\documentclass[a4paper,singlespace]{uopthesis}

\usepackage{amsmath}

\ifpdf%
  \hypersetup{
      pdftitle={Primordial Magnetic Fields in Cosmology},
      pdfauthor={Iain Brown (iain.brown@port.ac.uk)},
      pdfsubject={PhD, Thesis},
      pdfkeywords={CMB,MHD,magnetohydrodynamics,microwave background,magnetic},
      pdfpagemode=UseOutlines,
      pdfstartpage=1,
      pdfstartview=FitV,
      pdfview=FitV
  }
\fi

\title{Primordial Magnetic Fields in Cosmology}

\ifpdf
  \author{\href{mailto:iain.brown@port.ac.uk}{Iain A.~Brown}}
  \department{\href{http://www.tech.port.ac.uk/icg}{Institute of Cosmology and Gravitation}}
  \university{\href{http://www.port.ac.uk}{University of Portsmouth, UK}}
  \principaladvisor{\href{http://www.tech.port.ac.uk/staffweb/crittenr}{Dr.~Robert ~Crittenden}}
  \firstreader{First Reader Name}
  \secondreader{Second Reader Name}
\else
  \author{Iain A. Brown}
  \department{Institute of Cosmology and Gravitation}
  \university{University of Portsmouth, UK}
  \principaladvisor{Dr R. Crittenden}
\fi

\submitdate{March, 2006}
\copyrightyear{2006}

\figurespagefalse
\tablespagefalse

\usepackage{amsmath}
\usepackage{amssymb}
\usepackage{longtable}

\setlongtables

\newcommand{\be}{\begin{equation}}
\newcommand{\ee}{\end{equation}}
\newcommand{\bea}{\begin{eqnarray}}
\newcommand{\eea}{\end{eqnarray}}
\newcommand{\bdm}{\begin{displaymath}}
\newcommand{\edm}{\end{displaymath}}
\newcommand{\beas}{\begin{eqnarray*}}
\newcommand{\eeas}{\end{eqnarray*}}
\newcommand{\bml}{\begin{subequations}}
\newcommand{\eml}{\end{subequations}}

\newcommand{\bkr}{\overline{\rho}}        

\newcommand{\Ev}{\mathbf{E}}              

\newcommand{\Bv}{\mathbf{B}}              
\newcommand{\Ba}{B}                        

\newcommand{\jv}{\mathbf{J}}              

\newcommand{\vr}{\varrho_a}                 
\newcommand{\ca}{\sigma_a}                   

\newcommand{\ra}{\rho_{ba}}                    
\newcommand{\pa}{p_{ba}}                        

\newcommand{\bkrO}{\overline{\rho}_{ba}}       
\newcommand{\bkpO}{\overline{p}_{ba}}           

\newcommand{\fb}{B}                        

\newcommand{\xv}{\mathbf{x}}              
\newcommand{\hub}{\frac{\dot{a}}{a}}      
\newcommand{\kh}{\hat{k}}                 
\newcommand{\khv}{\hat{\mathbf{k}}}       
\newcommand{\kv}{\mathbf{k}}              
\newcommand{\nh}{\hat{n}}                 
\newcommand{\nhv}{\hat{\mathbf{n}}}   
\newcommand{\fgv}{\mathbf{v}}             

\newcommand{\Tr}[1]{\tilde{#1}}           

\newcommand{\Sc}[1]{{#1}^{S}}           
\newcommand{\V}[1]{{#1}^{V}}            
\newcommand{\T}[1]{{#1}^{T}}            

\begin{document}
\begin{preliminary}

\begin{abstract}
Magnetic fields have been observed in galaxies, clusters of galaxies and probably in superclusters. While mechanisms exist to generate these in the late universe, it is possible that magnetic fields have existed since very early times. A field existing before the formation of the cosmic microwave background will leave imprints from its impact on plasma physics that might soon be observable. This thesis is concerned with investigating methods to predict the form of such imprints.

In chapter \ref{Chapter-PerturbationTheory} we review in detail a standard, linearised cosmology based on a Robertson-Walker metric and a universe filled with photons, massless neutrinos, cold dark matter, a cosmological constant, and baryons. We work in a synchronous gauge for ease of implementation in a Boltzmann code, and keep our formalism general. We then consider the statistics of the cosmic microwave background radiation, assuming that only scalar (density) perturbations cause significant impact.

Chapter \ref{Chapter-MagnetisedCosmology} introduces an electromagnetic field of arbitrary size and presents the equations governing the magnetised cosmology, and the structure of the electromagnetic fields, in greater detail than has hitherto been shown. We then invoke a hierarchy of approximations, treating the conductivity of the universe as infinite and thus removing the electric field, and considering the energy density of the magnetic field to be small. We present the resulting system in a computationally useful form and end by reviewing previous studies into the damping scales induced by photon viscosity.

Chapter \ref{Chapter-SourceStats} considers the intrinsic statistics of the magnetic stresses. We approach this issue in two ways. Analytical methods are exact but reliant on an underlying Gaussian distribution function for the magnetic field, and simulating fields on a finite grid allows a great freedom in the form of the field but imposes an undesirable granularity and an unphysical infra-red cut-off. We construct the two-point moments in Fourier space, extending and improving the analytical results, some of which we present for the first time. There is excellent agreement between the results from analysis and those from the simulated fields. At the one- and three-point level we find significant intrinsic non-Gaussianities.

In chapter \ref{Chapter-CMB} we turn to the observable impacts a primordial magnetic field. We briefly review previous studies into constraints from the epoch of nucleosynthesis before turning to consider the cosmic microwave background. We briefly consider the potential impact of an evolving damping scale, which may cause decoherence in the sources. Assuming coherence, which is likely to be accurate for very infra-red fields, the statistics of the source can be mapped onto the cosmic microwave background in an extensible manner, modelling the source statistics and the photon evolution entirely seperately. We demonstrate that our approach is valid by reproducing the signals for Gaussian power law fields on the microwave sky. After outlining how we will improve our predictions by employing a Boltzmann code, we show that it is a purely technical matter to extend the method to the three-point function. With detector sensitivity increasing, the non-Gaussianity of the cosmic microwave background will be well constrained in the near future and our work allows us to employ a new probe into the nature of an early-universe magnetic field.
\end{abstract}

\begin{preface}
The work of this thesis was carried out at
the
\ifpdf%
  \href{http://www.port.ac.uk/icg}{Institute of Cosmology and Gravitation},
\else
Institute of Cosmology and Gravitation
\fi
University of Portsmouth, United Kingdom.

The work in this thesis is based on work in collaboration with Robert Crittenden (ICG, Portsmouth). Much of chapter \ref{Chapter-SourceStats} is based on the paper ``Non-Gaussianity from Cosmic Magnetic Fields'', I. Brown and R. Crittenden, Phys. Rev. {\bf D}72 063002 (2005).

I hereby declare that this thesis has not been submitted, either in the same or different form, to this or any other university for a degree, and that it represents my own work.

Iain Brown
\end{preface}

\begin{acknowledgements}
The work presented in this thesis could not have been performed without the help and collaboration of my supervisor, Robert Crittenden. One of the chapters of this thesis is based on and expanded from a paper written in collaboration with him. The support of the Institute of Cosmology and Gravitation, and in particular of Prof. Roy Maartens, has also been invaluable, as have been discussions with Kishore Ananda, Chris Clarkson, David Wands, David Parkinson, Marco Bruni, Konstantinos Dimopoulos, Antony Lewis and others too numerable to mention. I must also thank my family for their support as writing the thesis wore on, and the patience and tolerance of my friends, particularly of Mike, Ricki, Laura, Emma, Chris and Vivian. The largest thanks go to Kishore, Richard, Mat and Dan, who offered much appreciated hospitality and sat through innumerable "discussions" about the thesis and the state of the world, and to Kathryn, who rarely complained about the weight of emails and helped my sanity enormously. Thanks to them all, this was a lot less unpleasant than it might otherwise have been.
\end{acknowledgements}

\end{preliminary}

\startthesis

\setcounter{equation}{0}

\chapter{Introduction}
\label{Chapter-Introduction}
Magnetic fields are observed on many scales in the cosmos, from planetary scales with coherence lengths of a few thousand kilometres and strengths of a few Gauss, to galactic scales with coherence lengths on the order of kiloparsecs and a strength of approximately $B\approx\mu\mathrm{G}$. There are also fields with coherence lengths on the order of megaparsecs that lie between galaxies and have field strengths lying between nano- and micro-Gauss; fields larger yet within clusters are also likely to exist, with field strengths of a comparable size. While fields on supercluster scales are extraordinarily difficult to detect, there are suggestions that fields up to the order of micro-Gauss may exist even there. (See for example \cite{Kronberg94, KimKronbergTribble91, ZweibelHeiles97, GrassoRubenstein01, Widrow02, Giovannini04-Mag} for reviews.) It is no great leap to suggest that larger-scale fields yet may be present and, if so, we require a means of probing them.

The three chief observational probes for astronomical fields are the Zeeman effect, synchotron radiation, and the Faraday rotation. The Zeeman effect, in which the uniform component of the magnetic field separates molecular energy levels, is typically rather small (of the order of a few hertz for a micro-Gauss field) and thermal effects can readily induce a greater splitting than this. For the strong fields found in many astrophysical situations, the Zeeman effect can be a good probe; however, for fields on galactic scales and above it is unlikely to be useful. Synchotron radiation is emitted from electrons spiralling around magnetic field lines; this allows one to estimate the total magnetic field transverse to the electron motion. Synchotron emission is very useful for detections of magnetic fields in external galaxies, but unfortunately the relative scarcity of free electrons in clusters and greater scales limits its use to galactic scales. Moreover while synchotron measurements are useful for determining the transverse field strength they do not give an estimate of the total field strength and are also rather model dependent. For larger-scale fields and to determine a total field strength we resort to employing measures from Faraday rotation, in which a magnetic field rotates the plane of a light-beam's polarisation. A Faraday rotation signal is readily detected from its strong frequency dependence. Even with the Faraday rotation, fields on cluster and supercluster scales are notoriously difficult to calculate with any certainty; estimates range from the nano-Gauss to micro-Gauss levels.

The origin of these fields remains uncertain; many of the mechanisms suggested to generate the observed fields require a pre-existing seed field and are more accurately dubbed ``amplification'' mechanisms than generation mechanisms. For our purposes we separate creation and amplification mechanisms into processes occurring before, during and after recombination.


Popular post-recombination processes include the dynamo mechanism \cite{GrassoRubenstein01, ZeldovichRuzmaikinSolokoff80, KulsrudEtAl97} and the adiabatic compression of a previously-magnetised cloud \cite{GrassoRubenstein01,KingColes05}. The dynamo mechanism occurs when a rotating galaxy contains a pre-existing, small, seed field. The vorticity induced by the galactic rotation then ``winds up'' the magnetic field lines, boosting the field strength on an astrophysical timescale, the extra magnetic energy coming from rotational kinetic energy. The dynamo mechanism can likely boost a seed field of the order of $10^{-30}$ Gauss up to the observed level of around $10^{-6}$ Gauss, although this efficiency is still a matter of some debate (see for example \cite{KulsrudEtAl97}). The adiabatic compression of a previously magnetised cloud is also effective in boosting field strength; again one may visualise this as a compression of magnetic field lines, this time through an ever tighter packing rather than a tighter winding, and the extra magnetic energy comes from gravitational potential energy. Adiabatic compression is less efficient than is the dynamo mechanism; estimates are that a seed field of around $10^{-20}$ Gauss is necessary to produce the observed fields on a realistic timescale. Of course, there is little reason to believe that both these processes cannot be operating simultaneously, with the compression of a rotating, magnetised cloud.

Neither of these mechanisms generate fields from scratch; we still require some initial genesis mechanism to explain the observed fields. This could conceivably occur within clusters and galaxies themselves, by some battery mechanism for example; \cite{Giovannini04-Mag} provides some coverage of traditional approaches (the ``Biermann battery'' for example) to this matter and \cite{BiermannGalea03,HanayamaEtAl05} are a few modern treatments of magnetic fields produced in supernova batteries; such astrophysical sources do not here concern us. Instead we choose to consider the possibility that the seed fields, or at least a significant component of them, were relics of an earlier era.

Mechanisms certainly exist to generate this field at reionisation \cite{SubramanianEtAl94,GnedinFerraraZweibel00,LangerPugetAghanim03} or at recombination itself \cite{Hogan00, BerezhianiDolgov03}, and they might also have been created before recombination or even before nucleosynthesis. During the reionisation era magnetic fields can be created by the ``shadow'' an overdensity casts in the flux of an ionising source. While outside of the shadow an induced electric field is balanced by the radiation pressure, within the shadowed area the flux is weaker and the forces no longer balance; the ionised material is then induced into motion and the resulting current produces a magnetic field. Such fields can be as large as $10^{-14}$ or even $10^{-12}$ Gauss, and obey a power spectrum $\mathcal{P}\approx k^{n-2}$ where $n$ is the spectral index of the overdensities (assumed to be approximately $n\approx -1$ in the realm of applicability \cite{LangerPugetAghanim03}). These fields have coherence lengths on the order of around $1\mathrm{Mpc}$.

Fields originating at recombination will be relatively small-scale (though still large compared to galactic scales) and with strengths around $10^{-20}$ Gauss. They are generated by current flows induced between electrons and ionised hydrogen and helium during recombination by gravitational perturbations and radiation pressure.

There are many suggested mechanisms that can produce fields before recombination. One possibility arises from certain inflationary theories. In general, the density perturbations and gravitational waves produced during inflation are not conformally invariant, while magnetic fields are. Generating large-scale fields then depends on breaking the conformal invariance of the magnetic field, with highly model-dependent results. Possible generation mechanisms arise relatively naturally during the electroweak symmetry breaking phase, before or during inflation itself, or in a preheating stage; see for example \cite{GrassoRubenstein01,Giovannini04-Mag,TurnerWidrow88,TornkvistEtAl00,DimopoulosProkopecTornkvistDavis02,ProkopecPuchwein04,BambaYokoyama04,BassettEtAl01}. Such fields have power spectra ranging from $\mathcal{P}\propto k$ to $\mathcal{P}\propto k^{-1}$, or even further tilted to the red. These fields are often dubbed ``acausal'' since they are produced during or before inflation. It can be shown (e.g. \cite{CapriniDurrer02}) that a field produced by a ``causal'' mechanism must have an index $n\geq 2$. There are also recent studies into generic second-order phase transitions generating large-scale fields (e.g., \cite{MartinDavis95,HindmarshEverett97,BoyanovskyVegaSimionato03,BoyanovskyVega05}); such fields generally have power spectra with an index $n\approx 2$, heavily modified by the complex physics. Cosmic defects, both those models from GUT theories and more recent string-inspired models, might also be responsible \cite{Dimopoulos98,DaviesDimopoulos05}; these fields are generated by vorticity induced by the self-interaction of a string network.

More recently, attention has been given to the possibility that magnetic fields might be created continuously in the period between lepton decoupling and recombination, through the vorticity naturally occurring at higher order in perturbation theory \cite{BetschartDunsbyMarklund04, MatarreseEtAl04, GopalSethi04, TakahashiEtAl05}. Due to the nature of their production, such fields are necessarily rather weak and small-scale, though adequate for sourcing cluster fields; they have a complicated power spectrum that goes as $k^2$ for low $k$ before damping away.

Dolgov \cite{Dolgov03} provides a brief overview of many creation mechanisms. It is also worth commenting that, after production, their nature could evolve in the very early universe, perhaps as a result of hydromagnetic turbulence \cite{BrandenburgEnqvistOleson96} or the inverse cascade \cite{ChristenssonHindmarshBrandenburg00,HindmarshChristenssonBrandenburg02}. 

The exact magnetogenesis model is not our present focus; we are principally interested in studying the impact a primordial magnetic field might have on cosmological perturbation theory and, more specifically, the cosmic microwave background. It is to be hoped that studying this field will allow constraints on the possible strength of a primordial magnetic field -- or, indeed, determine whether such a field is incompatible with observations. Research in this area is not new; studies of magnetised universes of one form or another date back at least to the 1960s (see for example \cite{Thorne67, Jacobs68, Jacobs69}). However, in recent years much study has gone into the development of a theory of cosmological magnetohydrodynamics.

Limits from nucleosynthesis can be particularly powerful, and applicable to tangled as well as large-scale magnetic fields \cite{GrassoRubenstein01, Widrow02,CapriniDurrer02}. The direct impact of a magnetic field, through the induced splitting in electron energy levels and through the energy density it contributes to the universe can be used to constrain the current field strength to the order of micro-Gauss. There is also an indirect bound arising from the gravitational waves generated by a magnetic field; Caprini and Durrer \cite{CapriniDurrer02} demonstrate that the bounds on acausal fields generated in the extremely early universe are extraordinarily strong, to the level of $B\approx 10^{-30}$ Gauss for $n\approx 0$, and lessening only for strongly red spectra nearing $n\approx -3$. These limits are not entirely uncontested, however (\cite{KosowskyEtAl04,CapriniDurrer05}).

The cosmic microwave background (hereafter CMB) provides additional tools for investigating the properties of large-scale magnetic fields. Regardless of the time at which they were generated, all fields could be expected to leave a magnetised Sunyaev-Zel'dovich effect (e.g. \cite{HuLou04,Zhang03}) and Faraday rotations of the primordial CMB \cite{CampanelliEtAl04,ScoccolaHarariMollerach04,KosowskyEtAl04}. Fields present before reionisation will impact on the physics of the era (e.g. \cite{SethiSubramanian04}), and fields predating recombination will have a direct impact on the cosmological perturbations producing the primordial CMB \cite{ScannapiecoFerreira97, SubramanianBarrow98, DurrerKahniashviliYates98, KohLee00, DurrerFerreiraKahniashvili00, SubramanianBarrow02, SeshadriSubramanian00, MackKahniashviliKosowsky02, CapriniDurrerKahniashvili03, SubramanianSeshadriBarrow03, BereraEtAl03, Lewis04-Mag, Giovannini04-CMB, YamazakiIchikiKajino04}. It is with the final early-time possibility that this thesis is concerned.

The field strength of early-time fields is already constrained by limits from the CMB. By assuming the universe to be of Bianchi type VII, Barrow \emph{et. al.} \cite{BarrowFerreiraSilk97} demonstrated that, assuming the total anisotropy allowed by the 4-year COBE results \cite{COBE} to be due to a magnetic field, the field strength is constrained to a current value of $B\approx 10^{-9}$ Gauss. More recently, Clarkson \emph{et. al.} \cite{ClarksonEtAl02} place weak limits on the strength of the magnetic field by considering the impact on the CMB in a generic geometry; assuming a Robertson-Walker form tightens these bounds back to the order of nano-Gauss. It is worth stressing that these limits are for a large-scale, directional component to the field active on scales larger than the Hubble length. Limits for a ``tangled'' configuration require a closer study.

Primordial magnetic fields can have a significant impact on the CMB. While early treatments focused on the dynamics of a Bianchi universe \cite{Jacobs69, MilaneschiFabbri85}, more modern treatments \cite{ScannapiecoFerreira97, SubramanianBarrow98, DurrerKahniashviliYates98, KohLee00, DurrerFerreiraKahniashvili00, SubramanianBarrow02, SeshadriSubramanian00, MackKahniashviliKosowsky02, CapriniDurrerKahniashvili03, SubramanianSeshadriBarrow03, BereraEtAl03, Lewis04-Mag, Giovannini04-CMB, YamazakiIchikiKajino04, YamazakiEtAl06} consider either small perturbations around a large-scale homogeneous field or a tangled field configuration in which fields are taken to possess no net directionality above a certain scale, often associated with the cluster scale -- that is, the orientation of the net cluster scale fields is random. Both of these scenarios source CMB perturbations, directly through the scalar, vector and tensor stresses, and indirectly by the density and velocity perturbations they induce in the charged proton-electron fluids.  Code for calculating the vector and tensor anisotropies generated by primordial magnetic fields was recently added to the publicly-available Boltzmann code CAMB \cite{CAMB,Lewis04-Mag}, and that for scalars has been modelled independently \cite{KohLee00, Giovannini04-CMB, YamazakiIchikiKajino04}.

Considering the dynamics of the magnetised cosmological plasma, Brandenburg \emph{et. al.} \cite{BrandenburgEnqvistOleson96}, Tsagas and Barrow \cite{TsagasBarrow97}, Jedamzik \emph{et. al.} \cite{JedamzikKatalinicOlinto98} and Subramanian and Barrow \cite{SubramanianBarrow98-MHD} studied various aspects of magnetised cosmological perturbation theory and in particular the damping of magnetosonic and Alfv\'en waves within the magnetised cosmological fluid. They independently found that while fast magnetosonic waves undergo Silk damping in much the same way as standard acoustic waves, the slow magnetosonic and Alfv\'en waves are heavily overdamped and on some scales can survive Silk damping. This not only yields a power boost for small scales of the CMB but could also have a strong impact on cluster formation.

Subramanian and Barrow \cite{SubramanianBarrow98, SubramanianBarrow02} make semi-analytical estimates of the impact of a stochastic tangled magnetic field on the small-scale ($l\geq1000$) temperature perturbations on the CMB, extending this with Seshadri to the polarisation \cite{SeshadriSubramanian00,SubramanianSeshadriBarrow03}. They found that the Alfv\'en waves generated by a tangled field of size $B\approx 10^{-9}$ Gauss can contribute a signal to the temperature angular power spectrum that begins to dominate at $l\approx 3000$. The effect is more pronounced for bluer spectra. Durrer, Kahniashvili and collaborators (e.g. \cite{DurrerKahniashviliYates98, DurrerFerreiraKahniashvili00,MackKahniashviliKosowsky02}) investigate the impact a tangled magnetic field would have on the large-scale ($l\leq500$) CMB for both temperature and polarisation angular power spectra. They find a similar decrease of signal for low $n$ and that a field is constrained to be of the order of nano-Gauss if it has a spectral index $n\approx -3$, while a causal field with $n\approx 2$ is even more tightly constrained. Numerical codes are naturally the most accurate and previous findings have been confirmed and enhanced for scalar modes by Koh and Lee \cite{KohLee00}, Giovannini \cite{Giovannini04-CMB} and Yamazaki \emph{et. al.} \cite{YamazakiIchikiKajino04}, and for vector and tensor modes by Lewis \cite{Lewis04-Mag}.

The studies thus far have been limited to particular configurations of field statistics and power spectra. Even with the numerical models studied by Lewis, Koh and Lee and Yamazaki \emph{et. al.} much reliance for the primordial field's power spectrum and statistics is placed on the results of Mack \emph{et. al.} \cite{MackKahniashviliKosowsky02}. This limits the study to a power-law power spectrum; while likely sufficient for fields produced by inflation, this is unlikely to be accurate for more complex scenarios such as fields generated from phase transitions or plasma processes. Large-scale causal fields are restricted to power spectra $n\geq 2$; due to the power that this concentrates on small scales, such spectra are unphysical without some additional cutoff or turnover. Moreover the fields considered have been purely Gaussian and there is little reason to believe this to be a realistic assumption. Magnetogenesis mechanisms in the early universe tend to be exceptionally complicated and to merely assume a resulting Gaussian field is not warranted. Indeed, one of the few mechanisms for which the statistical nature has been derived is the second-order production of, for example, Matarrese \emph{et. al.} \cite{MatarreseEtAl04} which, due to its nature, produces $\chi^2$ fields with a damped causal power spectrum.

Moreover, the results in Mack \emph{et. al.} and the work based on them is limited to the accuracy of the approximations they employ for the damping scales, Alfv\'en velocity and normalisation of the spectral index. These approximations are again restricted to Gaussian power-law fields. Despite their efforts there is also still not a fully unified treatment of magnetic fields in cosmology. While full treatments are provided in principle in Tsagas \cite{Tsagas04} or Tsagas and Maartens \cite{TsagasMaartens99-Perts}, resulting naturally from their use of the covariant and gauge-invariant approach to cosmology, Lewis' application of this approach to CMB analysis \cite{Lewis04-Mag} neglected the scalar modes. In a more traditional metric-perturbation approach to cosmology, Mack \emph{et. al.} \cite{MackKahniashviliKosowsky02} work in a gauge-invariant formalism and also neglect scalar modes. Koh and Lee \cite{KohLee00} work in a conformal Newtonian gauge and thus consider only scalars. Giovannini \cite{Giovannini04-Mag}, enhancing the study of Koh and Lee and seeking to fill the gaps in Lewis' work, also employs conformal Newtonian gauge.

In this thesis we aim to review and extend the current research in this area, with a specific aim towards developing techniques by which a field with a significant tangled component could be constrained by CMB observations, including contributions from the scalar, vector and tensor components in a fully unified manner. In chapter \ref{Chapter-PerturbationTheory} we review in detail a linear-order cosmological perturbation theory of the classical type pioneered by Lifshitz (\cite{Lifshitz46, LifshitzKhalatnikov63, LandauLifshitz-ClassicalTheoryFields}). We generally restrict ourselves to synchronous gauge for ease of future integration with the CMBFast code \cite{ZaldarriagaSeljak97}. In chapter \ref{Chapter-MagnetisedCosmology} we introduce the modifications to the standard theory introduced by a magnetic field. Magnetic fields impact directly on the geometry through the magnetic energy density and isotropic pressure, but also through the anisotropic shear they introduce. Perhaps more importantly, however, the magnetic fields modify the conservation and Euler equations for the baryons, making them into a cosmological equivalent of the classical MHD equations. The fields then source slow magnetosonic waves and Alfv\'en waves, both of which become heavily overdamped and can to some measure survive Silk damping \cite{SubramanianBarrow98-MHD}. In the standard picture vector perturbations rapidly damp away (though Lewis has demonstrated that in principle a detectable signature might remain from primordial vector modes sourced by the neutrino anisotropic stress after neutrino decoupling \cite{Lewis04-Vector}); as with a system of cosmic defects, however, the magnetic field provides an active source.

It appears increasingly likely from cosmological data such as the CMB that the dominant source of perturbations in the universe was adiabatic and Gaussian in nature, consistent with the expectations of an inflationary universe \cite{WMAP-Bennett,WMAP-Komatsu}. However, this does not exclude the possibility that non-linear sources -- such as, but certainly not limited to, our magnetic fields -- might also have played a role in generating perturbations. While their impact on the power spectrum might be small, they could dominate any non-Gaussianities that we observe. In order to search for these optimally we need clear predictions for the nature of the non-Gaussian signals that they would source. In chapter \ref{Chapter-StatsOfCosmicMagneticFields} we investigate in detail the intrinsic statistics of an early-universe tangled magnetic field. The stress-energy tensor of the magnetic field is quadratic and so at the very least we expect the energy density of the field to obey a $\chi^2$ distribution. To test this suspicion, we analyse the intrinsic statistics of Gaussian-random magnetic fields, concentrating on the two- and three-point correlations between the scalar, vector and tensor components of the magnetic stresses. We do so by both generating realisations of Gaussian-random magnetic fields and through numerically integrating pure analytical results, most of which are presented here for the first time. While the integrations we present are reliant upon the fields being Gaussian-random, the code we use to generate our realisations is easily extended to include any type of non-Gaussian seed field and any form of power spectrum, not just the power-law typically considered. One motivating example is a field continuously sourced in the pre-recombination universe by the mechanism of, for example, Matarrese \emph{et. al.} \cite{MatarreseEtAl04} which, being produced by the combination of two Gaussian fields, is $\chi^2$ in nature and which has a significantly non-trivial power spectrum. As a first step towards considering this scenario, we model a Gaussian magnetic field with a causal power spectrum, exponentially damped on smaller scales. Unfortunately, the limited dynamic range of our realisations prevents us from directly modelling the end-result of the Matarrese \emph{et. al.} mechanism; however, the field we consider will provide us with clues as to its nature.

The impact of magnetic fields on the CMB temperature and ``gradient'' $E$-mode polarisation angular power spectra is likely to be relatively minor on the angular scales most easily resolved; they are expected to begin to dominate at multipoles of about $l\approx3000$, which remains unprobed by WMAP and unlikely to be detected with such accuracy for a long time yet, although measurements by the CBI, VSA (see e.g. \cite{RajguruEtAl05}), ACBAR \cite{ACBAR}, Boomerang \cite{MontroyEtAl05} and CAPMAP \cite{BarkatsEtAl04} probes are ever-improving, and the South Pole Telescope \cite{SPT04} is soon to come online. Moreover, angular scales this small are likely to be heavily contaminated by foregrounds that will be extremely hard to remove without washing away the effects of a magnetic field.

The ``rotational'' $B$-mode of the CMB polarisation is relatively unexcited; in the standard model $B$-modes are generated by gravitational waves produced by inflationary perturbations and by weak lensing, both with characteristic spectra. In general only vector and tensor perturbations excite the $B$-mode. Since a magnetic field produces both of these we have, in principle, a clean probe for its properties. Moreover, a cosmological magnetic field would naturally induce Faraday rotation in the CMB polarisation \cite{KosowskyLoeb96, ScoccolaHarariMollerach04, CampanelliEtAl04, KosowskyEtAl04}, which would convert an amount of the $E$-modes generated by all three classes of perturbations into the $B$-modes, in a characteristically frequency-dependant manner. In reality, unfortunately, a strong angular power spectrum for the $B$-mode is a long way off.

In chapter \ref{Chapter-CMB} we finally consider the observational impact of magnetic fields, briefly reviewing the known constraints from nucleosynthesis \cite{Widrow02,CapriniDurrer02} before turning in detail to the CMB temperature angular power spectrum. Exploiting the simple time-evolution of a magnetic field we can tackle the issue from a slightly different angle to previous authors, transferring the intrinsic magnetic statistics directly onto the CMB by integrating them across magnetised transfer functions. This approach requires pre-computed power- or bi-spectra, which could naturally be either purely analytic or resulting from the realisations we generated in chapter \ref{Chapter-SourceStats}. The transfer functions are derived entirely independently. The advantages of this approach are manifold. At the two-point level one might argue that the approach of Subramanian and Barrow \cite{SubramanianBarrow02}, Mack \emph{et. al.} \cite{MackKahniashviliKosowsky02}, Lewis \cite{Lewis04-Mag} and Yamazaki \emph{et. al.} \cite{YamazakiIchikiKajino04,YamazakiEtAl06} is sufficient and a new formalism entirely unnecessary. However, the techniques employed by these authors are restricted, both in the nature of the fields that it can model -- only purely Gaussian, power-law fields -- and in the heavy reliance on successive approximate solutions to a range of integrals. With our numerical approach, we are not at all restricted in this manner and can consider a far wider range of magnetic fields and their imprint on the CMB angular power spectrum. We consider the large-scale approximations to the magnetised transfer functions, which then enables us to demonstrate the validity of this approach. The extensibility is demonstrated using the damped causal field we considered in chapter \ref{Chapter-SourceStats}.

Magnetic impacts on the two-point CMB spectra are likely to be subdominant to other sources until very small scales. An alternative probe comes from the non-Gaussianity that any magnetic field -- including a Gaussian field -- will impart onto the CMB. From a large-scale magnetic field, this non-Gaussianity would be expected to show at relatively low multipoles. Given the quality of the current data at low mutipoles generated by the WMAP satellite and, in the near future, the Planck satellite, this would seem a sensible approach. Unfortunately, the analytical and semi-analytical approaches employed by, for example, Mack \emph{et. al.} and Subramanian and Barrow are unsuited to predictions for the CMB non-Gaussianity; the expressions become rapidly unwieldy, albeit probably not insoluble. We outline briefly how our approach can instead be employed to realistically predict the CMB angular bispectra arising from the various three-point correlations we derived in chapter \ref{Chapter-SourceStats}. This approach shares its advantages -- and disadvantages -- with the two-point case. It is severely compromised by the limited dynamic range, more so even than the two-point correlations, due to the strong mode-selection involved. Accurate intrinsic bispectra thus take a very long time and a large amount of memory to produce since we must use the largest possible dynamic range and average many different realisations. However, using such bispectra makes this approach highly extensible and the method we outline is entirely general. We conclude by outlining the procedure by which we will generate CMB angular bispectra using transfer functions generated from the CMBFast Boltzmann code \cite{SeljakZaldarriaga96}.



\setcounter{equation}{0}

\chapter{Standard Linearised FLRW Cosmology}
\label{Chapter-PerturbationTheory}
In this chapter we present a review, independently derived but, for the large part, recreating previously-known results, of the theory of linear cosmological perturbations of a Friedmann-Lema\^itre-Robertson-Walker universe. Sections believed original to this thesis include the discussions of vector perturbations.

\section{The FLRW Geometry}
\subsubsection{Cosmological Spacetimes and the Robertson-Walker Metric}
\label{Sec-Metric}
The Copernican principle states that the Earth is in no special part of the universe. If one assumes that the universe is isotropic about the Earth then applying the Copernican principle will immediately lead one to conclude that the universe is homogeneous and isotropic throughout. Perhaps the strongest evidence for an isotropic universe is the cosmic microwave background (CMB), which to a startling accuracy is observed to possess the same temperature (about $2.728\mathrm{K}$) in all directions \cite{WMAP-Bennett,COBE}. Applying the Copernican principle leads us to conclude that the universe is isotropic around every point -- \emph{i.e.}, homogeneous. We then expect the metric modelling the present universe to be maximally-symmetric -- that is, possessing the greatest possible numbers of isometries or, equivalently, possessing the maximum number of Killing vectors. Clearly this cannot be literally true, or we would not ourselves exist. However, if we consider the universe on a large enough scale that the clumping of matter is entirely negligible, then it becomes a very reasonable assumption. We should note that these are indeed merely assumptions, and one can work with mildly anisotropic cosmologies and reproduce the observed microwave background (e.g. \cite{TsagasMaartens99-Bianchi}) to within experimental error. Moreover, critics might point out that the direct observation of the CMB only makes a statement about the isotropy of the universe about the Earth; the homogeneous and isotropic universe follows only by declaring that isotropy complete and enforcing the Copernican principle. Non-homogeneous cosmologies are in general rather intractable, though there has been renewed work on the spherically-symmetric Lema\^itre-Tolman-Bondi metrics since some authors suggest such models may exhibit late-time acceleration without the need for dark energy (see \cite{AlnesEtAl06,VanderveldEtAl06,ParanjapeEtAl06} for three recent, conflicting, viewpoints). We assume the Copernican principle to hold.

In appendix \ref{Appendix-Perturbations} we derive the Robertson-Walker metric by foliating spacetime with maximally-symmetric three-spaces, and progress to a discussion of gauge issues in cosmology. Here we content ourselves with presenting our metric ansatz, in a flat synchronous gauge. We employ the line element
\be
  ds^2=-dt^2+a^2(t)\left(\gamma_{ij}+h_{ij}\right)dx^idx^j
    =a^2(\eta)\left(-d\eta^2+\left(\gamma_{ij}+h_{ij}\right)dx^idx^j\right).
\ee
where $\gamma_{ij}=\mathrm{diag}(1,1,1)$ and $h_{ij}$ is implicitly of order a small parameter $\varepsilon$ and includes contributions from components transforming under co-ordinate maps as scalar, vector and tensor parts. The separation will be performed in Fourier space. Synchronous gauge provides us with an unambiguous and familiar time co-ordinate $x^0$ and, as long as we take care when specifying initial conditions to remove the spurious gauge modes \cite{PressVishniac80}, is no worse than any other gauge choice for physical interpretation of results. Note that an observer measures the proper time and co-ordinates, $dt=d\eta/a$ and $dr^i=dx^i/a$. We will generally employ the conformal time $d\eta=dt/a$, where $a(\eta)$ is the scale factor of the FLRW geometry, swapping to co-ordinate time $t$ only in certain cases. We use the notation $\dot{A}=dA/d\eta$ and $A'=dA/dt$ for differentiation of an object $A$ with respect to the conformal and co-ordinate times respectively, and employ an overbar $\overline{A}$ to denote background quantities and $\delta A$ to denote perturbed quantities.

\subsubsection{Basic Definitions}
Consider a real-space covariant vector $B_i$, the 4-velocity, for example. Then the contravariant form will be
\bdm
  B^i=a^{-2}\gamma^{ij}B_j \Rightarrow \left(aB^i\right)=\gamma^{ij}\left(a^{-1}B_j\right)
\edm
and so one can define the comoving covariant vector $b_i$ and its contravariant counterpart
\bdm
  b^i=aB^i, \; b_i=a^{-1}B_i, \; b^i=\gamma^{ij}b_j.
\edm
In particular, this implies that
\bdm
  \partial^i=\gamma^{ij}\partial_j,
\edm
where $x^i$ is a comoving co-ordinate.

Similarly, for a tensor $B_{ij}$,
\bdm
  B^{ij}=a^{-4}\gamma^{im}\gamma^{jn}B_{mn} \Rightarrow \left(a^2B^{ij}\right)=\gamma^{im}\gamma^{jn}\left(a^{-2}B_{mn}\right)
\edm
leads to the definitions of the comoving tensors $b_{ij}$ and $b^{ij}$ as
\bdm
  b^{ij}=a^2B^{ij}, \; b_{ij}=a^{-2}B_{ij}, \; b^{ij}=\gamma^{im}\gamma^{jn}b_{mn} .
\edm
The mixed form $b^i_j$ is then simply
\bdm
  B^i_j=a^{-2}\gamma^{im}B_{mj}=\gamma^{im}b_{mj}=b^i_j .
\edm

The inverse metric \cite{Wald} is
\bdm
  g^{00}=-a^{-2}, \quad g^{0i}=0, \quad g^{ij}=a^{-2}\left(\gamma^{ij}- h^{ij}\right)
\edm
which takes this form to retain the normality of the metric; the metric perturbation is a comoving tensor
\bdm
  h_{ij}=a^{-2}\delta g_{ij},
\edm
and so
\bdm
  h^i_j=\gamma^{im}h_{mj}=\gamma_{ij}h^{mj} \quad \mathrm{and} \quad h^{ij}=a^2\delta g^{ij}=\gamma^{im}\gamma^{jn}h_{mn} .
\edm

The scale factor $a$ is normalised today to unity, \emph{i.e.} $a_0=a(\eta_0)=1$. This implies that comoving quantities become physical at the current epoch. The conformal time $\eta$ is then closely related to the Hubble distance.\footnote{The equations derived in \S\ref{Section-SingleFluidUniverses} imply that if we assume the current acceleration to have started only recently, we can approximate the current conformal time as $\eta\approx 2/H_0$ where $H_0$ is the current Hubble parameter.}

\subsubsection{Affine Connections and the Ricci Tensor}
The affine connections in a metric space without torsion are the Christoffel symbols,
\bdm
  \Gamma^{\lambda}_{\mu\nu}=\frac{1}{2}g^{\lambda\kappa}\left(g_{\mu\kappa,\nu}
   +g_{\nu\kappa,\mu}-g_{\mu\nu,\kappa}\right).
\edm
For a perturbed Robertson-Walker metric these are
\be
 \label{Christoffel}
 \begin{array}{rcl}
  \Gamma^{0}_{00}&=&\hub, \\
  \Gamma^{i}_{00}&=&0  \\
  \Gamma^{0}_{i0}&=&0  \\
  \Gamma^{0}_{ij}&=&\hub\left(\gamma_{ij}+ h_{ij}\right)+\frac{1}{2}\dot{h}_{ij} \\
  \Gamma^{i}_{j0}&=&\hub\delta^i_j+\frac{1}{2}\dot{h}^i_j \\
  \Gamma^{i}_{jk}&=&\frac{1}{2}\left(\partial_k h^i_j+\partial_jh^i_k-\partial^ih_{jk}\right).
 \end{array}
\ee
The Ricci tensor is then
\bdm
  R_{\mu\nu}=R^\alpha_{\phantom{\alpha}\mu\alpha\nu}=\Gamma^{\alpha}_{\mu\nu,\alpha}
  -\Gamma^{\alpha}_{\mu \alpha,\nu}+\Gamma^{\alpha}_{\beta\alpha}\Gamma^{\beta}_{\mu\nu}
  -\Gamma^{\alpha}_{\beta\nu}\Gamma^{\beta}_{\mu \alpha} .
\edm
Straightforward calculations then show that
\bea
\label{RicciTensor}
  a^2R^0_0&=&3\left(\frac{\ddot{a}}{a}-\left(\hub\right)^2\right)
   +\frac{1}{2}\left(\ddot{h}+\hub\dot{h}\right) \nonumber \\
  -2a^2R^i_0&=&\partial^i\dot{h}-\partial^j\dot{h}^i_j \\
  -a^2R^i_j&=&-\left(\frac{\ddot{a}}{a}+\left(\hub\right)^{2}\right)\delta^i_j
   -\frac{1}{2}\ddot{h}^i_j-\hub\dot{h}^i_j-\frac{1}{2}\hub\dot{h}\delta^i_j
   \nonumber \\ && \;\;
   +\frac{1}{2}\left(\partial^i\partial_jh+\nabla^2h^i_j
   -\partial^k\partial_ih^k_j-\partial^k\partial_jh^i_k\right). \nonumber
\eea
Here $h=\gamma^{ij}h_{ij}=h^i_i$ is the trace of the perturbation and we have raised one of the indices.

The Ricci scalar for this perturbed universe is then
\bea
 \label{RicciScalar}
  R&=&R^\mu_\mu=
   \frac{1}{a^2}\left(6\frac{\ddot{a}}{a}+\ddot{h}+3\hub\dot{h}
    -\nabla^2h+\partial_i\partial^jh^i_j\right).
\eea
This is in conformal time; were one to transfer back into co-ordinate time, one would find that the background term becomes the more familiar $\overline{R}=6\left(a''/a+a'^2/a^2\right)$.

\subsubsection{The Einstein Equations}
If one assumes only that the universe is composed of various fluids -- effective (photons and neutrinos, for example) and actual (baryons, CDM etc.), as well as other interacting or non-interacting components such as electromagnetic fields or networks of cosmic defects -- then one may express the matter content of the universe in a total stress-energy tensor $T^\mu_\nu$.

For intuitive purposes, this tensor can be defined in Minkowski space by
\bdm
  T^0_0=-\rho,\quad T^i_0=-\mathcal{E}^i,\quad T^i_j=\mathcal{P}^i_j
\edm
where $\rho$ is the mass/energy density of the matter component, $\mathcal{E}_i$ is the energy flux vector (which with $c=1$ is equivalent to the momentum density), and $\mathcal{P}^i_j$ is the momentum flux density (see \cite{LifshitzPitaevskii-FluidMechanics}). A more rigorous definition of a general stress-energy tensor is presented later. The momentum flux density can be seperated into its trace and traceless parts by
\be
  T^i_j=p\delta^i_j+\Pi^i_j \Rightarrow  3p=T^i_i, \quad \Pi^i_j=T^i_j-\frac{1}{3}\delta^i_jT^m_m
\ee
where $p$ is the isotropic pressure and $\Pi^i_j$ embodies the anisotropic stress and contains a traceless scalar degree of freedom along with two vector and two tensor degrees of freedom. This vanishes for a perfect fluid, for which we have
\be
  T^\mu_\nu=\left(\rho+p\right)u^\mu u_\nu+p\delta^\mu_\nu .
\ee

The Einstein equations can be written
\be
 \label{constraint}
  G^\mu_\nu=R^\mu_\nu-\frac{1}{2}\delta^\mu_\nu R=8\pi GT^\mu_\nu+\Lambda g^\mu_\nu.
\ee
Contracting these equations allows one to eliminate the Ricci scalar and express the Einstein equations in an alternative ``trace-reversed'' form,
\be
\label{evolution}
  R^\mu_\nu=8\pi G\left(T^\mu_\nu-\frac{1}{2}\delta^\mu_\nu T^\sigma_\sigma\right)-\Lambda\delta^\mu_\nu ,
\ee
which emphasises the vacuum equations $R^\mu_\nu=0$. There is obviously great redundancy amongst these equations. At zeroth order, the $G^0_0$ equation generates the Friedmann constraint equation and the $G^i_j$ equation generates the Raychaudhuri evolution equation. At first order, the $G^0_0$, $G^i_0$ and $R^i_0$ generate constraint equations while the $G^i_j$, $R^0_0$ and $R^i_j$ equations produce evolution equations. We shall select as evolution equations the $R^0_0$ for the scalar trace and the $R^i_j$ for the other components, and as constraints we shall select $G^0_0$ for the scalar trace and $G^i_0$ for the other components.

Employing the Ricci tensor and scalar (\ref{RicciTensor} -- \ref{RicciScalar}) found above, one readily finds that in the background we have
\bea
 \label{Friedmann}
 \scriptstyle{\mathrm{Friedmann; \; constraint, \;} \overline{G}^0_0 \; :} &&
 \mathcal{H}^2=\left(\hub\right)^2=-\frac{8\pi G}{3}\left(a^2\overline{T}^0_0\right)+\frac{a^2\Lambda}{3}, \\
 \label{Raychaudhuri}
 \scriptstyle{\mathrm{Raychaudhuri; \; evolution, \;} \delta^j_i\overline{G}^i_j \; :} &&
 2\frac{\ddot{a}}{a}-\left(\hub\right)^2=-\frac{8\pi G}{3}\left(a^2\overline{T}^i_i\right)+a^2\Lambda .
\eea
where $\mathcal{H}=aH$ is the Hubble factor in conformal time, $H$ being the observable Hubble parameter. The physical meaning of these equations can be rapidly illustrated with the perfect fluid form of the stress-energy tensor, for which we have
\be
  \left(\hub\right)^2=\frac{8\pi G}{3}a^2\left(\overline{\rho}+\frac{\Lambda}{8\pi G}\right), \;
  2\frac{\ddot{a}}{a}-\left(\hub\right)^2=-8\pi Ga^2\left(\overline{p}-\frac{\Lambda}{8\pi G}\right)
\ee
from which we may identify the cosmological constant as an effective perfect fluid with stress energy
\bdm
  \overline{T}^{(\Lambda)\mu}_\nu=\frac{\Lambda}{8\pi G}\mathrm{diag}\left(1,-\mathbf{1}\right)
\edm
or, to phrase it differently, an unperturbed perfect fluid of density $\rho_{\Lambda}=\Lambda/8\pi G$ and pressure $p_\Lambda=-\rho_\Lambda$. Using the typical definition of a cosmological equation of state $w=p/\rho$ we then have $w_\Lambda=-1$.

The Friedmann equation can also be written as
\be
  1=\frac{8\pi G}{3\mathcal{H}^2}a^2\overline{\rho}=\frac{8\pi G}{3H^2}\overline{\rho}=\Omega
\ee
where $\Omega$ is the ratio between the universe's total energy density and the critical density. For an individual species, then,
\be
  \Omega_i=\frac{8\pi G}{3H^2}\overline{\rho}_i, \quad \sum_i\Omega_i=1 .
\ee
These are normally presented as observed during the current era and the Friedmann equation can then be written
\be
  \sum_i\Omega_{i0}a^{-n_i}=1
\ee
with the scale-factor embodying the time-evolution. It is seen later that for radiation $n_i=4$, while for matter $n_i=3$ and for a cosmological constant $n_i=0$. Had we included a curvature term in our system then the effective energy density would evolve with $n_i=2$.

The linearised equations are
\bea
 \label{Evolution00}
  \scriptstyle{\mathrm{Evolution, \;} \delta R^0_0 \; :} &&
  \ddot{h}+\hub\dot{h}=16\pi Ga^2\left(\delta T^0_0-\frac{1}{2}\delta T^i_i\right), \nonumber \\
 \label{EvolutionIJ}
  \scriptstyle{\mathrm{Evolution, \;} \delta R^i_j \; :} &&
  \ddot{h}^i_j+2\hub\dot{h}^i_j+\hub\dot{h}\delta^i_j-\left(\partial^i\partial_jh+\nabla^2h^i_j
   -\partial^a\partial_jh^i_a-\partial_a\partial^ih^a_j\right) \nonumber \\ && \quad
   =16\pi Ga^2\left(\delta T^i_j-\frac{1}{2}\delta^i_j\delta T^\sigma_\sigma\right), \nonumber \\
 \label{Constraint00}
  \scriptstyle{\mathrm{Constraint, \;} \delta G^0_0 \; :} &&
  2\hub\dot{h}+\partial^j\partial_ih^i_j-\nabla^2h=-16\pi Ga^2\delta T^0_0, \nonumber \\
 \label{Constraint0i}
  \scriptstyle{\mathrm{Constraint, \;} \delta G^i_0 \; :} &&
  \partial^j\dot{h}^i_j-\partial^i\dot{h}=16\pi Ga^2\delta T^i_0 \nonumber .
\eea

These equations are entirely generic; henceforth, however, the matter sources we shall consider will be baryons, cold dark matter, photons, (massless) neutrinos, and in chapter \ref{Chapter-MagnetisedCosmology} an electromagnetic field. We could also add, for example, scalar fields and networks of cosmic defects. The explicit forms of the stress-energy tensor for the differing fluids employed in standard cosmology shall be considered in turn.

\subsubsection{Fourier Space and the Scalar-Vector-Tensor Split}
The system of equations presented above are most easily analysed by transferring them to Fourier space and separating the components into parts transforming as scalars, vectors and tensors under co-ordinate transformations. We work with the Fourier convention
\bdm
  \partial^i\rightarrow-ik\kh^i,
\edm
that is,
\bdm
  A(\eta,\mathbf{x})=\int A(\eta,\mathbf{k})e^{-i\mathbf{k}.\mathbf{x}}\frac{d^3\mathbf{k}}{(2\pi)^3}
\edm
for a scalar, vector or tensor quantity $A$. When necessary, co-ordinate axes will be aligned such that $\hat{x}_3\parallel \kh$.

A vector quantity in this space can be separated into a scalar and a solenoidal part as
\be
  B^i=i\kh^iB_S+B^i_V,\quad\kh_iB^i_V=0,
\ee
with $\kh_i$ a comoving wavevector, which implies that
\be
\label{VecSplit}
  B_S=-i\kh_iB^i, \quad B^i_V=B^i-\kh^i\kh_jB^j=\left(\delta^i_j-\kh^i\kh_j\right)B^j=P^i_j(\mathbf{k})B^j ,
\ee
where $P^i_j(\mathbf{k})=\delta^i_j-\kh^i\kh_j$ is a projection operator projecting a quantity onto a hypersurface defined by the Fourier
modes. Likewise, a tensor quantity can be split into
\be
  A^i_j=\frac{1}{3}\delta^i_jA+\left(\kh^i\kh_j-\frac{1}{3}\delta^i_j\right)A_S+\left(\kh^i\V{A}_j+\kh_jA^i_V\right)+A^{iT}_j,
\ee
with the various conditions
\bdm
  \kh^i\V{A}_i=0, \quad \kh^jA^{iT}_j=A^{iT}_i=0 ,
\edm
where $A=A^i_i$ is the trace of $A$, $A_S$ is the traceless scalar part of $A$, $\V{A}_i$ is the (divergenceless) vector part of $A$ and $A^{iT}_j$ is the (traceless, divergenceless) tensor part.

It is straightforward to demonstrate that the different components of $A^i_j$ are then recovered by applying various projection operators\footnote{Technically only $P^i_j(\mathbf{k})$ is a true projection operator since it obeys $P^i_i(\mathbf{k})=2$, $P^i_j(\mathbf{k})P^j_k(\mathbf{k})=P^i_k(\mathbf{k})$. Any sensible definition of a trace for the other projection operators vanishes; neither are they idempotent. However, we shall habitually refer to them all as projections.} to the full tensor:
\bea
\label{SVTProjections}
  A&=&\delta^j_iA^i_j, \nonumber \\
  A_S&=&\frac{3}{2}\kh^j\kh_iA^i_j-\frac{1}{2}A=\left(\delta^j_i-\frac{3}{2}P^j_i(\mathbf{k})\right)A^i_j
   =Q^j_i(\mathbf{k})A^i_j, \nonumber \\
  \V{A}_i&=&\kh_jA^j_i-\kh_i\kh_j\kh^kA^j_k=\kh_jP_i^k(\mathbf{k})A^j_k=\mathbf{P}^{kV}_{ij}(\mathbf{k})A^j_k(\mathbf{k}), \\
  A^{iT}_j&=&A^i_j-\frac{1}{3}\delta^i_jA-\left(\kh^i\kh_j-\frac{1}{3}\delta^i_j\right)A_S
  -\left(\kh^i\V{A}_j+\kh_jA^{iV}\right) \nonumber \\
   &=&\left(P^i_a(\mathbf{k})P_j^b(\mathbf{k})-\frac{1}{2}P^i_j(\mathbf{k})P_a^b(\mathbf{k})\right)A^a_b
    =\mathcal{P}^{ibT}_{ja}(\mathbf{k})A^a_b(\mathbf{k}). \nonumber
\eea
It is occasionally useful to work with fully symmetrised forms of the vector and tensor projection operators; when this is so we shall employ $\mathcal{P}^{kV}_{ij}=\kh_{(j}P_i^{k)}(\mathbf{k})$ as the vector projector to operate on $T^j_k$ and $\mathcal{P}^{ibT}_{ja}=\left(P^i_{(a}(\mathbf{k})P_j^{b)}(\mathbf{k})-\frac{1}{2}P^i_j(\mathbf{k})P_a^b(\mathbf{k})\right)$ for the tensors, where the curved brackets denote symmetrisation on the enclosed indices $A_{(i}B_{j)}=(1/2)(A_iB_j+A_jB_i)$.
We shall do this, for example, when we come to consider the statistics of the magnetic field.

Separating the metric perturbation, one can see that $h^{iT}_j$ will carry information about the gravitational waves while $\V{h}_i$ will carry information about vorticity.

We can now make identifications between our variables, separated using the above prescription, and the $\phi,B,S_i,\psi,E,F_i,G_{ij}$ employed in appendix \ref{Appendix-Perturbations} and by Mukhanov, Feldman and Brandenberger \cite{MukhanovFeldmanBrandenberger92}. Since we are working in synchronous gauge the lapse and shift functions vanish, implying $\phi=B=S_i=0$; the spatial perturbation is, in the two notations,
\bdm
  -2\psi\gamma_{ij}+2E_{,ij}+2F_{(i|j)}+G_{ij}=h_{ij}
\edm
which in Fourier space is
\bdm
  -2\psi\gamma_{ij}-2k^2\kh_i\kh_jE-2ik\kh_{(i}F_{j)}+G_{ij}=
   \frac{1}{3}h\gamma_{ij}+\left(\kh_i\kh_j-\frac{1}{3}\gamma_{ij}\right)h_S+2\kh_{(i}\V{h}_{j)}+h^T_{ij}
\edm
and so
\be
  \psi=\frac{1}{6}(h_S-h), \quad E=-\frac{1}{2k^2}h_S, \quad F_i=\frac{i}{k}h^V_i, \quad G_{ij}=h^T_{ij}
\ee
or
\be
  h=-\left(2k^2E+6\psi\right), \quad h_S=-2k^2E, \quad h^V_i=-ikF_i, \quad h^T_{ij}=G_{ij} .
\ee
In this representation of the gauge, then, the gauge-invariant Bardeen variables (\ref{BardeenSynchronous}) are
\be
\label{BardeenVariables}
  2k^2\Phi=\ddot{h}_S+\hub\dot{h}_S, \quad 2k^2\Psi=\hub\dot{h}_S+\frac{k^2}{3}\left(h_S-h\right), \quad
  k\tilde{V}_i=i\dot{h}^V_i .
\ee
This is little more physically lucid than in the alternative prescription and in this chapter we shall not employ the Bardeen variables though obviously we could rewrite the scalar and vector components of the Einstein equations below as equations for them. We return to the vector Bardeen variable in chapter \ref{Chapter-CMB}.

Transferring the Einstein equations into Fourier space and separating them across their scalar, vector and tensor components (noting that the $\delta R^i_j$ equation contains a redundant scalar trace term, which we remove) leads us to
\bea
\label{EinsteinEqs}
  \begin{array}{c}{}_{\mathrm{Background}} \\ \mathrm{{}^{Equations}} \end{array}&:&
   \left(\hub\right)^2=-\frac{8\pi G}{3}\left(a^2\overline{T}^0_0\right)+\frac{a^2\Lambda}{3}, \quad
   2\frac{\ddot{a}}{a}-\left(\hub\right)^2=-\frac{8\pi G}{3}\left(a^2\overline{T}^i_i\right)+a^2\Lambda , \nonumber \\
\cline{1-3}
  \begin{array}{c}\mathrm{{}_{Scalar\;Trace}}\\\mathrm{{}^{Evolution}} \end{array}&:&
    \ddot{h}+\hub\dot{h}=8\pi Ga^2\left(\delta T^0_0-\delta T^i_i\right), \nonumber \\
  \begin{array}{c}\mathrm{{}_{Traceless\;Scalar}}\\\mathrm{{}^{Evolution}} \end{array}&:&
    \ddot{h}_S+2\hub\dot{h}_S+\frac{1}{3}k^2\left(h-h_S\right)=16\pi Ga^2\delta T_S \nonumber \\
  \begin{array}{c}\mathrm{{}_{Vector}}\\\mathrm{{}^{Evolution}} \end{array}&:&
   \V{\ddot{h}}_i+2\hub\V{\dot{h}}_i=16\pi Ga^2\V{\delta T}_i, \\
  \begin{array}{c}\mathrm{{}_{Tensor}}\\\mathrm{{}^{Evolution}} \end{array}&:&
   \ddot{h}^{iT}_j+2\hub\dot{h}^{iT}_j+k^2h^{iT}_j=16\pi Ga^2\delta T^{iT}_j, \nonumber \\
  \begin{array}{c}\mathrm{{}_{Scalar}}\\\mathrm{{}^{Constraints}} \end{array}&:&
     \hub\dot{h}+\frac{1}{3}k^2\left(h-h_S\right)=-8\pi Ga^2\delta T^0_0 , \quad
     \dot{h}-\dot{h}_S=24\pi G\frac{a^2}{k}\delta T_{(0)}^{(i)S} , \nonumber \\
  \begin{array}{c}\mathrm{{}_{Vector}}\\\mathrm{{}^{Constraint}} \end{array}&:&
   \dot{h}^i_V=16i\pi G\frac{a^2}{k}\delta T_{(0)}^{iV} . \nonumber
\eea
with $\delta T_S$, $\V{\delta T}_i$ and $\delta T^{iT}_j$ defined as above and $\delta T^i_0$ separated as
\bdm
  \delta T^i_0=i\kh^i\delta T_{(0)}^{(i)S}+\delta T^{iV}_{(0)}.
\edm
Here all the variables are now in Fourier space and are functions of conformal time $\eta$ only.

We can then see that, in the absence of any source, the vector perturbations will damp away rapidly. The tensor evolution equation is a damped wave equation and in the absence of tensor sources it will generate damped plane waves. Although vector perturbations are neglected in standard models, they must be considered in models containing active sources -- models with networks of cosmic defects or magnetic fields, for example. Inflation does, however, generate an amount of power in gravitational waves and so tensor modes are taken into account in the most accurate pictures. It is, however, a good approximation to consider only scalar perturbations.

When these equations are studied for the separate components, it becomes necessary to discuss the bases we are working in for the vectors and the tensors. These bases are introduced in section \S\ref{BoltzmannVectorPert}.

We now turn to the standard components of the matter/energy content: cold dark matter (\S\ref{SecCDM}), the photons (\S\ref{SecPhotons}), the massless neutrinos (\S\ref{SecNeutrinos}) and the baryons (\S \ref{SecBaryons}). We also consider, for brevity as background sources only, scalar field matter (\S\ref{SecScalarFields}) which can drive both inflationary and late-time accelerating phases. The electromagnetic field is discussed in chapter \ref{Chapter-MagnetisedCosmology}.

\section{Fluid Matter}
\label{Sec-Fluids}
\subsection{Generic Stress-Energy Tensor}
\label{Sec-Fluids-SETensor}
Cold dark matter and baryons are perhaps the simplest matter that can be placed into the Einstein equations, especially if one neglects all viscous effects. Here we shall quote both shear and bulk viscosities and also heat conduction and their impacts on the stress-energy tensor; this will prove useful, for instance, when modelling photon viscosities. However when we consider the standard cosmological sources of cold dark matter (CDM) and baryons we shall neglect the imperfect elements.

To model imperfect fluids, seperate the stress-energy and particle current tensors into an ideal and a non-ideal part,
\bea
  T^\mu_\nu=T^{(\mathrm{Ideal})\mu}_\nu+T^{(\mathrm{Non-Ideal})\mu}_\nu
    &=&\left(\rho+p\right)u^\mu u_\nu+p\delta^\mu_\nu+T^{(\mathrm{Non-Ideal})\mu}_\nu \\
  N^\mu=N^\mu_{(\mathrm{Ideal})}+N^\mu_{(\mathrm{Non-Ideal})}&=&nu^\mu+N^\mu_{(\mathrm{Non-Ideal})}
\eea
where $\rho$ is the fluid's mass-energy density, $p$ its pressure, $n$ its number density and $u^\mu$ its 4-velocity. This decomposition introduces an ambiguity into the definitions of $\rho$, $p$, $n$ and $u^\mu$, so define
\bdm
   T^0_0\equiv-\rho, \qquad N^0\equiv-n
\edm
in a comoving frame where $u^\mu=(1,\mathbf{0})$, and $\rho$ is the total energy density in this frame. $p$ is defined as the ``intrinsic'' pressure when the non-ideal components vanish. The 4-velocity may be taken, as do Lifshitz and Pitaevskii \cite{LifshitzPitaevskii-FluidMechanics}, to be the velocity of energy transfer ($\Rightarrow T^{0i}=0$ in a comoving frame), or it can be taken, as do Eckart \cite{Eckart} and Weinberg \cite{Weinberg}, to be the velocity of particle transport ($\Rightarrow N^i=0$ in a comoving frame). We take Eckart and Weinberg's definition.

These definitions imply that $T^{(\mathrm{Non-Ideal})0}_0\equiv0$, $N^0_{(\mathrm{Non-Ideal})}\equiv 0$ and $N^i_{(\mathrm{Non-Ideal})}\equiv0$ in a comoving frame. Thus one can see that, in a comoving frame, the dissipative effects are held entirely within the stress-energy tensor; in a general frame,
\bea
  N^\mu_{\mathrm{(Non-Ideal)}}&\equiv&0 \\ u^\mu u^\nu T^{(\mathrm{Non-Ideal})}_{\mu\nu}&=&0 .
\eea
An arbitrary stress-energy can then be found by finding the most general $T^{\mu\nu}_{\mathrm{(Non-Ideal)}}$ allowed by both this equation and by the second law of thermodynamics; for a demonstration see Weinberg \cite{Weinberg}.

The resulting stress-energy tensor of a generic barotropic fluid in an arbitrary frame is
\beas
  T^\mu_\nu&=&\rho u^\mu u_\nu+pH^\mu_\nu-\chi\left(H^{\mu\alpha}u_\nu+H^\alpha_\nu u^\mu\right)Q_\alpha
  -\xi H^{\mu\alpha}H_{\nu\beta}W_\alpha^\beta-\zeta H^\mu_\nu u^{\alpha}_{\phantom{\alpha};\alpha}
   \\ &=&\rho u^\mu u_\nu+pH^\mu_\nu+\tilde{\chi}^\mu_\nu+\tilde{\xi}^\mu_\nu+\tilde{\zeta}^\mu_\nu
\eeas
with a supplementary thermodynamic equation of state usually taken in cosmology to be of the barotropic and isentropic form $p=w(\rho)\rho$. Here the 4-velocity is
\be
  u^\mu=\frac{dx^\mu}{d\sqrt{-ds^2}},
\ee
the projection tensor onto a hypersurface orthogonal to $u^\mu$ is
\bdm
  H^\mu_\nu=\delta^\mu_\nu+u^\mu u_\nu,
\edm
the heat-flow vector is defined as
\be
  Q_\mu=\Theta_{;\mu}+\Theta u_{\mu;\nu}u^\nu,
\ee
the shear tensor is defined as
\be
  W^\mu_\nu=u^\mu_{\phantom{\mu};\nu}+u_\nu^{\phantom{\nu};\mu}-\frac{2}{3}\delta^\mu_\nu u^\alpha_{\phantom{\alpha};\alpha},
\ee
and $\Theta$ is the temperature of the fluid. $\chi$, $\xi$ and $\zeta$ are the (positive) coefficients of heat-flow, shear viscosity and bulk viscosity respectively\footnote{Note the change of notation from that usual in the fluid literature, as we are reserving $\eta$ for the conformal time rather than the shear viscosity.}. For non-relativistic ``dust'' -- \emph{i.e.} non-relativistic, pressureless fluids, one then takes $w=0$. Both the baryons and the cold dark matter are taken to be non-relativistic since, even in the presence of a primordial magnetic field, the baryons remain highly non-relativistic; see \cite{BarrowFerreiraSilk97,SubramanianBarrow98-MHD, ClarksonEtAl02,Barrow97}.

This decomposition of the fluid stress-energy tensor is useful since it acts as a direct relativistic generalisation of the familiar components of the fluid stress-energy (see for example Landau and Lifshitz \cite{LifshitzPitaevskii-FluidMechanics} and Weinberg \cite{Weinberg}). An alternative is to decompose it according to the four-velocity of an observer; this is, for example, employed in the gauge-invariant and covariant approach to cosmology \cite{EllisBruni89,EllisHwangBruni89,EllisVanElst98}, developed from covariant fluid mechanics \cite{Hawking66,Ellis71}, wherein spacetime is decomposed into a $3+1$ split with the observer's four-velocity acting as a time parameter, and all tensorial objects are projected parallel to the velocity and onto a hypersurface perpendicular to it. In this formalism, the stress-energy tensor splits into $T_{\mu\nu}=\rho u_\mu u_\nu+pH_{\mu\nu}+q_\mu u_\nu+u_\mu q_\nu+\pi_{\mu\nu}$ with $q_\mu$ the momentum flux and $\pi_{\mu\nu}$ the anisotropic stress. This approach has many advantages but for the purposes of this thesis we shall retain the older, more familiar formalism. It should be noted, however, that the gauge-invariant and covariant approach has been used with great success in the field of magnetised cosmologies; see for example \cite{TsagasBarrow97,Tsagas04,TsagasMaartens99-Perts, TsagasMaartens99-Bianchi, Maartens00}. The great benefit of such an approach is that one formulates first the entirely non-linear equations and then curtails these to an approximation based around a Robertson-Walker background, as opposed to the metric-based approach we are employing wherein we set a (fictional) background metric and perturb up to a more realistic approximation. This then lets us draw qualitative conclusions about the non-linear behaviour of a system; we can also find a clear physical interpretation of each term.


The two main thermodynamic properties of a fluid that we shall be concerned with are the equation of state and the speed of sound, defined by
\be
  w=\frac{p}{\rho}, \quad c_s^2=\frac{\partial p}{\partial\rho}=w+\frac{\partial w}{\partial\rho}\rho .
\ee
and its temperature. We assume that the fluid is isentropic and barotropic, that is that $w=w(\rho)$. The perfect fluid stress-energy tensor then has the components
\be
  T^0_0=-\rho, \quad T^i_0=-\left(\rho+p\right)v^i, \quad T^i_j=p\delta^i_j\Rightarrow\Pi^i_j=0 .
\ee
We have not imposed any restrictions on $w=w(\rho)$, which is implicitly a function of space and time. To retain homogeneity, $v^i=\mathcal{O}(\varepsilon)$ and so we neglect any second-order combinations of $v^i$ and the metric perturbations. It is worth emphasising, however, that we have heavily restricted the physics of the fluid species we can consider; a more general fluid equation of state could involve terms $p=w(\rho,S)\rho$ with $S$ the entropy of the fluid, and a truly realistic equation of state would be of the form $p=p(\rho,S)$ derived from the micro-physics.

Expressing the temperature of the imperfect fluid as a background homogeneous component and a perturbation,\footnote{While in this section we rigorously employ this notation, we shall in the future omit the overline on the background temperature where the situation seems unambiguous.}
\bdm
  \Theta(\mathbf{x},\eta)=\overline{\Theta}(\eta)+\delta\Theta(\mathbf{x},\eta),
\edm
and assuming that the viscosities can be treated as quantities of zeroth order in $\varepsilon$, we can then find that, as one would expect, the energy flux is modified by both the heat conduction and also a drag from the bulk viscosity:
\be
  \tilde{\chi}^i_0=\chi_a\overline{\Theta}\left(\dot{v}^i
   +\left(\hub+\frac{\dot{\overline{\Theta}}}{\overline{\Theta}}\right)v^i
   +\frac{\partial^i\delta\Theta}{\overline{\Theta}}\right), \quad
  \tilde{\zeta}^i_0=3\zeta_a\hub v^i
\ee
while the stresses acquire contributions from the two viscosities,
\bea
  \tilde{\zeta}^i_j&=&-\zeta_a\left(3\hub+\nabla\cdot\mathbf{v}+\frac{1}{2}\dot{h}\right)\delta^i_j, \nonumber \\
  \tilde{\xi}^i_j&=&-\xi_a\left(\partial^iv_j+\partial_jv^i+\dot{h}^i_j
   -\frac{1}{3}\delta^i_j\left(2\nabla\cdot\mathbf{v}+\dot{h}\right)\right)
\eea
where we have scaled the coefficients of heat conductivity and bulk and shear viscosity by $\chi_a=\chi/a$ and similar. Note that, as one should expect given our background, the heat conduction and shear viscosity are purely first-order in $\varepsilon$. Were they to impact on the background they would impose a directionality on the universe by, for example, a net heat flow or a net momentum.
If we assume that a particle species is collisionless then we can derive the temperature change from the first law of thermodynamics;
\be
  dQ=\frac{3}{2}d\left(\frac{p}{\rho}\right)+pd\left(\frac{1}{\rho}\right)
\ee
where $dQ$ is the change in heat and $p$ the pressure. We can approximate the pressure as $p=N\Theta$ where $N$ is the particle number density ($\dot{n}=-3(\dot{a}/a)n$), the density as $\rho=Nm$, $m$ being the particle mass, and for a collisionless fluid set $dQ=0$. With these substitutions we quickly see that
\be
\label{CollisionlessFluidT}
  \dot{\Theta}+2\hub\Theta=0
\ee
and so
\be
  \Theta\propto \frac{1}{a^2} .
\ee

The background isotropic stress is, in full,
\be
\label{FullBackgroundPressure}
  \frac{1}{3}T^i_i=p-3\zeta_a\hub
\ee
and so we see that, insofar as the background dynamics are concerned, the presence of a bulk viscosity serves to reduce the pressure of the universe.

For simplicity, we now assume that any fluids in the universe are perfect and derive their dynamics. Non-ideal fluids are considered in appendix \ref{Appendix-ViscousFluids}.

\subsubsection{Real-Space Dynamics}
The stress-energy conservation laws for the ideal part of this fluid are given by
\bdm
  T^\nu_{\mu;\nu}+T^{\nu(\mathrm{I})}_{\mu;\nu}=\mathcal{C}_\mu,
\edm
where $T^{\nu(\mathrm{I})}_\mu$ is the summed stress-energy tensor of any interacting species (for example, the electromagnetic field will contribute an interaction term to the baryon fluid) and $\mathcal{C}_\mu$ is a collisional term providing a source or sink for stress-energy, a transfer of momentum between baryons and photons coupled by Thomson scattering, for example. For clarity we shall always present the contributions to the continuity equations as $-T^\nu_{0;\nu}$. The interaction and non-ideal contributions add linearly to the results of this section.

Evaluating the ideal fluid contribution, neglecting terms second order in $h$ or $v^i$ but without yet linearising thermodynamic properties yields
\bea
  -T^\mu_{0;\mu}&=&\dot{\rho}+\nabla\cdot\left(\left(\rho+p\right)\fgv\right)
  +\left(\rho+p\right)\left(3\hub+\frac{1}{2}\dot{h}\right), \\
  T^\nu_{i;\nu}&=&\frac{\partial\left(\left(\rho+p\right)v_i\right)}{\partial \eta}
  +\partial_ip+4\hub\left(\rho+p\right)v_i-h_i^j\partial_jp \nonumber .
\eea

Linearising the pressure will in general give
\bdm
  \delta p=\frac{\partial p}{\partial\rho}\delta\rho+\frac{\partial p}{\partial S}\delta S
\edm
and so for an isentropic fluid
\be
  \delta p=c_s^2\bkr\delta
\ee
where we have linearised the density with a dimensionless perturbation $\delta=\delta\rho/\bkr$ with a spatial average density $\bkr$ -- that is, $\rho=\bkr(1+\delta)$.

It is straightforward to now show that
\bea
\label{GenericFluidSEConsB}
  \dot{\bkr}+3\hub\bkr\left(1+w\right)&=&-\overline{\mathcal{C}}_0, \nonumber \\
  \partial^i\overline{p}&=&\overline{\mathcal{C}}^i
\eea
for the background dynamics, and
\bea
\label{GenericFluidSEConsF}
  \dot{\delta}+3\hub\left(c_s^{\phantom{s}2}-w\right)\delta+\left(1+w\right)
  \left(\nabla.\fgv+\frac{1}{2}\dot{h}\right)&=&-\delta\mathcal{C}_0, \\
  \dot{v}^i+\left(\frac{\dot{w}}{1+w}+\hub\left(1-3w\right)\right)v^i
  +\frac{c_s^2}{1+w}\partial^i\delta&=&\delta\mathcal{C}^i. \nonumber
\eea

\subsubsection{Fourier space}
Transferring into Fourier space and separating across the scalar, vector and tensor components is generally straightforward. Denoting the spatial trace of a tensor $T^i_j$ as $T=T^i_i$, we quickly see that, for the ideal sector of the stress-energy in the background,
\bdm
  \overline{T}^0_0=-\bkr, \quad \overline{T}^i_0=0, \quad \overline{T}^i_j=c_s^2\bkr\delta^i_j
\edm
and in the foreground,
\bdm
  \delta T^0_0=-\bkr\delta, \quad \delta T^i_0=-\bkr\left(1+w\right)v_i, \quad \delta T^i_j=c_s^2\bkr\delta\delta^i_j .
\edm
We separate across the scalar, vector and tensor parts to get
\bdm
  \overline{T}^0_0=-\bkr, \quad \overline{T}^i_0=0, \quad \overline{T}^i_i=3c_s^2\bkr, \quad \overline{\pi}^i_j=0
\edm
in the background and
\beas
  &\delta T^0_0=-\bkr\delta, \quad \delta T^{(i)S}_{(0)}=-\bkr\left(1+w\right)v_S,& \\
  &\delta T^{iV}_{(0)}=-\bkr\left(1+w\right)\V{v}_i, \quad \delta T=3c_s^2\bkr\delta, \quad \delta\pi^i_j=0 .&
\eeas
%
The background matter conservation equation is
\be
  \dot{\bkr}+3\hub\bkr\left(1+w\right)=-\overline{\mathcal{C}}_0 .
\ee
At linear order, the ideal components are
\bea
  \dot{\delta}+3\hub\left(c_s^{\phantom{s}2}-w\right)\delta+\left(1+w\right)
   \left(kv_S+\frac{1}{2}\dot{h}\right)&=&-\delta\mathcal{C}_0, \\
  \dot{v}_S+\left(\frac{\dot{w}}{1+w}+\hub\left(1-3w\right)\right)v_S
   -k\frac{c_s^{\phantom{s}2}}{1+w}\delta&=&\delta\mathcal{C}^{(i)}_S, \\
  \dot{v}^i_V+\left(\frac{\dot{w}}{1+w}+\hub\left(1-3w\right)\right)v^i_V&=&\delta\mathcal{C}^i_V .
\eea


\subsection{Cold Dark Matter}
\label{SecCDM}
Cold dark matter (CDM) is defined as a fluid that interacts with other species purely gravitationally, via the Einstein equations. Governing this fluid, in addition to the Einstein equations, will be the CDM specialisation of the conservation laws (\ref{GenericFluidSEConsB}, \ref{GenericFluidSEConsF}). ``Cold'' refers to the extreme non-relativistic nature of the dark matter; one assumes that it has a vanishing equation of state and speed of sound. Moreover, since CDM interacts only gravitationally, it may be used to define the synchronous frame of reference; hence we set $v_c^i=0$ identically.

CDM then obviously has a background stress-energy tensor
\be
  \overline{T}^0_0=-\bkr_c,\quad\overline{T}^i_0=\overline{T}^i_i=\overline{\Pi}^i_j=0
\ee
which contributes only an average density to the dynamics, and
\be
  \delta T^0_0=-\bkr_c\delta_c,\quad \delta T^i_0=\delta T^i_i=\delta \Pi^i_j=0,
\ee
in the foreground, with the CDM fluctuations providing potential wells that will seed the formation of large-scale baryonic structure. The stress-energy conservation equations become
\be
\label{StressEnergyCDM1}
  \dot{\bkr}_c+3\hub\bkr_c=0, \quad \dot{\delta}_c+\frac{1}{2}\dot{h}=0 .
\ee

In the simplest model there are no viscous effects that can be associated with cold dark matter; if there were, it would by definition not remain ``cold''. While naturally it would not take a particular leap of intuition to envisage a bulk viscous material acting as CDM it is standard to instead reduce the fluid to its very barest fundamentals.


\subsection{Single-Fluid Universes}
\label{Section-SingleFluidUniverses}
\subsubsection{An Ideal Fluid Universe}
Consider a universe dominated by some generic perfect fluid with (constant) equation of state $w$ and speed of sound $c_s^2$; this universe then evolves according to the (redundant) set of equations
\be
  \dot{\overline{\rho}}+3\hub\overline{\rho}(1+w)=0, \quad \mathcal{H}^2=\frac{8\pi G}{3}a^2\overline{\rho},
  \quad 2\frac{\ddot{a}}{a}-\left(\hub\right)^2=-8\pi Ga^2w\overline{\rho} .
\ee
We differentiate the Friedmann equation to obtain an expression for $\dot{\overline{\rho}}$ and substitute this back into the conservation equation, whence
\bdm
  2\mathcal{H}\left(\dot{\mathcal{H}}+(1+3w)\mathcal{H}^3\right)=0 .
\edm
This has the trivial solution $\mathcal{H}=0$, corresponding to an empty universe, and an ordinary differential equation for $\mathcal{H}$ which we quickly solve for both the Hubble parameter and the scale factor to find
\be
  a=a_0\eta^{\frac{2}{1+3w}}, \quad \mathcal{H}=\frac{2}{1+3w}\frac{1}{\eta}, \quad
  \frac{\ddot{a}}{a}=\frac{2}{1+3w}\frac{1-3w}{1+3w}\frac{1}{\eta^2} .
\ee
The background density is best found by direct integration of the continuity equation, whence
\be
  \overline{\rho}=\overline{\rho}_0\left(\frac{a_0}{a}\right)^{3(1+w)} .
\ee
We can then consider special cases of the equation of state:
\begin{itemize}
\item $w=1/3$:
A purely radiative or ultra-relativistic fluid has this equation of state, as demonstrated in the next section. With $w=1/3$ we see that
\be
\label{RadiationUniverse}
  a=a_0\eta, \quad \mathcal{H}=\frac{1}{\eta}, \quad \frac{\ddot{a}}{a}=0
\ee
and so a radiative universe does not accelerate with respect to the conformal time, and the scale factor grows linearly with $\eta$. The background density of the radiation fluid decays as $a^{-4}$, which can be interpreted as a stretching due to the volume expansion of the universe plus a stretching due to red-shifting.

\item $w=0$:
A fluid with an effectively vanishing pressure is labelled ``dust''; due to the extreme non-relativistic nature of their bulk distribution, baryons are assumed to be dust. Cold dark matter is also taken to be dusty. A dust-dominated universe evolves as
\be
\label{DustUniverse}
  a=a_0\eta^2, \quad \mathcal{H}=\frac{2}{\eta}, \quad \frac{\ddot{a}}{a}=\frac{2}{\eta^2} .
\ee

The background density of dust evolves, as one might expect, with $a^{-3}$ which is merely the volume expansion of the universe.

\item $w=-1/3$:
A fluid with such an equation of state causes (artificial) singularities to appear in the expressions for the Hubble and acceleration parameters. We will consider the importance of this type of fluid shortly.

\item $w=-1$:
As we saw earlier, a fluid with this equation of state behaves as a cosmological constant. The formalism above breaks down for a cosmological constant and we will consider it shortly.
\end{itemize}

In the light of the observations of distant type-Ia supernovae \cite{RiessEtAl04} implying the recent onset of an accelerating epoch, we might also ask when the universe appears to undergo an accelerating expansion; demanding that $\ddot{a}/a>0$ implies that the universe will expand for all fluids with $w<1/3$. Thus all fluids from the critical radiative through non-relativistic and into negative pressure accelerate the universal expansion with respect to conformal time. This is not, however, the physical acceleration observed by the astronomers as this is with respect to the conformal rather than proper time.

If we convert our system to co-ordinate time $dt=a(\eta)d\eta$, we can rewrite the Hubble and Raychaudhuri equations as
\be
\label{BackgroundH}
  \left(\frac{a'}{a}\right)^2=\frac{8\pi G}{3}\overline{\rho}, \quad
  \left(\frac{a''}{a}\right)=-\frac{8\pi G}{6}\overline{\rho}(1+3w)
\ee
with the matter continuity equation unchanged in form. We can immediately state from the Raychaudhuri equation that the universe will undergo accelerated expansion with respect to the observed co-ordinate time if $w<-1/3$. Should there be a significant amount of a fluid with a lower equation of state than this then the universe will ultimately enter a period of never-ending accelerated expansion.

If we repeat our analysis for co-ordinate time, with $H$ again denoting the observed Hubble parameter and $a'$ the derivative of the scale factor with respect to time, we find that
\be
  a=a_0t^{\frac{2}{3(1+w)}}, \quad H=\frac{2}{3(1+w)}\frac{1}{t}, \quad
  \frac{a''}{a}=-\frac{2}{9}\left(\frac{1+3w}{(1+w)^2}\right)\frac{1}{t^2} .
\ee

\begin{itemize}
\item $w=1/3$:
For radiation,
\be
  a=a_0\sqrt{t}, \quad H=\frac{1}{2t}, \quad \frac{a''}{a}=-\frac{1}{4t^2}
\ee
and so the universe is manifestly undergoing deceleration during a radiation-dominated era.

\item $w=0$:
For dust,
\be
  a=a_0t^{2/3}, \quad \mathcal{H}=\frac{2}{3t}, \quad \frac{a''}{a}=-\frac{2}{9t^2} .
\ee

\item $w=-1/3$:
Here
\be
  a=a_0t, \quad H=\frac{1}{t}, \quad a''=0
\ee
and any fluid with a more negative equation of state will undergo an accelerating expansion. According to this formalism the Hubble parameter will go singular for a cosmological constant and negative for a fluid with $w<-1$.
\end{itemize}

\subsubsection{de Sitter Spaces}
The case $w=-1$ cannot be analysed in the above way due to the vanishing of $\dot{\overline{\rho}}$. Instead we turn back to the Hubble equation for a universe dominated by the constant term; in conformal time this is
\be
  \left(\hub\right)^2=a^2\frac{\Lambda}{3}
\ee
which has the unedifying solution
\be
  a(\eta)=\left(\frac{1}{a_1}\pm\sqrt{\frac{\Lambda}{3}}\left(\eta-\eta_1\right)\right)^{-1}
\ee
where $a_1$ is the scale factor at some conformal time $\eta_1$. For $\Lambda>0$ this is called de Sitter space, and for $\Lambda<0$ it is anti-de Sitter space. This is not a very useful form for physical interpretation, and so we return again to co-ordinate time, wherein
\be
  \left(\frac{a'}{a}\right)^2=\frac{\Lambda}{3}
\ee
with the more lucid solution
\be
\label{deSitter}
  a(\eta)=a_1e^{\pm\sqrt{\frac{\Lambda}{3}}(t-t_1)} , \quad H=\pm\sqrt{\frac{\Lambda}{3}}, \quad
  \frac{a''}{a}=\frac{\Lambda}{3} .
\ee
de Sitter spaces are thus exponentially expanding or contracting spacetimes. Fluids with equations of state nearing $w=-1$ are quasi-de Sitter spacetimes; inflationary and quintessential models generate quasi-de Sitter spaces from universes dominated by scalar fields of some sort. Any inflating model, not necessarily of scalar fields, is characterised phenomenologically with $w$, and this can be found in the recent universe from observations of distant supernovae, as well as from the microwave background (wherein it is tangled in various degeneracies). Recently it was determined from large-scale structure with the observation of the baryon oscillations in the galactic distribution. The current equation of state of the inflating component of the universe -- whatever that may turn out to be -- is entirely consistent with a cosmological constant, as tested by groups analysing the WMAP+Supernovae datasets \cite{WMAP-Bennett,RiessEtAl04} and recent detections of the baryon oscillations in the large-scale galactic structure \cite{2dF,SDSS,SDSS2}.

\section{Boltzmann Fluids}
Here we will discuss a generic effective fluid governed by a Boltzmann equation. We then turn to consider in great detail a collisional photon fluid which exchanges energy-momentum with the baryon fluid, and finally briefly consider massless neutrinos.

While the approach in this section is our own, the basic material is well-known and can be drawn from various sources; see in particular Ma and Bertschinger \cite{MaBertschinger95} and Landriau and Shellard \cite{LandriauShellard03}, along with Peebles \cite{Peebles-LargeScaleStructure, Peebles-PrinciplesPhysicalCosmology}, Padmanabhan \cite{Padmanabhan}, Crittenden \cite{CrittendenThesis} and Crittenden \emph{et. al.} \cite{CrittendenEtAl93,CrittendenCoulsonTurok94}.

\subsection{General Boltzmann Fluids}
\subsubsection{The Boltzmann Equation and Stress-Energy Tensors}
\label{Sec-Boltzmann}
Let us work initially in a Minkowski spacetime. Liouville's theorem states that if a system is entirely satisfied by a Lagrangian formulation with co-ordinates $x^i$ and their conjugate momenta $P_i$, then the distribution function $f$ defined by
\be
  dN=f\left(x^i,P_i,t\right)d^3\mathbf{x}d^3\mathbf{P}
\ee
where $dN$ is the number of particles in a volume element does not change in transport -- \emph{i.e.}, $df=0$. If one takes the time derivative and includes a collisional term acting as a source or sink, the result is Boltzmann's equation,
\be
  \frac{df}{dt}=\frac{\partial f}{\partial t}+\frac{\partial f}{\partial x^i}\frac{dx^i}{dt}
  +\frac{\partial f}{\partial P_i}\frac{dP_i}{dt}=\frac{\partial f}{\partial t}
  +v^i\frac{\partial f}{\partial x^i}-\frac{\partial U}{\partial x^i}\frac{\partial f}{\partial P_i}
  =C[f]
\ee
for some external 3-potential $U$ and some collisional term $C[f]$. It is important to note that, since this arises in a Lagrangian system, $\left\{v^i,P_i\right\}$ are variables canonical to $x^i$.

Given that the mass (or number), momentum and energy of particles are conserved in collisional processes, we may state that
\bdm
  \int C[f]d^3\mathbf{P}=\int m_0C[f]d^3\mathbf{P}=\int m C[f]d^3\mathbf{P}=\int P_iC[f]d^3\mathbf{P}=0
\edm
where $m^2=P^2+m_0^2$ is the mass-energy of the particle concerned.

Now, it is clear that the number density of the distribution function is given at an event $(\mathbf{x},t)$ by
\bdm
  N=\int fd^3\mathbf{P}
\edm
and so we immediately see that
\be
  \rho=\int mfd^3\mathbf{P} .
\ee
It should also be clear that the average velocity of the system is given by
\be
  \langle v^i\rangle=\frac{\int v^ifd^3\mathbf{P}}{\int fd^3\mathbf{p}}
\ee
and that the average of two velocities is
\be
  \langle v^iv_j\rangle=\frac{\int v^iv_jfd^3\mathbf{P}}{\int fd^3\mathbf{p}} .
\ee
We can then immediately define the components of the stress-energy tensor (in Minkowski space) as
\bea
  T^0_0=-\rho&=&-\int mfd^3\mathbf{p} \nonumber \\
\label{MinkowskiBoltzmannSE}
  T^i_0=-\mathcal{E}^i=\rho\langle v^i\rangle&=&\int mv^ifd^3\mathbf{p} \\
  T^i_j=\mathcal{P}^i_j=\rho\langle v^iv_j\rangle&=&\int mv^iv_jfd^3\mathbf{p} . \nonumber
\eea

Using these one may show that the various conservation conditions imply mass continuity,
\be
  \frac{\partial\rho}{\partial t}+\frac{\partial}{\partial x^i}\left(\rho\langle v^i\rangle\right)=0,
\ee
the Euler equation
\be
  \frac{\partial}{\partial t}\left(\rho\left<v^i\right>\right)+\frac{\partial\mathcal{P}^i_j}{\partial x_j}+\rho\frac{\partial U}{\partial x_i}=0
\ee
and energy flow,
\be
  \frac{\partial}{\partial t}\left(N\left<m\right>\right)+\frac{\partial\mathcal{E}^i}{\partial x^i}
   +\rho\frac{\partial U}{\partial x^i}\left<v^i \right> .
\ee

The stress-energy tensor (\ref{MinkowskiBoltzmannSE}) is valid only in Minkowski space; however, we may define the momentum four-tensor $P^\mu=m_0U^{\mu}=m_0dx^\mu/\sqrt{-ds^2}$ and convert it to a tensorial expression valid in all reference frames. In this way one may express a covariant stress-energy tensor of a Boltzmann fluid,
\be
  T^\mu_\nu=\int\sqrt{-g}d^3\mathbf{P}\frac{P^\mu P_\nu}{P^0}f\left(x^\mu,P_\mu\right).
\ee
where the $\sqrt{-g}$ is included to ensure normality of the integral.

In a non-trivial metric, one has to take care to differentiate between the momentum $P_\mu$ appearing in the stress-energy tensor, canonical to the co-ordinate $x^\mu$, and the proper momenta $p_\mu$ measured in a Riemannian (locally-Minkowski) co-ordinate system ($p^\mu=(m,p^i)$, $p_\mu=(-m,p_i)$). Clearly the canonical momentum $P_\mu$ is gauge-dependant, while the proper momentum, $p_\mu$ is gauge-invariant. To link the two let us find the transform between the FLRW frame and the comoving Riemannian frame.

Using the tensor transformation law to compare the two conjugate metrics,
\be
  g^{\mu\nu}=a^{-2}\left(\begin{array}{cc}-1&\mathbf{0}\\\mathbf{0}&\gamma^{ij}-h^{ij}\end{array}\right)
   =\frac{\partial x^\mu}{\partial\overline{x}^\alpha}\frac{\partial x^\nu}{\partial\overline{x}^\beta}
   \overline{g}^{\alpha\beta}=\frac{\partial x^\mu}{\partial\overline{x}^\alpha}\frac{\partial x^\nu}{\partial\overline{x}^\beta}\mathrm{diag}(-1,1,1,1)
\ee
one finds that
\be
\label{ContravariantCoordinateTransformation}
  \frac{\partial x^0}{\partial\overline{x}^\alpha}=\frac{1}{a}\delta^0_\alpha, \quad
  \frac{\partial x^i}{\partial\overline{x}^0}=0, \quad
  \frac{\partial x^i}{\partial\overline{x}_j}=\frac{1}{a}\sqrt{\gamma^{ij}-\frac{1}{2}h^{ij}}
\ee
to the first order in $h^i_j$, and comparing the two metrics,
\be
  g_{\mu\nu}=a^2\left(\begin{array}{cc}-1&\mathbf{0}\\\mathbf{0}&\gamma_{ij}+h_{ij}\end{array}\right)
   =\frac{\partial x_\mu}{\partial\overline{x}_\alpha}\frac{\partial x_\nu}{\partial\overline{x}_\beta}
   \overline{g}_{\alpha\beta}=\frac{\partial x_\mu}{\partial\overline{x}_\alpha}\frac{\partial x_\nu}{\partial\overline{x}_\beta}\mathrm{diag}(-1,1,1,1)
\ee
gives
\be
\label{CovariantCoordinateTransformation}
  \frac{\partial x_0}{\partial\overline{x}_\alpha}=a\delta_0^\alpha, \quad
  \frac{\partial x_i}{\partial\overline{x}_0}=0, \quad
  \frac{\partial x_i}{\partial\overline{x}^j}=a\sqrt{\gamma_{ij}+\frac{1}{2}h_{ij}} .
\ee

Thus we can write the canonical momentum as
\be
  P^\mu=\frac{1}{a}\left(m,\left(\gamma^{ij}-\frac{1}{2}h^{ij}\right)p_j\right), \quad
  P_\mu=a\left(-m,\left(\gamma_{ij}+\frac{1}{2}h_{ij}\right)p^j\right)
\ee
which satisfy $P^\mu=g^{\mu\nu}P_\nu$ as one should hope. Taking the inner product of the four-momentum with itself again gives
\bdm
  P^\mu P_\mu=-m_0^2=p^2-m^2
\edm
as expected.

We remove the explicit dependence on $a$ by making a (non-canonical) transformation to $\epsilon=am$, $ap^j=q^j=q\hat{n}^j$ where $\hat{n}^i$ is the unit vector in the direction of the fluid's momentum -- that is, $\epsilon$ will be the mass-energy measured by a comoving FLRW observer and we write the comoving momentum as an amplitude and a direction. Doing this gives us the canonical momentum
\be
  P_\mu=\left(-\epsilon,q\left(\gamma_{ij}+\frac{1}{2}h_{ij}\right)\hat{n}^j\right) .
\ee
and the taking the amplitude of this momentum gives
\be
  \epsilon^2=q^2+a^2m_0^2 .
\ee

Working in polar co-ordinates in momentum space, the volume element is then
\bdm
  d^3\mathbf{P}=\left(1+\frac{1}{2}h\right)q^2dqd\Omega_{\mathbf{n}}
\edm
and we can rapidly calculate the determinant of the FLRW metric,
\bdm
  \sqrt{-g}=a^{-4}\left(1-\frac{1}{2}h\right).
\edm

We now perturb the distribution function around a homogeneous (thermal) background
\bdm
  f_0(q)=g_s\left(e^{m/\Theta_0}\pm 1\right)^{-1}
\edm
where $g_s$ -- two, for both photons and neutrinos -- is the statistical weight, $\Theta_0=a\Theta$ is the temperature of the particles today, and where fermions take the positive sign and bosons take the negative sign. We write the expansion as
\bdm
  f\left(x^i,q,n_j,\eta\right)=f_0(q)+f_0(q)f_1\left(x^i,q,n_j,\eta\right) .
\edm
Noting that
\bdm
  \int\hat{n}^i\hat{n}_jd\Omega_{\mathbf{n}}=\frac{4\pi}{3}\delta_{ij}, \quad
  \int\hat{n}_id\Omega_{\mathbf{n}}=\int\hat{n}_i\hat{n}_j\hat{n}_kd\Omega_{\mathbf{n}}=0 ,
\edm
one can show that the linearised stress-energy components are
\bea
  T^0_0&=&-\rho=-\frac{1}{a^4}\iint\sqrt{q^2+a^2m_0^2}f_0(1+f_1)q^2dqd\Omega_\mathbf{n},
   \nonumber \\
    &=&-\frac{4\pi}{a^4}\int\sqrt{q^2+a^2m_0^2}f_0q^2dq
     -\frac{1}{a^4}\iint\sqrt{q^2+a^2m_0^2}f_0f_1q^2dqd\Omega_\mathbf{n},\\
  T^i_0&=&-\frac{1}{a^4}\iint q\left(\hat{n}^i+\frac{1}{2}\hat{n}_jh^{ij}\right)f_0(1+f_1)q^2dqd\Omega_\mathbf{n}
   \nonumber \\
    &=&-\frac{1}{a^4}\iint q^2dqd\Omega q\hat{n}^if_0f_1 , \\
  T^i_j&=&\frac{1}{a^4}\iint\frac{q^2\hat{n}_a\hat{n}^b\left(\gamma^{ia}-\frac{1}{2}h^{ia}\right)
     \left(\gamma_{jb}+\frac{1}{2}h_{jb}\right)}{\sqrt{q^2+a^2m_0^2}}f_0(1+f_1)q^2dqd\Omega_\mathbf{n}
   \nonumber \\
    &=&\frac{4\pi}{3a^4}\delta^i_j\int\frac{q^2f_0}{\sqrt{q^2+a^2m_0^2}}q^2dq
     +\frac{1}{a^4}\iint\frac{q^2}{\sqrt{q^2+a^2 m_0^2}}\hat{n}^i\hat{n}_jf_0f_1q^2dqd\Omega_\mathbf{n} .
\eea

For massless particles, such as the photons and neutrinos we consider, we can set $m_0=0$ in the above equations which reduce to
\be
  -\overline{T}^0_0=\overline{T}^i_i=\frac{4\pi}{a^4}\int qf_0q^2dq , \quad \overline{T}^i_0=0,\quad \overline{\Pi}^i_j=0
\ee
in the background, and
\bea
  -\delta T^0_0=\delta T^i_i=\frac{1}{a^4}\iint qf_0f_1q^2dqd\Omega_\mathbf{n} ,\nonumber \\
  \delta T^i_0=-\frac{1}{a^4}\iint q\hat{n}^if_0f_1q^2dqd\Omega_\mathbf{n} , \\
  \delta\Pi^i_j=\frac{1}{a^4}\iint qf_0f_1\left(\hat{n}^i\hat{n}_j-\frac{1}{3}\delta^i_j\right)q^2dqd\Omega_\mathbf{n}
\eea
for the perturbations.

A massless Boltzmann fluid -- corresponding to a collection of photons or massless neutrinos, or to an extremely relativistic fluid --  then behaves to zeroth order as a barotropic perfect fluid with equation of state $\overline{p}=\frac{1}{3}\bkr$, as we previously stated. While this equation of state also holds in the foreground, the extant anisotropic stresses ensure that the effective fluid is far from perfect.


Returning to the Boltzmann equation (for a species not necessarily massless), we may now write it in terms of our new variables as
\be
  f_0\frac{\partial f_1}{\partial\eta}+f_0\frac{dx^i}{d\eta}\frac{\partial f_1}{\partial x^i}
  +\frac{dq}{d\eta}\frac{\partial f_0}{\partial q}+f_0\frac{dq}{d\eta}\frac{\partial f_1}{\partial q}
  +f_1\frac{dq}{d\eta}\frac{\partial f_0}{\partial q}+f_0\frac{d\hat{n}_i}{d\eta}\frac{\partial f_1}{\partial\hat{n}_i}=C[f] .
\ee
In this expression, the only quantities of unknown order are $dq/d\eta$ and $d\hat{n}_i/d\eta$. If we consider the geodesic equation
\bdm
  P^0\frac{dP^\mu}{d\eta}=-\Gamma^\mu_{\phantom{\mu}\alpha\beta}P^\alpha P^\beta
\edm
where $\sqrt{-ds^2}=d\eta$ since we are considering particles along their worldlines, then the time component will yield
\bdm
  \varepsilon\frac{d\varepsilon}{d\eta}=\hub\left(\varepsilon^2-q^2\right)-\frac{1}{2}q^2\nh_i\nh_j\dot{h}^{ij}
\edm
which, on substituting for the mass-energy relation, gives
\bdm
  q\frac{dq}{d\eta}=\hub\left(\varepsilon^2-q^2-a^2m_0^2\right)-\frac{1}{2}q^2\nh_i\nh_j\dot{h}^{ij}
\edm
and so
\be
  \frac{dq}{d\eta}=-\frac{1}{2}q\nh_i\nh_j\dot{h}^{ij} .
\ee
Similarly, considering the space component will give
\be
  2\frac{d\nh^i}{d\eta}=\nh^i\nh_m\nh_n\dot{h}^{mn}-\nh_j\dot{h}^{ij}-\frac{2q}{\varepsilon}\nh_m\nh_n\partial^m h^{in}
   +\frac{q}{\varepsilon}\nh_m\nh_n\partial^i h^{mn}\approx\mathcal{O}(h).
\ee
We may also write $dx^i/d\eta\approx(q/\varepsilon)\nh^i$ to zeroth order.

The Boltzmann equation then becomes in Fourier space
\bdm
  f_0\frac{\partial f_1}{\partial\eta}-ik\frac{q}{\varepsilon}\kh_i\nh^if_0f_1
  -\frac{1}{2}q\nh_i\nh_j\dot{h}^{ij}\frac{\partial f_0}{\partial q}=C[f]
\edm
to the first order. Dividing through by the background distribution function and setting $m_0=0$ we finally have, defining the angle cosine $\mu=\khv.\nhv$,
\be
  \frac{\partial f_1}{\partial\eta}-ik\mu f_1-\frac{1}{2}q\nh_i\nh_j\dot{h}^{ij}\frac{\partial\ln f_0}{\partial\ln q}=\frac{C[f]}{f_0} .
\ee

\subsubsection{Brightness Functions}
The ``brightness'' function of a Boltzmann fluid can be loosely seen as an analogue of the fractional density perturbation and is defined by
\be
  F\left(k^i,\hat{n}_i,\eta\right)=\frac{\int q^2dqf_0f_1q}{\int q^2dqf_0q}
\ee
One can then see that the exact analogue of the fractional density perturbation,
\bdm
  \delta=\frac{\delta\rho}{\rho}=\frac{\int q^2dqd\Omega qf_1f_0}{\int q^2 dq d\Omega qf_0}=\frac{1}{4\pi}\int d\Omega F
\edm
since $f_0$ has no angular dependence. One can then see the brightness function as the density contrast within some small solid angle.

It proves useful to expand the brightness function over the Legendre polynomials with
\be
\label{ExpBF}
  F\left(\mathbf{k},\hat{\mathbf{n}},\eta\right)=\sum_{l=0}^\infty(-i)^l
  \left(2l+1\right)F_l \left(\mathbf{k},\eta\right)P_l\left(\mu\right) .
\ee
The usefulness of this expansion is that the moments of the brightness function are then recovered by
\bdm
  F_l(\mathbf{k},\eta)=\frac{(-i)^{-l}}{4\pi}\int d\Omega P_l(\mu)F(\mathbf{k},\hat{\mathbf{n}},\eta),
\edm
and one sees that the zeroth moment is the density and the first moment is related to the velocity. Ignoring the possibility of vector and tensor perturbations for the moment, one can write
\be
  \delta=F_0, \quad v_S=-\frac{3i}{4}F_1 .
\ee
Curtailing the Legendre expansion at the first order will then give a rough approximation to the brightness function which is occasionally useful,
\be
  F\approx \delta+4\mu v_s .
\ee

To express the Boltzmann equation in terms of the brightness function it is easiest to relate the perturbation to the distribution function with the brightness function. Using the relation between a massless particle's density and temperature,
\bdm
  \frac{\delta\rho}{\overline{\rho}}=4\frac{\delta\Theta}{\Theta}
\edm
one may write the distribution function as the background function with a temperature perturbation,
\bdm
  f=2\left(\exp\left(\frac{q}{\Theta_0(1+\frac{1}{4}F)}\right)-1\right)^{-1} .
\edm
Since $F$ is a fully arbitrary function this retains the full generality of the distribution function. Expanding the exponentials using the binomial expansions one may now see that
\bdm
  f_1=F\left(\frac{q}{4\Theta_0}\frac{\exp(q/\Theta_0)}{\exp(q/\Theta_0)-1}\right) .
\edm
Using the form of $f_0$ we may now see that this is identical to
\be
  f_1=-\frac{1}{4}\frac{\partial(\ln f_0)}{\partial(\ln q)}F .
\ee
The Boltzmann equation, expressed in terms of the brightness function, then becomes
\be
  \frac{\partial F}{\partial\eta}-ik\mu F+2\nh_i\nh_j\dot{h}^{ij}=C\left[F\right] ,
\ee
where
\bdm
  C\left[F\right]=-\frac{4}{f_0}\left(\frac{\partial\left(\ln f_0\right)}{\partial\left(\ln q\right)}\right)^{-1}C[f].
\edm

Written in terms of the brightness function, the stress energy tensor is
\bea
  \frac{\delta T^0_0}{\bkr}&=&\frac{\delta T^i_i}{\bkr}=-\frac{1}{4\pi}\int d\Omega_{\mathbf{n}}F=-\delta, \nonumber \\
\label{BoltzmannStressEnergy1}
  \frac{\delta T^i_0}{\bkr}&=&-\frac{1}{4\pi}\int d\Omega_{\mathbf{n}}\hat{n}^iF=-\frac{4}{3}v^i, \\
  \frac{\delta T^i_j}{\bkr}&=&\frac{1}{4\pi}\int d\Omega_{\mathbf{n}}\hat{n}^i\hat{n}_jF=\Pi^i_j \nonumber 
\eea
with the solid angle $d\Omega_{\mathbf{n}}=\sin\theta d\theta d\phi$ and $\theta$ and $\phi$ are defined in a co-ordinate system with $x^3\parallel\mathbf{n}$.  We may separate the energy flux and momentum flux density into scalar, vector and tensor parts as
\bea
  \frac{\delta T^{(i)S}_{(0)}}{\bkr}&=&\frac{i}{4\pi}\int d\Omega_{\mathbf{n}}\mu F,  \nonumber \\
  \frac{\delta T^{iV}_{(0)}}{\bkr}&=&-\frac{1}{4\pi}\int d\Omega_{\mathbf{n}}\left(\hat{n}^i-\mu\hat{k}^i\right)F,  \nonumber \\
\label{BoltzmannStressEnergy2}
  \frac{\delta T_S}{\bkr}&=&\frac{1}{4\pi}\int d\Omega_{\mathbf{n}}\frac{F}{2}\left(3\mu^2-1\right), \\
  \frac{\delta T^i_V}{\bkr}&=&\frac{1}{4\pi}\int d\Omega_{\mathbf{n}}\mu\left(\hat{n}^i-\mu\hat{k}^i\right)F,  \nonumber \\
  \frac{\delta T^{iT}_j}{\bkr}&=&\frac{1}{4\pi}\int d\Omega_{\mathbf{n}}\left(\hat{n}^i\hat{n}_j+\frac{1}{2}\left(\mu^2-1\right)\delta^i_j
    +\frac{1}{2}\hat{k}^i\hat{k}_j\left(\mu^2+1\right)-\mu\left(\hat{k}^i\hat{n}_j+\hat{k}_j\hat{n}^i\right)\right)F . \nonumber 
\eea
While we could expand the brightness function across the Legendre polynomials and so reduce these to forms dependant on the moments of the brightness function we shall refrain from doing so until after we have considered the vector and tensor modes in further detail; the bases we are lead to employ to simplify the photon Boltzmann equation complicate the Legendre expansion.

While physically lucid, this approach is not sufficient for photons, since it does not take into account polarisation states and we now turn to a more detailed consideration of a collection of photons.
It is customary to notate the photon brightness function by $\Delta$. We shall retain $F$ for the neutrinos.

\subsection{The Photons}
\label{SecPhotons}
There are two main approaches to dealing with photons in cosmology, the more traditional approach in which the Boltzmann equation is separated across the Legendre polynomials to generate a hierarchy of coupled differential equations, and the more recent ``line-of-sight'' approach in which one expresses the brightness function as an integration over the past light-cone of the photons. While one can employ the traditional approach without the line-of-sight approach, to do so numerically is ruinously slow. The line-of-sight approach is not entirely independent since the sources are formulated in terms of the Legendre moments of the brightness function; however, one would only need to integrate a limited number of the coupled equations to generate these sources and then employ the line-of-sight approach for all higher photon moments.

A classical beam of light is fully specified by the four Stokes parameters, here labelled $I$, $Q$, $U$ and $V$ (see for example Jackson \cite{Jackson} or Chandrasekhar \cite{Chandrasekhar60}). The electric fields ($\mathbf{e}$) that make up the beam are separated along two axes, labelled $L$ and $R$. $I$ is the total intensity of the beam, and is the sum of the intensities along each axis. $Q$ is a measure of the linear polarisation and is the difference between the intensities along the two axes. $U$ and $V$ are measures of the phase differences along the two directions; $U$ is the cosine of the phase difference, while $V$ is its sine.

For an electric field instantaneously resolved along $L$ and $R$ as $\mathbf{e}=e_L\sin(\phi-\varepsilon_L)\mathbf{L}+e_R\cos(\phi-\varepsilon_R)
\mathbf{R}$,
\be \begin{array}{rl}
  I=\left| e_L\right|^2 + \left| e_R\right|^2, & Q=\left| e_L\right|^2
  - \left| e_R\right|^2, \\ U=2e_Le_R\cos(\varepsilon_L-\varepsilon_R),
  & V=2e_Le_R\sin(\varepsilon_L-\varepsilon_R). \end{array}
\ee
We can obviously also define the Stokes parameters as the set $\{I_l, I_r, U, V\}$. It can then be demonstrated (see, for example, \cite{Chandrasekhar60}) that the Boltzmann equation of transfer for a classical ray of light within a classically scattering atmosphere (in a cosmological context, \emph{i.e.} employing the brightness function $\Delta$ rather than the intensity $I$ and including the term $\nh^i\nh^j\dot{h}_{ij}$) is
\bea
  \frac{d}{d\eta}\left(\begin{array}{c} {}^L\Delta \\ {}^R\Delta \\ U \\ V\end{array}\right)
  -ik\mu\left(\begin{array}{c} {}^L\Delta \\ {}^R\Delta \\ U \\ V\end{array}\right)
  +\nh^i\nh^j\dot{h}_{ij}\left(\begin{array}{c}1\\1\\0\\0\end{array}\right)
  = -\dot{\tau}\left(
  \left( \begin{array}{c}{}^L\Delta \\ {}^R\Delta \\ U \\ V
   \end{array} \right) \right. \qquad \qquad \qquad \\
  \left.  \qquad \qquad \qquad
  - \frac{1}{4\pi}\int\mathbf{P}(\mu,\phi;\overline{\mu}, \overline{\phi})
  \left( \begin{array}{c}{}^L\Delta(\overline{\mu},\overline{\phi})\\
   {}^R\Delta(\overline{\mu},\overline{\phi})\\U(\overline{\mu},\overline{\phi})\\
   V(\overline{\mu},\overline{\phi})\end{array} \right)
  d\overline{\Omega}-2\hat{\mathbf{n}}\cdot\mathbf{v}\left( \begin{array}{c}1\\1\\0\\0\end{array} \right)
 \right) , \nonumber
\eea
where $d\overline{\Omega}=d\overline{\mu}d\overline{\phi}$ and $\dot{\tau}$ is the differential scattering cross-section. The scattering matrix $\mathbf{P}$ separates naturally into parts transforming as scalars, vectors and tensors (identified by their spin-0, spin-1 and spin-2 dependence on the azimuthal angle $\phi$):
\beas
  \Sc{\mathbf{P}}=\frac{3}{4}\left(\begin{array}{cccc}\mu^2\overline{\mu}^2+2\left(
   1-\mu^2\right)\left(1-\overline{\mu}^2\right) & \mu^2 & 0 & 0 \\ \overline{\mu}^2
   & 1 & 0 & 0 \\ 0&0&0&0 \\ 0&0&0&2\mu\overline{\mu} \end{array}\right) , \\
  \V{\mathbf{P}}=\frac{3}{4}\sqrt{1-\mu^2}\sqrt{1-\overline{\mu}^2}
   \left(\begin{array}{cccc}4\mu\overline{\mu}\cos\left(
   \overline{\phi}-\phi\right) & 0 & 2\mu\sin\left(\overline{\phi}-\phi\right) & 0 \\
   0&0&0&0 \\ -4\overline{\mu}\sin\left(\overline{\phi}-\phi\right)&0&2\cos\left(
   \overline{\phi}-\phi\right) & 0 \\ 0&0&0&2\cos\left(\overline{\phi}-\phi\right)
   \end{array}\right), \\
\eeas
\beas
  \T{\mathbf{P}}=\frac{3}{4}\left(\begin{array}{cccc}\mu^2\overline{\mu}^2\cos
   2\left(\overline{\phi}-\phi\right) & -\mu^2\cos2\left(\overline{\phi}-\phi\right)
   & \mu^2\overline{\mu}\sin2\left(\overline{\phi}-\phi\right) & 0 \\
   -\overline{\mu}^2\cos2\left(\overline{\phi}-\phi\right) &
   \cos2\left(\overline{\phi}-\phi\right) &
   -\overline{\mu}\sin2\left(\overline{\phi}-\phi\right) & 0 \\
   -2\mu\overline{\mu}^2\sin2\left(\overline{\phi}-\phi\right) &
   2\mu\sin2\left(\overline{\phi}-\phi\right)&2\mu\overline{\mu}\cos
   2\left(\overline{\phi}-\phi\right) & 0 \\ 0&0&0&0 \end{array}\right) .
\eeas
We have defined the Stokes parameters for the photons as
\be
  {}^{L,R}\mathbf{\Delta}=\left({}^L\Delta, {}^R\Delta, U, V\right)
\ee
and we have expanded $\nh$ across the usual polar co-ordinates,
\bdm
  \mathbf{\nh}=\sin\theta\cos\phi\hat{x}^1+\sin\theta\sin\phi\hat{x}^2
  +\cos\theta\kh
\edm
with $\mu=\cos\theta$ -- that is, we have defined a basis $\{\hat{x}^1, \hat{x}^2, \kh\}$.

Our approach will be closely related to that of Crittenden in his PhD thesis and work with Coulson and Turok \cite{CrittendenThesis,CrittendenCoulsonTurok94}; another useful reference (taking a slightly different approach) is that of Landriau and Shellard \cite{LandriauShellard03}. However, the details presented beneath are, except where noted, our own and in particular we employ different bases for the vector and tensor components to either of the above groups. When we consider the line-of-sight approach the vector components are, to our knowledge, entirely original to this thesis; while Lewis \cite{Lewis04-Vector} incorporated vector perturbations into the CAMB code he employs a different formalism and does not express the transfer functions as a line-of-sight integration.

The dominant scattering source will be Thomson scattering, with the differential cross-section
\be
  \dot{\tau}=an_e\sigma_T
\ee
where $n_e$ is the density of free electrons and $\sigma_T$ the Thompson cross-section \cite{Jackson,Chandrasekhar60}. Thompson scattering does not excite the $V$ Stokes parameter, and we shall henceforth neglect it as it can be set to zero without loss of generality.

To proceed, we shall also separate $\mathbf{\Delta}$, the metric contribution $\nh^i\nh^j\dot{h}_{ij}$ and the velocity term $\nh^iv_i$ into their respective components. We shall also wish to convert ${}^{L,R}\mathbf{\Delta}$ to $\mathbf{\Delta}=\{\Delta_T, Q=\Delta_P, U, V\}$ by means of the transformation
\be
  \mathbf{\Delta}=\left(\begin{array}{c}\Delta_T\\ \Delta_P\\U\end{array}\right)
  =\left(\begin{array}{ccc}1&1&0\\1&-1&0\\0&0&1\end{array}\right)
  \left(\begin{array}{c}{}^L\Delta\\{}^R\Delta\\U\end{array}\right)
  =\mathbf{A}\cdot{}^{L,R}\mathbf{\Delta}
\ee
which has the inverse transformation
\bdm
  {}^{L,R}\mathbf{\Delta}=\mathbf{A}^{-1}\cdot\mathbf{\Delta}, \quad
  \mathbf{A}^{-1}=\frac{1}{2}\left(\begin{array}{ccc}1&1&0\\1&-1&0\\0&0&2
  \end{array}\right).
\edm

\subsubsection{Scalar Perturbations}
The metric contribution from scalar perturbations is readily shown to be
\bdm
  \nh^i\nh^j\Sc{\dot{h}}_{ij}=\frac{1}{3}\dot{h}+\left(\mu^2-\frac{1}{3}\right)
  \dot{h}_S=\frac{1}{3}P_0(\mu)\dot{h}+\frac{2}{3}P_2(\mu)\dot{h}_S,
\edm
while the scalar velocity term is
\bdm
  \nh^i\Sc{v}_i=i\mu v_S=iP_1(\mu)v_S .
\edm

The Boltzmann equation for scalar perturbations to the photons (suppressing the $\Sc{}$ on $\mathbf{\Delta}$) is then
\beas
  \frac{d}{d\eta}\left(\begin{array}{c} \Delta_T \\ \Delta_P \\ U \end{array}\right)
  -ik\mu\left(\begin{array}{c} \Delta_T \\ \Delta_P \\ U \end{array}\right)
  +\left(\frac{2}{3}P_0(\mu)\dot{h}+\frac{4}{3}P_2(\mu)\dot{h}_S\right)
   \left(\begin{array}{c}1\\0\\0\end{array}\right)  \qquad \qquad \qquad \\
  = -an_e\sigma_T\left(
  \left( \begin{array}{c}\Delta_T \\ \Delta_P \\ U \end{array} \right)
  -4iP_1(\mu)v_S\left(\begin{array}{c}1\\0\\0\end{array}\right)
  - \frac{1}{4\pi}\int\mathbf{A}\cdot\Sc{\mathbf{P}}\cdot\mathbf{A}^{-1}
  \left( \begin{array}{c}\Delta_T(\overline{\mu},\overline{\phi})\\
   \Delta_P(\overline{\mu},\overline{\phi})\\U(\overline{\mu},\overline{\phi})\end{array} \right)
  d\overline{\Omega}
 \right) .
\eeas
A small amount of matrix algebra then demonstrates that the integrand is
\beas
  \mathbf{A}\cdot\Sc{\mathbf{P}}\cdot\mathbf{A}^{-1}.\mathbf{\Delta} = \frac{3}{8}\left(\begin{array}{c}
  \left[\left(\overline{\mu}^2+1\right)\left(\mu^2+1\right)
   +2\left(1-\overline{\mu}^2\right)\left(1-\mu^2\right)\right]\Delta_T \\
  \left[\left(\overline{\mu}^2+1\right)\left(\mu^2-1\right)
   +2\left(1-\overline{\mu}^2\right)\left(1-\mu^2\right)\right]\Delta_T \\
  0 \end{array} \right) \qquad \qquad \\ \qquad \qquad +\frac{3}{8}\left( \begin{array}{c}
  +\left[\left(\overline{\mu}^2-1\right)\left(\mu^2+1\right)
   +2\left(1-\overline{\mu}^2\right)\left(1-\mu^2\right)\right]\Delta_P \\
  +\left[\left(\overline{\mu}^2-1\right)\left(\mu^2-1\right)
   +2\left(1-\overline{\mu}^2\right)\left(1-\mu^2\right)\right]\Delta_P \\
  0 \end{array} \right) .
\eeas

One then sees that $U$ is unexcited by scalar perturbations, and we shall thus set $\Sc{\mathbf{\Delta}}=\{\Delta_T,\Delta_P\}$. There being no dependence on $\phi$ in the scalar Boltzmann transport equation, we can immediately integrate it out,
leaving us with the equation in an identical form but with a collisional term
\beas
  \Sc{\mathcal{C}}(\mathbf{\Delta})=\frac{3}{16}\left(\begin{array}{c}
  \left(\mu^2+1\right)\int\left[\left(\overline{\mu}^2+1\right)\Delta_T
   +\left(\overline{\mu}^2-1\right)\Delta_P\right]d\overline{\mu} \\
  \left(\mu^2-1\right)\int\left[\left(\overline{\mu}^2+1\right)\Delta_T
   +\left(\overline{\mu}^2-1\right)\Delta_P\right]d\overline{\mu}
   \end{array} \right. \qquad \qquad \\ \qquad \qquad \left. \begin{array}{c}
  +2\left(1-\mu^2\right)\int\left[\left(1-\overline{\mu}^2\right)\Delta_T
   +\left(1-\overline{\mu}^2\right)\Delta_P\right]d\overline{\mu} \\
  +2\left(1-\mu^2\right)\int\left[\left(1-\overline{\mu}^2\right)\Delta_T
   +\left(1-\overline{\mu}^2\right)\Delta_P\right]d\overline{\mu}
  \end{array} \right)d\overline{\mu} .
\eeas
Relating the powers of $\mu$ to the Legendre polynomials,
\bdm
  \mu^2+1=\frac{4}{3}P_0(\mu)+\frac{2}{3}P_2(\mu), \; \mu^2-1
  =\frac{2}{3}\left(P_2(\mu)-P_0(\mu)\right)
\edm
and we may expand $\Delta_{T,P}$ across the Legendre polynomials as we did the basic brightness function (\ref{ExpBF}). The orthogonality of the Legendre polynomials \ref{LegendreOrgogonality} gives the relation
\bdm
  \int P_a(\overline{\mu})\Delta_{T,P}(\overline{\mu})d\overline{\mu}=2(-i)^a\Delta_{T,Pa}.
\edm
Obviously, no summation is implied in the above statement.

Using these definitions, one may readily demonstrate that
\bdm
  \Sc{\mathcal{C}}(\mathbf{\Delta})=\left(\begin{array}{c}
  P_0(\mu)\Delta_{T0}-\frac{1}{2}P_2(\mu)\left(\Delta_{T2}
   +\Delta_{P0}+\Delta_{P2}\right) \\
  \frac{1}{2}P_0(\mu)\left(\Delta_{T2}+\Delta_{P0}+\Delta_{P2}\right)
   -\frac{1}{2}P_2(\mu)\left(\Delta_{T2}+\Delta_{P0}+\Delta_{P2}\right)
  \end{array}\right)
\edm
and so we may write the two transport equations (for the photon energy density and polarisation respectively) as
\beas
  \lefteqn{\dot{\Delta}_T-ik\mu\Delta_T} \\
   &&=-\frac{2}{3}\left(P_0(\mu)\dot{h}+2P_2(\mu)\dot{h}_S\right)
    -an_e\sigma_T \left(\Delta_T-P_0(\mu)\Delta_{T0}-4iP_1(\mu)v_S+\frac{1}{2}P_2(\mu)\Phi_S\right), \\
  \lefteqn{\dot{\Delta}_P-ik\mu\Delta_P
    =-an_e\sigma_T\left(\Delta_P-\frac{1}{2}\left(P_0(\mu)-P_2(\mu)\right)\Phi_S\right)}
\eeas
where
\bdm
  \Phi_S=\Delta_{T2}+\Delta_{P0}+\Delta_{P2} .
\edm

Expanding these equations across the Legendre polynomials and employing the recursion relation for Legendre polynomials (\ref{LegendreRecursion}) yields the equation
\beas
  \lefteqn{\sum_l(-i)^l\left((2l+1)P_l(\mu)\dot{\Delta}_{Tl}-ik\left((l+1)P_{l+1}(\mu)+lP_{l-1}(\mu)\right)\Delta_{Tl}\right)} \\ &&
  =-\frac{2}{3}\left(P_0(\mu)\dot{h}+2P_2(\mu)\dot{h}_S\right)
  -an_e\sigma_T\Big(\sum_l(-i)^l(2l+1)P_l(\mu)
 \\  &&
  \qquad \qquad \qquad \qquad \qquad \qquad -P_0(\mu)\Delta_{T0}-4iP_1(\mu)v_S+\frac{1}{2}P_2(\mu)\Phi_S\Big).
\eeas
Applying the integral operator $\int_{-1}^{1}d\mu P_a(\mu)$ and employing the orthogonality of the Legendre polynomials (\ref{LegendreOrgogonality}) leads rapidly to the general transfer equation for the moments of the intensity,
\beas
  \dot{\Delta}_{Ta}=\frac{k}{2a+1}\left((a+1)\Delta_{Ta+1}-a\Delta_{Ta-1}
   \right)-\frac{2}{3}\dot{h}\gamma_{0a}+\frac{4}{15}\dot{h}_S\gamma_{2a} \\
  -an_e\sigma_T\left(\Delta_{Ta}-\Delta_{T0}\gamma_{0a}+\frac{4}{3}v_S\gamma_{1a}
   +\frac{1}{10}\Phi_S\gamma_{2a} \right).
\eeas
If we undertake a similar process for the polarisation, we get
\beas
  \dot{\Delta}_{Pa}=\frac{k}{2a+1}\left((a+1)\Delta_{Pa+1}-a\Delta_{Pa-1}
   \right)-an_e\sigma_T\left(\Delta_{Pa}-\frac{1}{2}\Phi_S\gamma_{0a}
   -\frac{1}{10}\Phi_S\gamma_{2a}\right).
\eeas

Expanding these equations into hierarchies, we finally reach the equations of motion for scalar perturbations to a photon fluid,
\bea
  \dot{\Delta}_{T0}&=&k\Delta_{T1}-\frac{2}{3}\dot{h}, \nonumber \\
\label{PhotonsScalarHierarchyT}
  \dot{\Delta}_{T1}&=&\frac{1}{3}k\left(2\Delta_{T2}-\Delta_{T0}\right)-an_e\sigma_T
   \left(\Delta_{T1}+\frac{4}{3}v_S\right), \\
  \dot{\Delta}_{T2}&=&\frac{1}{5}k\left(3\Delta_{T3}-2\Delta_{T1}\right)+\frac{4}{15}\dot{h}_S
   -an_e\sigma_T\left(\Delta_{T2}+\frac{1}{10}\Phi_S\right) , \nonumber \\
  \dot{\Delta}_{Tl}&=&\frac{1}{2l+1}k\left((l+1)\Delta_{Tl+1}-l\Delta_{Tl-1}\right)
   -an_e\sigma_T\Delta_{Tl}, \quad l\geq 3 ; \nonumber \\
  \nonumber \\
  \dot{\Delta}_{P0}&=&k\Delta_{P1}-an_e\sigma_T\left(\Delta_{P0}-\frac{1}{2}\Phi_S\right), \nonumber \\
\label{PhotonsScalarHierarchyP}
  \dot{\Delta}_{P1}&=&\frac{1}{3}k\left(2\Delta_{P2}-\Delta_{P1}\right)-an_e\sigma_T\Delta_{P1}, \\
  \dot{\Delta}_{P2}&=&\frac{1}{5}k\left(3\Delta_{P3}-2\Delta_{P1}\right)-an_e\sigma_T\left(
   \Delta_{P2}+\frac{1}{10}\Phi_S\right), \nonumber \\
  \dot{\Delta}_{Pl}&=&\frac{1}{2l+1}k\left((l+1)\Delta_{Pl+1}-l\Delta_{Pl-1}\right)
   -an_e\sigma_T\Delta_{Pl}, \quad l\geq 3 . \nonumber
\eea

\subsubsection{Vector Perturbations}
\label{BoltzmannVectorPert}
The vector perturbations depend on the azimuthal angle $\phi$, and their analysis is thus rendered more complicated than that for the scalars. The metric contribution is
\bdm
 \nh^i\nh^j\V{\dot{h}}_{ij}=2\mu\nh^i\V{\dot{h}_i} .
\edm
Employing the basis $\{\hat{x}^1,\hat{x}^2,\kh\}$ and the expansion of a generic vector quantity as $\mathbf{V}=iV_S\kh+V_\epsilon\hat{x}^\epsilon, \; \epsilon\in\{1,2\}$, one sees that the vector part of the metric perturbation may be written as
\bdm
  \V{h}_{ij}=\kh_i\V{h}_j+\kh_j\V{h}_i=\gamma_{i3}\V{h}_j+\gamma_{j3}\V{h}_i=
  \left(\begin{array}{ccc}0&0&\V{h}_1 \\ 0&0&\V{h}_2 \\
  \V{h}_1 & \V{h}_2 & 0 \end{array} \right) .
\edm
The metric contribution is then
\be
  \nh^i\nh^j\V{\dot{h}}_{ij}=2\mu\sqrt{1-\mu^2}\left(\cos\phi\V{\dot{h}}_1
  +\sin\phi\V{\dot{h}}_2 \right),
\ee
and the vector velocity term in the Boltzmann transport equation is
\be
  \nh^i\V{v}_i=\sqrt{1-\mu^2}\left(\V{v}_1\cos\phi+\V{v}_2\sin\phi\right) .
\ee
The Boltzmann equation for vector perturbations is thus
\beas
  \frac{d}{d\eta}\left(\begin{array}{c} \Delta_T \\ \Delta_P \\ U \end{array}\right)
  -ik\mu\left(\begin{array}{c} \Delta_T \\ \Delta_P \\ U \end{array}\right)
  +4\mu\sqrt{1-\mu^2}\left(\V{\dot{h}}_1\cos\phi+\V{\dot{h}}_2\sin\phi\right)
   \left(\begin{array}{c}1\\0\\0\end{array}\right) \qquad \qquad \\
  = -an_e\sigma_T\left(
  \left( \begin{array}{c}\Delta_T \\ \Delta_P \\ U \end{array} \right)
  -4\sqrt{1-\mu^2}\left(\V{v}_1\cos\phi+\V{v}_2\sin\phi\right)\left(\begin{array}{c}1\\0\\0\end{array}\right)
\right. \qquad \\ \left.
  - \frac{1}{4\pi}\int\mathbf{A}\cdot\V{\mathbf{P}}\cdot\mathbf{A}^{-1}
  \left(\begin{array}{c}\Delta_T(\overline{\mu},\overline{\phi})\\
   \Delta_P(\overline{\mu},\overline{\phi})\\U(\overline{\mu},\overline{\phi})\end{array}\right)
  d\overline{\Omega}
 \right) .
\eeas
The scattering matrix is, after some matrix algebra, seen to be
\bdm
  \mathbf{A}\cdot\V{\mathbf{P}}\cdot\mathbf{A}^{-1}=
  \frac{3}{2}\sqrt{1-\mu^2}\sqrt{1-\overline{\mu}^2}\left(\begin{array}{ccc}
  \mu\overline{\mu}C & \mu\overline{\mu}C & \mu S \\
  \mu\overline{\mu}C & \mu\overline{\mu}C & \mu S \\
  -\overline{\mu}S & -\overline{\mu}S & C \end{array} \right)
\edm
where
\bdm
  \begin{array}{c}
    C=\cos\left(\overline{\phi}-\phi\right)=\cos\overline{\phi}\cos\phi+\sin\overline{\phi}\sin\phi \\
    S=\sin\left(\overline{\phi}-\phi\right)=\sin\overline{\phi}\cos\phi-\cos\overline{\phi}\sin\phi .
  \end{array}
\edm
If
\bdm
  I_C(T,P)=
    \int\overline{\mu}\sqrt{1-\overline{\mu}^2}\Delta_{T,P}\cos\overline{\phi}d\overline{\Omega}, \quad
  I_C(U)=
    \int\sqrt{1-\overline{\mu}^2}U\cos\overline{\phi}d\overline{\Omega}
\edm
with $I_S$ corresponding to the analogous integrations over $\sin\phi$, the collisional term is
\beas
  \V{\mathcal{C}}\left(\mathbf{\Delta}\right)=\frac{3}{8\pi}\sqrt{1-\mu^2}
  \left\{
  \left(\begin{array}{ccc}
   \mu I_C(T)+\mu I_C(P)+\mu I_S(U) \\
   \mu I_C(T)+\mu I_C(P)+\mu I_S(U) \\
   I_C(U)-I_S(T)-I_S(P)
  \end{array}\right)\cos\phi \right.
  \\ \left. +
  \left(\begin{array}{ccc}
   \mu I_S(T)+\mu I_S(P)-\mu I_C(U) \\
   \mu I_S(T)+\mu I_S(P)-\mu I_C(U) \\
   I_S(U)+I_C(T)+I_C(P)
  \end{array}\right)\sin\phi
  \right\}.
\eeas

It proves useful to now transform the variables to
\bdm
  \left(\begin{array}{c}\Delta_T\\ \Delta_P\\U\end{array}\right)=\sqrt{1-\mu^2}
  \left(\begin{array}{c}-i\Delta_T^1\cos\phi\\\mu\Delta_P^1\cos\phi\\
  -U^1\sin\phi\end{array}\right)+\sqrt{1-\mu^2}\left(\begin{array}{c}
  -i\Delta_T^2\sin\phi\\\mu\Delta_P^2\sin\phi\\U^2\cos\phi\end{array}\right)=\mathbf{\Delta}^{(1)}
  +\mathbf{\Delta}^{(2)}.
\edm
An understanding of why this expansion is chosen can be gained by taking the cue from $\nh^i\nh^j\dot{h}_{ij}$ and expanding $\mathbf{\Delta}$ across the basis $\{\cos\phi, \sin\phi\}$, upon which one finds it necessary to impose further redefinitions to reduce the equation to a tractable form. Note that our transformation agrees with Landriau and Shellard \cite{LandriauShellard03} and differs from that of Crittenden and Coulson \cite{CrittendenThesis,CrittendenCoulsonTurok94} by a power of $\mu$ in the temperature redefinition; the difference arises because they did not include the velocity term. The sign convention on $U$ and the presence of $-i$ on $\Delta_T$ produce a neater form for the interaction term $\V{\Phi}_\epsilon$.

With this transformation, the Boltzmann transfer equation separates to
\beas
 \lefteqn{
  \frac{d}{d\eta}\left(\begin{array}{c}\Delta_T\\ \Delta_P\\U\end{array}\right)^\epsilon
  -ik\mu\left(\begin{array}{c}\Delta_T\\ \Delta_P\\U\end{array}\right)^\epsilon+4i\mu\V{\dot{h}}_\epsilon
  \left(\begin{array}{c}1\\0\\0\end{array}\right)
 } \\ &&
  =-an_e\sigma_T\left(\left(\begin{array}{c}\Delta_T\\ \Delta_P\\U\end{array}\right)^\epsilon-4i\V{v}_\epsilon
  \left(\begin{array}{c}1\\0\\0\end{array}\right)
  -\V{\Phi}_\epsilon\left(\begin{array}{c}i\mu\\1\\-1\end{array}\right)\right)
\eeas
where
\be
\label{VectorPhi1}
  \V{\Phi}_\epsilon=\frac{3}{8}\int\left(1-\overline{\mu}^2\right)\left\{\overline{\mu}^2\Delta_P^\epsilon
   -i\overline{\mu}\Delta_T^\epsilon-U^\epsilon\right\}d\overline{\mu} .
\ee

We now briefly define a variable $\alpha^\epsilon=\Delta_P^\epsilon+U^\epsilon$, which has the equation of motion
\bdm
  \dot{\alpha}^\epsilon-ik\mu\alpha^\epsilon=-an_e\sigma_T\alpha^\epsilon
\edm
with the simple solution
\bdm
  \alpha^\epsilon(\eta)=\alpha^\epsilon(\eta_i)\exp\left(ik\mu\left(\eta-\eta_i\right)
   -\sigma_T\int_{\eta_i}^\eta a(\overline{\eta})n_e(\overline{\eta})d\overline{\eta}\right)
\edm
where $\eta_i$ is the initial conformal time. Thus, for a system initially unpolarised with $\alpha^\epsilon(\eta_i)=0$,
\bdm
  \Delta_P^\epsilon=-U^\epsilon
\edm
holds for all times. This then means that we need only evolve two parameters for the vector modes, rather than three, giving us the evolution equation
\beas
\lefteqn{
   \frac{d}{d\eta}\left(\begin{array}{c}\Delta_T\\ \Delta_P\end{array}
   \right)^\epsilon-ik\mu\left(\begin{array}{c}\Delta_T\\ \Delta_P\end{array}\right)^\epsilon
   +4i\mu\V{\dot{h}}_\epsilon\left(\begin{array}{c}1\\0\end{array}\right)
} \\ &&
   =-an_e\sigma_T\left(\left(\begin{array}{c}\Delta_T\\ \Delta_P\end{array}\right)^\epsilon
   -4i\V{v}_\epsilon\left(\begin{array}{c}1\\0\end{array}\right)-\V{\Phi}_\epsilon
   \left(\begin{array}{c}i\mu \\ 1\end{array}\right) \right)
\eeas
The interaction term reduces to
\bdm
  \V{\Phi}_\epsilon=\frac{3}{8}\int\left\{\left(1-\overline{\mu}^4\right)\Delta_{^\epsilon}
  -i\overline{\mu}\left(1-\overline{\mu}^2\right)\Delta_T^{\epsilon}\right\}
\edm
and the physical Stokes parameters are recovered with
\be
  \left(\begin{array}{c}\Delta_T\\ \Delta_P\\U\end{array}\right)
  =\sqrt{1-\mu^2}\left(\begin{array}{c}
   -i\left(\Delta_T^1\cos\phi+\Delta_T^2\sin\phi\right) \\
   \mu\left(\Delta_P^1\cos\phi+\Delta_P^2\sin\phi\right) \\
   \Delta_P^1\sin\phi-\Delta_P^2\cos\phi
  \end{array}\right) .
\ee

We now expand $\Delta_T$ and $\Delta_P$ across the Legendres in the usual way, and in a process exactly analogous to that for scalars find the hierarchies
\bea
  \dot{\Delta}^\epsilon_{T0}&=&k\Delta^\epsilon_{T1}-an_e\sigma_T\left(\Delta^\epsilon_{T0}-
   4i\V{v}_\epsilon\right), \nonumber \\
\label{PhotonsVectorHierarchyT}
  \dot{\Delta}^\epsilon_{T1}&=&\frac{1}{3}k\left(2\Delta_{T2}^\epsilon-\Delta_{T0}^\epsilon\right)
   +\frac{4}{3}\V{\dot{h}}_\epsilon-an_e\sigma_T\left(\Delta_{T1}^\epsilon+\frac{1}{3}\V{\Phi}_\epsilon\right), \\
  \dot{\Delta}^\epsilon_{Tl}&=&\frac{1}{2l+1}k\left((l+1)\Delta_{Tl+1}^\epsilon-l\Delta_{Tl-1}^\epsilon\right)
   -an_e\sigma_T\Delta_{Tl}^\epsilon , \quad l\geq 2 \nonumber \\
  \nonumber \\
\label{PhotonsVectorHierarchyP}
  \dot{\Delta}_{P0}^\epsilon&=&k\Delta_{P1}^\epsilon-an_e\sigma_T\left(\Delta_{P0}^\epsilon-\V{\Phi}_\epsilon\right), \\
  \dot{\Delta}_{Pl}^\epsilon&=&\frac{1}{2l+1}k\left((l+1)\Delta_{Pl+1}^\epsilon-l\Delta_{Pl-1}^\epsilon\right)
   -an_e\sigma_T\Delta_{Pl}^\epsilon, \quad l\geq1 , \nonumber
\eea
with
\be
  \V{\Phi}_\epsilon=\frac{3}{5}\Delta_{P0}^\epsilon+\frac{3}{7}\Delta_{P2}^\epsilon
  -\frac{6}{35}\Delta_{P4}^\epsilon-\frac{3}{10}\Delta_{T1}^\epsilon-\frac{3}{10}\Delta_{T3}^\epsilon .
\ee

In the context of CMB polarisation, unfortunately, the $U$ and $V$ Stokes parameters are not the variables one would prefer to measure due to their heavy rotational dependence. In appendix \ref{Appendix-EandB} we construct rotationally independent analogues of $U$ and $V$. These are the $E$ (or gradient) and $B$ (or curl) modes of the polarisation actually employed in studies of CMB polarisation.

\subsubsection{Tensor Perturbations}
\label{BoltzmannTensorPert}
Again following Landriau and Shellard, we shall expand the tensor perturbations along the symmetric and trace-free basis
\beas
  \mathcal{M}_+=\mathbf{\hat{x}}^1 \otimes \mathbf{\hat{x}}^1
  -\mathbf{\hat{x}}^2 \otimes \mathbf{\hat{x}}^2 & = & \left(
  \begin{array}{ccc}1&0&0\\0&-1&0\\0&0&0\end{array}\right),\\
  \mathcal{M}_\times=\mathbf{\hat{x}}^1 \otimes \mathbf{\hat{x}}^2
  +\mathbf{\hat{x}}^2 \otimes \mathbf{\hat{x}}^1 & = & \left(
  \begin{array}{ccc}0&1&0\\1&0&0\\0&0&0\end{array}\right) .
\eeas
A symmetric, traceless tensor $\Tr{A}_{ij}$ is then expressed as
\bdm
  A_{ij}=A_+ \mathcal{M}_{+ij}+A_\times\mathcal{M}_{\times ij} .
\edm
In particular, one may expand the metric perturbation to $\T{h}_{ij}=h_+\mathcal{M}_{+ij}+h_\times\mathcal{M}_{\times ij}$ and show that the metric contribution to the tensor Boltzmann transport equation is
\bdm
  \nh^i\nh^j\T{\dot{h}}_{ij}=\left(1-\mu^2\right)\left(\T{\dot{h}}_+
  \cos 2\phi + \T{\dot{h}}_\times \sin 2\phi\right) .
\edm

The Boltzmann transport equation then becomes
\beas
  \frac{d}{d\eta}\left(\begin{array}{c} \Delta_T \\ \Delta_P \\ U \end{array}\right)
  -ik\mu\left(\begin{array}{c} \Delta_T \\ \Delta_P \\ U \end{array}\right)
  +2\left(1-\mu^2\right)\left(\T{\dot{h}}_+\cos2\phi+\T{\dot{h}}_\times\sin2\phi\right)
   \left(\begin{array}{c}1\\0\\0\end{array}\right) \qquad \qquad \\
  = -an_e\sigma_T\left(
  \left( \begin{array}{c}\Delta_T \\ \Delta_P \\ U \end{array} \right)
  - \frac{1}{4\pi}\int\mathbf{A}\cdot\T{\mathbf{P}}\cdot\mathbf{A}^{-1}
  \left( \begin{array}{c}\Delta_T(\overline{\mu},\overline{\phi})\\
   \Delta_P(\overline{\mu},\overline{\phi})\\U(\overline{\mu},\overline{\phi})\end{array} \right)
  d\overline{\Omega}
 \right) .
\eeas
The scattering matrix reduces to
\bdm
  \mathbf{A}\cdot\T{\mathbf{P}}\cdot\mathbf{A}^{-1}=
  \frac{3}{8}\left(\begin{array}{ccc} \left(1-\overline{\mu}^2\right)\left(1-\mu^2\right)C
  & \left(\overline{\mu}^2+1\right)\left(\mu^2-1\right)C & 2\overline{\mu}\left(\mu^2-1\right)S \\
  \left(\overline{\mu}^2-1\right)\left(\mu^2+1\right)C & \left(\overline{\mu}^2+1\right)\left(\mu^2+1\right)C
  & 2\overline{\mu}\left(\mu^2+1\right)S \\ 2\left(1-\overline{\mu}^2\right)\mu S &
  -2\left(\overline{\mu}^2+1\right)\mu S & 4\overline{\mu}\mu C \end{array} \right)
\edm
where now we have for clarity defined
\beas
  &C=\cos2\left(\overline{\phi}-\phi\right)=\cos2\overline{\phi}\cos2\phi+\sin2\overline{\phi}\sin2\phi,& \\
  &S=\sin2\left(\overline{\phi}-\phi\right)=\sin2\overline{\phi}\cos2\phi-\cos2\overline{\phi}\sin2\phi.&
\eeas
This implies that, defining
\beas
  &I_C(T)=\int\left(1-\overline{\mu}^2\right)\Delta_T\cos2\overline{\phi}d\overline{\Omega}, \;
  I_C(P)=\int\left(1+\overline{\mu}^2\right)\Delta_P\cos2\overline{\phi}d\overline{\Omega}, &\\
  &I_C(U)=2\int\overline{\mu}\sin2\overline{\phi}d\overline{\Omega}&
\eeas
and analogous expressions for integrals over $\sin2\overline{\phi}$, the collisional term is
\beas
  \T{\mathcal{C}}\left(\mathbf{\Delta}\right)=\frac{3}{32\pi} \left\{
  \left(\begin{array}{c} \left(1-\mu^2\right)I_C(T)+\left(\mu^2-1\right)I_C(P)+\left(\mu^2-1\right)I_S(U) \\
   -\left(1+\mu^2\right)I_C(T)+\left(\mu^2+1\right)I_C(P)+\left(\mu^2+1\right)I_S(U) \\
   2\mu I_S(T)-2\mu I_S(P)+2\mu I_C(U) \end{array}\right)\cos2\phi \qquad \right. \\ \qquad \left.
  +\left(\begin{array}{c} \left(1-\mu^2\right)I_S(T)+\left(\mu^2-1\right)I_S(P)+\left(1-\mu^2\right)I_C(U) \\
   -\left(1+\mu^2\right)I_S(T)+\left(\mu^2+1\right)I_S(P)-\left(\mu^2+1\right)I_C(U) \\
   -2\mu I_C(T)+2\mu I_C(P)+2\mu I_S(U) \end{array}\right)\sin2\phi \right\} .
\eeas

We now change variables to
\bdm
  \left(\begin{array}{c}\Delta_T\\ \Delta_P\\U\end{array}\right)=
  \left(\begin{array}{c}\left(1-\mu^2\right)\Delta_T^+\cos2\phi\\
   \left(1+\mu^2\right)\Delta_P^+\cos2\phi\\ 2\mu U^+\sin2\phi\end{array}\right)
  +\left(\begin{array}{c}\left(1-\mu^2\right)\Delta_T^\times\sin2\phi\\
   \left(1+\mu^2\right)\Delta_P^\times\sin2\phi\\-2\mu U^\times\cos2\phi\end{array}\right)
  =\mathbf{\Delta}^{(+)}+\mathbf{\Delta}^{(\times)}.
\edm
This transformation may be motivated by expanding $\mathbf{\Delta}$ across the basis $\{\cos2\phi, \sin2\phi\}$ and applying a further transformation to render the Boltzmann transport equation as lucid as possible. See for example Polnarev \cite{Polnarev85}, Crittenden \cite{CrittendenThesis} or Landriau and Shellard \cite{LandriauShellard03} for further discussion.

Under this transformation, the Boltzmann equation separates to
\bdm
  \frac{d}{d\eta}\left(\begin{array}{c}\Delta_T\\ \Delta_P\\U\end{array}\right)^\star
  -ik\mu\left(\begin{array}{c}\Delta_T\\ \Delta_P\\U\end{array}\right)^\star+2\T{\dot{h}}_\star
  \left(\begin{array}{c}1\\0\\0\end{array}\right)
  =-an_e\sigma_T\left(\left(\begin{array}{c}\Delta_T\\ \Delta_P\\U\end{array}\right)^\star
  -\T{\Phi}_\star\left(\begin{array}{c}1\\-1\\-1\end{array}\right)\right)
\edm
where
\be
  \T{\Phi}_\star=\frac{3}{32}\int\left\{\left(1-\overline{\mu}^2\right)^2\Delta_T^\star
  -\left(1+\overline{\mu}^2\right)^2\Delta_P^\star-\left(2\overline{\mu}\right)^2U^\star
  \right\}d\overline{\mu}
\ee
and $\star\in\left\{+,\times\right\}$.

Consider first the $+$ mode and select a ``temperature'' vector to act as a basis vector,
\bdm
  \mathbf{a}_+=\left(1-\mu^2\right)\cos2\phi\left(\begin{array}{c}1\\0\\0\end{array}\right) ;
\edm
when transformed to the transformed variables, this is $\mathbf{A}_+=(1,0,0)$. Applying the integral operator
\bdm
  \hat{I}_\phi=\frac{1}{4\pi}\int_0^{2\pi}d\phi\left(2\mathbf{A}^{-1}\cdot\T{\mathbf{P}}\right)
\edm
and expressing the result as a linear combination of the basis vectors necessary for the excited system, we find
\bdm
  \frac{3}{16}\overline{\mu}\left(1-\overline{\mu}^2\right)\left(\begin{array}{c}
  \left(\mu^2-1\right)\cos2\phi \\ \left(\mu^2+1\right)\cos2\phi \\ 2\mu\sin2\phi
  \end{array}\right)=\frac{3}{32}\overline{\mu}^2\left(1-\overline{\mu}^2\right)
  \left(C_a\mathbf{a}_++C_b\mathbf{b}_+\right)
\edm
which (with $C_a=-2, \; C_b=2$) gives us our ``polarisation'' vector
\bdm
  \mathbf{b}_+=\left(\begin{array}{c}0\\ \left(1+\mu^2\right)\cos2\phi \\ 2\mu\sin2\phi
  \end{array}\right);
\edm
its transformed analogue is $\mathbf{B}_+=(0,1,1)$. We may now expand
\bdm
  \mathbf{\Delta}^{(+)}=\alpha^+\mathbf{a}_++\beta^+\mathbf{b}_+
\edm
where $\mathbf{\Delta}^{(+)}$ is the contribution to the brightness function from the $+$ mode.

Performing the same process on the $\times$ mode yields the basis vectors
\bdm
  \mathbf{a}_\times=\left(1-\mu^2\right)\sin2\phi\left(\begin{array}{c}1\\0\\0\end{array}\right), \;
  \mathbf{b}_\times=\left(\begin{array}{c}0\\\left(1+\mu^2\right)\sin2\phi\\-2\mu\cos2\phi\end{array}\right);
\edm
the physical Stokes parameters are thus
\bea
  \Delta_T&=&\left(1-\mu^2\right)\left(\alpha^+\cos2\phi+\alpha^\times\sin2\phi\right), \nonumber \\
  \Delta_P&=&\left(1+\mu^2\right)\left(\beta^+\cos2\phi+\beta^\times\sin2\phi\right), \\
  U&=&2\mu\left(\beta^+\sin2\phi-\beta^\times\cos2\phi\right),\nonumber
\eea
with the equations of motion
\bea
  \dot{\alpha}^\star-ik\mu\alpha^\star+2\T{\dot{h}}_\star&=&-an_e\sigma_T\left(\alpha^\star
    -\T{\Phi}_\star\right), \nonumber \\
  \dot{\beta}^\star-ik\mu\beta^\star&=&-an_e\sigma_T\left(\beta^\star+\T{\Phi}_\star\right) .
\eea
Expanding $\alpha^\star$ and $\beta^\star$ as with the basic brightness function (\ref{ExpBF}) reduces $\T{\Phi}_\star$ to
\be
  \T{\Phi}_\star=\frac{1}{10}\alpha_0^\star-\frac{3}{5}\beta_0^\star
  +\frac{1}{7}\left(\alpha_2^\star+6\beta_2^\star\right)+\frac{3}{70}
  \left(\alpha_4^\star-\beta_4^\star\right) .
\ee
Reducing the equations of motion to hierarchies in exactly the same way as before yields
\bea
  \dot{\alpha}_0^\star&=&k\alpha_1^\star-2\T{\dot{h}}_\star-an_e\sigma_T
   \left(\alpha_0^\star-\T{\Phi}_\star\right) \nonumber \\
\label{PhotonsTensorHierarchyA}
  \dot{\alpha}_l^\star&=&\frac{1}{2l+1}k\left(\left(l+1\right)\alpha_{l+1}^\star-
   l\alpha_{l-1}^\star\right)-an_e\sigma_T\alpha_l^\star \\
  \nonumber \\
  \dot{\beta}_0^\star&=&k\alpha_1^\star-an_e\sigma_T\left(\beta_0^\star+\T{\Phi}_\star\right) \nonumber \\
\label{PhotonsTensorHierarchyB}
  \dot{\beta}_l^\star&=&\frac{1}{2l+1}k\left(\left(l+1\right)\beta_{l+1}^\star-
   l\beta_{l-1}^\star\right)-an_e\sigma_T\beta_l^\star .
\eea

\subsubsection{Stress-Energy Tensor}
\label{PhotonStressEnergyTensor}
We are now finally in a position to expand the photon (and thus the massless neutrino) stress-energy tensor, (\ref{BoltzmannStressEnergy1}, \ref{BoltzmannStressEnergy2}) across the modes of the brightness function.

For the scalars, $\Delta^S_T=\sum_l(-i)^l(2l+1)P_l(\mu)\Delta_{Tl}(\kv,\eta)$, $\partial_\phi\Delta_T=0$ and so $\int d\Omega \rightarrow 2\pi\int d\mu$; thus
\be
  -\frac{\delta T^0_0}{\bkr}=\frac{\delta T^i_i}{\bkr}
  =\frac{1}{2}\int d\mu\Delta^S_TP_0(\mu)=\Delta^S_{T0},
\ee
which is the density perturbation,
\be
  \frac{\delta T^{(i)S}_{(0)}}{\bkr}=-\frac{i}{2}\int d\mu\Sc{\Delta}_TP_1(\mu)=-\Sc{\Delta}_{T1}
\ee
which is to be related to the photon scalar velocity field, and
\be
  \frac{\delta T_S}{\bkr}=\frac{1}{2}\int d\mu\Sc{\Delta}_TP_2(\mu)=-\Sc{\Delta}_{T2}
\ee
which represents the scalar anisotropic stress.

For the vectors we work explicitly within our basis. Considering the first component, we see that $\Delta_T=-i\sqrt{1-\mu^2}\left(\Delta_T^1\cos\phi+\Delta_T^2\sin\phi\right)$ and
\beas
  \frac{\Delta T^{1V}_{(0)}}{\bkr}&=&-\frac{1}{4\pi}\int d\Omega\V{\Delta}_T\sin\theta\cos\phi
  =\frac{i}{6}\int d\mu\left(P_0(\mu)+P_2(\mu)\right)\Delta_T^1 \\
  &=&\frac{i}{3}\left(\Delta^1_{T0}-\Delta_{T2}^1\right) .
\eeas
The second component gives identical forms, so
\be
  \frac{\delta T^{\epsilon V}_{(0)}}{\bkr}=\frac{i}{3}\left(\Delta_{T0}^\epsilon-\Delta_{T2}^\epsilon\right) ,
\ee
which is related to the vector velocity field. We derive the velocity fields shortly.

Similarly,
\beas
  \frac{\delta T_V^1}{\bkr}&=&\frac{1}{4\pi}\int d\Omega\V{\Delta}_T\mu\sin\theta\cos\phi
  =\frac{-i}{10}\int d\mu\left(P_1(\mu)+P_3(\mu)\right)\Delta_T^1 \\
  &=&\frac{1}{5}\left(\Delta_{T3}^1-\Delta_{T1}^1\right)
\eeas
and so
\be
  \frac{\delta T_V^\epsilon}{\bkr}=\frac{1}{5}\left(\Delta_{T3}^\epsilon-\Delta_{T1}^\epsilon\right)
\ee
which is a measure of vorticity induced by the photons.

For the tensors, we must again be explicit in our basis. The tensor stress-energy tensor (\ref{BoltzmannStressEnergy2}) contains the term
\beas
  \lefteqn{\nh_i\nh_j-\frac{1}{3}\gamma_{ij}+\frac{1}{2}\mu^2\left(\kh_i\kh_j+\gamma_{ij}\right)
  +\frac{1}{2}\left(\kh_i\kh_j-\frac{1}{3}\gamma_{ij}\right)-\mu\left(\kh_i\nh_j+\kh_j\nh_i\right)} \\ &&
  \qquad \qquad \qquad \qquad \qquad \qquad \qquad
  =\frac{1}{2}\sin^2\theta\left(\cos2\phi\mathcal{M}_{+ij}+\sin2\phi\mathcal{M}_{\times ij}\right).
\eeas
This enables us to separate out $\delta T_+$ and $\delta T_\times$. Consider the $+$ mode, with $\T{\Delta}_T=(1-\mu^2)(\alpha^+\cos2\phi+\alpha^\times\sin\phi)$:
\beas
  \frac{\delta T_+}{\bkr}&=&\frac{1}{8\pi}\int d\Omega\sin^2\theta\cos2\phi\T{\Delta_T}
  \frac{1}{8}\int d\mu\left(\frac{8}{15}P_0(\mu)-\frac{16}{21}P_2(\mu)+\frac{8}{35}P_4(\mu)\right)\alpha^+ \\
  &=&\frac{2}{15}\alpha^+_0+\frac{4}{21}\alpha^+_2+\frac{2}{35}\alpha^+_4 .
\eeas
This also holds for the $\times$ mode and so
\be
  \frac{\delta T_*}{\bkr}=\frac{2}{15}\alpha_0^*+\frac{4}{21}\alpha_2^*+\frac{2}{35}\alpha_4^*
\ee
are the sources of gravitational radiation from the photons.

\subsubsection{Velocity Fields}
Relating our energy flux to that for a radiative fluid we can say
\bdm
  \delta T^i_0=-\frac{4}{3}\bkr(i\kh^iv_S+v^i_V)
\edm
Then
\bdm
  \frac{4}{3}v_S=\frac{\delta T^{(i)S}_{(0)}}{\bkr}, \quad
  \frac{4}{3}v^V_\epsilon=-\frac{\delta T^{\epsilon V}_{(0)}}{\bkr} ,
\edm
whence we may define the effective velocity fields of the photons to be
\be
\label{PhotonVelocities}
  v_S=\frac{3}{4}\Delta^S_{T1}, \quad
  v^V_{\epsilon}=\frac{i}{4}\left(\Delta^{\epsilon V}_{T2}-\Delta^{\epsilon V}_{T0}\right) .
\ee

We can then rewrite the evolution equations of the photon density perturbation (scalar $l=0$) and velocity (scalar $l=1$ and vector $l=0,2$) as
\bea
  \dot{\delta}_\gamma+\frac{4}{3}v_{S\gamma}+\frac{2}{3}\dot{h}&=&0, \nonumber \\
\label{PhotonDensityVelocityEquations}
  \dot{v}_{S\gamma}+\frac{1}{4}k\left(2\Delta^0_{T2}-\delta_\gamma\right)-an_e\sigma_T\left(v_{Sb}-v_{S\gamma}\right)&=&0, \\
  \dot{v}^V_{\gamma\epsilon}-\frac{ik}{20}\left(3\Delta_{T3}^{\epsilon V}-7\Delta_{T1}^{\epsilon V}\right)
   +an_e\sigma_T\left(v^V_{\gamma\epsilon}-v^V_{b\epsilon}\right)&=&0
\eea
which are useful when we wish to approximate the photons as a fluid.

\subsection{E and B Modes}
The polarisation bases we have thus far employed are far from ideal for study of the sky, due to their heavy rotation dependence. While a detailed study is somewhat tangential to the direct field of this thesis we present an introduction to the $E$- and $B$- modes in appendix \ref{Appendix-EandB}. An intuitive way to consider $E$ and $B$ modes is to think of the $E$ modes as the gradient component of the polarisation and the $B$ mode as the curl component -- that is, the $E$ component makes patterns radial or normal to the radial around a centre, while the $B$ component makes curling patterns set at $45^o$ to the $E$ patterns, reminiscent of a catherine wheel. In a small-angle limit the $E$ and $B$ modes reduce to the $Q$ and $U$ polarisation respectively.

\subsection{The Line-of-Sight Approach}
It is very slow to na\"{\i}vely implement a basic Boltzmann approach by integrating the system of equations detailed above. The introduction by Seljak and Zaldarriaga \cite{SeljakZaldarriaga96} of fast methods employing the line-of-sight integration approach make it stubborn to insist on any alternative. This method, which we review below, reduces the necessary maximum multipole number $l_{\mathrm{max}}$ from the region of $1,500$ to the order of $10$s or even less. The saving in computational time is extremely significant.

We consider the line-of-sight approach to be split into two sections -- first expressing the evolution equations in terms of some convenient and simply evaluable line-of-sight integral, and second in terms of the statistics on the CMB. This second part will be discussed in full in \S\ref{Line-of-Sight-CMB} and we shall merely here present the pertinent results. We define the optical depth between us at $\eta_0$ and an arbitrary time $\eta$ to be
\be
  \tau=\int_{\eta}^{\eta_0}\dot{\tau}(\overline{\eta})d\overline{\eta}
\ee
and the visibility function
\be
  g(\eta)=\dot{\tau}\exp(-\tau) .
\ee
We often use the co-ordinate $x\equiv k(\eta-\eta_0)$ rather than $\eta$. We consider here only the temperature perturbations.

\subsubsection{Scalars}
Consider first the scalar perturbations. Here
\bdm
  \dot{\Delta}_T-\left(ik\mu-\dot{\tau}\right)\Delta_T=-\frac{2}{3}\left(\dot{h}+\left(3\mu^2-1\right)\dot{h}_S\right)+\dot{\tau}\left(\Delta_{T0}+4i\mu v_S-\frac{1}{4}\left(3\mu^2-1\right)\Phi_S\right)
\edm
where
\bdm
  \Phi_S=\Delta_{T2}+\Delta_{P0}+\Delta_{P2}.
\edm
We can formally integrate this between $\eta=0$ and $\eta=\eta_0$ as a standard first-order differential equation to give
\bea
  \lefteqn{\Delta_T(\eta_0)=\int_0^{\eta_0}e^{-ix\mu}\left(-\frac{2}{3}\left(\dot{h}+\left(3\mu^2-1\right)\dot{h}_S\right)e^{-\tau(\eta)}\right.} \nonumber \\ && \left. \qquad \qquad
  +g(\eta)\left(\Delta_{T0}+4i\mu v_S-\frac{1}{4}\left(3\mu^2-1\right)\Phi_S\right)\right)d\eta.
\eea

While this equation is none too edifying, it is simple to integrate it by parts; we can take any term
\be
  \int_0^{\eta_0}i\mu e^{-ik\mu(\eta-\eta_0)}A(\eta)d\eta=\int_0^{\eta_0}\frac{\dot{A}(\eta)}{k}e^{-ik\mu(\eta-\eta_0)}d\eta
\ee
where the boundary term is taken to vanish since we assume it disappears at $\eta=0$ and at the current time is a monopole which we are not interested in. Effectively this means we can replace any term proportional to $i\mu$ by its time derivative divided by the wavenumber. Doing so leads us after a bit of manipulation to
\beas
  \lefteqn{\Delta_T(\eta_0)=\int_0^{\eta_0}d\eta e^{-ix\mu}\left[
   \frac{2}{3}e^{-\tau}\left(\dot{h}_S-\dot{h}+\frac{3}{k^2}\frac{\partial}{\partial\eta}\ddot{h}_S\right) \right.} \\ && \left.
   +g\left(\Delta_{T0}+4\frac{\dot{v}_S}{k}-4\frac{\ddot{h}_S}{k^2}+\frac{1}{4}\Phi_S
  +\frac{3}{4}\frac{\ddot{\Phi}_S}{k^2}\right)
   +\dot{g}\left(4\frac{v_S}{k}-2\frac{\dot{h}_S}{k^2}+\frac{3}{2}\frac{\dot{\Phi}_S}{k^2}\right)
   +\frac{3}{4}\ddot{g}\frac{\Phi_S}{k^2}
   \right] .
\eeas
We can thus express
\be
  \Delta_T(\eta_0)=\int_0^{\eta_0}d\eta e^{-ix\mu}S^{(S)}_T(k,\eta)
\ee
with
\bea
  S^{(S)}_T(k,\eta)&=&\frac{2}{3}e^{-\tau}\left(\dot{h}_S-\dot{h}+\frac{3}{k^2}\frac{\partial}{\partial\eta}\ddot{h}_S\right)
   +g\left(\Delta_{T0}+4\frac{\dot{v}_S}{k}-4\frac{\ddot{h}_S}{k^2}+\frac{1}{4}\Phi_S+\frac{3}{4}\frac{\ddot{\Phi}_S}{k^2}\right)
   \nonumber \\ && \;
   +\dot{g}\left(4\frac{v_S}{k}-2\frac{\dot{h}_S}{k^2}+\frac{3}{2}\frac{\dot{\Phi}_S}{k^2}\right)
   +\frac{3}{4}\ddot{g}\frac{\Phi_S}{k^2}
\eea

In section \S\ref{Line-of-Sight-CMB} on CMB anisotropies, we shall first expand the plane waves across the spherical Bessel functions and then take ensemble averages; in doing so we can ultimately identify
\bea
\label{ScalarLineOfSightCls}
  \Delta^S_{Tl}(k,\eta)=\int_0^{\eta_0}S^{S}_{T}(k,\eta)j_l(x)d\eta .
\eea

We have thus separated the brightness function and polarisation modes into a part $S_T(k,\eta)$, dependant on the hierarchies with $l\leq2$ and encoding the total fluid dynamics of the system between earliest times and the present day, and a geometrical part $j_l(x)$ which doesn't vary with differing cosmological models. We then immediately see that we need only evaluate the Boltzmann hierarchies up to some low $l$ -- about $l=7$, say -- and, so long as we have precomputed the spherical Bessel functions we can integrate equation (\ref{ScalarLineOfSightCls}) to determine the modes up to an arbitrarily high-$l$ with relatively little computation.

\subsubsection{Vector Perturbations}
Here we had that the physical Stokes parameter is
\bdm
  \Delta_T=-i\sqrt{1-\mu^2}\left(\Delta_T^1\cos\phi+\Delta_T^2\sin\phi\right)
\edm
and that this obeys the evolution equation
\bdm
  \dot{\Delta}_T^\epsilon-\left(ik\mu-\dot{\tau}\right)\Delta_T^\epsilon=
    -4i\mu\dot{h}_\epsilon^{(V)}+i\dot{\tau}\left(4v_\epsilon^{(V)}+\mu\Phi_\epsilon^{(V)}\right)
\edm
with
\bdm
  \V{\Phi}_\epsilon=\frac{3}{5}\Delta_{P0}^\epsilon+\frac{3}{7}\Delta_{P2}^\epsilon
  -\frac{6}{35}\Delta_{P4}^\epsilon-\frac{3}{10}\Delta_{T1}^\epsilon-\frac{3}{10}\Delta_{T3}^\epsilon .
\edm
Following the same process as we did for the scalars then quickly yields the source term
\be
  S_T^\epsilon(k,\eta)=-\frac{4}{k}\ddot{h}_\epsilon^{(V)}e^{-\tau}+g\left(4i\V{v}_\epsilon+\frac{1}{k}\V{\dot{\Phi}}_\epsilon\right)
\ee
However, we shall not directly write the line-of-sight integrations since they are not in an ideal basis for this formalism. Instead we rotate the basis such that the trigonometric functions become exponentials; rewriting the trigonometric functions as exponentials gives us
\bdm
  \Delta_T=-\frac{i}{2}\sqrt{1-\mu^2}\left(\left(\Delta_T^1-i\Delta_T^2\right)e^{i\phi}
   +\left(\Delta_T^1+i\Delta_T^2\right)e^{-i\phi}\right) .
\edm
We can thus make the co-ordinate transformation
\be
  2\tilde{\Delta}_T=\Delta_T^1-i\Delta_T^2, 2\breve{\Delta}_T=\Delta_T^1+i\Delta_T^2 .
\ee
The statistics of the vector perturbations are characterised by two variables $\tilde{\zeta}$ and $\breve{\zeta}$ with the properties
\be
  \langle\tilde{\zeta}(\mathbf{k})\tilde{\zeta}(\mathbf{k}')\rangle=\langle\breve{\zeta}(\mathbf{k})\breve{\zeta}(\mathbf{k}')\rangle=\frac{1}{2}\mathcal{P}_V(k)\delta\left(\mathbf{k}-\mathbf{k}'\right), \quad \langle\tilde{\zeta}(\mathbf{k})\breve{\zeta}(\mathbf{k}')\rangle=0 .
\ee
Assuming that our modes are thus uncorrelated but similar, we can employ the source term generated from, for example, the $1$ mode, and write our line-of-sight integral as
\be
  \Delta_T=-i\sqrt{1-\mu^2}\left(\tilde{\zeta}(\mathbf{k})e^{i\phi}+\breve{\zeta}(\mathbf{k})e^{-i\phi}\right)\int_0^{\eta_0}e^{-ix\mu}\V{S}_T(k,\eta)d\eta
\ee
with
\be
  \V{S}_T(k,\eta)=-\frac{4}{k}\V{\ddot{h}}_1e^{-\tau}+g(\eta)\left(4i\V{v}_1+\frac{1}{k}\V{\dot{\Phi}}_1\right) .
\ee

The line-of-sight moments are a bit protracted to calculate; again we only present the result here. We find that
\be
  \Delta^V_{Tl}=\sqrt{\frac{(l+1)!}{(l-1)!}}\int_\eta\V{S}_T(k,\eta)\frac{j_l(x)}{x}d\eta .
\ee

\subsubsection{Tensor Perturbations}
For the tensors,
\bdm
  \Delta_T=\left(1-\mu^2\right)\left(\alpha^+\cos2\phi+\alpha^\times\sin2\phi\right),
\edm
with the evolution equation
\bdm
  \dot{\alpha}^*-\left(ik\mu-\dot{\tau}\right)\alpha^*=-2\T{\dot{h}}_*+\dot{\tau}\T{\Phi}
\edm
and
\bdm
  \T{\Phi}_*=\frac{1}{10}\alpha_0^*-\frac{3}{5}\beta_0^*+\frac{1}{7}\left(\alpha_2^*+6\beta_2^*\right)+\frac{3}{70}\left(\alpha_4^*-\beta_4^*\right) .
\edm
From here we rapidly find the source term
\be
  S_\alpha^*(k,\eta)=-2\T{\dot{h}}_*e^{-2\tau}+g\T{\Phi}_* .
\ee

As with the vector case, things will run more smoothly if we rotate our basis to convert the trigonometric functions into exponentials; following the process we did with the vectors we see that the new variables
\be
  \alpha^1=\frac{\alpha^+-i\alpha^\times}{2}, \; \alpha^2=\frac{\alpha^++i\alpha^\times}{2}, \quad
\ee
puts our Stokes parameter into the form
\be
  \Delta_T=\left(1-\mu^2\right)\left(\alpha^1e^{2i\phi}+\alpha^2e^{-2i\phi}\right) .
\ee
Following Zaldarriaga and Seljak we also choose to characterise the statistics of the gravity waves with variables $\xi^1$ and $\xi^2$ and assume that our two modes are similar but uncorrelated, which means that we can evolve one set ($+$, for example) and use that for our source term, with the solution
\be
  \Delta_T=\left(1-\mu^2\right)\left(\xi^1(\mathbf{k})e^{2i\phi}+\xi^2(\mathbf{k})e^{-2i\phi}\right)\int_0^{\eta_0}e^{-ix\mu}S^{(T)}_T(k,\eta)d\eta ,
\ee
with
\be
  \langle\xi^1(\mathbf{k})\xi^{1*}(\mathbf{k}')\rangle=\langle\xi^2(\mathbf{k})\xi^{2*}(\mathbf{k}')\rangle=\frac{1}{2}\mathcal{P}_T(k)\delta\left(\mathbf{k}-\mathbf{k}'\right), \quad \langle\xi^1(\mathbf{k})\xi^{2*}(\mathbf{k})\rangle=0
\ee
and
\be
  \T{S}_T(k,\eta)=-2\T{\dot{h}}_+e^{-\tau(\eta)}+g(\eta)\T{\Phi}_+ .
\ee

The tensor moments can then be shown to be
\be
  \Delta^T_{Tl}=\sqrt{\frac{(l+2)!}{(l-2)!}}\int_\eta S^T_T(k,\eta)\frac{j_l(x)}{x^2}d\eta
\ee
which is pleasing in light of the vector and scalar moments; the spin-2 tensors are undefined for the dipole and the quadrupole and the geometrical Bessel term is divided through by $x^2$, while in the spin-1 vector case the moments are undefined for the dipole and the geometrical term is merely divided through by $x$. For the spin-0 scalars, the moments are cleanly defined for all $l>0$ and the geometrical term is divided by $x^0$. These symmetries between the separate types are exploited in far greater depth in the total angular momentum formalism of CMB perturbations developed by Hu and White \cite{HuWhite97}.

\subsection{The Neutrinos}
\label{SecNeutrinos}
Recent neutrino oscillation observations strongly suggest that, in fact, the neutrino mass is small but non-vanishing (a combined mass of $\sum_\nu m_\nu <1eV$ is a recent estimate from cosmology; see for example \cite{Fukugita05} for a cosmological presentation) but considering them to be massless will remain a good approximation. While it is possible to add mass in this formalism (see for example \cite{MaBertschinger95}), we shall not do so.

We start our system long after the neutrinos have decoupled and so assume they do not interact with matter; however, it is worth noting that the magnetic field introduced in chapter \ref{Chapter-MagnetisedCosmology} could well have been sourced \emph{before} neutrino decoupling; if this was the case then we will have additional initial conditions generated by the unmodified anisotropic stresses of the magnetic field. Lewis \cite{Lewis04-Mag} demonstrated that, after neutrino decoupling, the neutrino anisotropic stress generally acts to cancel that of the magnetic field and so the greatest impact of such a primordial magnetic field is through the sourcing of perturbations in the extremely early universe, before neutrino decoupling.

We model each species of neutrino as a radiative fluid in exactly the same way as the intensity of the photons -- thus to recover the formalism for neutrinos, we should take the equations for $\Delta_T$ and set all interaction terms to zero. Some modification is required for the tensor modes; here we do not expand across the basis employed for the photons but rather stop with the change of variables (denoting the brightness function for the neutrinos as $F$)
\bdm
  \T{F}=\left(1-\mu^2\right)F^+\cos2\phi+\left(1-\mu^2\right)F^\times\sin2\phi .
\edm

The resulting Boltzmann hierarchies are
\bea
  \Sc{\dot{F}}_0 & = & k\Sc{F}_1-\frac{2}{3}\dot{h}, \nonumber \\
  \Sc{\dot{F}}_1 & = & \frac{1}{3}k\left(2\Sc{F}_2-\Sc{F}_0\right), \nonumber \\
\label{NeutrinoScalarHierarchy}
  \Sc{\dot{F}}_2 & = & \frac{1}{5}k\left(3\Sc{F}_3-2\Sc{F}_1\right)
   +\frac{4}{15}\dot{h}_S , \\
  \Sc{\dot{F}}_l & = & \frac{k}{2l+1}\left(\left(l+1\right)\Sc{F}_{l+1}
   -l\Sc{F}_{l-1}\right) , \quad l \geq 3 \nonumber ,
\eea
\bea
 \V{\dot{F}}_{\epsilon 0}&=&k\V{F}_{\epsilon 1} , \nonumber \\
\label{NeutrinoVectorHierarchy}
 \V{\dot{F}}_{\epsilon 1}&=&\frac{1}{3}k\left(2\V{F}_{\epsilon 2}-\V{F}_{\epsilon 0}\right)
  -\frac{4}{3}\V{\dot{h}}_\epsilon, \\
 \V{\dot{F}}_{\epsilon l}&=&\frac{1}{2l+1}k\left(\left(l+1\right)\V{F}_{\epsilon l+1}
  -l\V{F}_{\epsilon l-1}\right), \quad l\geq 2 \nonumber ,
\eea
\bea
  \T{\dot{F}}_{\star0}&=&k\T{F}_{\star1}-2\T{\dot{h}}_\star , \nonumber \\
\label{NeutrinoTensorHierarchy}
  \T{\dot{F}}_{\star l}&=&\frac{1}{2l+1}k\left(\left(l+1\right)\T{F}_{\star l+1}
   -l\T{F}_{\star l}\right), \quad l\geq1 .
\eea

The stress-energy tensor is identical in form to that for the photons, separating out as
\bea
  -\frac{\delta T^0_0}{\bkr_\nu}=\frac{\delta T^i_i}{\bkr_\nu}=\Sc{F}_{0}, \quad
  \frac{\delta T^{(i)S}_{(0)}}{\bkr_\nu}=-\Sc{F}_{T1}, \quad
  \frac{\delta T_S}{\bkr_\nu}=-\Sc{F}_{T2}, \nonumber \\
  \frac{\delta T^{\epsilon V}_{(0)}}{\bkr_\nu}=\frac{i}{3}\left(F_{T2}^\epsilon-F_{T0}^\epsilon\right), \quad
  \frac{\delta T^\epsilon_V}{\bkr_\nu}=\frac{1}{5}\left(F_{T3}^\epsilon-F_{T1}^\epsilon\right), \\
  \frac{\delta T_*}{\bkr_\nu}=\frac{2}{15}F_0^*+\frac{4}{21}F_2^*+\frac{2}{35}F_4^* \nonumber
\eea
and it has an equivalent density fluctuation and velocity fields
\be
  \delta_\nu=F^S_0, \quad v_S^\nu=-\frac{3}{4}F_1^S, \quad v^V_\epsilon=\frac{i}{4}\left(F^V_2-F^V_0\right)
\ee

We would require the neutrino hierarchies only to a level equivalent to the photon ones; the neutrinos will merely provide sources in the Einstein equations, requiring moments only as high as the fourth.

\subsection{Truncation}
In the line-of-sight integration approach, we need only integrate up a limited number of modes of the Boltzmann hierarchies to provide the sources for the line-of-sight integrals; these can be cut off at, say, $l\approx7-10$ \cite{SeljakZaldarriaga96}. However, a na\"{\i}ve sudden truncation will lead to dreadful errors propagating through the code from the error induced in the evolution equation for $\Delta_{l+1}$ cascading down to $\Delta_{0}$ in a finite time (and back up to the cut-off again). The resulting errors, without some more sensible truncation approach than the harsh cutoff, will be major. Inspired by the line-of-sight integral approach we derive an improved approximation for $\Delta_{l+1}$ (equivalent to that in Ma and Bertschinger \cite{MaBertschinger95}; it might be emphasised that although the approach is our own this is not an original result.)

Consider first the scalars. From the line-of-sight integration we saw that their oscillatory character is governed for the most part by the spherical Bessel functions (see Zaldarriaga and Seljak \cite{SeljakZaldarriaga96} for a detailed discussion on the oscillations of the two parts of the integral). In the absence of scattering and time-varying metrics, the result is the integral of a spherical Bessel function. Employing now the recursion relation for the spherical Bessel functions (\ref{SphericalBesselRecursion}) we can write the exact solution
\bdm
  \Delta^S_{l+1}=(2l+1)\int_\eta S_T^S(k,\eta)\frac{j_l(x)}{x}d\eta-\Delta_{l-1}(x)
\edm
which we approximate to
\be
  \Delta^S_{l+1}\approx(2l+1)\frac{\Delta^S_l(x)}{x}-\Delta^S_{l-1}(x)
\ee
since we are concerned primarily with the oscillatory behaviour over time, which holds equally well for both temperature and polarisation. Cutting short the hierarchy at $l_\mathrm{max}$, we can employ this to write the evolution equation for $\Delta_{l_\mathrm{max}}$ as
\be
\label{TruncatedHierarchy}
  \dot{\Delta}^S_{l_\mathrm{max}}=k\left(\frac{l_\mathrm{max}+1}{x}\Delta^S_{l_\mathrm{max}}(x)-\Delta^S_{l_\mathrm{max}-1}\right)
   -an_e\sigma_T\Delta^S_{l_\mathrm{max}}(x) .
\ee

While the vector line-of-sight solution does not strictly apply to the vector hierarchies, since they are formulated in different bases, it is apparent that the oscillatory behaviour will still be dominated by the spherical Bessels, here appearing as $j_l(x)/x$. If one works through, it is clear that the above scheme also then holds for the vectors and the final equation in the hierarchy has an identical form to (\ref{TruncatedHierarchy}). The tensors' oscillations will be dominated by $j_l(x)/x^2$ and, again, the same truncation scheme will hold.

The neutrino solutions do not contain the scattering sources but otherwise the solutions will be identical in behaviour to the photons; removing the scattering term in (\ref{TruncatedHierarchy}) will generate a scheme equally as applicable to the neutrinos as to the photons.

\section{Baryonic Matter}
\label{SecBaryons}
We turn now to the (unmagnetised) baryon fluid; we shall assume that, as with the cold dark matter, the baryonic matter can be represented by a perfect, pressureless fluid. However, we must include a coupling term that transfers momentum between the baryons and the photons -- recalling that there was no impact from the photons onto the baryonic mass continuity (\ref{PhotonsScalarHierarchyT}).

If $C^i_{b\rightarrow\gamma}$ represents the transfer of momentum into the photons, we can express the equations governing the baryons (\ref{GenericFluidSEConsB}, \ref{GenericFluidSEConsF}) as
\bea
  \dot{\bkr}+3\hub\bkr&=&0, \nonumber \\
  \partial^i\overline{p}+\overline{C}^i_{b\rightarrow\gamma}&=&0
\eea
in the background and
\bea
\label{BaryonStandardCosmology}
  \dot{\delta}+\left(\nabla\cdot\fgv+\frac{1}{2}\dot{h}\right)&=&0, \\
  \dot{v}^i+\hub v^i+c_s^2\partial^i\delta+C^i_{b\rightarrow\gamma}&=&0
\eea
where we have set $w=c_s^2=0$ except where they multiply a spatial derivative.

To evaluate the transferral of the momentum, consider the scalar $l=1$ (\ref{PhotonsScalarHierarchyT}) and vector $l=0$ (\ref{PhotonsVectorHierarchyT}) moments of the photon brightness function hierarchy coupling the photons to the baryons. Since the photons have the velocity fields $v_{S\gamma}=-(3/4)\Sc{\Delta}_{T1}$ and $\V{v}_{\gamma\epsilon}=(i/4)(\V{\Delta}_{T2}-\V{\Delta}_{T0})$, we see that (in addition to the vector $l=2$ equation) these are effectively the photonic equivalents of the Euler equation.

For a generic fluid, the contribution to momentum conservation comes from
\bdm
  \frac{1}{(1+w)\bkr}\delta T^{i\nu S}_{\phantom{i\nu};\nu}=\dot{v}^i
  +\left(\hub(1-3w)+\frac{\dot{w}}{1+w}\right)v^i-k\frac{c_s^2}{1+w}\delta
\edm
(c.f. (\ref{GenericFluidSEConsF}).) We can then construct the photon analogue of this for the scalar modes,
\bdm
  \delta T^{(i)\nu S}_{\gamma;\nu}=\frac{4\overline{\rho}_\gamma}{3}\left(
  \dot{v}_{S\gamma}+\frac{1}{4}k\left(2\Delta_{T2}-\Delta_{T0}\right)-an_e\sigma_T
  \left(v_S-v_{S\gamma}\right)\right)
\edm
and compare it with the corresponding baryon contribution,
\bdm
  \delta T^{(i)\nu S}_{b;\nu}=\bkr_b\left(
  \dot{v}_S+\hub v_S-kc_s^{\phantom{s}2}\delta_b+C^S_{b\rightarrow\gamma}\right) .
\edm
Thus we see that to account for the loss of momentum from the photons we need to take
\be
  C^S_{b\rightarrow\gamma}=\frac{4}{3}\frac{\overline{\rho}_\gamma}{\bkr_b}an_e\sigma_T\left(v_S-v_{S\gamma}\right)
\ee
as the term in the baryon Euler equations accounting for the transferral of momentum and from the photons.

Similarly, in the vector case we combine the vector $l=0$ and $l=2$ equations to form the photon equation equivalent to the Euler equation,
\bdm
  \delta T^{\nu V}_{\gamma\epsilon;\nu}=\frac{4\overline{\rho}_\gamma}{3}\left(
  \V{\dot{v}}_{\gamma\epsilon}-\frac{ik}{20}\left(3\Delta_{T3}^\epsilon-7\Delta_{T1}^\epsilon\right)
  +an_e\sigma_T\left(\V{v}_{\gamma\epsilon}-\V{v}_\epsilon\right)\right)
\edm
and compare it with the baryon term
\bdm
  \delta T^{\epsilon\nu V}_{b;\nu}=\bkr_b\left(
  \V{\dot{v}}_\epsilon+\hub\V{v}_\epsilon+{}^{(V)}C_\epsilon^{b\rightarrow\gamma}\right)
\edm
to identify
\be
  {}^{(V)}C_{b\rightarrow\gamma}=\frac{4}{3}\frac{\overline{\rho}_\gamma}{\bkr_b}an_e\sigma_T
  \left(\V{v}_\epsilon-\V{v}_{\gamma\epsilon}\right) .
\ee

Converting these back into the moments of the brightness function,
\bea
  C^S_{b\rightarrow\gamma}&=&\frac{4}{3}\frac{\overline{\rho}_\gamma}{\bkr_b}an_e\sigma_T
   \left(v_S+\frac{3}{4}\Sc{\Delta}_{T1}\right), \nonumber \\
  {}^{(V)}C_\epsilon^{b\rightarrow\gamma}&=&\frac{4}{3}\frac{\overline{\rho}_\gamma}{\bkr_b}an_e\sigma_T
   \left(\V{v}_\epsilon+\frac{i}{4}\Delta^{\epsilon V}_{T0}-\frac{i}{4}\Delta^{\epsilon V}_{T2}\right) .
\eea

The sound speed is evaluated \cite{MaBertschinger95} from
\be
\label{BaryonCs2}
  c_s^2=\frac{\Theta_b}{\mu}\left(1-\frac{1}{2}\frac{d\ln\Theta_b}{d\ln a}\right)
\ee
where $\mu$ is the (approximately constant) mean molecular weight and $\Theta_b$ is the temperature of the baryon fluid
\be
  \frac{1}{2}m\langle v^2\rangle\approx\frac{3}{2}\Theta_b
\ee
where we work in units with the Boltzmann constant $k_B=1$. The temperature evolves by \cite{MaBertschinger95}
\be
  \dot{\Theta}_b+2\hub\Theta_b-\frac{8}{3}\frac{\mu}{m_e}\frac{\bkr_\gamma}{\bkr_b}
  an_e\sigma_\Theta\left(\Theta_\gamma-\Theta_b\right) =0.
\ee
This is derived from the first law of thermodynamics in the same way as for a collisionless fluid (\ref{CollisionlessFluidT}), but with the heating rate \cite{MaBertschinger95,Peebles-PrinciplesPhysicalCosmology,LiddleLyth}
\be
  \frac{dQ}{d\eta}=4\frac{\bkr_\gamma}{\bkr_b}an_e\sigma_T\left(\Theta_\gamma-\Theta_b\right) .
\ee

In Fourier space we then have for the baryons the stress-energy tensor
\be
  \overline{T}^0_0=-\bkr, \quad \overline{T}^i_0=\overline{T}^i_j=\Pi^i_j=0 \\
\ee
in the background and
\bea
  &\delta T^0_0=-\bkr\delta, \quad
  \delta T^{iS}_{(0)}=\bkr v_S, \quad
  \delta T^{iV}_{(0)}=-\bkr v^V_i,& \nonumber \\
  &\delta T^i_i=\delta T_S=\delta T^V_i=\delta T^T_*=0 .&
\eea
The equation of mass continuity is
\be
\label{BaryonMassContinuity}
  \dot{\delta}+kv_S+\frac{1}{2}\dot{h}=0 .
\ee
The scalar component of the Euler equation is
\be
\label{BaryonScalarEuler}
  \dot{v}_S+\hub v_S-kc_s^{\phantom{s}2}\delta+\Sc{c}_{b\rightarrow\gamma}=0
\ee
where $v_S=-i\kh^iv_i$ and $\Sc{c}_{b\rightarrow\gamma}$ is similarly the scalar component of the momentum transfer into the photons in Fourier space. The vector component is
\be
\label{BaryonVectorEuler}
  \V{\dot{v}}_i+\hub\V{v}_i+{}^{(V)}c_i^{b\rightarrow\gamma}=0.
\ee

\subsection{Tight-Coupling}
\label{Sec-TightlyCoupled}
Throughout almost the entire period we are considering, the Thomson scattering term $t_c^{-1}=an_e\sigma_T$ will be extremely large due to the high density of free electrons. It is only for a short period before recombination that the free electron density and thus the scattering term drops. We shall shortly demonstrate that recombination occurs well within the matter-dominated era and so for the entire time that the universe is dominated by radiative species the photons and baryons are tightly coupled together. Even within the era of matter domination a tight-coupling approximation will hold until relatively near recombination.

To build up our tight-coupling approximation, take the interaction time $t_c\ll 1$ and consider the scalar photon and baryon evolution equations separated into zeroth and higher orders:
\begin{itemize}
  \item $\mathcal{O}(t_c^0)$:
  \beas
    \dot{\delta}_b+kv_S+\frac{1}{2}\dot{h}=0; \quad \Delta_{T1}^S=-\frac{4}{3}v_S; \\
    \dot{\Delta}_{T0}^S+\frac{2}{3}\dot{h}=k\Delta_{T1}^S; \quad \Delta_{Tl}^S=-\frac{1}{10}\Phi_S; \quad
    \Delta_{Tl}^S=0; \; l\geq 3 ; \\
    \Delta_{P0}^S=\frac{1}{2}\Phi_S; \quad \Delta^S_{P1}=0; \quad \Delta_{P2}^S=\frac{1}{10}\Phi_S; \quad
    \Delta_{Pl}^S=0; \; l\geq3
  \eeas

  \item $\mathcal{O}(t_c^1)$:
  \beas
    \dot{v}_s+\hub v_S-kc_s^2\delta_b=0; \quad
    \dot{\Delta}_{T1}^S=\frac{1}{3}k\left(2\Delta_{T2}^S-\Delta_{T0}^S\right); \\
    \dot{\Delta}_{T2}^S=\frac{1}{5}k\left(3\Delta_{T3}^S-2\Delta_{T1}^S\right)+\frac{4}{15}\dot{h}_S; \quad
    \dot{\Delta}_{Tl}^S=\frac{1}{2l+1}k\left((l+1)\Delta_{Tl+1}^S-l\Delta_{Tl-1}^S\right), \; l\geq 3 ; \\
    \dot{\Delta}_{P0}^S=k\Delta_{P1}^S; \quad
    \dot{\Delta}_{Pl}^S=\frac{1}{2l+1}k\left((l+1)\Delta_{Pl+1}^S-l\Delta_{Pl-1}^S\right), \; l\geq 1 ; \\
  \eeas
\end{itemize}
Employing the form of $\Phi_S$ then lets us see that at zeroth order in $t_c$ polarisation is driven to zero in the tight-coupled era. So, therefore, during tight-coupling the only significant photon moments are the zeroth and the first, corresponding to the density fluctuation and the velocity respectively, with the anisotropic stresses and polarisation of at least first-order in the small $t_c$.

\subsubsection{The First Moment: Tight-Coupled Euler Equations}
If we consider the first moment of the photon hierarchies, we can write a formal solution for $\Delta_{T1}^S$:
\bdm
  \Delta_{T1}^S=-\frac{4}{3}v_S+t_c\left(\frac{1}{3}k\left(\Delta_{T0}^S-2\Delta_{T2}^S\right)-\dot{\Delta}_{T1}^S\right) .
\edm
Applying this recursively to itself we can express this as
\be
  \Delta_{T1}^S=-\frac{4}{3}v_S+t_c\left(\frac{4}{3}\dot{v}_S-\frac{1}{3}k\Delta_{T0}^S\right)+\mathcal{O}(t_c^2)
\ee
Substituting this in the baryon Euler equation then gives us
\bdm
  \dot{v}_S+\hub v_S-kc_s^2\delta_b
  +\frac{\overline{\rho}_\gamma}{\overline{\rho}_b}\left(\frac{4}{3}\dot{v}_S-\frac{1}{3}k\Delta^S_{T0}\right)+\mathcal{O}(t_c)
  =0 .
\edm

\subsubsection{The Zeroth Moment: Entropy Perturbations}
If we consider now the zeroth moment of the photon hierarchy, we have that
\bdm
  \dot{\Delta}_{T0}^S=-\frac{2}{3}\dot{h}-\frac{4}{3}kv_S+\mathcal{O}(t_c)
\edm
implying with the baryon continuity equation that
\bdm
  \dot{\Delta}_{T0}^S=\dot{\delta}_\gamma=\frac{4}{3}\dot{\delta}_b+\mathcal{O}(t_c) .
\edm
Integrating gives us the condition that
\be
  \frac{3}{4}\delta_\gamma-\delta_b=\delta S_{\gamma b}=\mathrm{const}+\mathcal{O}(t_c)
\ee
where we have unidentified this difference as the entropy perturbation between the photons and the baryons. We may define similar relations between the baryons and the cold dark matter, or between the photons and the neutrinos,
\bdm
  \delta S_{\gamma\nu}=\delta_\gamma-\delta_\nu, \quad \delta S_{bc}=\delta_b-\delta_c .
\edm
We may thus state that if a system is set an initial entropy perturbation then to first order in the tight-coupling parameter it will retain this perturbation until at the least the epoch of recombination. Stated differently, we may separate the initial conditions for the universe into \emph{adiabatic} modes with $\delta S_{\gamma b}=\delta S_{bc}=\delta S_{\gamma\nu}=0$, or into \emph{isocurvature} modes which possess entropy perturbations but can be shown to have vanishing spatial curvature on a given spatial slicing. Adiabatic and isocurvature perturbations provide a basis for cosmological initial conditions. CMB analyses are entirely consistent with adiabatic initial conditions with no isocurvature component (\cite{WMAP-Bennett}), although a certain level of isocurvature perturbation is allowed.

With the relation between the density fluctuations, we may then express the Euler equation for adiabatic perturbations as
\bea
  \left(\overline{\rho}_b+\frac{4}{3}\overline{\rho}_\gamma\right)\dot{v}_S+\hub\overline{\rho}_bv_S
   &=&k\left(c_s^2\overline{\rho}_b\delta_b+\frac{1}{3}\overline{\rho}_\gamma\Delta^S_{T0}\right) \nonumber \\
   &=&k\left(c_{sb}^2\overline{\rho}_b+\frac{4}{3}c_{s\gamma}^2\overline{\rho}_\gamma\right)\delta_b \nonumber \\
   &\approx&\frac{1}{3}k\overline{\rho}_\gamma\delta_\gamma
\eea
labelling the effective speeds of sound as $c_{s\gamma}^2\gg c_{sb}^2$, which identifies the tightly-coupled fluid as a perfect fluid with density $\overline{\rho}_{t_c}=\overline{\rho}_b+(4/3)\overline{\rho}_\gamma$ and pressure $p_{t_c}\approx(1/3)\overline{\rho}_\gamma\delta_\gamma$. An isocurvature perturbation adds an extra effective pressure term.

\section{Scalar Field Matter}
\label{SecScalarFields}
General Relativity can be formulated as a variational theory; the relevant vacuum Lagrangian density is
\be
  \mathcal{L}_{EH}=\sqrt{-g}R .
\ee
Including a matter Lagrangian, the action is then
\be
  S=\int\left(R+\mathcal{L}_{M}\right)\sqrt{-g}d^4x .
\ee
The Einstein equations can then be found from this action by variation with respect to the metric and an assumption about the form of stress-energy tensor for the matter. A Klein-Gordan field of mass $m$, for example, has the Lagrangian density
\be
  \mathcal{L}_{KG}=-\frac{1}{2}\left(\partial^\mu\phi\partial_\mu\phi+m^2\phi^2\right) .
\ee
The form of the stress-energy tensor for matter is somewhat arbitrary; we refer the reader to Weinberg \cite{Weinberg} for details. Here we shall take the stress-energy tensor to be
\be
\label{ScalarFieldStressEnergy}
  T^\mu_\nu=\delta^\mu_\nu\mathcal{L}_M-2g^{\mu\sigma}\frac{\delta\mathcal{L}_M}{\delta g^{\nu\sigma}} .
\ee

If we let the universe contain a dominating massless, spatially uniform scalar field $\phi$ which is minimally coupled to gravity and has some arbitrary associated potential $U(\phi)$ then the relevant action is
\be
  S=\int d^4x\sqrt{-g}\left(R-\frac{1}{2}\partial^\mu\phi\partial_\mu\phi-U(\phi)\right)
\ee
where $\phi$ obeys the massless Klein-Gordan equation
\bdm
  \frac{1}{\sqrt{-g}}\partial_\mu\sqrt{-g}\partial^\mu\phi=\frac{dU}{d\phi} .
\edm

Considering only the background metric and expanding out the Klein-Gordan equation then gives us
\be
  \ddot{\phi}+2\hub\dot{\phi}+a^2\frac{dU}{d\phi}=0 ;
\ee
converting this to co-ordinate time will yield the more familiar
\bdm
  \phi''+3\hub\phi'+\frac{dU}{d\phi}=0 .
\edm
This is merely an harmonic equation with a friction term due to the expansion of space.

The stress-energy tensor (\ref{ScalarFieldStressEnergy}) gives us a vanishing energy flux and the pressure and density
\be
  \rho_\phi=\frac{1}{2a^2}\dot{\phi}^2+U(\phi), \quad
  p_\phi=\frac{1}{2a^2}\dot{\phi}^2-U(\phi) .
\ee
The Friedmann and Raychaudhuri equations (\ref{Friedmann}, \ref{Raychaudhuri}), including a curvature $\mathcal{K}$, are then
\be
  \frac{8\pi G}{3}\left(\frac{1}{2}\dot{\phi}^2+a^2V\right)=\mathcal{H}^2-\frac{\mathcal{K}}{a^2}, 
  \quad 2\frac{\ddot{a}}{a}-\mathcal{H}^2=-8\pi G\left(\frac{1}{2}\dot{\phi}^2-a^2V\right) .
\ee
For a de Sitter or quasi-de Sitter state, then, when $a$ expands exponentially, the curvature term $\mathcal{K}$ is effectively driven to zero regardless of its initial value.

In co-ordinate time, inflation occurs when $w<-1/3$; for a scalar field this condition is
\be
  \dot{\phi}^2<a^2V
\ee
and for extremal inflation, normally dubbed ``slow-roll'',
\be
  \dot{\phi}^2\ll a^2V
\ee
and therefore
\be
  \rho\approx -p,
\ee
that is, a slowly-rolling scalar field will mimic a cosmological constant and the universe will be in a quasi-de Sitter state. In slow-roll inflation we have the Friedmann and Klein-Gordan equations
\be
  \frac{8\pi Ga^2}{3}U=\mathcal{H}^2-\frac{\mathcal{K}}{a^2}, \quad
  \mathcal{H}\dot{\phi}=-\frac{a^2}{2}\frac{dU}{d\phi}
\ee
from which it may be shown that, given the exponential expansion of $a$, $\mathcal{K}$ is effectively driven rapidly to zero.

Finally we define the slow-roll parameters, introduced by Liddle and Lyth; if we define
\be
  \varepsilon_\phi=\frac{8\pi G}{3}\frac{1}{U^2}\left(\frac{dU}{d\phi}\right)^2, \quad
  \eta_\phi=\frac{8\pi G}{3}\frac{1}{U}\frac{d^2U}{d\phi^2}
\ee
then it can be demonstrated that
\be
  \varepsilon_\phi \ll 1, \quad |\eta_\phi |\ll 1
\ee
are necessary, though not sufficient, conditions for inflation.

All the above applies to a universe dominated by a scalar field minimally-coupled to gravity, regardless of whether it is in the early universe or the present universe. Both theories of inflation and theories of quintessence employ scalar fields of this type, with various potentials, as well as employing multiple-field models or models with modified kinetic terms in the action. However, while it is relatively easy to justify the existence of a scalar field in the exceptionally early universe -- high-energy physics predicting many such fields -- it is relatively hard to justify their existence in the late-time universe when the energy is so very much lower, if only because we can directly probe such energies in the laboratory. Moreover, the extremely shallow gradients on the potential must be seen to argue against the simplest quintessence models. As a candidate for an early-universe inflaton a single scalar field model (with a scale-invariant spectral index) is entirely consistent with CMB observations \cite{WMAP-Bennett} although other, more complicated, are not ruled out (the two-field curvaton model \cite{LythWands02} for example); these models are in principle distinguishable by the relative amplitudes of the tensor and scalar modes they produce, or their primordial non-Gaussianity. For the late-time acceleration we observe, scalar field candidates are popular but not unrivalled, with a standard cosmological constant entirely consistent with CMB and large-scale structure observations.

\section{The Evolution of the Universe}
\subsection{Periods in the Universal Evolution}
\label{Sec-PeriodsInEvolution}
\begin{figure}\begin{center}
\includegraphics{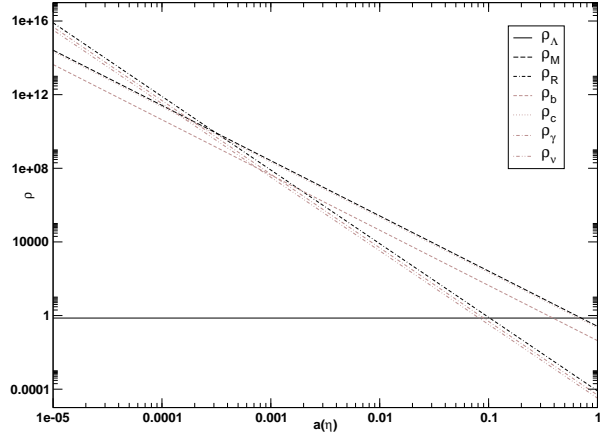}
\caption{Schematic fluid evolution for a $\Lambda$CDM cosmology.}\label{Figure-FluidEvolution}
\end{center}\end{figure}

Because the fluids filling the universe evolve at different rates, there will be different periods in the universal history at which a different form of matter dominates; the situation will be similar to the schematic in figure \ref{Figure-FluidEvolution}. The most rapidly-decaying species, radiation (photons and neutrinos), will have dominated the universe at some early time. Following some brief era of approximate matter-radiation equality, the radiation density will have become subdominant to the dust species (baryons and cold dark matter). As we shall shortly show, the CMB last-scattered some way into matter domination. If the universal expansion is indeed beginning to accelerate \cite{WMAP-Bennett,RiessEtAl04}, then in the simplest model we are sitting at the beginning of the period of vacuum energy domination. This bleak epoch begins when the matter densities decay below the intrinsic energy of the cosmological constant and the universe tends asymptotically towards an eternal de Sitter state. More complex dark energetic models naturally have differing future states. It may be commented that our exact position at the beginning of vacuum energy domination is a ``coincidence problem''; why should the vacuum energy density and matter density have grown roughly equivalent in the very recent past ($z\approx 0.36$)? There are various suggestions ranging from arbitrary tracker potentials for dark energy models that tend to similar universal evolutions independent of initial conditions (e.g. \cite{SteinhardtWangZlatev99}), to equally arbitrary modifications of the gravitational action (see \cite{CarrollEtAl05,NojiriOdinstov06} for recent reviews); other suggestions include modifications of the averaging procedure employed to derive our background densities \cite{Rasanen04} or other back-reactions which derive effective dark energies from non-linear terms in the Friedmann equations \cite{IshibashiWald05,MartineauBrandenberger05}.

\subsection{Vacuum Domination -- Inflation}
The epoch of inflation occurs at some point when the energy of the universe lies between the Planck scale and the electroweak scale (at about $100\mathrm{GeV}$). We do not discuss this era in detail, and merely wish to note that, without a period of vacuum domination, there are large fine-tuning problems in cosmology, along with the ``horizon problem''. The main fine-tuning problem is the universe's flatness -- a curvature term in the Friedmann equation decays as $a^2$ and, to be as flat as it is observed to be today, this implies an extremely finely-tuned initial condition. The ``horizon problem'' is most readily observed from the microwave sky; two points in opposing directions are at the same temperature and yet, obviously, are two horizon distances apart. It seems incredible to believe that the entire surface of last scattering formed at the same time without it having been in causal contact at some point, and yet it cannot have been. Inflation remedies these problems; a small patch of space can have an arbitrary curvature and be well within causal contact and in thermal equilibrium, and then inflated until every perturbation is well outside the Hubble distance. The curvature is driven to zero and the horizon problem is solved. It can be shown that in a standard model around 60 e-foldings are necessary to solve the problems.

Moreover, the perturbations of simple inflationary models predict initial conditions for standard cosmology that are adiabatic, Gaussianly distributed with an almost scale-invariant (Harrison-Zel'dovich) power spectrum. This is entirely consistent with the observations of the CMB. We refer the reader further to, for example, Liddle and Lyth \cite{LiddleLyth} for a relatively recent overview of simple inflation and assume now that some inflationary, or vacuum dominated, epoch has occurred and left us with Gaussian and adiabatic initial conditions.

\subsection{Radiation Domination}
Technically the universe is radiation dominated as far back as the validity of general relativity and the standard model of physics and in the usual history radiation-dominated back to the end of the inflationary epoch; however, the phrase is more usually used to describe the universe after the electroweak phase transition or even after neutrino decoupling at about $1\mathrm{MeV}$. Cosmological codes such as CAMB and CMBFast begin deep in radiation domination but typically after neutrinos have decoupled. We assume the phrase to mean this.

During radiation domination as we have defined it the Friedmann equation is effectively
\be
  \mathcal{H}^2=\frac{8\pi Ga^2}{3}\left(\bkr_\nu+\bkr_\gamma\right) \Rightarrow \Omega_\nu+\Omega_\gamma\approx 1
\ee
As we saw earlier (\ref{RadiationUniverse}) a universe dominated by a radiative fluid evolves as
\bdm
  a=a_0\eta, \quad \mathcal{H}=\frac{1}{\eta}, \quad \frac{\ddot{a}}{a}=0 .
\edm
The universe remains tightly-coupled throughout radiation domination and adiabatic perturbations thus remain adiabatic. Employing the convenient unit $y=k\eta$, the system of tightly-coupled equations that we developed in \S\ref{Sec-TightlyCoupled}, along with the Einstein equation for the evolution of the scalar trace, are
\be
  \frac{d\delta_\gamma}{dy}=-\frac{2}{3}\frac{dh}{dy}-\frac{4}{3}v_S, \quad 4\frac{dv_S}{dy}=\delta_\gamma, \quad
  y^2\frac{d^2h}{dy^2}+y\frac{dh}{dy}+6\left(R_\nu\delta_\gamma+(1-R_\nu)\delta_\nu\right)=0
\ee
where we have defined
\be
  R_\nu=\frac{\bkr_\gamma}{\bkr_\gamma+\bkr_\nu} .
\ee
We can then find the evolution of the baryon and CDM densities by the adiabatic condition. This equation also reduces the source term in the Einstein equation to $-6\delta_\gamma$. Using the Euler equation to eliminate the velocity and the mass continuity equation to eliminate the metric perturbation gives the evolution equation for $\delta_\gamma$,
\be
  y^3\frac{d^3\delta_\gamma}{dy^3}+2y^2\frac{d^2\delta_\gamma}{dy^2}
  +\left(\frac{1}{3}y^2-4\right)y\frac{d\delta_\gamma}{dy}+\left(\frac{2}{3}y^2+4\right)\delta_\gamma=0 .
\ee
We study this equation in two limits, the super-horizon limit when $y=k\eta\ll 1$ and the sub-horizon limit where $y\gg 1$.

\begin{itemize}
\item Super-horizon:
The general evolution equation in the super-horizon limit is
\bdm
  y^3\frac{d^3\delta_\gamma}{dy^3}+2y^2\frac{d^2\delta_\gamma}{dy^2}-4y\frac{d\delta_\gamma}{dy}+4\delta_\gamma=0
\edm
which is an Euler-Cauchy equation for $\delta_\gamma$, easily solved by
\bdm
  \delta_\gamma=\delta_0y^2+\delta_1y+\frac{\delta_2}{y^2} .
\edm
Press and Vishniac \cite{PressVishniac80} demonstrated that the decaying mode is a symptom of the lack of specification in the synchronous gauge and we remove it; we also neglect the slower-growing mode. Then we can see that
\be
  \delta_\gamma=\delta_0y^2
\ee
and so, always requiring that the initial perturbations are small or vanishing,
\be
  h=-\frac{3}{2}\delta_0y^2=-\frac{3}{2}\delta_\gamma, \quad
  v_S=\frac{1}{12}\delta_0y^3
\ee
The velocity is thus sub-dominant on super-horizon scales. We then employ the first Einstein scalar constraint to find
\be
  h_S=h+\frac{3}{y}\frac{dh}{dy}-\frac{9}{y}\delta_\gamma=h^0_S-\frac{3}{2}\delta_0\eta^2 .
\ee

Outside of the horizon, then, the density and metric perturbations variables grow simply as the scale factor, while the velocity grows slightly faster.

\item Sub-horizon:
The equation for $\delta_\gamma$ has the general solution
\bea
  \delta_\gamma&=&\frac{D_0}{y^2}
  +D_1\left(\cos\left(\frac{y}{\sqrt{3}}\right)-2\frac{\sqrt{3}}{y}\sin\left(\frac{y}{\sqrt{3}}\right)
    -\frac{6}{y^2}\cos\left(\frac{y}{\sqrt{3}}\right)\right) \\ && \quad
  +D_2\left(\sin\left(\frac{y}{\sqrt{3}}\right)+2\frac{\sqrt{3}}{y}\cos\left(\frac{y}{\sqrt{3}}\right)
    -\frac{6}{y^2}\sin\left(\frac{y}{\sqrt{3}}\right)\right) .
\eea
For regularity as $y\rightarrow 0$ we remove $D_2$, and $D_0$ is a gauge mode. Taking the sub-horizon limit $y\gg 1$ then gives us the solution
\be
  \delta_\gamma=D_1\cos\left(\frac{y}{\sqrt{3}}\right) .
\ee
We can then rapidly find that
\be
  v_S=\frac{\sqrt{3}}{4}D_1\sin\left(\frac{y}{\sqrt{3}}\right)
\ee
and the metric solutions are
\be
  h\approx h_S\approx -\delta_\gamma .
\ee
The oscillations in the densities and coupled photon-baryon velocity found once modes have entered the horizon produce the characteristic patterns on the CMB. They are merely sound waves -- hence the terminology ``acoustic peak'' for the peaks of the CMB angular power spectrum -- in the tightly-coupled fluid, driven by gravitational attraction and self-pressure. They are all the more important since, as we shall shortly show, oscillations cease once the universe enters matter domination.

\end{itemize}

\subsection{Matter-Radiation Equality}
At matter-radiation equality, we can say
\be
  \bkr_R=\bkr_m\Rightarrow \frac{\bkr_{0R}}{a^4}=\frac{\bkr_{0m}}{a^3}\Rightarrow
  (1+z_{\mathrm{eq}})=a_{\mathrm{eq}}^{-1}=\frac{\bkr_{0m}}{\bkr_{0R}}=\frac{\Omega_{0m}}{\Omega_{0R}}
\ee
where $R$ denotes photons plus three species of massless neutrino, $z$ denotes the redshifting of light, and $m$ denotes CDM and baryons. From the temperature of the CMB it can be shown \cite{LiddleLyth} that the density ratio of radiation, assuming three species of massless neutrinos, is
\bdm
  \Omega_{0R}=4.17\times 10^{-5}h^{-2}
\edm
where $h$ is the observed correction to an assumed Hubble parameter, $H=100h\mathrm{kms}^{-1}\mathrm{Mpc}^{-1}$. With this we see that
\be
  z_{\mathrm{eq}}\approx (2.4\times 10^4)\Omega_{m0}h^2
\ee
and thus, taking the concordance values of
\bdm
  \Omega_{m0}\approx 0.26, \quad h\approx 0.72
\edm
we have
\be
  z_{\mathrm{eq}}\approx 4500 .
\ee

\subsection{Matter Domination}
Once the universe has entered a matter-dominated era the situation is changed. The Friedmann equation is now given by
\be
  \mathcal{H}^2=\frac{8\pi G}{3}a^2\left(\bkr_b+\bkr_c\right) ,
\ee
which has the solutions (\ref{DustUniverse})
\bdm
  a=a_0\eta^2, \quad \mathcal{H}=\frac{2}{\eta}, \quad \frac{\ddot{a}}{a}=\frac{2}{\eta^2} ,
\edm
and we work with the system
\be
  \frac{d\delta_b}{dy}+v_S+\frac{1}{2}\frac{dh}{dy}=0, \quad \frac{dv_S}{dy}+\frac{2}{y}v_S=c_s^2\delta_b, \quad
  \frac{d\delta_c}{dy}+\frac{1}{2}\frac{dh}{dy}=0 .
\ee

\begin{itemize}
\item On the super-horizon scale the adiabatic condition holds between the CDM and the baryons, as well as between the baryons and the photons, because the modes are well outside of the acoustic horizon. We choose to employ the CDM density as a tracer of the universe's evolution. Employing again the scalar trace evolution equation and working in a manner analogous to that for radiation domination, we can quickly find that
\be
  y^2\frac{d^2\delta_c}{dy^2}+2y\frac{d\delta_c}{dy}-6\delta_c=0
\ee
with the solution
\be
  \delta_c=d_0y^2
\ee
where we have again removed a decaying gauge mode. This implies we have the metric trace
\be
  h=H_0-2d_0y^2
\ee
and so the traceless metric perturbation
\be
  h_s=h+\frac{3}{y}\frac{dh}{dy}-\frac{18}{y^2}\delta_c=(H_0-30d_0)-2d_0y^2.
\ee
We can match these solutions to those from the previous section to fix the integration constants if we so desire. The velocity in the super-horizon limit couples to nothing and decays as the scale factor, that is,
\be
  v_S=\frac{v_0}{\eta^2} .
\ee

\item The sub-horizon scale is again more complicated. From (\ref{BaryonCs2}), the baryon speed of sound, in a tight-coupling limit, evolves approximately as
\be
  c_s^2=\frac{c_0^2}{\eta^4} .
\ee

The CDM density, naturally, is unaffected by the system entering the sound horizon and evolves as before,
\be
  \delta_c=d_1\eta^2 .
\ee
However, the baryons oscillate weakly; one may derive that the baryon density evolves as
\be
  y^2\frac{d^2\delta_b}{dy^2}+2y\frac{d\delta_b}{dy}+y^2c_s^2\delta_b=6d_1y^2
\ee
which is a forced damped wave equation producing waves with a frequency of $c_s$.

\end{itemize}

\subsection{Recombination -- Formation of the CMB}
Recombination occurs when the temperature of the universe has dropped low enough that hydrogen atoms can form. Once this occurs, obviously the density of free electrons plummets and the tight-coupling approximation breaks. During this brief period higher moments of the photon hierarchy are excited, and the presence of a quadrupole in turn excites polarised modes. Since recombination is relatively brief and photons free-stream when it is finished (excepting foreground effects such as the Sunyaev-Zel'dovich effect and the reionisation of the universe) polarisation will then give us a unique picture of the fluid dynamics at the formation of the CMB.

It is not our intention here to provide an overview of recombination physics; for such we recommend Peebles \cite{Peebles-LargeScaleStructure,Peebles-PrinciplesPhysicalCosmology} or Seager \emph{et. al.} \cite{RECFast-1,RECFast-2}; the latter programmed the widely-used RECFast code employed by both CMBFast and CAMB. We shall merely calculate the redshift at which recombination occurs; it is readily seen that this is well into the matter dominated era and thus that the largest scale sound waves imprinted upon the CMB were frozen in a long time previously. For the photons, the temperature is related to the density by
\be
  \Theta\propto\rho^4
\ee
which implies that the bulk temperature evolves by
\be
  \Theta=\frac{\Theta_0}{a}=\Theta_0\left(1+z\right)
\ee
by the definition of redshift.

Recombination occurred when the average temperature of the universe was approximately $3000\mathrm{K}$ -- that is, when the average energy was of the order of tenths of electron-volts \cite{Peebles-LargeScaleStructure}. The binding energy of hydrogen is approximately $13\mathrm{eV}$; however, sufficient energy remains in the tail of the photon distribution to keep the universe ionised until the temperature has dropped well below this amount. The CMB today is at a temperature of about $2.7\mathrm{K}$; this then gives a recombination redshift of about
\be
  z_{\mathrm{rec}}\approx 1100 .
\ee
With $z_{\mathrm{eq}}\approx 4500$, it is clear that the CMB was formed well into the epoch of matter domination.

It is likely that the existence of a magnetic field might alter the recombination process, most likely to impede it. However, there has been little study in this area.

\subsection{Late-Time Acceleration}
If the universe should contain a source of effective vacuum energy, such as a cosmological constant or some lingering remnant of a slowly-rolling inflaton, then the matter-dominated era will give way to a second period of vacuum domination. Observations of type Ia supernova, generally regarded as standard candles\footnote{The use of Type Ia supernova as standard candles is debated; possible alternative explanations include dust extinction (see e.g. \cite{SullivanEtAl03} for a study into this possibility), and an evolution of Type Ia supernovae, perhaps with metallicity \cite{Filippenko03}.}, suggests that the universe did indeed begin to enter such an accelerating phase at a redshift of approximately $z\approx 0.36$ \cite{RiessEtAl04}. The nature of this ``dark energy'', and the coincidence that it has appeared at the same broad time as humanity, is not yet understood. Suggestions range from the relatively orthodox, such as ``quintessence'' models which are effectively inflatons operating in the current universe \cite{CaldwellDaveSteinhardt97}, through ``k-essence'' \cite{ChibaOkabeYamaguchi99}, which are scalar fields with modified kinetic terms, ``Chaplygin gases'' with inverse equations of state \cite{KamenshchikEtAl01}, and through to the more extreme, such as modifying our methods of taking the averages employed in the Friedmann equation \cite{Rasanen03}, or direct modifications to Einstein's law of gravity \cite{CarrollEtAl05,NojiriOdinstov06}. Each approach has a certain logic behind it and each has flaws and it is not the purpose of this thesis to explore these. It should be noted, however, that a straight cosmological constant -- or, indeed, some field mimicking a cosmological constant so closely as to be indistinguishable, at which point Occam's razor can probably be invoked -- is an excellent fit for the observed CMB and large-scale structure observations, as well as the supernova observations.

A simple alternative that has reoccurred occasionally in the literature is the existance of a bulk viscosity acting at zeroth order -- see \cite{MaartensMendez96,MaartensTriginer97,MaartensTriginer98,RenMeng05,HuMeng05,Giovannini05-Viscous} for a few recent studies. If we assume that the universe is dominated by a fluid with a bulk viscosity and express the effective pressure (\ref{FullBackgroundPressure}) using real rather than conformal time, we have that
\be
  p_{\mathrm{eff}}=p-3\zeta H .
\ee
Then we have the condition for acceleration
\bdm
  w_{\mathrm{eff}}=\frac{p_{\mathrm{eff}}}{\rho}<-\frac{1}{3}\Rightarrow \zeta>\frac{\rho}{9H}(1+3w) .
\edm
Using the Friedmann equation, and working briefly in SI units, we can then express this condition as
\bdm
  \zeta>\Omega_mH\frac{(1+3w)}{24\pi G}c^2
\edm
for any particular epoch where $\Omega_m=(8\pi G/3H^2)\rho_m$ is the relative density for the viscous component. If we substitute the observed values for our current epoch, assuming $w=0$ and $\Omega_m\approx0.3$, we find a numerical bound of
\be
  \zeta\gtrsim 6\Omega_mh\times 10^7\mathrm{kgm}^{-1}\mathrm{s}^{-1}\approx 1.3\times10^7\mathrm{kgm}^{-1}\mathrm{s}^{-1}
\ee
This is highly viscous and a fluid description for such a material is perhaps not warranted, although glass does possess a higher bulk viscosity. However, one may see directly from the expression for the effective pressure (with $w=0$) that this fluid is unphysical:
\be
  p_{\mathrm{eff}}\approx-3\zeta H<-\frac{1}{3}\rho
\ee
shows that $\left|\zeta H\right|$ is of the order of $\rho$. However, the thermodynamics is valid \cite{MaartensMendez96} only for
\be
  \left|\zeta H\right|\ll\rho .
\ee
Employing a bulk viscosity to entirely drive the current expansion of the universe, while neat, is not valid on thermodynamical grounds. This, of course, does not rule out the existence of a bulk viscosity contributing to the observed acceleration without being the major cause.

\section{The Cosmic Microwave Background}
\subsection{The Sachs-Wolfe Equation}
\label{Sec-SachsWolfe}
While a full treatment of photons is complex, insight into the structure of the last-scattering surface can be found by formally integrating the Boltzmann equation for a species of collisionless photons between the surface of last scattering and the present day. The ``surface of last scattering'' formed when the collisional term in the Boltzmann equation became negligible and the photons began to free-stream. From the collisionless Boltzmann equation we have
\be
  \frac{\partial\Delta_{\mathbf{k}}}{\partial\eta}-ik\mu\Delta_{\mathbf{k}}=-2\nh_i\nh_j\dot{h}^{ij} .
\ee
Formally integrating this between the surface of last scattering at a conformal time $\eta_{\mathrm{rec}}$ and the present day at $\eta_0$ we have
\bdm
  \Delta_{\mathbf{k}}\left(\eta_0\right)=\Delta_{\mathbf{k}}(\eta_{\mathrm{rec}})e^{ik\mu\left(\eta_0-\eta_{\mathrm{rec}}\right)}
  -2\int^{\eta_0}_{\eta_{\mathrm{rec}}}\dot{h}^{ij}\nh_i\nh_je^{ik\mu\left(\eta_0-\eta\right)}d\eta .
\edm
This equation assumes an infinitesimal thickness of the surface of last scattering.

Using again $\rho\propto\Theta^4\Rightarrow\Delta=4\delta\Theta/\Theta$ and truncating the Legendre expansion at $l=1$, one recovers the Sachs-Wolfe equation,
\be
  \frac{\delta\Theta}{\Theta}(\eta_0,\mathbf{k})=\left(\frac{1}{4}\delta_\gamma+\mu v\right|_{\eta=\eta_{\mathrm{rec}}}e^{ik\mu\left(\eta_0-\eta_{\mathrm{rec}}\right)}
  -2\int^{\eta_0}_{\eta_{\mathrm{rec}}}\dot{h}^{ij}\nh_i\nh_je^{ik\mu\left(\eta_0-\eta\right)}d\eta .
\ee

The separate contributions to this term are easily interpreted; the photon density perturbation contributes the ``intrinsic'' temperature perturbation on the surface of last scattering, while the $\mu v$ term is a Doppler-type effect caused by the red- or blue-shifting of photons scattered from the surface. The integral term is a gravitational contribution, more readily intuitive in the conformal Newtonian gauge in a universe filled with perfect fluids, where one sees that it reduces merely to the time-derivative of the gravitational potential. A photon passing through a time-dependant potential well will obviously emerge with a different energy; had the well deepened the light will be redshifted and had the well shallowed the light would be blueshifted. In a dust-dominated universe the gravitational potential is constant and this term can be neglected. However, the same is not true in a universe dominated by a dark energy and so the integrated term, known as the integrated Sachs-Wolfe effect, is an important test of dark energy theories. It is in the Doppler term that we expect the majority of the impacts of a magnetic field on scalar and vector modes to be felt, through velocity perturbations induced by the magnetic field. Tensor modes might be expected to be dominated by the integrated Sachs-Wolfe term.

Obviously, in reality we do not assume an infinitesimal surface of last scattering. For practical purposes, CMB codes such as CMBFast, CMBEasy and CAMB \cite{CAMB,SeljakZaldarriaga96, CMBEasy} tend to employ the RECFast recombination code \cite{RECFast-1,RECFast-2}. We now consider statistical measures of the CMB in rather more detail.

\subsection{Statistical Measures on the Sky}
\label{Sec-CMB}
From $\rho_\gamma\propto\Theta^4$, where $\Theta$ is here the ambient photon temperature,
\bdm
  \frac{\delta\Theta}{\Theta}(\mathbf{x},\eta,\nhv)=\frac{1}{4}\int\Delta(\mathbf{k},\eta,\nhv)
   e^{-i\mathbf{k}.\mathbf{x}}\frac{d^3\mathbf{k}}{\left(2\pi\right)^3}.
\edm
Since we observe these projected onto our celestial sphere, it makes sense to expand the temperature perturbation across the spherical harmonics as
\bdm
  \frac{\delta\Theta}{\Theta}(\mathbf{x},\eta,\nhv)=\sum_{l,m}a_{lm}Y_{lm}(\nhv)
\edm
which, with the orthogonality of the spherical harmonics (\ref{SphericalHarmonicsOrthogonality}), gives the harmonic coefficients $a_{lm}$ as
\bdm
  a_{lm}=\int\frac{\delta\Theta}{\Theta}(\mathbf{x},\eta,\nhv)Y^*_{lm}(\nhv)d\Omega_{\mathbf{n}} .
\edm

Assuming the statistical map of the sky to be rotationally invariant immediately leads to the definition of the expectation of multipole moments as
\be
  \langle a^{}_{lm}a^*_{pn}\rangle=C_l\gamma_{lp}\gamma_{mn}, \quad C_l=\frac{1}{2l+1}\sum_m\langle a^{}_{lm}a^*_{lm}\rangle .
\ee
$C_l$ is known as the angular power spectrum of the CMB.

The simplest statistical measure of the sky is the two-point correlation function in two different photon directions $\nhv$ and $\nhv'$. For the correlations of a map with itself, this is called the \emph{auto-correlation}. For temperature, the auto-correlation is
\bea
  C(\nhv\cdot\nhv')&=&\left<\frac{\delta\Theta}{\Theta}(\nhv)\frac{\delta\Theta}{\Theta}(\nhv')\right>
   =\sum_{lm}\sum_{pn}\left<a^{}_{lm}a^*_{lm}\right>Y^{}_{lm}(\nhv)Y^*_{pn}(\nhv') \noindent \\
  &=&\sum_{lm}C_lY^{}_{lm}(\nhv)Y^*_{lm}(\nhv')=\frac{1}{4\pi}\sum_l\left(2l+1\right)C_lP_l\left(\nhv\cdot\nhv'\right)
\eea
where we have used the relation between the Legendre polynomials and the spherical harmonics (\ref{LegendreSphericalHarmonicRelation}). Since the auto-correlation is linear, we may separate $C(\nhv\cdot\nhv')$ or, equivalently, $C_l$ into the contributions from scalar, vector and tensor perturbations of the brightness function.

\subsection{Temperature Auto-Correlation - Traditional Approach}
\label{Sec-CMB-TradAutoCorrelation}
\subsubsection{Scalar}
For purposes of clarity, and to demonstrate some results that will be useful in the future, we present the derivations, in the traditional Boltzmann approach, of the scalar, vector and tensor temperature auto-correlations. For scalars,
\bdm
  \Sc{\frac{\delta\Theta}{\Theta}}\left(\mathbf{x},\nhv,\eta\right)=\frac{1}{4}
  \sum_l(-i)^l(2l+1)\int\Delta_{Tl}P_l(\mu)e^{-i\mathbf{k}.\mathbf{x}}\frac{d^3\mathbf{k}}{(2\pi)^3}
\edm
and so the temperature auto-correlation function is
\beas
   C^S(\nhv\cdot\nhv')&=&\frac{1}{16}\sum_{l,p}(-i)^li^p(2l+1)(2p+1)\iint P_l(\khv\cdot\nhv)P_p(\khv'\cdot\nhv') \\&&\qquad
   \qquad \qquad \qquad \qquad \times
  \left<\Delta_{Tl}(\mathbf{k})\Delta^*_{Tp}(\mathbf{k}')\right>e^{i(\mathbf{k}'-\mathbf{k}).\mathbf{x}}
  \frac{d^3\mathbf{k}}{(2\pi)^3}\frac{d^3\mathbf{k}'}{(2\pi)^3}
\eeas
Since the scalar hierarchy is independent of directions of $\kv$, we can set the expectation value of $\Delta(\kv)$ to be
\be
  \langle\Delta_{Tl}(\kv)\Delta^*_{Tp}(\kv)\rangle=\mathcal{P}_S(k)\Delta_{Tl}(k)\Delta^*_{Tp}(k')\delta(\kv'-\kv)
\ee
where $\mathcal{P}_S(k)$ is the initial power spectrum of the fluctuations. Substituting this into the auto-correlation function, expanding the Legendres across the spherical harmonics and integrating over $\kv'$ gives us
\beas
  C^S(\nhv\cdot\nhv')&=&\frac{(4\pi)^2}{16(2\pi)^6}\sum_{l,m}\sum_{p,n}(-i)^li^p\int Y^*_{pn}(\nhv')
  Y_{lm}(\nhv)\left(\int Y^*_{lm}(\khv)Y_{pn}(\khv)d\Omega_\kv\right) \\ &&
  \qquad \qquad \qquad \qquad \qquad \qquad \qquad \qquad \qquad \times \mathcal{P}_S(k)\Delta_{Tl}(k)\Delta_{Tp}(k)
  k^2dk
\eeas
which, employing the orthogonality of the spherical harmonics, leads rapidly back to
\bdm
  C^S(\nhv\cdot\nhv')=\frac{4\pi}{16(2\pi)^6}\sum_l(2l+1)\int\mathcal{P}_S(k)\left|\Delta_{Tl}\right|^2k^2dk
\edm
and hence
\be
  C^S_l=A\int\mathcal{P}_S(k)\left|\Delta_{Tl}\right|^2k^2dk
\ee
where we have rolled the prefactors into one term that can be normalised to the results of COBE or WMAP.

From this expression we see the meaning of the terminology ``transfer functions'' for $\Delta_{Tl}$; these functions literally wrap the initial conditions -- here the scalar power spectrum -- onto the microwave background sky. In effect the transfer functions embody the physics of the problem and, once found, we can determine the effect of any initial state onto the microwave background sky. While in this traditional approach the vector and tensor terms will not yield such a neat result, it is part of the power of the line-of-sight approach that we can, in fact, express each contribution to the $C_l$s in the above form.

\subsubsection{Vector}
\label{VectorCMBSpectra}
The standard contribution to the temperature shift in the CMB from vector perturbations was
\bdm
  \V{\frac{\delta\Theta}{\Theta}}\left(\mathbf{x},\nhv,\eta\right)=\frac{1}{4}
  \sqrt{1-\mu^2}\sum_l(-i)^l(2l+1)P_l(\mu)\left(\alpha_l^{(1)}(\kv,\eta)\cos\phi
   +\alpha_l^{(2)}(\kv,\eta)\sin\phi\right) .
\edm
Deriving the vector contribution to the temperature auto-correlation is moderately unwieldy; here we merely summarise the steps and present the full derivation in appendix \ref{VectorCMBSpectraAppendix}.

Since the photon hierarchies are symmetric with respect to the change of vector mode, we shall assume that the two are uncorrelated but similar -- that is,
\bea
  \left<\alpha_l^{(1)}(\kv)\alpha_p^{*(1)}(\kv')\right>&\approx&
  \left<\alpha_l^{(2)}(\kv)\alpha_p^{*(2)}(\kv')\right>, \nonumber \\
  \left<\alpha_l^{(1)}(\kv)\alpha_p^{*(2)}(\kv')\right>&=&
  \left<\alpha_l^{(2)}(\kv)\alpha_p^{*(1)}(\kv')\right>^*\approx0.
\eea
As in the scalar case, the independence of the evolution equations on the direction of $\kv$ then allows us to define
\be
  \left<\alpha_l^{(1)}(\kv)\alpha_p^{*(1)}(\kv')\right>=\mathcal{P}_V(k)\alpha_l(k)\alpha^*_p(k')\delta
  (\kv'-\kv).
\ee
Substituting these and using 
\beas
  \nhv'.\nhv&=&\sqrt{1-\mu^2}\sqrt{1-\overline{\mu}^2}\cos\left(\overline{\phi}-\phi\right)+\mu\overline{\mu}
\eeas
to remove the azimuthal angle $\phi$, and then expanding the Legendre polynomials over the spherical harmonics and integrating out ultimately gives
\be
  \V{C}_l=A\frac{l(l+1)}{(2l+1)^2}\int\mathcal{P}_V(k)\left|\alpha_{l-1}+\alpha_{l+1}\right|^2k^2dk.
\ee
Note that this differs dramatically from that presented by Crittenden and Coulson because of the difference in the temperature vector that we chose as part of our basis. To our knowledge this expression has not appeared elsewhere.

\subsubsection{Tensor}
\label{TensorCMBSpectra}
Assuming as with the vectors that the two tensor modes are uncorrelated but similar, the temperature auto-correlation is
\beas
  C^T(\nhv\cdot\nhv')&=&\frac{1}{16}\sum_l\sum_p(-i)^li^p(2l+1)(2p+1)\iint
  (1-\mu^2)(1-\overline{\mu}^2)P_l(\mu)P_p(\overline{\mu}) \\ && \qquad \qquad \quad \times
  \left<\alpha_l^+(\kv)\alpha_p^{*+}(\kv')\right>\cos2\left(\overline{\phi}-\phi\right)
  e^{i(\kv'-\kv).\xv}\frac{d^3\kv}{(2\pi)^3}\frac{d^3\kv'}{(2\pi)^3} .
\eeas
From here we may proceed in a manner entirely analogous to the approach to the vector $C_l$; the derivation is presented in appendix \ref{TensorCMBSpectraAppendix}. $\phi$ is removed with the square of $\nhv'.\nhv$, which leads after expansions and integrations to
\be
  C^T_l=A(l-1)l(l+1)(l+2)\int\mathcal{P}_T(k)\left|\tilde{\alpha}_{l-2}
  +\tilde{\alpha}_l+\tilde{\alpha}_{l+2}\right|^2k^2dk
\ee
where the scaled variables are
\be
  \tilde{\alpha}_{l-2}=\frac{\alpha_{l-2}}{(2l-1)(2l+1)}, \
  \tilde{\alpha}_l=\frac{2\alpha_l}{(2l-1)(2l+3)}, \
  \tilde{\alpha}_{l+2}=\frac{\alpha_{l+2}}{(2l+1)(2l+3)} .
\ee
This agrees with the result of Crittenden \emph{et. al.} \cite{CrittendenEtAl93}, although our derivation is substantially different.

While it would in principle be possible to repeat this derivation for the polarisation as derived from a traditional Boltzmann approach, to do so is not particularly useful. We could either simply determine the power spectra of the Stokes parameters $Q$ and $U$, the interpretation of which necessitates the use of a small-angle approximation due to the ambiguities induced by the rotational variance of the parameters, or reconstruct the $E$ mode polarisation induced by the scalar perturbations from the standard approach, but an attempt to do the same with the vector and tensor perturbations leads to some unpleasant and convoluted integrations and, on the whole, it is simpler by far to rebuild the $E$ and $B$ modes from a line-of-sight approach. We now finally turn to building the power spectra for this approach.

\subsection{Angular Power Spectra from the Line-of-Sight Approach}
\label{Line-of-Sight-CMB}
\subsubsection{Scalar Modes}
Here we found that
\be
  \Delta_T\left(\mathbf{k},\eta_0\right)=\int_\eta d\eta\Sc{S}_T\left(k,\eta\right)e^{-ix\mu} .
\ee
Instead of aiming for the real-space auto-correlation function we can derive the CMB angular power spectra from the (real-space) spherical moments $a_{lm}$ and reconstruct the relevant angular power spectra with
\bdm
  C_{ab,l}=\frac{1}{2l+1}\sum_m\langle a_{a,lm}a^*_{b,lm}\rangle
\edm
where in the full generality $\{a,b\}\in\{T,E,B\}$ (and, of course, here $B=0$). Polarisation is introduced in appendix \ref{Appendix-EandB}. Setting ourselves at the centre of the co-ordinate system (that is, $\mathbf{x}_0=0$), we see that for the temperature auto-correlation we have
\be
\label{Scalar_aTlm}
  a_{T,lm}=\int_{\mathbf{k}}\left(\frac{1}{4}\int_{\Omega_\mathbf{n}}Y^*_{lm}(\mathbf{n})\int_\eta\Sc{S}_T(k,\eta)
  e^{-ix\mu}d\eta d\Omega_\mathbf{n}\right)\frac{d^3\mathbf{k}}{(2\pi)^3} .
\ee

With the usual definition of the scalar power spectrum $\mathcal{P}_S(k)$, we can then write the auto-correlation function of $a_{T,lm}$ as
\beas
  \langle a^*_{T,lm}a_{T,lm}\rangle&=&\frac{1}{(2\pi)^3}\int_\mathbf{k}\mathcal{P}_S(k)\left|
   \int_{\Omega_\mathbf{n}}Y^*_{lm}(\mathbf{n})\frac{1}{4}\int_\eta\Sc{S}_T(k,\eta)e^{-ix\mu}d\eta
   d\Omega_\mathbf{n}\right|^2\frac{d^3\mathbf{k}}{(2\pi)^3}
\eeas
and then expand the exponential across the spherical harmonics and spherical Bessel functions (\ref{ExponentialSphericalHarmonics}) . Integrating over the directions of $\mathbf{n}$ reduces our expression to
\beas
  \langle a^*_{T,lm}a_{T,lm}\rangle&=&\frac{1}{(2\pi)^3}\int_\mathbf{k}\mathcal{P}_S(k)\left|
    4\pi(-i)^l\frac{1}{4}\int_\eta\Sc{S}_T(k,\eta)j_l(x)Y^*_{lm}(\mathbf{k})d\eta\right|^2
    d^3\mathbf{k}
\eeas
which we can then convert to
\be
\label{StandardApproachCl}
  C_{T,l}=\frac{2}{\pi}\int_k\mathcal{P}_S(k)\left|\frac{1}{4}\int_\eta\Sc{S}_T(k,\eta)j_l(x)d\eta\right|^2k^2dk
\ee
and, by comparison with the earlier expression we can then identify
\be
\label{ScalarTemperatureTransferFunctions}
  \Delta_{Tl}(\mathbf{x},\eta_0,\mathbf{n})=\int_0^{\eta_0}\Sc{S}_T(k,\eta)j_l(x)d\eta
\ee
as the scalar contribution to the temperature transfer function -- the scalar contribution to the temperature moments, as we quoted earlier, and
\be
  C^S_{T,l}=\frac{1}{8\pi}\int_k\mathcal{P}_S(k)\left|\Delta^S_{Tl}(k,\eta_0)\right|^2k^2dk .
\ee

\subsubsection{Vector Modes}
We approach the vector modes in an analogous manner to the scalars. Considering first the temperature case, we had
\bdm
  \Delta^V_T=-i\sqrt{1-\mu^2}\left(\tilde{\zeta}e^{i\phi}+\breve{\zeta}e^{-i\phi}\right)\int_\eta S^V_T(k,\eta)e^{-ix\mu}d\eta .
\edm
The correlation of two multipole moments is then
\bdm
  \langle a_{T,lm}a^*_{T,lm}\rangle=\int_\mathbf{k}\mathcal{P}_V(k)\left|\int_{\Omega_\mathbf{n}}Y^*_{lm}(\mathbf{n})
   \sqrt{1-\mu^2}e^{i\phi}\frac{1}{4}\int_\eta S^V_T(k,\eta)e^{-ix\mu}d\eta    d\Omega_{\mathbf{n}}\right|^2\frac{d^3\mathbf{k}}{(2\pi)^3}
   .
\edm
Integrating over the directions of $\mathbf{k}$ and using the definition of the spherical harmonics (\ref{SphericalHarmonicsDefinition}) we can convert this to
\beas
  \lefteqn{\langle a_{T,lm}a^*_{T,lm}\rangle=}
  \\ && \;
  \frac{4\pi}{(2\pi)^3}\int_k\mathcal{P}_V(k)\left|
   \int_{\Omega_\mathbf{n}}\frac{1}{4}\int_\eta\sqrt{\frac{2l+1}{4\pi}\frac{(l-m)!}{(l+m)!}}P^m_l(\mu)
    \right. \\ && \qquad \times \left.
   \sqrt{1-\mu^2}e^{i(1-m)\phi}
   \int_\eta S^V_T(k,\eta)e^{-ix\mu}d\eta d\Omega_{\mathbf{n}}\right|^2k^2dk .
\eeas
But the integral over $\phi$ will yield a Kronecker delta,
\bdm
  \int_\phi e^{(1-m)i\phi}d\phi=2\pi\delta^1_m ,
\edm
and so if we sum across the index $m$, divide by $2l+1$ and employ the definition of the associated Legendre function (\ref{AssociatedLegendreDefinition}) we can find that
\bdm
  C^V_{T,l}=\frac{1}{2\pi}\frac{(l-1)!}{(l+1)!}\int_k\mathcal{P}_V(k)\left|\frac{1}{4}
  \int_\mu\int_\eta -\frac{\partial P_l(\mu)}{\partial\mu}S^V_T(k,\eta)\left(1-\mu^2\right)e^{-ix\mu}d\eta d\mu\right|^2k^2dk .
\edm
We now notice that we can convert $\mu\rightarrow i\partial_x$ operating on the exponential and so we can integrate this expression once by parts (neglecting a boundary term that would affect the monopole) to leave us with
\bdm
  C^V_{T,l}=\frac{1}{2\pi}\frac{(l-1)!}{(l+1)!}\int_k\mathcal{P}_V(k)\left|\frac{1}{4}
  \int_\mu\int_\eta P_l(\mu)S^V_T(k,\eta)\left(1+\partial_x^2\right)\left(xe^{-ix\mu}\right)d\eta d\mu\right|^2k^2dk .
\edm
Expanding the exponential across the Legendre polynomials and spherical Bessel functions and integrating over $\mu$, we end up with
\be
 C^V_{T,l}=\frac{2}{\pi}\frac{(l+1)!}{(l-1)!}\int_k\mathcal{P}_V(k)\left|\frac{1}{4}\int_\eta
   S^V_T(k,\eta)\frac{j_l(x)}{x}d\eta\right|^2k^2dk
\ee
where we have also used
\be
  \left(1+\partial_x^2\right)\left(xj_l(x)\right)=\frac{(l+1)!}{(l-1)!}\frac{j_l(x)}{x}
\ee
as can be shown by direct computation and substitution of the spherical Bessel equation (\ref{SphericalBesselEquation}). We can thus identify the temperature transfer function from vector perturbations,
\be
  \Delta^V_{Tl}(k,\eta_0)=\sqrt{\frac{(l+1)!}{(l-1)!}}\int_\eta S^V_T(k,\eta)\frac{j_l(x)}{x}d\eta
\ee
as was stated earlier, and
\be
 C^V_{T,l}=\frac{1}{8\pi}\int_k\mathcal{P}_V(k)\left|\Delta^V_{Tl}(k,\eta_0)\right|^2k^2dk .
\ee
The derivation and presentation of the vector transfer function in the line-of-sight approach is clearly enormously simplified compared to the standard approach.

\subsubsection{Tensor Modes}
Deriving the tensor temperature transfer function is very analogous to the vector case; here we had
\bdm
  \Delta^T_T(\mathbf{k},\eta)=\left(1-\mu^2\right)\left(\xi^1(\mathbf{k})e^{2i\phi}+\xi^2(\mathbf{k})e^{-2i\phi}\right)
  \int_\eta S^T_T(k,\eta)e^{-ix\mu}d\eta .
\edm
Following much the same process as for the vectors, we expand this across the spherical harmonics, employ the statistical properties of the variables $\xi^{1,2}(\mathbf{k})$, integrate over the directions of $\mathbf{k}$ and express the spherical harmonic in terms of an associated Legendre polynomial and an exponential to give us
\beas
  \lefteqn{\langle a_{T,lm}a^*_{T,lm}\rangle=
  \frac{4\pi}{(2\pi)^3}\int_k\mathcal{P}_T(k)\left|\frac{1}{4}
  \int_\mu\int_\phi\int_\eta\sqrt{\frac{2l+1}{4\pi}\frac{(l-m)!}{(l+m)!}}S^T_T(k,\eta)\right.}
   \\ && \qquad \times \left.
  P^m_l(\mu)e^{(2-m)i\phi}\left(1-\mu^2\right)e^{-ix\mu}d\eta d\phi d\mu\right|^2k^2dk .
\eeas
Integrating over $\phi$, summing over $m$ and dividing by $2l+1$ leads us to
\bdm
  C^T_{T,l}=\frac{1}{2\pi}\frac{(l-2)!}{(l+2)!}\int_k\mathcal{P}_T(k)\left|\frac{1}{4}
  \int_\mu\int_\eta P^2_l(\mu)S^T_T(k,\eta)\left(1-\mu^2\right)e^{-ix\mu}d\eta d\mu\right|^2k^2dk .
\edm
The definition of the associated Legendre polynomial (\ref{AssociatedLegendreDefinition}) allows us to write
\bdm
  C^T_{T,l}=\frac{1}{2\pi}\frac{(l-2)!}{(l+2)!}\int_k\mathcal{P}_T(k)\left|\frac{1}{4}
  \int_\mu\int_\eta\frac{\partial^2}{\partial\mu^2}P_l(\mu)S^T_T(k,\eta)\left(1+\partial_x^2\right)^2e^{-ix\mu}d\eta d\mu\right|^2k^2dk .
\edm
We can then integrate twice over $\mu$ by parts (again neglecting boundary terms), expand the exponential in terms of the Legendre polynomials and Bessel functions, and then integrate out the dependence on $\mu$, which finally gives us
\bea
  C^T_{T,l}&=&\frac{2}{\pi}\frac{(l-2)!}{(l+2)!}\int_k\mathcal{P}^T(k)\left|\frac{1}{4}\int_\eta S^T_T(k,\eta)\left(1+\partial_x^2\right)^2
   \left(x^2j_l(x)\right)d\eta\right|^2k^2dk \nonumber \\
   &=&\frac{2}{\pi}\frac{(l+2)!}{(l-2)!}\int_k\mathcal{P}^T(k)\left|\int_\eta S^T_T(k,\eta)\frac{j_l(x)}{x^2}d\eta\right|^2k^2dk
\eea
using
\be
\label{x^2j_l(x)}
  \left(1+\partial_x^2\right)^2\left(x^2j_l(x)\right)=\frac{(l+2)!}{(l-2)!}\frac{j_l(x)}{x^2} .
\ee
From here we can immediately identify the tensor contribution to the temperature transfer function,
\be
  \Delta^T_{Tl}(k,\eta_0)=\sqrt{\frac{(l+2)!}{(l-2)!}}\int_\eta S^T_T(k,\eta)\frac{j_l(x)}{x^2}d\eta
\ee
and 
\be
  C^T_{T,l}=\frac{1}{8\pi}\int_\eta\mathcal{P}^T(k)\left|\Delta^T_{Tl}(k,\eta_0)\right|^2k^2dk .
\ee

We thus see (as we commented earlier) that there is a simple symmetry on the temperature transfer functions for scalars, vectors and tensors; if scalars have spin $s=0$, vectors $s=1$ and tensors $s=2$ we have
\bdm
  \Delta^s_{Tl}(k,\eta_0)=\sqrt{\frac{(l+s)!}{(l-s)!}}\int_\eta S^s_T(k,\eta)\frac{j_l(x)}{x^s}d\eta .
\edm
This connection between spins is incorporated rigorously in the spin-weighted approach to cosmological perturbation theory, explored in detail in the spin-weighted formalism \cite{HuWhite97,HuEtAl97,Durrer01}.

As with the vector case, the simple transfer functions in this model, as opposed to the traditional approach, are more than apparent. Moreover, the transfer functions emphasise the split between the geometrical properties embodied in the spherical Bessel functions, and the input from the physics which is entirely encoded within the source terms.

\subsection{The CMB Angular Power Spectrum}
It can be useful to derive crude approximations for the CMB anisotropies, based on an instantaneous epoch of recombination. In this case we take the visibility function $g=\delta(\eta-\eta_\mathrm{rec})$.
Then we have for scalar perturbations
\bea
\label{DeltaTS-Approx}
  \Delta_{Tl}^S&\approx& \left(\Delta_{T0}+\frac{4\dot{v}_S}{k}-4\frac{\ddot{h}_S}{k^2}
   +\frac{1}{4}\Phi_S+\frac{3}{4}\frac{\ddot{\Phi}_S}{k^2}\right|_{\eta_{\mathrm{rec}}}
   j_l(x_{\mathrm{rec}}) \nonumber \\ && \qquad \qquad
   +\frac{2}{3}\int_{\eta_{\mathrm{rec}}}^{\eta_0}\left(\dot{h}_S-\dot{h}
   +\frac{3}{k^2}\frac{\partial}{\partial\eta}\ddot{h}_S\right)j_l(x)d\eta ,
\eea
while for vectors we have
\bea
\label{DeltaTV-Approx}
  \Delta^V_{Tl}&\approx&\sqrt{\frac{(l+1)!}{(l-1)!}}
   \left(\left(4iv^v+\frac{1}{k}\dot{\Phi}^V\right|_{\eta_{\mathrm{rec}}}
   \frac{j_l(x_{\mathrm{rec}})}{x_{\mathrm{rec}}}
   +\frac{4}{k}\int_{\eta_{\mathrm{rec}}}^{\eta_0}\ddot{h}^V\frac{j_l(x)}{x}d\eta\right)
\eea
and for tensors
\bea
\label{DeltaTT-Approx}
  \Delta^T_{Tl}&\approx&\sqrt{\frac{(l+2)!}{(l-2)!}}\left(\left.\Phi^T\right|_{\eta_\mathrm{rec}}
  \frac{j_l(x_{\mathrm{rec}})^2}{x_{\mathrm{rec}}}^2
   +2\int_{\eta_{\mathrm{rec}}}^{\eta_0}\dot{h}^T\frac{j_l(x)}{x^2}d\eta\right) .
\eea
These instantaneous-recombination approximations simplify further when one, for example, assumes a tightly-coupled limit throughout. In the next section we shall employ only the scalar contributions to $\Delta_{Tl}$ and return to the vector and tensor contributions in sections \S\ref{Sec-VectorCl} and \S\ref{Sec-TensorCl}.

\subsubsection{The Sachs-Wolfe Plateau}
On very large scales in the matter dominated era, considering only scalar perturbations, we have
\bdm
  \Delta_{Tl}\approx -4d_0j_l(x_\mathrm{rec})=-4\frac{\delta_0}{k^2}j_l(x_\mathrm{rec})
\edm
as the contribution from the $\ddot{h}_S$ term; the others are subdominant to this. The integrated Sachs-Wolfe term has vanished. The angular power spectrum of the CMB is then approximately
\be
  C_l\approx \frac{2}{\pi}\int\mathcal{P}_S(k)\left|\frac{\delta_0}{k^2}\right|^2j_l(x_{\mathrm{rec}})k^2dk .
\ee
Assuming a spectral index
\be
  \mathcal{P}_S(k)=Ak
\ee
as is produced by many inflationary models, we then have
\bdm
  C_l\approx \frac{2}{\pi}A\left|\delta_0\right|^2\int j_l(k(\eta_{\mathrm{rec}}-\eta))\frac{dk}{k} .
\edm
This is a standard integral \cite{AbramowitzStegun} and the result gives
\be
  l(l+1)C_l\approx \frac{1}{\pi}A\left|\delta_0\right|^2=\mathrm{const}.
\ee
This is called the Sachs-Wolfe plateau, and explains why a Harrison-Zel'dovich spectral index is known as a scale-invariant spectrum. Different initial conditions and different spectral indices will obviously not produce a Sachs-Wolfe plateau. While the real large-scale CMB sky is not entirely flat, this behaviour is observed to a fair approximation by WMAP.

\subsubsection{The Acoustic Peaks}
The oscillations in the coupled photon-baryon fluid also leave their imprint. With a phase of $kc_s\eta$ where $c_s\approx1/\sqrt{3}$, and with the density perturbations going as the cosine, one may predict that the first acoustic peak will be at $k=0$ -- that is, large scales. However, modes on these scales had yet to enter the horizon when the CMB formed and on these scales we instead see the Sachs-Wolfe plateau. The next extremum, technically a compression peak, will be at
\be
  k\eta_\mathrm{rec}c_s=\pi
\ee
and the separation between peaks is
\be
  \Delta k=\frac{\pi}{\eta_{\mathrm{rec}}c_s} .
\ee
This corresponds to a comoving wavelength
\bdm
  \Delta\lambda=\frac{2\pi}{\Delta k} .
\edm
We can map comoving separations onto angles at our own time and convert angles approximately into multipole moments by
\be
  \Delta l\approx \frac{2\pi}{\Delta\theta} \Rightarrow \Delta l\approx F\Delta k
\ee
where
\bdm
  F=\left(\frac{1}{H_0}\int_0^{z_{\mathrm{rec}}}\frac{1}{\sqrt{(1+z)^4\Omega_\gamma+(1+z)^3\Omega_m+\Omega_\Lambda}}dz\right)^2
\edm
is a function transferring the wavelength to an angle. This corresponds in a flat universe with adiabatic initial conditions \cite{Durrer01} to
\be
  \Delta l\approx 220
\ee
and so the first acoustic peak is expected to be at around this multipole number. Since different initial conditions affect the phase of the perturbations, this separation remains constant regardless and is chiefly a test of universal curvature; however, the absolute position of peaks in multipole-space obviously varies.

Despite their name, these acoustic peaks are not actually Doppler peaks since their extrema are at points when the velocity vanishes. The extrema of the velocity perturbations act to level the acoustic peaks somewhat.

\subsubsection{The Full CMB Angular Power Spectrum}
\begin{figure}\begin{center}
\includegraphics{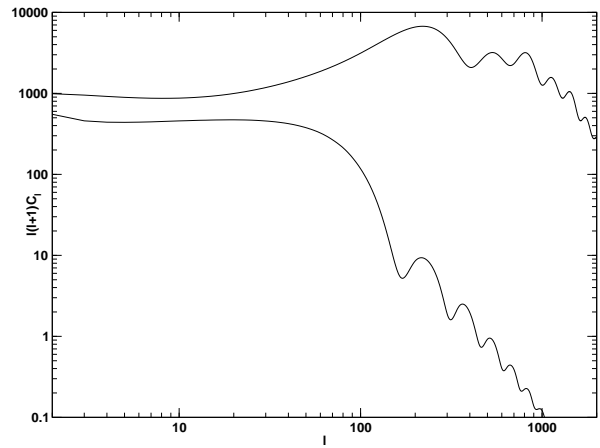}
\caption{CMB angular power spectra for scalar modes (top) and tensor modes (bottom) for the concordance cosmology. Plots generated by CAMB \cite{CAMB}.}\label{CMB_CAMB}
\end{center}\end{figure}

\begin{figure}\begin{center}
\includegraphics{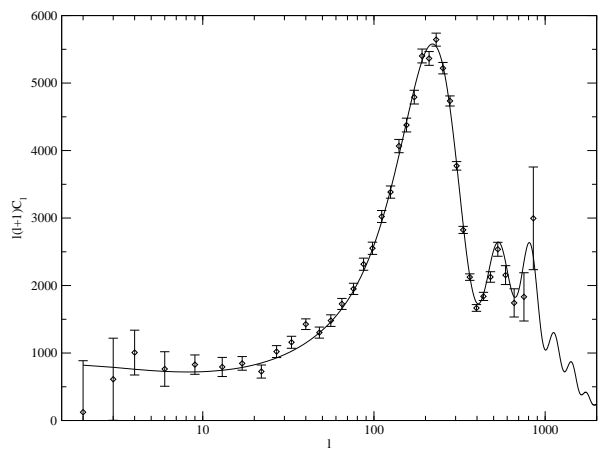}
\caption{CMB angular power spectra for scalar modes (generated by CAMB) plotted against the 1-year WMAP data.}\label{CMB_WMAP}
\end{center}\end{figure}

Figure \ref{CMB_CAMB} plots the scalar and tensor angular power spectrum predicted by the ``concordance model'' of cosmology, with $\Omega_{m0} h^2=0.135$ the total matter energy density, $\Omega_{b0}h^2=0.0224$ the energy density in baryons, a flat universe and a Hubble rate of $h=0.71$. Initial conditions were generated by a Harrison-Zel'dovich power spectrum with no running. Figure \ref{CMB_WMAP} plots the scalar spectrum against the 1-year WMAP data \cite{WMAP-Data}. One can clearly see at the low end the ``Sachs-Wolfe'' plateau. The full theoretical Sachs-Wolfe plateau rises slightly as one reaches the largest scales due to an integrated Sachs-Wolfe effect from the cosmological constant. The observed quadrupole (and, to a lesser extent, the octopole) \cite{WMAP-Bennett,GaztanagaEtAl03} is extraordinarily low, outside even the cosmic variance caused by our limited sample. There are many attempts to justify this low quadrupole -- such as a sharp cut-off in the primordial power spectrum \cite{Efstathiou03,ClineEtAl03}, topologically finite universes \cite{UzanEtAl04,PhillipsKogut04}, foregrounds such as the galactic plane \cite{GaztanagaEtAl03,Efstathiou04} or a thermal Sunyaev-Zel'dovich effect \cite{AbramoSodre03} or considered more generically \cite{SlosarSeljak04}, correlated adiabatic and isocurvature modes \cite{ValiviitaEtAl03,GordonHu04} or inflation powered by two inflatons \cite{FengEtAl03} -- and the matter has been much studied, especially in connection with curious alignments also found in the CMB \cite{TegmarkEtAl03,LandMagueijo05-CubicAnomaly}. It could, however, be no more than a statistical artifact (e.g. \cite{Efstathiou04,Efstathiou03-Stat}).

Beyond the Sachs-Wolfe plateau is the first acoustic peak, at an $l\approx 220$ as predicted. Beyond this peak the amplitude of the waves dies away; this is due to ``Silk damping'' \cite{Silk68} which is a small-scale (non-linear) effect contributing a drag to the oscillation of the waves. WMAP is cosmic-variance limited throughout the first acoustic peak and other than a few glitches at the base and peak of the first acoustic peak the agreement is superb. WMAP does not have good resolution much beyond the base of the first peak; datasets covering this smaller-$l$ region include those of CBI \cite{RajguruEtAl05}, Boomerang \cite{MontroyEtAl05} and ACBAR \cite{ACBAR}. Observations are consistent with the concordance model; however, there seems to be a systematic increase of power on small scales. Amongst many other possibilities, this could be at least partially due to the effects of a magnetic field \cite{SubramanianBarrow98,SubramanianBarrow02,YamazakiIchikiKajino04}.

\chapter{Magnetised Cosmology}
\label{Chapter-MagnetisedCosmology}
\section{The Electromagnetic Field}
\label{SecEMField}
In this chapter we consider the additions to the standard linearised cosmology that a tangled primordial magnetic field will introduce. After reducing more general expressions to a suitable cosmological approximation we then consider in some detail the transferral of the magnetic terms into Fourier space. We conclude with a brief review of the damping of magnetohydrodynamical modes by photon shear viscosity.

While much of the material presented in this chapter is known, the approach is entirely individual and the equations formulated more generally than has previously been seen in the metric based approach. Tsagas and Maartens \cite{Tsagas04,TsagasMaartens99-Perts} and Lewis \cite{Lewis04-Mag} work within the gauge-invariant and covariant formalism, and so in principle present a greater generality and consider first the full non-linear equations before truncating them to some expansion order. However, Lewis does not consider the scalar perturbations and the work Tsagas and Maartens is limited to a homogeneous field. Giovannini \cite{Giovannini04-Mag} works within a metric-based approach and presents equations to a similar generality but considers only scalar perturbations. For future ease we present equations with the electromagnetic fields held to an arbitrary perturbation order before linearising their effect; this will aid a two-parameter perturbation approach.

\subsection{Field and Stress-Energy Tensors}
\label{FieldAndStressEnergyTensors}
In an inertial frame, the electromagnetic field tensor, or Faraday tensor, can be defined by
\bdm
  F'_{\mu\nu}=2A_{[\nu,\mu]}=A_{\nu,\mu}-A_{\mu,\nu}
\edm
where
\bdm
  A_\mu=\left(-\phi,\mathbf{A}\right)
\edm
is the 4-potential and $\phi$ and $\mathbf{A}$ the scalar and vector electromagnetic potentials respectively. The electric field $\vec{\epsilon}$ and magnetic field $\vec{\beta}$ can be found from
\be
  \vec{\beta}=\nabla\times\mathbf{A}, \quad \vec{\epsilon}=-\nabla\phi-\frac{\partial\mathbf{A}}{\partial t}
\ee
where for the moment we are employing co-ordinate time. Since these are in Minkowski space we have $\beta_i=\beta^i$ and $\epsilon_i=\epsilon^i$. This gives the field tensor in a matrix format as
\be
  F'_{\mu\nu}=\left(\begin{array}{cccc}
   0 & -\epsilon_1 & -\epsilon_2 & -\epsilon_3 \\ \epsilon_1 & 0 & \beta_3 & -\beta_2 \\
   \epsilon_2 & -\beta_3 & 0 & \beta_1 \\ \epsilon_3 & \beta_2 & -\beta_1 & 0
  \end{array}\right) .
\ee
Extending Jedamzik, Katalini\'c and Olinto \cite{JedamzikKatalinicOlinto98}, we obtain the FLRW form of this from the locally-Minkowski tensors above by transforming from Minkowski space to a perturbed FLRW metric using (\ref{ContravariantCoordinateTransformation}, \ref{CovariantCoordinateTransformation}). This gives a field tensor
\beas
  F_{\mu\nu}&=&a^2\left( \begin{array}{cccc}
   0 & -\epsilon_1 & -\epsilon_2 & -\epsilon_3 \\ \epsilon_1 & 0 & \beta_3 & -\beta_2 \\
   \epsilon_2 & -\beta_3 & 0 & \beta_1 \\ \epsilon_3 & \beta_2 & -\beta_1 & 0
  \end{array}\right) \\ && \quad
  +\frac{1}{2}a^2\left(\begin{array}{cccc}
   0 & -h^a_1\epsilon_a & -h^a_2\epsilon_a & -h^a_3\epsilon_a \\
   h^a_1\epsilon_a & 0 & h\beta_3-h^a_3\beta_a & -\left(h\beta_2-h^a_2\beta_a\right) \\
   h^a_2\epsilon_a & -\left(h\beta_3-h^a_3\beta_a\right) & 0 & h\beta_1-h^a_1\beta_a \\
   h^a_3\epsilon_a & h\beta_2-h^a_2\beta_a & -\left(h\beta_1-h^a_1\beta_a\right) & 0 \end{array}\right), \\
  F^{\mu\nu}&=&\frac{1}{a^2}\left( \begin{array}{cccc}
   0 & \epsilon_1 & \epsilon_2 & \epsilon_3 \\ -\epsilon_1 & 0 & \beta_3 & -\beta_2 \\
   -\epsilon_2 & -\beta_3 & 0 & \beta_1 \\ -\epsilon_3 & \beta_2 & -\beta_1 & 0
  \end{array} \right) \\ && \quad
  +\frac{1}{2a^2}\left(\begin{array}{cccc}
   0 & -h^1_a\epsilon^a & -h^2_a\epsilon^a & -h^3_a\epsilon^a \\
   h_a^1\epsilon^a & 0 & -\left(h\beta_3-h_a^3\beta^a\right) & h\beta_2-h^a_2\beta_a \\
   h^a_2\epsilon_a & h\beta_3-h^a_3\beta_a & 0 & -\left(h\beta_1-h^a_1\beta_a\right) \\
   h^a_3\epsilon_a & -\left(h\beta_2-h^a_2\beta_a\right) & h\beta_1-h^a_1\beta_a & 0 \end{array}\right) .
\eeas
The electric and magnetic fields observed by a comoving observer with $u^\mu=a^{-1}(1,\mathrm{0})$ can then be recovered \cite{TsagasBarrow97,Tsagas04,TsagasMaartens99-Perts} by
\be
  e^\mu=u_\nu F^{\mu\nu}, \quad b^\mu=\frac{1}{2}\varepsilon^{\mu\alpha\beta}F_{\alpha\beta}
\ee
with corresponding covariant expressions. Note that we are actually computing the electric field and magnetic field $h$; our use of the magnetic induction in this case is habitual and the approximations we shall soon apply make the field and induction vectors equal. $\varepsilon^{\mu\alpha\beta}=\sqrt{-|g_{\sigma\rho}|}\epsilon^{\mu\alpha\beta}$ is the Levi-Civita tensor density for this space and $\epsilon_{\mu\alpha\beta}$ the totally anti-symmetric tensor density.

The co- and contra-variant forms of the electric and magnetic fields are
\bea
  e^{\mathrm{FLRW}}_\mu=a\left(0,\epsilon_i+\frac{1}{2}h_i^a\epsilon_a\right), &
  b^{\mathrm{FLRW}}_\mu=a\left(0,\beta_i+\frac{1}{2}h^a_i\beta_a\right), \\
  e_{\mathrm{FLRW}}^\mu=a^{-1}\left(0,\epsilon^i-\frac{1}{2}h^i_a\epsilon^a\right), &
  b^\mu_{\mathrm{FLRW}}=a^{-1}\left(0,\beta^i-\frac{1}{2}h^i_a\beta^a\right) .
\eea
We shall work with the comoving vectors $e_i=\epsilon_i+(1/2)h_i^a\epsilon_a$ and $b_i=\beta_i+(1/2)h_i^a\beta_a$, and it is tacitly assumed that they contain these internal structures. However, it is worth commenting that these forms for the electric and magnetic fields are more general than those presented in \cite{JedamzikKatalinicOlinto98}, for example. In previous work, authors have employed the conformal equivalence of Minkowski and Robertson-Walker magnetohydrodynamics \cite{SubramanianBarrow98} to map the locally-observed Minkowski fields directly onto the Robertson-Walker fields, that is $e^{\mathrm{FLRW}}_i=a\epsilon_i$, $b^{\mathrm{FLRW}}_i=a\beta_i$. This is sufficient for fields present in the background but not for a perturbed Robertson-Walker metric, which is no longer conformally Minkowski. The results found by \cite{DurrerFerreiraKahniashvili00,SubramanianBarrow02,MackKahniashviliKosowsky02} are no less valid, since the fields are assumed small and the coupling terms implicitly to be of order $\mathcal{O}(\varepsilon^{3/2})$. The forms presented above,  however, should be employed if one wishes to consider fields of a different order or, indeed, fields in a Robertson-Walker metric perturbed to second order.

The stress-energy tensor of the magnetic fields is then given \cite{TsagasBarrow97,Weinberg} by
\be
  4\pi T^\mu_\nu=F^{\mu\sigma}F_{\nu\sigma}-\frac{1}{4}\delta^\mu_\nu F^{\sigma\rho}F_{\sigma\rho} .
\ee
Noting, then, that
\bdm
  F^{\sigma\rho}F_{\sigma\rho}=2\left(b^2-e^2\right) ,
\edm
one can construct the energy density, momentum flux and momentum flux density of an electromagnetic field:
\bea
  4\pi T^0_0&=&F^{0\sigma}F_{0\sigma}-\frac{1}{4}F^{\sigma\rho}F_{\sigma\rho}=-\frac{1}{2}\left(e^2+b^2\right) ,
   \\
  4\pi T^i_0&=&F^{i\sigma}F_{0\sigma}=F^{ij}F_{0j}=-\left(\mathbf{e}\times\mathbf{b}\right)^i
   \\
  4\pi T^i_j&=&F^{i\sigma}F_{j\sigma}-\frac{1}{4}\delta^i_jF^{\sigma\rho}F_{\sigma\rho} .
   =\frac{1}{2}\left(b^2+e^2\right)\delta^i_j-e^ie_j-b^ib_j .
\eea
We have recovered the familiar energy density of an electromagnetic field, the Poynting vector, and the electromagnetic stresses. The latter separate into the isotropic pressure
\be
  4\pi T^i_i=4\pi\left(3p\right)=\frac{1}{2}\left(b^2+e^2\right),
\ee
which gives the expected equation of state $p=(1/3)\rho$, and the anisotropic stress
\be
  4\pi\Pi^i_j=\frac{1}{3}\left(b^2+e^2\right)\delta^i_j-e^ie_j-b^ib_j .
\ee

\subsection{Maxwell Equations and Stress-Energy Conservation}
\label{Sec-MaxwellEquations}
The Einstein equations are not the only equations governing the evolution of the electromagnetic field. One also has the Maxwell equations, Ohm's law and stress-energy conservation.


The electromagnetic 4-current is taken to be
\be
  j^\mu=a^{-1}\left(\varrho,\mathbf{j}\right)
\ee
where $\varrho$ is the comoving charge density and $\mathbf{j}$ the comoving charge current density. This is a 4-current as observed in a frame comoving with universal expansion. In a locally-Minkowski frame, one would expect to observe a current $\mathbf{j}_\mathrm{Mink}=\mathbf{j}-\varrho\fgv$.\footnote{This takes the opposite sign to that one might na\"ively expect since our velocity is defined with reference to the comoving expansion rather than with reference to a Minkowski observer.}

The Maxwell equations are then
\be
  F^{\mu\nu}_{\phantom{\mu\nu};\nu}=j^{\mu}, \quad F_{[\lambda\mu;\sigma]}=0 .
\ee
The skew-symmetry of the electromagnetic field tensor implies charge conservation
\be
  j^\mu_{\phantom{\mu};\mu}=0 .
\ee
The relativistic equivalent to the familiar Ohm's law arises when one projects the (invariant) 4-current into the observer's 3-space and sets this observed current proportional to the electric field viewed in this frame \cite{TsagasBarrow97,Tsagas04,Maartens00} -- that is
\be
  H_{\mu\nu}j^\nu=\left(g_{\mu\nu}+u_\mu u_\nu\right)j^\nu
   =\sigma F_{\mu\nu}u^\nu\Rightarrow j_\mu+u_\mu u_\nu j^\nu=\sigma F_{\mu\nu}u^\nu .
\ee
$\sigma$ is the electrical conductivity.

One may rapidly demonstrate that the Maxwell equations in a synchronous FLRW metric perturbed to first order, with electromagnetic fields held to arbitrary order, are
\bea
  \scriptstyle{F^{0\nu}_{\phantom{0\nu};\nu}=j^0} & : &
   \nabla\cdot\mathbf{e}+\frac{1}{2}\left(\mathbf{e}\cdot\nabla\right)h=a\varrho, \nonumber \\
\label{MaxwellInCosmology}
  \scriptstyle{F^{i\nu}_{\phantom{0\nu};\nu}=j^i} & : &
   \nabla\times\mathbf{b}-\left(\dot{\mathbf{e}}+2\hub\mathbf{e}\right)-\frac{1}{2}\dot{h}\mathbf{e}
    -\frac{1}{2}\left(\mathbf{b}\times\nabla\right)h=a\mathbf{j} , \\
  \scriptstyle{F_{[0ij]}=0} & : &
   \dot{\mathbf{b}}+2\hub\mathbf{b}+\nabla\times\mathbf{e}=0 , \nonumber \\
  \scriptstyle{F_{[ijk]}=0} & : & \nabla\cdot\mathbf{b}=0 . \nonumber
\eea
The requirement of charge continuity implies that
\be
\label{ChargeContInCosmology}
  \dot{\varrho}+3\hub\varrho+\nabla\cdot\mathbf{j}+\frac{1}{2}\dot{h}\varrho
   +\frac{1}{2}\left(\mathbf{j}\cdot\nabla\right)h=0.
\ee
Interpreting this in the absence of metric perturbations, we see that there is an effective ``sink'' of charge (as viewed by a comoving observer) caused by the universal expansion -- charge decays as $a^{-3}$.

The spatial part of Ohm's law gives
\be
\label{OhmInCosmology}
  \mathbf{j}-\varrho\fgv=\sigma\left(\mathbf{e}+\fgv\times\mathbf{b}\right)
\ee
which is the familiar Ohm's law from electrodynamics. The temporal component yields, to $\mathcal{O}(\epsilon)$, $\fgv\cdot\mathbf{j}=\sigma\fgv\cdot\mathbf{e}$ which is merely the scalar product of the standard Ohm's law with the observer's velocity.

Generalising the results of (for example) Jedamzik \emph{et. al.} \cite{JedamzikKatalinicOlinto98} to the case of finite conductivity and a perturbed universe, the contributions of electromagnetic fields to stress-energy conservation are
\bea
  -4\pi T^\nu_{0;\nu} & = &
   \mathbf{e}\cdot\dot{\mathbf{e}}+\mathbf{b}\cdot\dot{\mathbf{b}}+2\hub\left(e^2+b^2\right)
   +\nabla\cdot\left(\mathbf{e}\times\mathbf{b}\right)+\frac{1}{2}\left(e^2+b^2\right)\dot{h}\nonumber \\ && \quad
   +\frac{1}{2}\left(\left(\mathbf{e}\times\mathbf{b}\right)\cdot\nabla\right)h
   -\frac{1}{2}\dot{h}^i_j\left(e_ie^j+b_ib^j\right) \nonumber \\
   &=& -a\mathbf{e}\cdot\mathbf{j}-\frac{1}{2}\left(\dot{h}b^2+
   \left(\dot{h}^i_j+2\hub h^i_j\right)\left(e_ie^j+b_ib^j\right)\right),
    \\ && \nonumber \\
  4\pi T^\nu_{i;\nu} & = &
   \partial_\eta \left(\mathbf{e}\times\mathbf{b}\right)_i
   +\frac{1}{2}\partial_i\left(b^2\right)+\frac{1}{2}\partial_i\left(e^2\right)
   +\left(4\hub+\frac{1}{2}\dot{h}\right)\left(\mathbf{e}\times\mathbf{b}\right)_i
    \nonumber \\ && \quad
   -\frac{1}{2}e_i\left(\left(\mathbf{e}\cdot\nabla\right)h\right)
   -\frac{1}{2}b_i\left(\left(\mathbf{b}\cdot\nabla\right)h\right)
   +\hub h_i^j\left(\mathbf{e}\times\mathbf{b}\right)_j \nonumber
   \\ && \quad
   +\frac{1}{2}\left(e_je^k+b_jb^k\right)\partial_ih^j_k
   -\partial_j\left(e_ie^j+b_ib^j\right)
     \nonumber \\ & = &
   -a\left(\mathbf{j}\times\mathbf{b}\right)_i
   -a\varrho e_i
   +\frac{1}{2}\left(\mathbf{b}\times\left(\mathbf{b}\times\nabla\right)\right)_ih
   -\frac{1}{2}b_i\left(\mathbf{b}\cdot\nabla\right)h \nonumber \\ &&
   \quad
   +\hub h_i^j\left(\mathbf{e}\times\mathbf{b}\right)_j
   +\frac{1}{2}\left(e_je^k+b_jb^k\right)\partial_ih^j_k .
\eea
The first of these is the contribution an electromagnetic field will make to the conducting fluid's mass continuity; the field creates fluctuations in the fluid's current which then impacts on the matter conservation. The second of these is the contribution the field makes to the Euler equation (momentum conservation). Comparing with the form for non-relativistic magnetohydrodynamics (e.g. \cite{Jackson}) we can pick out the important contributions to the Euler equation
\be
  \mathbf{l}=a\left(\mathbf{j}\times\mathbf{b}+\varrho\mathbf{e}\right)
\ee
which is the standard Lorentz force. All other terms in the Euler contribution are couplings to the metric perturbations, representing the scattering of the fields from the underlying geometry. This coupling is readily apparent in covariant approaches to the problem (see for example \cite{TsagasBarrow97,Tsagas04,TsagasMaartens99-Perts,TsagasMaartens99-Bianchi}) in the full non-linear equations, where the fields couple directly to the Weyl tensor.

\subsection{The Electromagnetic Field: Robertson-Walker and Bianchi Cosmologies}
The Robertson-Walker cosmology is a cosmology homogeneous and isotropic about every event, at least to zeroth order, which forbids the addition in the background of a directional field. The electromagnetic field generally produces both an energy flux and a full anisotropic shear, neither of which are compatible with a Robertson-Walker metric.

To maintain the standard cosmology, it is necessary to ensure that the energy flux and anisotropic shear vanish at zeroth order in the perturbation parameter $\varepsilon$. To retain the FLRW form we thus set the electromagnetic fields to be ``half-order''; that is, that we should set $\mathcal{O}(e,b)\approx\mathcal{O}(\sqrt{\varepsilon})$. While this has the benefit of removing electromagnetic shear from the bulk cosmology, it is restrictive for two main reasons: firstly that a linearisation of the magnetic field employing the same parameter will push the perturbed fields to one-and-a-half order and our field will be correspondingly bland, and secondly that the scattering of the fields from gravity is also pushed to one-and-a-half order. This neglects many of the more interesting effects of a magnetic field, such as the direct sourcing of tensor perturbations, and removes almost the entire structure of MHD. (See \cite{DurrerFerreiraKahniashvili00} for some further discussion on this matter.) It does, however, mean that the comoving fields in the Robertson-Walker metric can now be identified with the Minkowski fields.

However restrictive in linear perturbation theory our approach may be, it allows a great freedom in the exact form of the fields that one may add. The modern approach in CMB physics is to employ a half-order tangled stochastic magnetic field, (with or without a homogeneous component) under the assumption that the initial spectrum is Gaussian and the tangled component has one or more length scales (e.g. \cite{SubramanianBarrow98,KohLee00,DurrerFerreiraKahniashvili00,SubramanianBarrow02,MackKahniashviliKosowsky02,YamazakiIchikiKajino04}). The fields are usually assumed to be tangled on cluster scales -- implying no directionality on supercluster scales -- and the observed strength on such scales can be used to fix the amplitude of the magnetic power spectrum.

The net effect of these assumptions, along with taking the limit of infinite conductivity, is merely to add to the standard Euler equation a Lorentz force. The magnetic field influences not only the temperature maps for the CMB but also the polarisation -- indeed, it is likely that the effect on the temperature-temperature power spectrum will be sub-dominant across most of the realistically observable range, and the effects of a magnetic field might be observable primarily by its imprint on the polarisation (e.g. \cite{KosowskyLoeb96, SeshadriSubramanian00, SubramanianSeshadriBarrow03, MackKahniashviliKosowsky02}) or perhaps its non-Gaussianity \cite{SubramanianBarrow02,BrownCrittenden05,Giovannini05-Mag}.

An alternative approach, while still remaining within the constraints of a Robertson-Walker cosmology, would be to employ a two-parameter linearisation approach, with a magnetic field too small to affect the bulk cosmology linearised by a free parameter $\varepsilon_B$. One would neglect terms of order $\mathcal{O}(\varepsilon^2,\varepsilon_B^2)$ but retain those of order $\mathcal{O}(\varepsilon\varepsilon_B)$ A brief study of two-parameter perturbation theory (in the context of but not restricted to rotating, relativistic stars) is presented by Bruni \emph{et. al.} \cite{BruniGualtieriSopuerta03} and the referenced works therein.

Were one to drop the condition of isotropy and introduce a mildly isotropic Bianchi cosmology rather than the maximally symmetric FLRW, then one could instead employ a magnetic field that gives the universe a preferred direction. It would also be possible to include the shear viscosity of the baryons (and possibly also of the CDM) at zeroth order which could potentially model a realistic fluid more than the perfect fluid approximations. Indeed, the limits that Barrow, Ferreira and Silk place on the strength of a bulk field \cite{BarrowFerreiraSilk97}, based on the 4-year COBE results, are derived by limiting the anisotropy introduced by a Bianchi (type VII) model and attributing it entirely to a magnetic field. The idea of imposing a preferred direction on the universe through an ordered magnetic field, while certainly not original (research in magnetised anisotropic cosmologies goes back to at least the late 60s; see \cite{Thorne67, Jacobs68, Jacobs69} and references therein) could correspond to Tegmark \emph{et. al.}'s speculative explanation \cite{TegmarkEtAl03} for the low power observed in the quadrupole by WMAP \cite{WMAP-Bennett, WMAP-Spergel} resulting from a preferred direction. It is also interesting to note that Tsagas and Maartens have demonstrated \cite{TsagasMaartens99-Bianchi} that, for a relatively weak field, the Robertson-Walker approximation is equivalent at first order to a Bianchi I approach.




\section{Baryonic Matter: Cosmological Magnetohydrodynamics}
\label{MHD}
\subsection{Real Space}
We now consider a magnetised baryon fluid. In the presence of a magnetic field, the mass continuity and Euler equations, retaining for generality the pressure terms and including again the collisional term $C^i_{b\rightarrow\gamma}$ transferring momentum between the baryons and the photons, become
\beas
  \lefteqn{\dot{\rho}+\nabla.\left(\left(\rho+p\right)\fgv\right)+\left(3\hub+\frac{1}{2}\dot{h}\right)
    \left(\rho+p\right)-\frac{1}{4\pi}a\mathbf{e}\cdot\mathbf{j}}
   \\ &&
    -\frac{1}{8\pi}\left(\dot{h}b^2+\dot{h}^i_j\left(e_ie^j+b_ib^j\right)\right)=0, \\
  \lefteqn{\frac{\partial}{\partial \eta}\left(\left(\rho+p\right)v^i\right)
    +\partial^ip+4\hub\left(\rho+p\right)v^i
    -\frac{a}{4\pi}\left(\mathbf{j}\times\mathbf{b}\right)^i-\frac{a}{4\pi}\varrho e^i}
   \\ &&
    +\frac{1}{8\pi}\left(\mathbf{b}\times\left(\mathbf{b}\times\nabla\right)\right)^ih
    -\frac{1}{8\pi}b^i\left(\mathbf{b}\cdot\nabla\right)h
    +\frac{1}{4\pi}\hub\left(\mathbf{e}\times\mathbf{b}\right)^jh^i_j
   \\ &&
    +\frac{1}{8\pi}\left(e_je^k+b_jb^k\right)\partial^ih^j_k +C^i_{b\rightarrow\gamma}=0 .
\eeas
These are supplemented by the Maxwell equations (\ref{MaxwellInCosmology}), charge continuity (\ref{ChargeContInCosmology}) and Ohm's law (\ref{OhmInCosmology}).

Clearly these equations are not in an ideal form either for further analytical study or for numerical implementation. Instead, we shall follow the lead of \cite{DurrerKahniashviliYates98,BrandenburgEnqvistOleson96, JedamzikKatalinicOlinto98, SubramanianBarrow98-MHD} in rescaling the magnetic fields to remove the Hubble terms in the Maxwell equations.

One may rapidly demonstrate that the Maxwell equations reduce (in the absence of metric perturbations) to the form of classical electrodynamics if one employs the scalings
\be
  \left\{\Ev,\Bv,\vr,\jv,\ca\right\}=\left\{a^2\mathbf{e},a^2\mathbf{b},a^3\varrho,a^3\mathbf{j},a^{-1}\sigma\right\}
\ee
with no scaling for the comoving co-ordinates (and hence none for $\partial^i$ and $\nabla$), the baryon velocity or the metric perturbation. These are generally physically motivated; the electric and magnetic fields scale with $a^2$ because their energy density must be that of radiation and decay as $a^4$. The charge and current densities, associated with matter, are scaled by $a^3$. The conductivity has dimensions of length and so scales with $a$.

With these new variables, the electrodynamical equations reduce to
\be
\begin{array}{rcl}
  \nabla\cdot\Ev&=&\vr-\frac{1}{2}\left(\Ev\cdot\nabla\right)h \\
  \nabla\times\Bv-\dot{\Ev}&=&\jv+\frac{1}{2}\dot{h}\Ev+\frac{1}{2}\left(\Bv\times\nabla\right)h \\
  \dot{\Bv}+\nabla\times\Ev&=&0 \\
  \nabla\cdot\Bv&=&0\\
  \dot{\varrho}_a+\nabla\cdot\jv&=&-\frac{1}{2}\left(\vr\dot{h}-\left(\jv\cdot\nabla\right)h\right) \\
  \jv-\vr\fgv&=&\ca\left(\Ev+\fgv\times\Bv\right) .
\end{array}
\ee

In the equations of cosmological MHD presented above, we see we should then apply the scaling
\be
 \left\{\ra,\pa\right\}=a^4\left\{\rho_b, p_b\right\}
\ee
to the mass density and pressure respectively. This scaling is not performed to erase Hubble terms but instead to ensure that each term is scaled by the same factor; our scalings were employed to reduce the unperturbed cosmological Maxwell equations to the Minkowski form, rather than remove the time-dependence of each variable. Consistency of scaling then requires that the matter density and pressure are scaled as the magnetic density and pressure are -- as $a^4$ rather than as $a^3$ as one would na\"{\i}vely expect. (Subramanian and Barrow \cite{SubramanianBarrow98-MHD} present a more formal and detailed analysis of the conformal relation between magnetohydrodynamics in an unperturbed Robertson-Walker metric and in a Minkowski metric.) Note also that
\bdm
  \pa=w_a\left(\ra\right)\ra\Rightarrow p_b=w_a\left(\ra\right)\rho_b
\edm
which implies that
\be
  w_a\left(\ra\right)=w\left(\rho_b\right) .
\ee
We also have
\be
  c_s^2=\frac{\partial p}{\partial\rho}=c_{sa}^2 .
\ee

Employing these substitutions, the MHD equations quickly become
\bea
  \lefteqn{\dot{\rho}_{ba}+\nabla\cdot\left(\left(\rho_{ba}\left(1+w\right)\right)\fgv\right)
   +\frac{1}{2}\left(\rho_{ba}\left(1+w\right)\right)\dot{h}} \nonumber \\ &&
   =\hub\rho_{ba}\left(1-3w\right)+\frac{1}{4\pi}\Ev\cdot\jv+\frac{1}{8\pi}\left(\dot{h}\Ba^2+ 
    \dot{h}^i_j\left(E_i(\mathbf{x})E^j(\mathbf{x})+B_i(\mathbf{x})B^j(\mathbf{x})\right)\right),
\\
  \lefteqn{\dot{v}^i+\left(\frac{\dot{w}}{1+w}+\frac{\dot{\rho}_{ba}}{\ra}\right)v^i
    +\frac{c_s^{\phantom{s}2}}{\ra\left(1+w\right)}\partial^i\ra+\frac{1}{1+w}C^i_{b\rightarrow\gamma}} \nonumber \\ &&
    =\frac{1}{4\pi\left(1+w\right)\ra}\left(\jv\times\Bv+\vr\Ev\right)^i
    +\frac{1}{8\pi\left(1+w\right)\ra}\left.\bigg(B^i(\mathbf{x})\left(\Bv\cdot\nabla\right)h
    -\left(\Bv\times\left(\Bv\times\nabla\right)\right)^ih  \right.
     \nonumber \\ && \left.
    -2\hub\left(\Ev\times\Bv\right)^jh^i_j
    +\left(E_j(\mathbf{k})E^k(\mathbf{x})+B_j(\mathbf{x})B^k(\mathbf{x})\right)\partial^ih^j_k\right).
\eea

\subsubsection{Linearisation}
We expand around the FLRW background by setting
\bdm
  \ra=\bkrO\left(1+\varepsilon\delta\right), \quad  \mathcal{O}(v^i)\approx\mathcal{O}(\varepsilon), \quad
  \left\{\Bv, \Ev, \jv, \vr\right\}\rightarrow \sqrt{\varepsilon}\left\{\Bv,\Ev,\jv,\vr\right\} .
\edm
The (dimensionless) perturbation to the density $\delta$ is naturally unscaled. The comoving electric and magnetic fields can then be identified with their Minkowski equivalents,
\bdm
  e_i=\epsilon_i, \quad b_i=\beta_i
\edm
and the contra- and co-variant components are equal. Neglecting all powers of $\varepsilon$ above unity and separating into background and perturbation leads to the Maxwell equations
\be
\begin{array}{rcl}
  \nabla\cdot\Ev&=&\vr, \\
  \nabla\times\Bv-\dot{\Ev}&=&\jv, \\
  \dot{\Bv}+\nabla\times\Ev&=&0, \\
  \nabla\cdot\Bv&=&0, \\
  \dot{\varrho}_a+\nabla\cdot\jv&=&0, \\
  \jv&=&\ca\Ev ,
\end{array}
\ee
supplementing the magnetohydrodynamical equations
\bdm
  \dot{\overline{\rho}}_{ba}=\hub\bkrO\left(1-3w\right),\quad \nabla \bkrO = 0
\edm
in the background and
\beas
  \dot{\delta}+\left(1+w\right)\nabla\cdot\fgv+\frac{1}{2}\left(1+w\right)\dot{h}&=&\frac{1}{4\pi\bkrO}\Ev\cdot\jv , \\
  \dot{v}^i+\left(\frac{\dot{w}}{1+w}+\hub\left(1-3w\right)\right)v^i \\ \qquad\qquad
    +\frac{c_s^{\phantom{s}2}}{\left(1+w\right)}\partial^i\delta+\frac{1}{1+w}C^i_{b\rightarrow\gamma} &=&
    \frac{1}{4\pi\bkrO\left(1+w\right)}\left(\jv\times\Bv+\vr\Ev\right)^i .
\eeas
at first order.

For baryonic matter we again set $w=c_s^2=0$ except where they multiply a spatial derivative giving
\be
  \dot{\overline{\rho}}_{ba}=\hub\bkrO\rightarrow \bkrO\propto a; \quad \nabla\bkrO=0
\ee
and
\bea
  \dot{\delta}_b+(1+w)\nabla\cdot\fgv_b+\frac{1}{2}\dot{h}&=&\frac{1}{4\pi\bkrO}\Ev\cdot\jv , \\
  \dot{v}^i+\hub v^i+c_s^2\partial^i\delta_b+C^i_{b\rightarrow\gamma}&=&
    \frac{1}{4\pi\bkrO}\left(\jv\times\Bv+\vr\Ev\right)^i\equiv\frac{1}{\bkrO}L^i.
\eea
One can now clearly see that the only difference in the Euler equation obeyed by the baryons with a magnetic field of this type, compared with the standard cosmological picture (\ref{BaryonStandardCosmology}), is the existence of the Lorentz force $\mathbf{L}$. The greater structure of the interactions between the magnetic fields, the geometry, and the perturbations is at a higher-order. This complex structure gives rise to, for example, direct coupling between the magnetic fields and the Weyl tensor (\cite{Tsagas04}) and magnetic fields automatically sourced by second-order vorticity (\cite{MatarreseEtAl04,GopalSethi04}). The vast bulk of cosmological MHD remains unexplored.

\subsubsection{Early-Universe Approximations}
It is usual to assume that the conductivity $\ca$ is extremely high until after recombination (see for example \cite{GrassoRubenstein01} or the appendix of \cite{TurnerWidrow88}); this loose statement may be quantified by considering the dimensionless magnetic Reynolds number
\be
  R_B=4\pi\mu_0\sigma\frac{\left|\nabla\times\left(\mathbf{v}\times\mathbf{B}\right)\right|}{\left|\nabla^2\mathbf{B}\right|}
   \approx 4\pi\mu_0\sigma vL
\ee
where $L$ is the typical length-scale of the system. To a good approximation, the universe possesses a low Reynold's number for much of its evolution. Using Ohm's law, assuming a vanishing conductivity while retaining a finite current density then removes the electric fields. This is sometimes dubbed the magnetohydrodynamical limit. It is also standard in MHD (see \cite{Jackson}) to take the displacement current as negligible for non-relativistic scenarios; the situation here is highly non-relativistic, and so we shall also employ this approximation.

The equations now become
\be
\label{ScaledMaxwellEq}
  \jv=\nabla\times\Bv, \quad \dot{\Bv}=0, \quad \vr=0, \quad \nabla\cdot\Bv=0, \quad \dot{\overline{\rho}}_{ba}=\hub\bkrO,
\ee
and so
\bea
  \dot{\delta}_b+\left(1+w\right)\nabla\cdot\fgv_b+\frac{1}{2}\dot{h}&=&0, \\
  \dot{v}^i_b+\hub v^i_b+c_s^2\partial^i\delta_b
   +C^i_{b\rightarrow\gamma}=&=&\frac{1}{\bkrO}L^i. \nonumber
\eea
The total stress-energy tensor for these two fluids is
\bea
  \overline{T}^0_0=\bkr_b, \quad \overline{T}^i_0=\overline{T}^i_j=0,  \\
  \delta T^0_0=\bkr_b\delta_b+\frac{1}{8\pi a^4}B^2(\mathbf{x}), \quad
   \delta T^i_0=-\bkr_bv^i_b, \\
  \delta T^i_i=\frac{1}{8\pi a^4}B^m(\mathbf{x})B_m(\mathbf{x}) , \quad
   \delta \Pi^i_j=\frac{1}{4\pi a^4}\left(\frac{1}{3}B^m(\mathbf{x})B_m(\mathbf{x})\delta^i_j
   -B^i(\mathbf{x})B_j(\mathbf{x})\right) .
\eea
The scaled magnetic field $B(\mathbf{x})$ is independent of time and so, if one postulates a strength for the magnetic field in the present epoch and normalises the scale factor so that $a(\eta_0)=1$, the present value can be employed throughout. This assumes that the conductivity of the universe remains negligible after recombination; while the conductivity is likely to remain extraordinarily high, this is not going to be an entirely valid assumption. Moreover, this assumption neglects any non-linear effects on the magnetic field. In effect we have neglected the action of both gravity and, indeed, magnetohydrodynamics, and any conclusions based on the above model after recombination should be treated with caution.

\subsection{Fourier Space}
\label{MHD-Fourier}
Even with our simplifying assumptions, the cosmological MHD equations form a complicated set of equations; however, despite the non-linear nature of the Lorentz force and the stress-energy tensor of the magnetic fields, we shall find the system is rendered somewhat more tractable in Fourier space.

Consider first the scaled Maxwell equations and conservation of charge, which are linear and so will not be convolved. In Fourier space we see that these are
\be
\label{MaxwellCosmological}
  \dot{\mathbf{B}}(\mathbf{k})=0, \quad \kv\cdot\mathbf{B}(\mathbf{k})=0
\ee
where we have neglected the trivial equations. The magnetic stress-energy tensor in real space has the isotropic and anisotropic stresses
\beas
  4\pi a^4T^i_i=4\pi\tau^i_i=\frac{1}{2}B^m(\mathbf{x})B_m(\mathbf{x}), \\
  4\pi a^4\Pi^i_j=\frac{1}{3}B^m(\mathbf{x})B_m(\mathbf{x})\delta^i_j-B^i(\mathbf{x})B_j(\mathbf{x}),
\eeas
with a vanishing Poynting vector and the energy density equal to a third of the isotropic stress. We employ $\tau_{\mu\nu}$ to denote the scaled stress-energy tensor. The isotropic pressure is transferred to Fourier space by
\beas
  4\pi\tau^i_i(\mathbf{k})
   &=&\frac{1}{2}\int B^m(\mathbf{x})B_m(\mathbf{x})e^{i\mathbf{k}.\mathbf{x}}d^3\mathbf{x} \\
   &=&\frac{1}{2}\iint B^m(\mathbf{k}')B_m(\mathbf{x})e^{i(\mathbf{k}-\mathbf{k}').\mathbf{x}}
    d^3\mathbf{x}\frac{d^3\mathbf{k}'}{2\pi^3} \\
   &=&\frac{1}{2}\int B^m(\mathbf{k}')B_m(\mathbf{k-k}')\frac{d^3\mathbf{k}'}{2\pi^3}
\eeas
which is the expected convolution. Note that the reality of the fields in co-ordinate space requires $B(\mathbf{k})=-B^*(\mathbf{k})$. For the anisotropic stresses we have
\be
  4\pi a^4\Pi^i_j(\mathbf{k})=\int\left\{\frac{1}{3}B^m(\mathbf{k}')B_m(\mathbf{k-k}')\delta^i_j
   -B^i(\mathbf{k}')B_j(\mathbf{k-k}')\right\}\frac{d^3\mathbf{k}'}{2\pi^3}.
\ee
The scalar, vector and tensor parts are derived from this by applying combinations of $P^i_j(\mathbf{k})$, $Q^i_j(\mathbf{k})$, $\mathcal{P}^{Vi}_{jk}(\mathbf{k})$ and $\mathcal{P}^{Tij}_{ab}(\mathbf{k})$ as in equations (\ref{SVTProjections}). We find
\bea
  4\pi\tau_S(\mathbf{k})&=&4\pi Q^j_i(\mathbf{k})\left(\frac{1}{3}\tau^a_a(\mathbf{k})\delta^i_j+\Pi^i_j\right)
   \nonumber \\
   &=&\frac{3}{2}\kh^j\kh_i\left(4\pi\Pi^i_j(\mathbf{k})\right)
   \nonumber \\
   &=&\frac{1}{2}\int\left\{B^m(\mathbf{k}')B_m(\mathbf{k-k}')
   -3\left(\khv\cdot\mathbf{B}(\mathbf{k}')\right)\left(\khv\cdot\mathbf{B}(\mathbf{k-k}')\right)\right\}
   \frac{d^3\mathbf{k}'}{2\pi^3}
\eea
for the anisotropic pressure. For the magnetic vorticity we employ the symmetrised vector projection which gives us
\bea
  4\pi\tau_a^V(\mathbf{k})&=&4\pi\mathcal{P}^{Vij}_a(\mathbf{k})\left(\frac{1}{3}\tau^a_a(\mathbf{k})\delta^i_j+a^4\Pi^i_j\right)
   \\
   &=&\kh_{(i}P^{j)}_a(\mathbf{k})\left(4\pi a^4\Pi^i_j(\mathbf{k})\right)
   \nonumber \\
   &=&\int\bigg\{\left(\khv\cdot\mathbf{B}(\mathbf{k}')\right)\left(\khv\cdot\mathbf{B}(\mathbf{k-k}')\right)\kh_a
    -\frac{1}{2}\left(\khv\cdot\mathbf{B}(\mathbf{k}')\right)B_a(\mathbf{k-k}') \nonumber \\
   &&-\frac{1}{2}\left(\khv\cdot\mathbf{B}(\mathbf{k-k}')\right)B_a(\mathbf{k}')\bigg\}\frac{d^3\mathbf{k}'}{2\pi^3} ,
\eea
and the symmetrised tensor projection likewise gives us
\bea
  4\pi\tau^{aT}_b&=&4\pi a^4\mathcal{P}^{ajT}_{bi}\Pi^i_j \nonumber \\
  &=&\frac{1}{2}\int\bigg\{B^m(\mathbf{k}')B_m(\mathbf{k-k}')\left(\delta^a_b-\kh^a\kh_b\right)
    -2B^{(a}(\mathbf{k}')B_{b)}(\mathbf{k-k}')
   \nonumber \\ &&
    -\left(\hat{\mathbf{k}}\cdot\mathbf{B}(\mathbf{k}')\right)
     \left(\hat{\mathbf{k}}\cdot\mathbf{B}(\mathbf{k-k}')\right)\left(\delta^a_b+\kh^a\kh_b\right)
   \nonumber \\ &&
    +2\left(\hat{\mathbf{k}}\cdot\mathbf{B}(\mathbf{k}')\right)\kh^{(a}B_{b)}(\mathbf{k-k}')
    +2\left(\hat{\mathbf{k}}\cdot\mathbf{B}(\mathbf{k-k}')\right)\kh^{(a}B_{b)}(\mathbf{k}')
    \bigg\}\frac{d^3\mathbf{k}'}{2\pi^3}
\eea
We shall study this stress-energy tensor in much greater detail in the next chapter and content ourselves here with commenting that this highly-nonlinear form will naturally introduce to the system a significant measure of non-Gaussianity, even for a magnetic field which is itself Gaussian in nature. It should also be commented that while our simplified system will retain the Gaussianity of these fields the preservation of Gaussianity of a magnetic field in general is not necessarily to be expected.

Turning now to the Lorentz forces, in co-ordinate space we had that
\bdm
  4\pi\mathbf{L}(\mathbf{x})=\left(\nabla\times\mathbf{B}(\mathbf{x})\right)\times\mathbf{B}(\mathbf{x})
\edm
for the infinitely conductive universe. Following the same process to push this into Fourier space, we have
\bea
  4\pi\mathbf{L}(\mathbf{k})&=&
  \iint\left(\nabla\times\mathbf{B}(\mathbf{k}')\right)\times\mathbf{B}(\mathbf{x})
   e^{i(\mathbf{k-k}').\mathbf{x}}d^3\mathbf{x}\frac{d^3\mathbf{k}'}{2\pi^3}
  \nonumber \\
  &=&-i\iiint\left(\mathbf{k}'\times\mathbf{B}(\mathbf{k}')\right)\times\mathbf{B}(\mathbf{k}'')
   e^{i(\mathbf{k-k'-k''}).\mathbf{x}}d^3\mathbf{x}\frac{d^3\mathbf{k}'}{2\pi^3}\frac{d^3\mathbf{k}''}{2\pi^3}
  \nonumber \\
  &=&-i\iint\left(\mathbf{k}'\times\mathbf{B}(\mathbf{k}')\right)\times\mathbf{B}(\mathbf{k}'')
   \delta(\mathbf{k-k'-k''})\frac{d^3\mathbf{k}'d^3\mathbf{k}''}{2\pi^3}
\eea
where we have used $(2\pi)^3\delta(\mathbf{k-k'})=\int e^{i(\mathbf{k-k}').\mathbf{x}}d^3\mathbf{x}$. Employing
\be
  \left(\left(\mathbf{A}\times\mathbf{B}\right)\times\mathbf{C}\right)
  =\mathbf{C}(\mathbf{A}\cdot\mathbf{B})-\mathbf{B}(\mathbf{A}\cdot\mathbf{C})
\ee
we can write
\be
  4\pi\mathbf{L}(\mathbf{k})=-i\int\mathbf{B}(\mathbf{k}')\left(\kv'\cdot\mathbf{B}(\mathbf{k-k}')\right)
  \frac{d^3\mathbf{k}'}{2\pi^3}
\ee
where we have also used Maxwell's third equation, $\kv\cdot\mathbf{B}(\mathbf{k})=0$. If we now use the freedom we have in our integration wavemode to write $\mathbf{k}'\rightarrow\mathbf{k}-\mathbf{k}'$ then $d^3\mathbf{k}'\rightarrow -d^3\mathbf{k}'$ and the Lorentz force becomes the simple
\be
  4\pi L_j(\mathbf{k})=ik\int B_j(\mathbf{k-k}')\left(\khv\cdot\mathbf{B}(\mathbf{k}')\right)
  \frac{d^3\mathbf{k}'}{2\pi^3}
\ee
where again we have used Maxwell's third equation. We can now split this into a scalar part,
\bdm
  4\pi L_S=-i\khv\cdot\mathbf{L}(\mathbf{x})=k\int\left(\khv\cdot\mathbf{B}(\mathbf{k}')\right)
  \left(\khv\cdot\mathbf{B}(\mathbf{k-k}')\right)\frac{d^3\mathbf{k}'}{2\pi^3}
\edm
which, comparing with the scalar pressures is
\be
\label{ScalarLorentzForce}
  L_S=\frac{2k}{3}\left(\tau^i_i-\tau_S\right)=\frac{2k}{3}P^i_j(\mathbf{k})\tau^j_i(\mathbf{k}) ,
\ee
and an (unsymmetrised) vector part
\bdm
  4\pi L_i^V(\mathbf{k})=P^a_i(\mathbf{k})L_a(\mathbf{k})
  =ik\int\left(\khv\cdot\mathbf{B}(\mathbf{k'})\right)\left(B_i(\mathbf{k-k}')
   -\left(\khv\cdot\mathbf{B}(\mathbf{k-k}')\right)\kh_i\right)\frac{d^3\mathbf{k}'}{2\pi^3}
\edm
which is readily seen to be
\be
\label{VectorLorentzForce}
  L_i^V(\mathbf{k})=ik\tau_i^V(\mathbf{k}) .
\ee
That is, the scalar contribution to the Lorentz force is directly proportional to the difference of the iso- and aniso-tropic pressures, while the vortical contribution to the Lorentz force is directly proportional to the magnetic vorticity itself, when seen in Fourier space.

Considering the full baryon fluid, then, aligning the co-ordinate axes along the Fourier mode and employing the vector and tensor bases defined in \S\ref{BoltzmannVectorPert} results in the stress-energy tensor
\be
  \overline{T}^0_0=-\bkr_b, \quad \overline{T}^0_i=\overline{T}^i_j=\Pi^i_j=0 \\
\ee
in the background and
\bea
  \delta T^0_0=-\bkr_b\delta_b-\frac{1}{a^4}\tau^i_i, &&
  \delta T^{i}_{(0)S}=-\bkr_bv^b_S, \quad
  \delta T^{i}_{(0)V}=-\bkr_bv_{bV}^i, \nonumber \\
  \delta T^i_i=\frac{1}{a^4}\tau^i_i, &&
  \delta T_S=\frac{1}{a^4}\tau_S , \quad
  \delta T^V_i=\frac{1}{a^4}\tau^V_i , \\
  \delta \T{T}_+&=&\frac{1}{8\pi a^4}\int\left\{\fb_2(\mathbf{k}')\fb_2(\mathbf{k}-\mathbf{k}')
   -\fb_1(\mathbf{k}')\fb_1(\mathbf{k}-\mathbf{k}')\right\}\frac{d^3\mathbf{k}'}
   {\left(2\pi\right)^3}, \nonumber \\
  \delta \T{T}_\times&=&\frac{-1}{8\pi a^4}\int\left\{\fb_1(\mathbf{k}')\fb_2(\mathbf{k-k}')
   +\fb_1(\mathbf{k-k}')\fb_2(\mathbf{k}')\right\}\frac{d^3\mathbf{k}'}{\left(2\pi\right)^3}. \nonumber
\eea
The equation of mass continuity becomes
\be
\label{MassContinuityMHD}
  \dot{\delta}_b+\left(1+w\right)kv_S+\frac{1}{2}\dot{h}=0 .
\ee
The scalar component of the Euler equation is
\be
\label{ScalarEulerMHD}
  \dot{v}^b_S+\hub v^b_S-kc_s^2\delta+\Sc{c}_{b\rightarrow\gamma}=\frac{2}{3\bkrO}k\left(\tau^i_i-\tau_S\right)
\ee
and the vector component is
\be
\label{VectorEulerMHD}
  \V{\dot{v}}_i+\hub\V{v}_i+{}^{(V)}c_i^{b\rightarrow\gamma}=\frac{i}{\bkrO}k\tau^V_i .
\ee

This then completes the system of equations that one would need to include in a Boltzmann integrator -- CMBFast \cite{ZaldarriagaSeljak97} would be ideal due to its use of synchronous gauge -- to include a primordial magnetic field along with a standard cosmology.

\section{Damping of Cosmological Magnetic Fields}
\label{Sec-DampingScaleMHD}
In this section we shall review the pertinent results of \cite{BrandenburgEnqvistOleson96,DurrerFerreiraKahniashvili00,MackKahniashviliKosowsky02,JedamzikKatalinicOlinto98,SubramanianBarrow98-MHD} governing the scales at which magnetohydrodynamical modes in the universe are damped by viscous processes. This scale, $k_D$, it should be emphasised, is distinct from an inherent magnetic damping scale arising from some particular magnetogenesis model, being viscous in origin and arising from the shear drag imposed on the baryons by the photon fluid. The primordial damping scale is likely to be very small and constant, while the viscous damping scale will be somewhat larger and time-dependent. The effective damping scale of the magnetic field will then be
\be
  k_c(\eta)=\mathrm{min}\left(k_c^{\mathrm{prim}},k_D(\eta)\right).
\ee
At some epoch, therefore, the effective magnetic damping scale will become time-dependent.

Mode damping is a small-scale effect of the photon (or neutrino) viscous drag on the coupled photon-baryon fluid; to analyse it we return to real space and, for ease of mathematics, linearise the magnetic field around an ordered background with
\bdm
  \mathbf{B}=\mathbf{B}_0+\mathbf{B}_1, \quad \dot{\mathbf{B}}_0=0, \quad \mathbf{B}_0=B_0\hat{z} .
\edm
We employ a shear viscosity to model the photon viscosity (see appendix \ref{Appendix-ViscousFluids}), impose infinite conductivity and, since we are working on scales much less than the Hubble scale, neglect $\dot{a}/a$. We have the system of equations
\bea
  \dot{\delta}+\nabla\cdot\mathbf{v}=0, \quad \frac{\dot{\rho}_{ba}}{\ra}\approx 0, \\
  \dot{\mathbf{v}}+c_s^2\nabla\delta
   =\frac{1}{4\pi(\bkrO+\bkpO)}\left(\mathbf{B}_0\times\left(\nabla\times\mathbf{B}_1\right)\right)
   +\xi_a\left(\nabla^2\mathbf{v}+\frac{1}{3}\nabla\left(\nabla\cdot\mathbf{v}\right)\right), \\
  \dot{\mathbf{B}}_1=\nabla\times\left(\mathbf{v}\times\mathbf{B}_0\right) .
\eea
We may then derive a wave equation for $\mathbf{v}$,
\be
  \ddot{\mathbf{v}}-c_s^2\nabla\left(\nabla.\mathbf{v}\right)=
   \mathbf{v}_A\times\left(\nabla\times\left(\nabla\times\left(\mathbf{v}\times\mathbf{v}_A\right)\right)\right)
   +\xi_a\left(\nabla^2\dot{\mathbf{v}}+\frac{1}{3}\nabla\left(\nabla\cdot\dot{\mathbf{v}}\right)\right) .
\ee
We have defined the Alfv\'en velocity
\be
  \mathbf{v}_A=\frac{\mathbf{B}_0}{\sqrt{4\pi(\bkrO+\bkpO)}} .
\ee

In Fourier space this becomes
\bea
 \lefteqn{\ddot{\mathbf{v}}+\left(c_s^2+v_A^2\right)\left(\mathbf{k}\cdot\mathbf{v}\right)\mathbf{k}
  } \nonumber \\ &&
  =\left(\mathbf{k}\cdot\mathbf{v}_A\right)\left(\mathbf{v}(\mathbf{k}\cdot\mathbf{v}_A)
     -\mathbf{k}(\mathbf{v}\cdot\mathbf{v}_A)
   -\mathbf{v}_A(\mathbf{k}\cdot\mathbf{v})\right)
   -\xi_a\left(k^2\mathbf{v}+\frac{1}{3}\mathbf{k}(\mathbf{k}\cdot\mathbf{v})\right) .
\eea

We align the magnetic field with the $z$-axis and define $\mathbf{k}\cdot\mathbf{v}_A=kv_A\cos\theta$ and $\mathbf{v}\cdot\mathbf{v_A}=v_z$. Then we can consider various perturbation modes:

\begin{itemize}
\item Incompressible waves: Alfv\'en Waves

If we take $\nabla\cdot\mathbf{v}=0\Rightarrow\mathbf{k}\cdot\mathbf{v}=0$ then by the continuity equation we are considering perturbations that do not support density perturbations. In the wave equation, we then see that
\be
  \ddot{\mathbf{v}}+\frac{3\xi_a}{4\bkrO}k^2\dot{\mathbf{v}}+k^2v_A^2\cos^2\theta\mathbf{v}=0 .
\ee
These waves will then oscillate when $kv_A\cos\theta\gg 3\xi_ak^2/4\bkrO$, and will overdamp when $kv_A\cos\theta\ll 3\xi_ak^2/4\bkrO$. For photons, we have the shear viscosity \cite{SubramanianBarrow98-MHD}
\be
  \xi_a=\frac{8}{15}\frac{\pi^2}{30}\Theta^4l_\gamma
\ee
where $\Theta$ is the temperature and $l_\gamma=(n_e\sigma_T)^{-1}$ the mean photon diffusion length. From here it can be shown that the ratio of the friction term to the natural frequency is approximately
\be
  \frac{D}{\omega_0}\approx 530\frac{k}{a(\eta)}\frac{l_\gamma(\eta)}{B_{-9}}
\ee
where $B_{-9}$ is the background field in units of nano-Gauss. Many modes will thus be overdamped. Of the two solutions of a heavily damped oscillator we expect one, with a large initial velocity, to damp rapidly, and another with a small initial velocity to effectively freeze. It is this second mode that will survive Silk damping.

\item Magnetosonic Waves

Consider now the compressible waves; defining $v_k=\khv.\hat{\mathbf{v}}$ and taking inner products of the wave equation with first $\khv$ and then $\hat{\mathbf{z}}$ we can find the two equations
\bea
  \ddot{v}_k+\left(c_s^2+v_A^2\right)k^2v_k-k^2v_A^2\cos^2\theta v_z+\frac{\xi_a}{\overline{\rho}_{ba}}k^2\dot{a}&=&0, \\
  \ddot{v}_z+c_s^2k^2\cos\theta v_k+\frac{3\xi_a}{4\overline{\rho}_{ba}}k^2\dot{v}_z+
   \frac{\xi_a}{4\bkrO}k^2\cos\theta\dot{v}_k .
\eea

Taking a mode parallel to the magnetic field ($\mathbf{k}\parallel\mathbf{v}_A$) gives
\be
  \ddot{v}_k+\frac{\xi_a}{\bkrO}k^2\dot{v}_k+c_s^2k^2v_k=0
\ee
which is a standard, Silk-damped, sound wave. If we consider a wavevector perpendicular to the magnetic field, $\mathbf{k}\perp\mathbf{v}_A$, we instead find
\be
  \ddot{v}_k+\frac{\xi_a}{\bkrO}k^2\dot{v}_k-\left(c_s^2+v_A^2\right)k^2v_k=0
\ee
which is a damped fast magnetosonic wave. This damps in the same way as a standard sound wave.

Finally if we consider a wavevector tilted with respect to the field, $\mathbf{k}.\mathbf{v}_A=kv_A\cos\theta$, we have a slow magnetosonic wave. Subramanian and Barrow showed that this behaves approximately as the Alfv\'en waves, though the analysis is slightly more involved. A different approach with the same conclusions was presented in \cite{JedamzikKatalinicOlinto98}.

\end{itemize}

To briefly summarise, then, we have seen that, during the tightly-coupled epoch, fast magnetosonic waves sourced by a magnetic field will damp as sound waves, while one mode apiece of the slow magnetosonic waves and Alfv\'en waves will become strongly overdamped and freeze, thus surviving Silk damping.

Free-streaming occurs when the mean free path of the photon $l_\gamma=(n_e\sigma_T)^{-1}$ grows larger than the wavelength of a given mode. At decoupling this is around $3$Mpc, and after recombination all of the modes will freely stream. Accurate evaluation of modes in the free-streaming r\'egime requires a full treatment of the photon and baryon fluids; however, Subramanian and Barrow provide an approximate treatment, considering the Euler equation
\be
  \dot{\mathbf{v}}+\hub\mathbf{v}+(\mathbf{v}\cdot\nabla)\mathbf{v}=-\frac{1}{\rho_b}\nabla p_b+\frac{1}{\rho_b}\mathbf{j}\times\mathbf{b}-\frac{1}{a}\nabla\phi-\frac{4}{3}\frac{\rho_\gamma}{\rho_b}an_e\sigma_T\mathbf{v}
\ee
where $\phi$ is the gravitational potential from Poisson's equation. We then have two chief r\'egimes, one in which the magnetic pressure dominates over the baryon pressure, and the opposite. When the baryon pressure dominates, one may derive an evolution equation for the velocity
\be
  \ddot{\mathbf{v}}+\left(\hub+\frac{4\rho_\gamma}{3\rho_b}an_e\sigma_T\right)\dot{\mathbf{v}}+\frac{4\rho_\gamma}{3\rho_b}k^2v_A^2\mathbf{v}=0 .
\ee
The Hubble damping will be much smaller than the viscous damping and so the ratio between the damping and natural frequency is
\be
  \frac{D}{\omega_0}=\sqrt{\frac{4\rho_\gamma}{3\rho_b}}\frac{an_e\sigma_T}{kv_A}
   \approx (3\times 10^3)\sqrt{\frac{\rho_\gamma}{\rho_b}}((k/a)l_\gamma(\eta)B_{-9})^{-1}
\ee
which, given that we are free-streaming and $(k/a)l_\gamma\approx 1$, and that we are considering particularly weak fields, is strongly overdamped.

Employing a ``terminal-velocity'' approximation for the low-amplitude mode which is effectively frozen -- that is, neglecting the acceleration -- one can find rough solutions to the equations. In this r\'egime the solution is rapidly
\be
  v(\eta)=v(\eta_f)e^{-\left(\int_{\eta_f}^\eta\frac{\omega_0^2}{D}d\eta\right)}
   =v(\eta_f)e^{-v_A^2\cos^2\theta\int {k^2}{a^2}l_\gamma dt}=v(\eta_f)e^{-\frac{k^2}{k_D^2}} .
\ee
where we have swapped the integral to co-ordinate time for neatness. This allows us to identify the damping scale
\be
  k^{-2}_D=v_A^2\cos^2\theta\int\frac{l_\gamma(t)}{a^2(t)}dt .
\ee
This is at a minimum at recombination, and evaluated at recombination and comparing with the Silk scale we have
\be
\label{FreeStreamDampingScale}
  k^{-1}_D\approx\sqrt{\frac{3}{5}}v_AL_S .
\ee
The analysis in the case of a stronger field is more complicated and unlikely to be necessary given the relative weakness of the primordial cosmological field.

In general we will take the damping scale to be approximately
\be
  k^{-1}_D\approx\sqrt{\frac{3}{5}}v_Al_\gamma
\ee
where $l_\gamma$ is the photon diffusion length. The Alfv\'en modes that survived Silk damping will be converted into fast and slow magnetosonic waves. The damping scales are similar to that above for fast waves, while for slow waves the baryon speed of sound dominates over the Alfv\'en velocity. The time-dependence of the viscous damping scale arises from the photon diffusion length; for the more complicated case of a tangled field configuration, addressed in chapter \ref{Chapter-CMB}, the Alfv\'en velocity is also time-dependent.

For further discussion on damping scales see appendix A of \cite{CapriniDurrer02}.

\section{Conclusions}
We have presented, in a synchronous gauge, the sources and Lorentz forces arising from a magnetic field of arbitrary size and type. We have then reduced the resulting system to a realistically tractable approximation reminiscent of standard MHD and found that the only difference between a magnetised and a standard cosmology at linear order is the existence of a Lorentz force. This Lorentz force is proportional to the stresses when viewed in Fourier space. This system is in an ideal form to transfer to some Boltzmann code; our motivation is to construct a formalism suitable for CMBFast, although we could also employ CMBEasy \cite{CMBEasy}. We also briefly considered the damping scales relevent to the problem, working in an altered formalism with a large-scale homogeneous field and a small perturbation to address issues that are otherwise only present in second-order magnetised theories. We derived the form of the viscous damping scale that will determine both the scale on which Alfv\'en and slow magnetosonic perturbations are cut-off, and also the scale on which the magnetic field will ultimately damp.

\chapter{Statistics of Cosmic Magnetic Sources}
\label{Chapter-StatsOfCosmicMagneticFields}
\label{Chapter-SourceStats}
\label{Chapter-SourceStatistics}
\section{Tangled Magnetic Fields}
\label{Sec-Tangled}
In this chapter we consider in detail the nature and statistics of the magnetic sources through a combination of simulated realisations and analytical study. Both have advantages and disadvantages; the simulated fields can be vastly more general but are limited in dynamic range and computational power, while the analytic fields are heavily restricted by mathematical limitations. In line with the modern treatment, we do not assume that the magnetic field is a large-scale, ordered field with a small arbitrary perturbation, but instead consider entirely inhomogeneous fields tangled on some length scale (see for example Mack \emph{et. al.} \cite{MackKahniashviliKosowsky02}). For simplicity (and to compare with the previous literature), we assume that the magnetic fields are random variables obeying a Gaussian probability distribution function; however, our realisations may be formulated more generally than this. It should however surprise no-one that a non-Gaussian magnetic field generates a non-Gaussian stress-energy, and so the Gaussian case is likely to be the most intuitively informative. Analytically we are currently restricted to Gaussian fields.
From equation (\ref{MaxwellInCosmology}), $\mathbf{b}\propto a^{-2}$; we will be modelling the time-independent scaled field $B$. The only time-dependence then arises from the damping scale. Since this will be associated with the grid-size of the realisations this is not here an issue, though it may need to be addressed when we come to wrap our statistics onto the CMB -- for some configurations this will lead to time-dependent stresses, thus introducing decoherence into the sourced perturbations.

Magnetic fields, being a non-linear source with a full anisotropic stress, naturally provide sources for scalar, vector and tensor perturbations. For scalar perturbations one does not expect a significant effect on large scales. However, the contribution to the temperature auto-correlation on the microwave sky can begin to dominate at an $l$ of $1000-3000$ depending on spectral index (see \S\S\ref{Sec-CMB-AnalyticLiterature},\ref{Sec-CMB-NumericalLiterature} or \cite{SubramanianBarrow02, Lewis04-Mag, YamazakiIchikiKajino04}). This effect comes both from the impact of the magnetic energy and anisotropic pressure directly onto the spacetime geometry and from the Lorentz forces imparted onto the coupled proton-electron fluid. One also expects a significant impact on the vector perturbations as compared to the standard picture since the magnetic fields will both directly generate vector perturbations in the spacetime and contribute vortical Lorentz forces. Prior to neutrino decoupling a primordial magnetic field acts as a source for gravitational waves; following neutrino decoupling there will also be a contribution from the neutrinos, which Lewis has shown serves to cancel much of the magnetic stress \cite{Lewis04-Mag}. Since the Lorentz forces and the stress-energy tensor are both quadratic in the magnetic field, we also expect a level of non-Gaussianity to be imprinted onto the fluid and perturbations, even for a magnetic field that is itself Gaussian.

The features of the magnetic field -- the Lorentz forces and the direct sourcing of geometric fluctuations -- are all contained within the stress-energy tensor.  We can then consider the statistics of the stress-energy tensor alone and be hopeful of characterising the majority of the non-Gaussian effects that might impact on the CMB. The non-Gaussianity predicted from our analysis and our simulations can be projected onto the CMB sky by folding them with the transfer functions generated by the modified CAMB code \cite{Lewis04-Mag,CAMB} or a modification of CMBFast \cite{KohLee00,YamazakiIchikiKajino04,ZaldarriagaSeljak97}. The great benefit of our formalism derived in the previous chapter is the ease with which we may implement it into CMBFast. Care would have to be taken, however, with the initial conditions -- see Lewis \cite{Lewis04-Mag} and Giovannini \cite{Giovannini04-CMB}. In principle, the $B$-mode polarisation will give the purest magnetic signal, being sourced purely by vector and tensor perturbations; however, care would have to be taken to disentangle these from any $B$-modes caused by the gravitational lensing of the dominant $E$-mode and from other early-universe sources of gravitational wave.\footnote{Magnetic fields will also cause Faraday rotation from $E$ modes into $B$ modes \cite{CampanelliEtAl04,ScoccolaHarariMollerach04,KosowskyEtAl04,KosowskyLoeb96}; support for scalar modes in a uniform magnetic field was added to CMBFast by \cite{ScoccolaHarariMollerach04} and also to CMBFast for tangled fields by \cite{KosowskyEtAl04}, both in the case of small rotations.} Nonetheless, even on the $B$-mode the contaminants will be myriad. The inherently non-Gaussian nature of the sources, both gravitational and the Lorentz forces, offers an alternative route to characterising the nature of a primordial magnetic field.

We shall concentrate our study of the statistics on the space-space part of the magnetic stress-energy tensor, the stresses, since the magnetic energy density is proportional to the isotropic pressure, and in our cosmological setting the Poynting vector vanishes. The Lorentz forces are directly proportional to components of the stress-energy; loosely speaking, their properties can be found by applying prefactors of the wavemodes to the stress-energy results.


The (scaled) stress-energy tensor can be written
\be
\label{ScaledStressEnergy}
  \tau_{ab}(\mathbf{k})=\frac{1}{2}\gamma_{ab}\tilde{\tau}^i_i(\mathbf{k})-\tilde{\tau}_{ab}(\mathbf{k}).
\ee
where
\bdm
  \tilde{\tau}_{ab}(\mathbf{k})=\int B_a(\mathbf{q})B_b(\mathbf{k}-\mathbf{q})d^3\mathbf{q}.
\edm

This chapter of the thesis is derived from work by myself and Robert Crittenden \cite{BrownCrittenden05}.

\subsection{Underlying Statistics of the Fields}
\label{Sec-Stats-Underlying}
We begin by specifying the underlying statistics of the tangled magnetic fields. In Fourier space, Maxwell's second law implies that
\be
  \left<\mathbf{B}_a(\kv)\mathbf{B}_b^*(\mathbf{k}')\right>
  =\left(\mathcal{P}(k)P_{ab}(\mathbf{k})+\frac{i}{2}\mathcal{H}(k)\epsilon_{abc}\hat{k}^c\right)\delta(\kv-\mathbf{k}')
\ee
where $\mathcal{P}({k})$ is the magnetic field power spectrum, $P_{ab}$ is the operator projecting vectors and tensors onto a plane orthogonal to $\kh_a$ and $\kh_b$,
\be
  P_{ab}(\mathbf{k})=\gamma_{ab}-\kh_a\kh_b ,
\ee
and $\mathcal{H}(k)$ is the power spectrum of the anti-symmetric helical term (see for example \cite{CapriniDurrerKahniashvili03,PogosianEtAl01,KahniashviliRatra05}.) Here we have assumed the fields are statistically isotropic and homogeneous. If the magnetic fields are Gaussianly distributed, then all their statistics are determined by their power spectrum. In the interests of simplicity we henceforth assume that the helical component of the field vanishes; however we should note that helicity may well be a generic feature of magnetic fields generated in the very early universe (e.g. the non-conservation of helicity during the lepton stage of the universe \cite{SemikozSokoloff05} and from the magnetic-axion coupling \cite{CampanelliGiannotti05}).


The power spectrum is often taken to be a simple power law,
\be
  \mathcal{P}({k})=Ak^n.
\ee
To avoid divergences, these power law spectra are generally assumed to have some small scale cutoff usually associated with the photon viscosity damping scale; see \S\ref{Sec-DampingScaleMHD} and \cite{SubramanianBarrow98,MackKahniashviliKosowsky02,JedamzikKatalinicOlinto98}.  Durrer and Caprini \cite{CapriniDurrer02, DurrerCaprini03} demonstrate that, for a causally-generated magnetic field, the spectral index must be $n\geq 2$; it must also in any event be $n\geq -3$ to avoid over-production of long-range coherent fields.
They also demonstrate that nucleosynthesis limits on the gravitational waves produced by the magnetic fields place extremely strong bounds on magnetic fields, to the level of $10^{-39}$G for inflation-produced fields with $n=0$, although this has been contested (see \S\ref{Nucleosynthesis-Indirect}). For spectra that might realistically imprint on the microwave background we must consider spectra with $n<-2$, which are much less tightly constrained. For purposes of comparison we shall usually take either $n=0$ for a flat spectrum -- where ``flat'' refers to the spectrum of primordial magnetic fields themselves -- in which each mode contributes equally, or $n=-2.5$ for a spectrum nearing the ``realistically observable'' $n\approx-3$, which we shall refer to as ``steep''. Such fields can be produced by inflation (for example, \cite{BambaYokoyama04,Ratra92}).

It is worth emphasising that, while we generally restrict ourselves to power-law spectra, this is not a necessity. For the analytical results we will numerically integrate the formulae rather than attempt approximations, while for our realisations the form of the power spectrum can be entirely arbitrary; in both cases it is as easy to employ a non-power law spectrum as it is a power-law. It is also a simple matter to employ a non-sharp damping scale -- we could, for example, employ an exponentially or Gaussian-damped tail for greater freedom in modelling the effective microphysics. We do not explore such spectra much, but we do present some few results for an exponentially-damped causal field, qualitatively similar to the spectrum derived by Matarrese \emph{et. al.} \cite{MatarreseEtAl04} for a field sourced by second-order vorticity in the electron-baryon plasma (see also \cite{BetschartDunsbyMarklund04,MatarreseEtAl04,GopalSethi04,TakahashiEtAl05}.) However, the results presented should not be taken to be much more than indicative of the potential nature of such a field; fields produced by the product of two Gaussian variables will possess a $\chi^2$ probability distribution function, while our results are for a Gaussian field. Moreover, the dynamic range presented in the simulations is significantly shorter for our results than should be for that type of field. The formalism we are employing will also not apply to such a field -- not least due to its time-dependence -- and their amplitude is very small. Similarly fields sourced at recombination \cite{BerezhianiDolgov03} or reionisation \cite{GnedinFerraraZweibel00}, for example, arise at a very small scale. The question of the statistical nature of both recombination and very early universe fields has not been well explored in the literature. Here we present results from a Gaussian field as an illustrative example; any conclusions about the non-Gaussian signatures of magnetic fields will depend sensitively on this assumption, and we plan to explore more general scenarios in future work.

The normalisation of a magnetic power spectrum is typically fixed by reference to a particular comoving smoothing scale $\lambda$ and the variance of the field strength at this scale, $B_\lambda$. Specifically, we smooth the field by convolving it with the Gaussian filter
\be
  f(k)=\mathrm{exp}\left(-\frac{\lambda^2k^2}{2}\right)
\ee
and define the variance of the field strength at the scale $\lambda$ by
\be
  B_\lambda^2=\left<B^a(\mathbf{x})B_a(\mathbf{x})\right>,
\ee
implying that the power spectrum and $B_\lambda$ are related by
\be
\label{SpectrumNormalisation}
  B_\lambda^2=\int d^3\mathbf{k}\mathcal{P}({k})e^{-\lambda^2k^2}.
\ee
This allows us to relate the astronomically observed field strengths at, say, cluster scales, to the amplitude of the magnetic power spectrum. We shall sometimes write
\be
\label{AmplitudePowerSpectrum}
  \mathcal{P}(k)=A\mathcal{Q}(k) .
\ee

\subsection{Realisations of Magnetic Fields}
\label{Sec-MagneticRealisations}
To aid our study of the non-Gaussian properties of tangled magnetic fields, we create static realisations of the fields numerically. The great advantage of doing so is that we are not bound by analytic difficulties; for example, while we can derive analytical forms for the power and bispectra for Gaussian fields it is not clear how we should approach the integrations for non-Gaussian fields. Realisations do not have this limitation. We create the fields on a grid in Fourier space of size $l_{\mathrm{dim}}^3$, where $l_{\mathrm{dim}}$ is typically 100-200. Since the fields are solenoidal we can generate the three magnetic field components for each $k$-mode using two complex Gaussian uncorrelated random fields with unit variance,
\bdm
 \mathbf{C}=\left(\begin{array}{c}C_1\\C_2\end{array}\right).
\edm
We then determine the magnetic field Fourier components by applying a rotation matrix,
\bdm
  \mathbf{B}=\left(\begin{array}{c}B_x\\B_y\\B_z\end{array}\right)=\mathbf{R}\cdot\mathbf{C},
\edm
where $\mathbf{R}$ is a $3 \times 2$ matrix. From the definition of the magnetic field power spectrum, we see that to get the proper statistical properties, we require
\bdm
  \left<B_{a}B^*_{b}\right>=R_{am}\left<C^mC^n\right>R^T_{nb}=(\mathbf{R}\cdot\mathbf{R}^T)_{ab}=\mathcal{P}(k)P_{ab}(\mathbf{k}).
\edm
While this does not specify the rotation matrix uniquely, it is straightforward to show that choosing the rotation matrix as
\be
\label{R}
 \mathbf{R}=\frac{\mathcal{P}(k)^{1/2}}{\sqrt{\kh^2_x+\kh^2_y}}\left(\begin{array}{cc}
   \kh_x\kh_z & \kh_y \\
   \kh_y\kh_z & -\kh_x \\
   -\left(\kh_x^2+\kh_y^2\right) & 0
  \end{array}\right)
\ee
will produce fields with the correct statistical properties. This rotation is well defined except in the case when $\kh_x=\kh_y=0.$  Here, $B_z=0$ and the other components are uncorrelated, so we instead choose
\be
  \label{R0} \mathbf{R}_0=\mathcal{P}(k)^{1/2}\left(\begin{array}{cc}1&0\\0&1\\0&0\end{array}\right) .
\ee
The reality of the fields is ensured by requiring $\mathbf{B}(\mathbf{-k})=\mathbf{B}^*(\mathbf{k})$.

Throughout, we are careful to avoid creating modes with frequencies higher than the Nyquist frequency of the grid, which could be aliased into power on other frequencies; this frequency is half the grid frequency. Since the quantity of greatest interest, the stress-energy, is a quadratic function of the fields, it will typically have power up to twice the cutoff frequency of the magnetic fields. To avoid aliasing these fields, we generally require that the magnetic field cutoff frequency be less than half the Nyquist frequency. We also have an infra-red cut-off which is the inevitable result of working on a finite grid. For steeply-red spectra the natural suppression at small-scales will allow us to expand our dynamic range up to $\sim l_{\mathrm{dim}}$.

\begin{center}
\begin{figure}
\includegraphics[width=\textwidth]{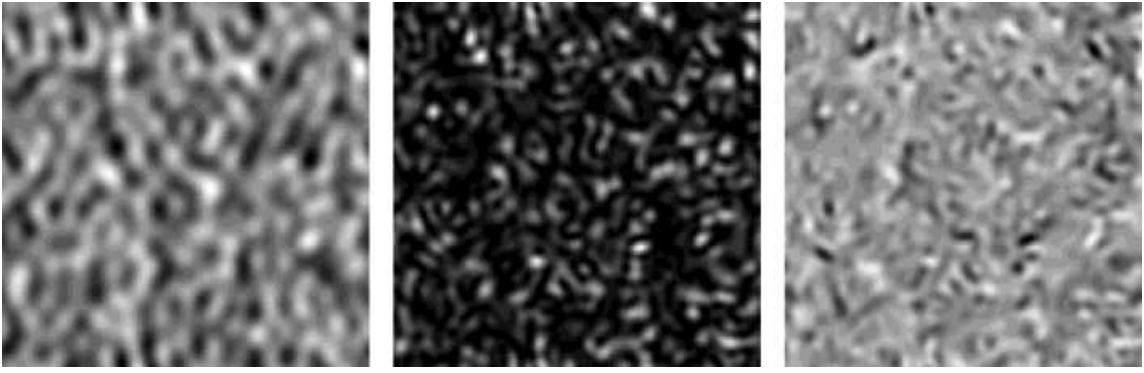}
\caption{Sample realisations of the magnetic field for a spectrum $n=0$. Left: Gaussian magnetic field slice, $B_{x}|_{z=0}$; centre: the (very non-Gaussian) isotropic pressure $\tau|_{z=0}$; right: the (slightly non-Gaussian) anisotropic pressure $\tau_S|_{z=0}$.  Compared to the magnetic field, non-linearity transfers power to smaller scales in the sources. (Figures by Robert Crittenden.)}\label{FieldSlices}
\end{figure}
\end{center}

Figure \ref{FieldSlices} shows a sample Gaussian realisation of one component of the magnetic field along a slice through the realisation, as well as the resulting trace and traceless components of the stress-energy (the isotropic and anisotropic pressures). Both the isotropic and anisotropic pressures show power on smaller scales than the fields themselves, a direct result of their non-linearity.  In addition, the isotropic pressure is darker, reflecting a paucity of positive fluctuations and a significant deviation from Gaussianity. The anisotropic pressure appears to be more similar to the magnetic field -- that is, relatively Gaussian. These observations will be made concrete in the next section when we consider the one-point statistics of the pressures.

\section{One-Point Moments}
\label{Sec-OnePoint}
There are many ways to characterise non-Gaussianity, particularly given such a strongly non-linear stress-energy term. In this section we briefly consider the skewness and kurtosis of the one-point probability distributions of the isotropic and anisotropic pressures. In this section the results we present are the mean of twenty realisations with a grid-size of $l_{\mathrm{dim}}=192$, and the errors quoted are one standard deviation.

The simplest to consider is the distribution of the trace part, since it is simply the square of the magnetic field. Despite the divergence-free condition, the three components of the magnetic field at a single point in space are uncorrelated and Gaussian. The product of $n$ Gaussian fields is a $\chi^2$ distribution with $n$ degrees of freedom, and so we expect the trace of the stress-energy to have a $\chi^2$ distribution with three degrees of freedom.

All the moments of a one-point distribution may be given by its moment generating function as
\be
  \mu'_n \equiv \langle X^n\rangle=\frac{\partial^n}{\partial t^n}\left.M(t)\right|_{t=0}
\ee
where, for a $\chi^2$ distribution with $p$ degrees of freedom
\be
  M(t)=\frac{1}{\left(1-2t\right)^{p/2}}.
\ee
The central moments are then readily calculated and the normalised skewness and kurtosis are defined to be
\be
  \gamma_1=\frac{\mu_3}{\mu_2^{3/2}}, \quad \gamma_2=\frac{\mu_4}{\mu_2^{2}}-3 .
\ee
We quickly find that, for the $\chi^2$ distribution, the normalised skewness and kurtosis are
\be
  \gamma_1=\sqrt{\frac{8}{p}}\approx 1.633, \quad \gamma_2=\frac{12}{p}=4
\ee
where the numerical results are for a distribution with $3$ degrees of freedom. The results from the realisations can be seen to be in agreement with the predictions; with a flat spectrum we find that, for the isotropic pressure, $\gamma_1=1.63\pm 0.01$ and $\gamma_2=3.99\pm0.05$. For a more realistically observable field, with a power spectrum of $n=-2.5$, say, we find $\gamma_1=1.61\pm0.01$ and $\gamma_2=3.92\pm0.05$. It is apparent that the statistics for the isotropic pressure are, as expected, relatively insensitive to the spectral index one employs.

The anisotropic stress is harder to characterise because it is not a local function of the fields, but contains derivatives of them. However, it is effectively the sum of the products of two Gaussian fields which are, for the most part, independent of each other. The distribution of the product of two independent Gaussians is non-Gaussian but is symmetric (actually following a modified Bessel distribution, as shown in the appendix of \cite{BoughnCrittenden05}.) Thus the effect of adding such terms is to dilute the skewness. That is, while the isotropic stress is the sum of three very skewed $\chi^2$ variables, the anisotropic stress is the sum of $\chi^2$ terms and symmetric modified Bessels, making the result less skewed.

\begin{figure}
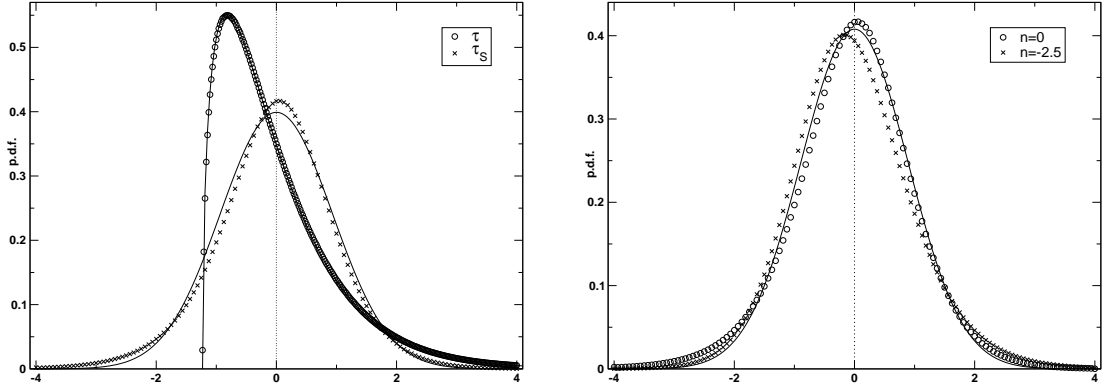
\begin{center}
\includegraphics[width=0.46\textwidth]{PDFs_n0.eps}\qquad\includegraphics[width=0.46\textwidth]{PDFs_comparison.eps}
\caption{The left figure shows the probability distribution of the isotropic and anisotropic pressures for a spectrum $n=0$, with a Gaussian distribution shown for comparison. The isotropic distribution is well fit by a $\chi^2$ curve. On the right we compare the anisotropic pressure distribution for different spectral indices. The $x$-axis is in units of the root mean square amplitude of the relevant field.}
\label{PDFs}
\end{center}\end{figure}

The probability distributions of the isotropic and anisotropic stresses for a flat spectrum are plotted in the left-hand panel of figure \ref{PDFs} along with a Gaussian and a $\chi^2$ distribution. The damping scale we employed was $k_c=l_{\mathrm{dim}}/4.1$ -- \emph{i.e.} just beneath half the Nyquist frequency. For the steep magnetic spectra, plotted in the right-hand panel, we again used twenty realisations and a grid-spacing of $l_{\mathrm{dim}}=192$ but with a damping scale at the size of the grid to ensure a reasonable mode coverage in the low-$k/k_c$ r\'egime. The anisotropic pressure distribution has quite different properties when the spectral index is changed, including a switch in sign of the skewness. For the flat spectrum we find $\gamma_1=-0.24\pm0.003$ and $\gamma_2=1.10\pm0.01$, while with a steep spectrum with a power spectrum of $n=-2.5$, we find  $\gamma_1=0.38\pm0.01$ and $\gamma_2=0.86\pm0.02$. The distributions are plotted in the right-hand side of the figure, again with a sample Gaussian, and the change in the skewness is readily apparent.

\section{Two-Point Moments}
\label{Sec-TwoPoint}
We next calculate the two point power spectra of the various perturbation types. These give a useful example of how the higher order calculations will proceed and provide a means of testing our realisations. Some of these have been previously calculated, such as the vector \cite{MackKahniashviliKosowsky02} and tensor \cite{CapriniDurrer02,DurrerFerreiraKahniashvili00, MackKahniashviliKosowsky02} power spectra, while the trace and traceless scalar auto-correlations and their cross correlation, to our knowledge, were first presented in our publication \cite{BrownCrittenden05}.

By the nature of the scalar-vector-tensor decomposition, we do not expect any cross correlations except between the trace and traceless scalar pieces. Thus, we consider five power spectra: one cross spectrum and the auto-spectra of the four pieces of the stress-energy.  We focus on constructing rotationally invariant spectra which will contain all the information in the general correlations.

In general, the power spectra will involve expectations of four magnetic fields. Since these are assumed to be Gaussian, they can be evaluated using Wick's theorem which, for four Gaussian fields, may be expressed as
\bdm
  \left<ABCD\right>=\left<AB\right>\left<CD\right>+\left<AC\right>\left<BD\right>+\left<AD\right>\left<BC\right> .
\edm
It is most useful to begin with the general two point correlation,
\bea
  \mathcal{B}_{abcd}=\lefteqn{\left<\tilde{\tau}_{ab}(\mathbf{k})\tilde{\tau}_{cd}^*(\mathbf{p})\right>=\delta(\mathbf{k}-\mathbf{p}) \int d^3\mathbf{k}'\mathcal{P}(k')\mathcal{P}(\left|\mathbf{k-k'}\right|)} \\ && \qquad
 \times \left(P_{ac}(\mathbf{k'})P_{bd}(\mathbf{k-k'})
 +P_{ad}(\mathbf{k'})P_{bc}(\mathbf{k-k'})\right) \nonumber
\eea
with indices $\{a,b,c\}$ running from 1 to 3. Note that there are two terms rather than three since we are interested in the perturbations from the mean value of the field. The power spectra of the various stresses may be obtained from this by applying an operator $\mathcal{A}^{abcd}$ formed from the relevant combinations of the projection operators (\ref{SVTProjections}) to yield
\be
\label{MagneticSourcePowerSpectra}
  \left<\tau_A(\mathbf{k})\tau_B(\mathbf{p})\right>=\delta(\mathbf{k}-\mathbf{p})
  \int d^3\mathbf{k}'\mathcal{P}(k')\mathcal{P}(\left|\mathbf{k-k'}\right|)
  \mathcal{F}_{AB}
\ee
with $A$ and $B$ denoting the two stress components and $\mathcal{F}_{AB}=\mathcal{F}_{AB}(\gamma,\mu,\beta)$ denoting the relevant angular integrand.  The relevant angles possible between the wavevectors have been defined as
\be
  \gamma=\hat{\mathbf{k}}\cdot\hat{\mathbf{k}}', \quad \mu=\hat{\mathbf{k}}'\cdot\widehat{\mathbf{k-k'}}, \quad \beta=\hat{\mathbf{k}}\cdot\widehat{\mathbf{k-k'}} ,
\ee
where $\widehat{\mathbf{k-k'}}$ denotes the unit vector in the direction of $\mathbf{k-k'}$.

The trace-trace correlation is found by applying the operator $\mathcal{A}^{abcd}=(1/4)\gamma^{ab}\gamma^{cd}$ whence we obtain
\be
  \mathcal{F}_{\tau\tau}=\frac{1}{2}\left(1+\mu^2\right) .
\ee
Similarly, we obtain the traceless scalar auto-correlation function by applying\newline $\mathcal{A}^{abcd}=(-1)^2Q^{ab}(\mathbf{k})Q^{cd}(\mathbf{k})$; some algebra yields the result
\be
  F_{\tau_S\tau_S}=2+\frac{1}{2}\mu^2-\frac{3}{2}\left(\gamma^2+\beta^2\right)-{3}\gamma\mu\beta+\frac{9}{2}\gamma^2\beta^2.
\ee
The cross correlation between the trace and traceless scalar pieces requires the operator\newline $\mathcal{A}^{abcd}=-(1/2)\gamma^{ab}Q^{cd}(\mathbf{k})$. This gives
\be
  F_{\tau\tau_S}=-1+\frac{3}{2}\left(\gamma^2+\beta^2\right)+\frac{1}{2}\mu^2-\frac{3}{2}\mu\gamma\beta.
\ee

For the vector and tensor contributions, it is useful to construct rotationally invariant combinations that can be relatively easily mapped onto the CMB. The divergenceless condition on the vectors implies that their correlation function can be written as
\be
\label{SpectrumTv}
  \left<\tau^V_a(\mathbf{k})\tau^{*V}_b(\mathbf{p})\right>=\frac{1}{2}
  \mathcal{P}^V(k)P_{ab}(\mathbf{k})\delta(\mathbf{k}-\mathbf{p}) .
\ee
where our definition differs by a factor of two from \cite{MackKahniashviliKosowsky02}. All the information is condensed in the rotationally invariant vector isotropic spectrum $\mathcal{P}^V(k)=\left<\tau^{iV}(\mathbf{k})\tau_i^{*V}(\mathbf{k})\right>.$ The operator necessary to recover this is
\be
  \mathcal{A}^{abcd}=\kh^{(a}P^{b)i}(\mathbf{k})\kh^{(c}P^{d)}_i(\mathbf{k})=\kh^a\kh^{(c}P^{d)b}(\mathbf{k})+\kh^d\kh^{(a}P^{b)c}(\mathbf{k}) .
\ee
The resulting angular term can then be shown to be
\be
  F_{\tau^V\tau^V}=1-2\gamma^2\beta^2+\mu\gamma\beta.
\ee

Similar arguments apply for the tensor correlations. The full tensor two-point correlation is
\be
\label{SpectrumTt}
  \left<\tau_{ab}^T(\mathbf{k})\tau_{cd}^{*T}(\mathbf{p})\right>=\frac{1}{4}\mathcal{P}^T(k)\mathcal{M}_{abcd}(\mathbf{k})\delta(\mathbf{k}-\mathbf{p}),
\ee
where $\mathcal{M}_{abcd}(\mathbf{k})=P_{ac}(\mathbf{k})P_{bd}(\mathbf{k})+P_{ad}(\mathbf{k})P_{bc}(\mathbf{k})-P_{ab}(\mathbf{k})P_{cd}(\mathbf{k})$ which, as can be readily shown, satisfies the transverse-traceless condition on the tensors, and $\gamma^{ac}\gamma^{bd}\mathcal{M}_{abcd}(\mathbf{k})=4$. We focus on the rotationally invariant tensor isotropic spectrum $\mathcal{P}^T(k)=\left<\tau^{ijT}(\mathbf{k})\tau_{ij}^{*T}(\mathbf{k})\right>$. Using the tensor projection operators and simplifying, the relevant operator is
\beas
  \mathcal{A}^{abcd}&=&\left(P^{i(a}(\mathbf{k})P^{b)j}(\mathbf{k})-(1/2)P^{ij}(\mathbf{k})P^{ab}(\mathbf{k})\right)\left(P_i^{(c}(\mathbf{k})P^{d)}_j(\mathbf{k})-(1/2)P_{ij}(\mathbf{k})P^{cd}(\mathbf{k})\right)
  \\ &=&P^{c(a}(\mathbf{k})P^{b)d}(\mathbf{k})-(1/2)P^{ab}(\mathbf{k})P^{cd}(\mathbf{k}) .
\eeas
This leads to a simple angular term of
\be
\label{TensorTwoPoint}
  F_{\tau^T\tau^T}=(1+\gamma^2)(1+\beta^2).
\ee

The vector and tensor results differ from those otherwise presented \cite{CapriniDurrer02,DurrerFerreiraKahniashvili00,MackKahniashviliKosowsky02} by $\pm(\beta^2-\gamma^2)$. It is straightforward, however, to see that if one redefines the integration wavemode as $\mathbf{k}'=\mathbf{k}-\mathbf{k}''$ then one maps $\mu'\rightarrow\mu,\;\beta'\rightarrow\gamma,\;\gamma'\rightarrow\beta$. On integration, then, the product $\gamma^2\beta^2$ is invariant while $\gamma^2-\beta^2$ may be taken to vanish. Our results are thus in agreement with those previously presented.

\begin{figure}\begin{center}\includegraphics{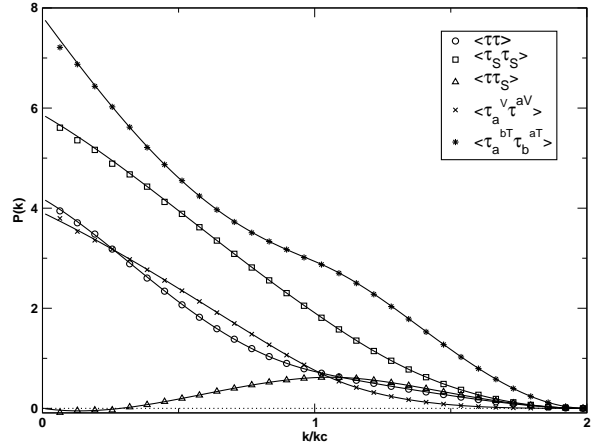}
\caption{Magnetic stress-energy power spectra ($n=0$) -- the realisations agree well with the analytic predictions. The error bars from the realisations are small except at very low $k$.}\label{Spectra_n0}\end{center}\end{figure}

We can compare numerical integrations of these power spectra with the results arising from the realised magnetic fields. Our results for a flat power spectrum ($n=0$) are presented in figure \ref{Spectra_n0} where $P(k)$ denotes the various spectra (all presented on the same scale). The agreement between the analytic results (lines) and a simulated field (data points) is striking. We have plotted the power spectra averaged over twenty realisations with a grid-size of $l_{\mathrm{dim}}=192$, a damping scale at $k_c=l_{\mathrm{dim}}/4.1$. We evaluated the 1-$\sigma$ error at each point but have not plotted it as it is negligible for all but the lowest modes.

We could also consider magnetic fields with $n\geq 2$ corresponding to causally-generated fields \cite{CapriniDurrer02}; the features for such fields are qualitatively similar to those for a flat spectra and there are no difficulties in evaluating the theoretical predictions in this r\'egime. We present a few results for a causal field damped to avoid piling power on an unrealistically small scale in \S\ref{Section-DampedCausal}.

\begin{figure}\begin{center}\includegraphics{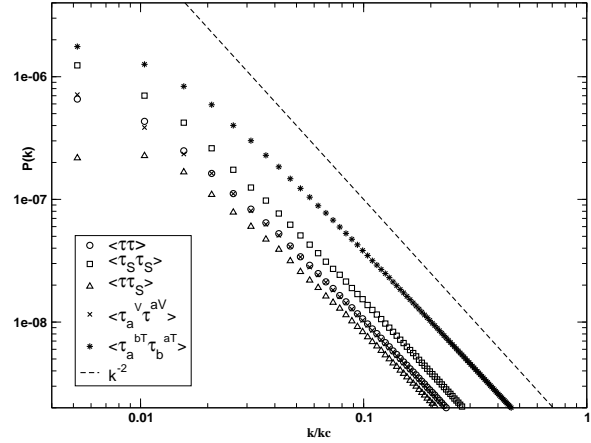}
\caption{Magnetic stress-energy power spectra ($n=-2.5$). The dashed line shows the spectral dependence expected naively, $\propto k^{2n+3}$. The turnover at low $k$ reflects the finite size of the grid.}
\label{Spectra_n-2.5}\end{center}\end{figure}

Analytic results can be found in certain limits. There are two r\'egimes of interest for the spectral index, as shown by Durrer et. al. \cite{DurrerFerreiraKahniashvili00}. For $n>-3/2$ the integrations are dominated by the cutoff scale. In this ultraviolet r\'egime, if $k\ll k_c$, the angular integrations are straight forward ($\mu\simeq-1,\; \beta\simeq-\gamma$). Relative to the trace correlation, the amplitudes of the other correlations are
\be
\label{LargestScaleTwoPoint}
  \frac{\langle\tau_S\tau_S\rangle}{\langle\tau\tau\rangle}=\frac{7}{5}, \quad
  \frac{\langle\tau\tau_S\rangle}{\langle\tau\tau\rangle}=0, \quad
  \frac{\langle\tau^V\tau^V\rangle}{\langle\tau\tau\rangle}=\frac{14}{15}, \quad
  \frac{\langle\tau^T\tau^T\rangle}{\langle\tau\tau\rangle}=\frac{28}{15}
\ee
respectively, agreeing with the large-scale results of both the realisations and the integrations. For $n<-3/2$ the situation is considerably more complex and we content ourselves with the results of our simulations in figure \ref{Spectra_n-2.5}. In this infra-red r\'egime the cutoff is unimportant and the spectra quickly approach the power law $P_A(k)\propto k^{2n+3}$ na\"{\i}vely expected from the $k$-integration. Also notable is the change of behaviour of the scalar cross-correlation; whereas this vanishes on large scales in the $n>-3/2$ r\'egime, it remains finite on large scales for $n<-3/2$ and so in principle might be observable on the sky. There are also the effects of the unphysical infra-red cut-off causing a suppression at low-$k$.

\section{Three-Point Moments}
\label{Sec-ThreePoint}
 In this section we focus on the three point moments in Fourier space, for which it is possible (if laborious) to obtain analytic expressions. The magnetic field realisations provide a way of exploring other kinds of non-Gaussianities which may arise.

At the three point level, it is no longer guaranteed that correlations between the scalar, vector and tensor pieces will vanish, and we present some of the first calculations of these here. There are many possible three point moments, but here we consider only the rotationally invariant combinations $\langle\tau\tau\tau\rangle$, $\langle\tau\tau\tau_S\rangle$, $\langle\tau\tau^{S}\tau^{S}\rangle$ and $\langle\tau^{S}\tau^{S}\tau^{S}\rangle$, the two scalar-vector correlations $\langle\tau\tau^{aV}\tau_a^V\rangle$ and $\langle\tau_S\tau^{aV}\tau_a^V\rangle$, the two scalar-tensor correlations $\langle\tau\tau^{abT}\tau_{ab}^T\rangle$ and $\langle\tau_S\tau^{abT}\tau_{ab}^T\rangle$, the vector-tensor cross-correlation $\langle\tau_a^V\tau^{aT}_b\tau^{bV}\rangle$ and the tensor auto-correlation $\langle\tau_{ab}^T\tau^{bT}_c\tau^{acT}\rangle$. We work throughout in Fourier space, where the three-point moments are known as the bispectra. One advantage of working in Fourier space is that the transfer functions, which fold in the fluid dynamics and describe the impact on the microwave background, are local.

\subsection{General Considerations}
\begin{figure}\begin{center}
\includegraphics{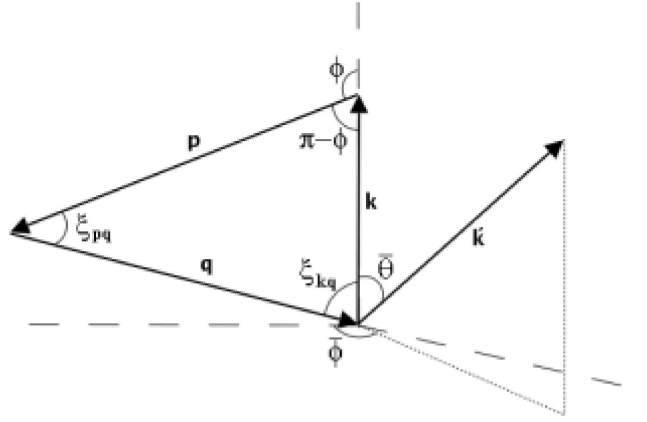}
\caption{The geometry for the bispectra calculations; $\mathbf{k},\mathbf{p},\mathbf{q}$ are the wavevectors for the three legs, while $\mathbf{k'}$ is an integration mode.}
\label{BispectraGeometry}\end{center}\end{figure}

In principle we can calculate all the three point statistics described above in Fourier space. The bispectra involve three wave modes, and since we assume the fields are homogeneous and isotropic, the sum of the three modes must be zero. Thus the bispectra are a function of the amplitudes of the modes alone (or alternatively, two amplitudes and the angle between them.) We denote different geometries by selecting a baseline $\mathbf{k}$ and a vector $\mathbf{p}$ making an angle $\phi$ with $\mathbf{k}$ and having an amplitude $p=rk$ (see figure \ref{BispectraGeometry}). We may then calculate $\mathbf{q}=-\mathbf{k}-\mathbf{p}$. For simplicity, we here concentrate on the colinear (degenerate) case in which $\mathbf{p}=\mathbf{k}$ implying $\mathbf{q}=-2\mathbf{k}$ -- that is, $r=1$ and $\phi=0$ -- though evaluating the contributions the primordial non-Gaussianities will have on the microwave background would require the full bispectra.

We calculate the bispectra analogously to the power spectra, although matters are complicated by the need to deal with expectations of six fields rather than four. The object of most general interest is $\mathcal{B}_{ijklmn}(\mathbf{k},\mathbf{p},\mathbf{q}) \equiv\left<\tilde{\tau}_{ij}(\mathbf{k})\tilde{\tau}_{kl}(\mathbf{q})\tilde{\tau}_{mn}(\mathbf{p})\right>,$ which is related to the expectation value of six magnetic fields,
\bea
  \lefteqn{\mathcal{B}_{ijklmn}(\mathbf{k},\mathbf{p},\mathbf{q})=\iiint d^3\mathbf{k}'d^3\mathbf{p}'d^3\mathbf{q}'} \nonumber \\ &&
   \times\left<B_i(\mathbf{k}')B_j(\mathbf{k}-\mathbf{k}')B_k(\mathbf{p}')B_l(\mathbf{p}-\mathbf{p}')B_m(\mathbf{q}')
    B_n(\mathbf{q}-\mathbf{q}')\right> .
\eea
As in the two-point case, all three-point moments of interest may be found by applying the relevant projection operator, $\mathcal{A}^{ijklmn}$, to this. Expanding this six-point correlation with Wick's theorem generates fifteen terms, eight of which contribute to the reduced bispectrum, that is, the bispectrum neglecting the one-point terms proportional to $\delta(\mathbf{k})$, $\delta(\mathbf{p})$ or $\delta(\mathbf{q})$. This leads eventually to
\bea
\label{SixFieldCorrelation}
  \lefteqn{\mathcal{B}_{ijklmn}(\mathbf{k},\mathbf{p},\mathbf{q})=
  \delta(\mathbf{k}+\mathbf{p}+\mathbf{q})\int d^3\mathbf{k}'\mathcal{P}(k')\mathcal{P}(\left|\mathbf{k}-\mathbf{k}'\right|)
  \mathcal{P}(\left|\mathbf{p}+\mathbf{k}'\right|)} \\ && \times P_{ik}(\mathbf{k}')\left(P_{jm}(\mathbf{k}-\mathbf{k}')
  P_{ln}(\mathbf{p}+\mathbf{k}')\right)
  +\left\{i\leftrightarrow j, p\rightarrow q\right\}, \{k\leftrightarrow l, m\leftrightarrow n.\} \nonumber
\eea
These eight terms reduce to the same contribution if the projection tensor $\mathcal{A}^{ijklmn}$ that recovers a set bispectrum is independently symmetric in $\{ij\}$, $\{kl\}$ and $\{mn\}$.

In the power spectra calculations, the geometry was straight forward; here it is considerably more complicated. The three wavevectors of the bispectrum are constrained by homogeneity to obey $\mathbf{k+q+p}=0$. Combined with the dummy integration wavevector, these define a four sided tetrahedron.  This has six edges, $\mathbf{k,q,p,k',k-k'}$ and $\mathbf{p+k'}$.  From these, we can generate fifteen unique angles which could arise in the bispectra calculations.  This is to be compared to just three edges and three angles required for the power spectra.  Clearly these angles are not all independent; they are, in fact, functions of just five underlying angles.

For our purposes, it is easiest to work with the fifteen which we separate into four hierarchies; those between the set wavevectors $\mathbf{k}$, $\mathbf{p}$ and $\mathbf{q}$, angle cosines of these vectors with $\mathbf{k}'$, cosines with $\mathbf{k}-\mathbf{k}'$, and cosines with $\mathbf{p}+\mathbf{k}'$. The final group are defined below. We take the angles, with $\left\{\mathbf{a},\mathbf{b}\right\}\subset\left\{\mathbf{k},\mathbf{p},\mathbf{q}\right\}$, to be
\bea
  &\theta_{ab}=\hat{\mathbf{a}}\cdot\hat{\mathbf{b}}, \quad \alpha_a=\hat{\mathbf{a}}\cdot\hat{\mathbf{k}}', \quad \beta_a=\hat{\mathbf{a}}\cdot\widehat{\mathbf{k}-\mathbf{k}'}, \quad \gamma_a=\hat{\mathbf{a}}\cdot\widehat{\mathbf{p}+\mathbf{k}'},&
 \nonumber \\
  &\bar{\beta}=\mathbf{\hat{k}'}\cdot\mathbf{\widehat{k-k'}} \quad \bar{\gamma} = \mathbf{\hat{k}'}\cdot\mathbf{\widehat{p+k'}}\quad \bar{\mu} = \mathbf{\widehat{k - k'}}\cdot\mathbf{\widehat{p+k'}}.&
\eea
In terms of the angles $\xi_{kq}$ and $\xi_{pq}$ in figure \ref{BispectraGeometry} we obviously have $\theta_{kq}=-\cos\xi_{kq}$ and similarly for $\theta_{pq}$. We also have $\alpha_k=\cos\bar{\theta}$. $\gamma_a$ and $\overline{\gamma}$ are, of course, entirely unrelated to the metric on our spatial hypersurface $\gamma_{ab}$ or its determinant.

In a manner entirely analogous to the two-point case we find the different bispectra by applying to (\ref{SixFieldCorrelation}) different projection operators to extract the relevant scalar, vector or tensor parts and, given that we ensure that $\mathcal{A}^{ijklmn}$ has the required symmetries, we may express the bispectra as
\be
\label{GenericBispectra}
  \left<\tau_A(\mathbf{k})\tau_B(\mathbf{q})\tau_C(\mathbf{p})\right>
    =\delta(\mathbf{k}+\mathbf{p}+\mathbf{q})\int d^3\mathbf{k}'\mathcal{P}(k')
    \mathcal{P}(\left|\mathbf{k}-\mathbf{k}'\right|)\mathcal{P}(\left|\mathbf{p}+\mathbf{k}'\right|)
    \left(8\mathcal{F}_{ABC}\right)
\ee
where $\left\{ABC\right\}$ denote denote different parts of the stress-energy tensor and $\mathcal{F}_{ABC}$ is the relevant angular component.

\subsection{Scalar Bispectra}
 We begin with the simplest case, the bispectrum of the magnetic pressure. This is found by defining $\mathcal{A}^{ijklmn}=(1/8)\gamma^{ij}\gamma^{kl}\gamma^{mn}$ which gives us
\be
  8\mathcal{F}_{\tau\tau\tau}=\bar{\beta}^2+\bar{\gamma}^2+\bar{\mu}^2-\bar{\beta}\bar{\gamma}\bar{\mu} .
\ee

The first scalar cross-correlation will be between the square of the pressure and the anisotropic pressure, found by using $\mathcal{A}^{ijklmn}=(-1/4)\gamma^{ij}\gamma^{kl}Q^{mn}(\mathbf{q})$ to give
\bea
  \lefteqn{-8\mathcal{F}_{\tau\tau\tau_S}=3\Big(1-\alpha_q^2-\gamma_q^2-\beta_q^2-\frac{1}{3}(\bar{\beta}^2+\bar{\gamma}^2+\bar{\mu}^2)}
   \nonumber \\ && \qquad \qquad
   +\alpha_q(\beta_q\bar{\beta}+\gamma_q\bar{\gamma})+\bar{\mu}(\beta_q\gamma_q+\frac{1}{3}\bar{\beta}\bar{\gamma})
   -\bar{\beta}\bar{\gamma}\beta_q\gamma_q\Big)
\eea

Similarly the second scalar cross-correlation, with $\mathcal{A}^{ijklmn}=(1/2)\gamma^{ij}Q^{kl}(\mathbf{p})Q^{mn}(\mathbf{q})$, gives
\be
  \mathcal{F}_{\tau\tau_S\tau_S}=\sum_{n=0}^5\mathcal{F}_{\tau\tau_S\tau_S}^{n}
\ee
where
\bea
  8\mathcal{F}_{\tau\tau_S\tau_S}^0&=&-6, \nonumber \\
  8\mathcal{F}_{\tau\tau_S\tau_S}^1&=&0, \nonumber \\
  8\mathcal{F}_{\tau\tau_S\tau_S}^2&=&\bar{\beta}^2+\bar{\gamma}^2+\bar{\mu}^2+3\left(\alpha_p^2+\alpha_q^2+\beta_p^2+\beta_q^2
   +\gamma_p^2+\gamma_q^2\right)+9\theta_{pq}^2, \\
  8\mathcal{F}_{\tau\tau_S\tau_S}^3&=&-\Big(\bar{\beta}\bar{\gamma}\bar{\mu}+3\bar{\mu}(\beta_p\gamma_p+\beta_q\gamma_q)
   +\bar{\gamma}(\alpha_p\gamma_p+\alpha_q\gamma_q)+\bar{\beta}(\alpha_p\beta_p+\alpha_q\beta_q)
    \nonumber \\ && \quad +9\theta_{pq}(\alpha_p\alpha_q+\beta_p\beta_q+\gamma_p\gamma_q)\Big) \nonumber \\
  8\mathcal{F}_{\tau\tau_S\tau_S}^4&=&3\left(\bar{\beta}(\bar{\mu}\alpha_p\gamma_p+\bar{\gamma}\beta_q\gamma_q+3\alpha_p\beta_q\theta_{pq})
   +3\left(\alpha_p\gamma_p\alpha_q\gamma_q+\beta_p\gamma_p\beta_q\gamma_q\right)\right) \nonumber \\
  8\mathcal{F}_{\tau\tau_S\tau_S}^5&=&-9\bar{\beta}\alpha_p\gamma_p\beta_q\gamma_q . \nonumber
\eea

Finally, the anisotropic scalar bispectrum is found by applying $A^{ijkmln}=(-1)^3Q^{ij}(\mathbf{k})Q^{kl}(\mathbf{p})$ $\times Q^{mn}(\mathbf{q})$ which results in
\be
  \mathcal{F}_{\tau_S\tau_S\tau_S}=\sum_{n=0}^6\mathcal{F}_{\tau_S\tau_S\tau_S}^{n}
\ee
with
\bea
  -8\mathcal{F}_{\tau_S\tau_S\tau_S}^0&=&9 \nonumber \\
  -8\mathcal{F}_{\tau_S\tau_S\tau_S}^1&=&0 \nonumber \\
  -8\mathcal{F}_{\tau_S\tau_S\tau_S}^2&=&-\Big(
   \bar{\beta}^2+\bar{\gamma}^2+\bar{\mu}^2
   +9(\theta_{kp}^2+\theta_{kq}^2+\theta_{pq}^2)
    \nonumber \\ && \quad
   +3(\alpha_k^2+\alpha_p^2+\alpha_q^2+\beta_k^2+\beta_p^2+\beta_q^2+\gamma_k^2+\gamma_p^2+\gamma_q^2)
   \Big) \nonumber \\
  -8\mathcal{F}_{\tau_S\tau_S\tau_S}^3&=&3\bigg(
   \bar{\mu}(\beta_k\gamma_k+\beta_p\gamma_p+\beta_q\gamma_q+\frac{1}{3}\bar{\beta}\bar{\gamma})
   +\bar{\gamma}(\alpha_k\gamma_k+\alpha_p\gamma_p+\alpha_q\gamma_q)
    \nonumber \\ && \quad
   +\bar{\beta}(\alpha_k\beta_k+\alpha_p\beta_p+\alpha_q\beta_q)
   +3\theta_{kp}(\alpha_k\alpha_p+\beta_k\beta_p+\gamma_k\gamma_p)
    \nonumber \\ && \quad
   +3\theta_{kq}(\alpha_k\alpha_q+\beta_k\beta_q+\gamma_k\gamma_q)
   +3\theta_{pq}(\alpha_p\alpha_q+\beta_p\beta_q+\gamma_p\gamma_q)
    \nonumber \\ && \quad
   +9\theta_{kp}\theta_{kq}\theta_{pq}
   \bigg) \nonumber \\
  -8\mathcal{F}_{\tau_S\tau_S\tau_S}^4&=&-3\bigg(
   \bar{\gamma}\bar{\mu}\alpha_k\beta_k+\bar{\beta}\bar{\mu}\alpha_p\gamma_p+\bar{\beta}\bar{\gamma}\beta_q\gamma_q
     \nonumber \\ && \quad
   +3\big(\bar{\mu}\theta_{kp}\beta_k\gamma_p+\bar{\gamma}\theta_{kq}\alpha_k\gamma_q+\bar{\beta}\theta_{pq}\alpha_p\beta_q\big)
     \nonumber \\ && \quad
   +3\big(\alpha_k\beta_k(\alpha_p\beta_p+\alpha_q\beta_q)
   +\alpha_p\gamma_p(\alpha_k\gamma_k+\alpha_q\gamma_q)
   +\beta_q\gamma_q(\beta_k\gamma_k+\beta_p\gamma_p)\big)
     \nonumber \\ && \quad
   +9(\theta_{kp}\theta_{kq}\gamma_p\gamma_q+\theta_{kp}\theta_{pq}\beta_k\beta_q+\theta_{kq}\theta_{pq}\alpha_k\alpha_p)
   \bigg)
\eea
\bea
  -8\mathcal{F}_{\tau_S\tau_S\tau_S}^5&=&9\bigg(
   \bar{\mu}\alpha_k\beta_k\alpha_p\gamma_p
   +\bar{\gamma}\alpha_k\beta_k\beta_q\gamma_q+\bar{\beta}\alpha_p\gamma_p\beta_q\gamma_q
    \nonumber \\ && \quad
   +3(\theta_{kp}\beta_k\gamma_p\beta_q\gamma_q+\theta_{kq}\alpha_k\alpha_p\gamma_p\gamma_q+\theta_{pq}\alpha_k\beta_k\alpha_p\beta_q)
   \bigg) \nonumber \\
  -8\mathcal{F}_{\tau_S\tau_S\tau_S}^6&=&-27\alpha_k\beta_k\alpha_p\gamma_p\beta_q\gamma_q. \nonumber
\eea

\subsection{Vector and Tensor Cross Bispectra}
For the vector and tensor correlations we restrict ourselves to the various rotationally-invariant quantities, which can be identified with cross-correlations between the scalar pressures and either the vector or tensor moduli. The first of these, the correlation between the scalar pressure and the vorticity, we recover with the operator $\mathcal{A}^{ijklmn}=(1/2)\gamma^{ij}\hat{p}^{(k}P^{l)}_a(\mathbf{p})\hat{q}^{(m}P^{n)a}(\mathbf{q})$. The eventual result is
\be
  \mathcal{F}_{\tau\tau^V\tau^V}=\sum_{n=1}^6\mathcal{F}_{\tau\tau^V\tau^V}^{n}
\ee
\noindent with
\bea
  8\mathcal{F}_{\tau\tau^V\tau^V}^1&=&-3\theta_{pq} \nonumber \\
  8\mathcal{F}_{\tau\tau^V\tau^V}^2&=&\gamma_p\gamma_q \nonumber \\
  8\mathcal{F}_{\tau\tau^V\tau^V}^3&=&
   +2\theta_{pq}\left(\alpha_p^2+\beta_p^2+\gamma_p^2+\alpha_q^2+\beta_q^2+\gamma_q^2+\frac{1}{2}\bar{\beta}^2+2\theta_{pq}^2\right)
    \nonumber \\ && \quad
   \bar{\mu}\left(\beta_p\gamma_q+\beta_q\gamma_p\right)+\bar{\gamma}\left(\alpha_p\gamma_q+\alpha_q\gamma_p\right)
   \nonumber \\
  8\mathcal{F}_{\tau\tau^V\tau^V}^4&=&-\bar{\beta}\left(\bar{\mu}\alpha_p\gamma_q+\bar{\gamma}\gamma_p\beta_q\right)
   -2\bar{\beta}\theta_{pq}\left(\alpha_q\beta_q+\alpha_p\beta_p\right)
   -2\left(\alpha_p\alpha_q+\beta_p\beta_q)(\gamma_p^2+\gamma_q^2+2\theta_{pq}^2\right)
    \nonumber \\ && \quad
   -2\gamma_p\gamma_q\left(\alpha_p^2+\beta_p^2+\alpha_q^2+\beta_q^2+\frac{1}{2}\bar{\beta}^2+2\theta_{pq}^2\right)
    \nonumber \\
  8\mathcal{F}_{\tau\tau^V\tau^V}^5&=&4\theta_{pq}\gamma_p\gamma_q\left(\alpha_p\alpha_q+\beta_p\beta_q\right)
   +2\bar{\beta}\gamma_p\gamma_q\left(\alpha_p\beta_p+\alpha_q\beta_q\right)
   +2\bar{\beta}\alpha_p\beta_q\left(\gamma_p^2+\gamma_q^2+2\theta_{pq}^2\right)
    \nonumber \\
  8\mathcal{F}_{\tau\tau^V\tau^V}^6&=&-4\bar{\beta}\theta_{pq}\alpha_p\gamma_p\beta_q\gamma_q. \nonumber
\eea

The cross-correlation with the anisotropic pressure is recovered with the operator $\mathcal{A}^{ijklmn}=-Q^{ij}(\mathbf{k})\hat{p}^{(k}P^{l)}_a(\mathbf{p})\hat{q}^{(m}P^{n)a}(\mathbf{q})$, giving
\be
  \mathcal{F}_{\tau_S\tau^V\tau^V}=-\sum_{n=1}^7\mathcal{F}_{\tau_S\tau^V\tau^V}^{n}
\ee
with
\bea
    8\mathcal{F}_{\tau_S\tau^V\tau^V}^1&=&6\theta_{pq} \nonumber \\
    -8\mathcal{F}_{\tau_S\tau^V\tau^V}^2&=&4\gamma_p\gamma_q
 \nonumber \\
    -8\mathcal{F}_{\tau_S\tau^V\tau^V}^3&=&
      \big(
      \overline{\beta}^2+2\left(\alpha_p^2+\beta_p^2+\gamma_p^2+\alpha_q^2+\beta_q^2+\gamma_q^2\right)
      +4\theta_{pq}^2+3\left(\alpha_k^2+\beta_k^2\right)
        \nonumber \\ && \quad
      +3\left(\theta_{kp}\gamma_q+\theta_{kq}\gamma_p\right)\gamma_k
      +\theta_{pq}\left(6\left(\theta_{kp}^2+\theta_{kq}^2\right)
      \beta_p\gamma_q+\beta_q\gamma_p\right)\overline{\mu}
      +(\alpha_p\gamma_q+\alpha_q\gamma_p)\overline{\gamma}
      \big)
\nonumber \eea
\bea
    8\mathcal{F}_{\tau_S\tau^V\tau^V}^4&=&
      4\theta_{pq}^2\left(3\theta_{kp}\theta_{kq}+\alpha_p\alpha_q+\beta_p\beta_q+\gamma_p\gamma_q\right)
      +\overline{\beta}\theta_{pq}\left(3\alpha_k\beta_k+2\left(\alpha_p\beta_p+\alpha_q\beta_q\right)\right)
        \nonumber \\ && \quad
      +6\theta_{pq}\left(\theta_{kp}\left(\alpha_k\alpha_p+\beta_k\beta_p\right)
        +\theta_{kq}\left(\alpha_k\alpha_q+\beta_k\beta_q\right)\right)
        \nonumber \\ && \quad
      +\gamma_p\gamma_q\left(3\alpha_k^2+3\beta_k^2+2\alpha_p^2+2\beta_p^2+2\alpha_q^2+2\beta_q
        +6\theta_{kp}^2+6\theta_{kq}^2+\overline{\beta}^2\right)
        \nonumber \\ && \quad
      +2\left(\gamma_p^2+\gamma_q^2\right)\left(\alpha_p\alpha_q+\beta_p\beta_q+3\theta_{kp}\theta_{kq}\right)
      +\overline{\beta}\left(\alpha_p\gamma_q\overline{\mu}+\beta_q\gamma_p\overline{\gamma}\right)
        \nonumber \\ && \quad
      +3\left(\alpha_k\gamma_p\overline{\gamma}\theta_{kq}+\beta_k\gamma_q\overline{\mu}\theta_{kp}\right)
      +3\gamma_k\left(\alpha_k\alpha_p\gamma_q+\beta_k\beta_q\gamma_p\right)
 \nonumber \\
    -8\mathcal{F}_{\tau_S\tau^V\tau^V}^5&=&
      4\theta_{pq}^2\left(\alpha_p\beta_q\overline{\beta}+3\theta_{kq}\alpha_k\alpha_p
        +3\theta_{kp}\beta_k\beta_q\right)
      +4\left(\theta_{pq}\left(\alpha_p\alpha_q+\beta_p\beta_q+3\theta_{kp}\theta_{kq}\right)\gamma_p\gamma
        \right. \nonumber \\ && \quad \left.
        +3\alpha_k\beta_k\left(\alpha_p\beta_p+\alpha_q\beta_q\right)\right)
      +6\theta_{kp}\left(\left(\alpha_k\alpha_p+\beta_k\beta_p\right)\gamma_p\gamma_q
        +\beta_k\beta_q\left(\gamma_p^2+\gamma_q^2\right)\right)
        \nonumber \\ && \quad
      +6\theta_{kq}\left(\left(\alpha_k\alpha_q+\beta_k\beta_q\right)\gamma_p\gamma_q
        +\alpha_k\alpha_p\left(\gamma_p^2+\gamma_q^2\right)\right)
      +2\alpha_p\beta_q\overline{\beta}\left(\gamma_p^2+\gamma_q^2\right)
        \nonumber \\ && \quad
      +\left(\alpha_p\beta_p+\alpha_q\beta_q+3\alpha_k\beta_k\right)\gamma_p\gamma_q\overline{\beta}
      +3\alpha_k\beta_k\left(\alpha_p\gamma_q\overline{\mu}+\beta_q\gamma_p\overline{\gamma}\right)
      \nonumber \\
    8\mathcal{F}_{\tau_S\tau^V\tau^V}^6&=&
      12\theta_{pq}^2\alpha_k\alpha_p\beta_k\beta_q
      +4\theta_{pq}\gamma_p\gamma_q\left(\alpha_p\beta_q\overline{\beta}+3\theta_{kq}\alpha_k\alpha_p
       +3\theta_{kp}\beta_k\beta_q\right)
        \nonumber \\ && \quad
      +6\alpha_k\beta_k\gamma_p\gamma_q\left(\alpha_p\beta_p+\alpha_q\beta_q\right)
      +6\alpha_k\beta_k\alpha_p\beta_q\left(\gamma_p^2+\gamma_q^2\right) \nonumber \\
    -8\mathcal{F}_{\tau_S\tau^V\tau^V}^7&=&12\theta_{pq}\alpha_k\alpha_p\beta_k\beta_q\gamma_p\gamma_q. \nonumber
\eea

The cross-correlation between the scalar trace and the tensor modulus, $\langle\tau\tau^{Tab}\tau_{ab}^T\rangle$, is found by the application of $\mathcal{A}^{ijklmn}=(1/2)\gamma^{ij}\mathcal{P}^{Tklab}(\mathbf{p})\mathcal{P}^{Tmn}_{ab}(\mathbf{q})$. After a lengthy calculation one sees that
\beas
  8\mathcal{F}^0_{\tau\tau^T\tau^T}&=&2
   \\
  8\mathcal{F}^1_{\tau\tau^T\tau^T}&=&0
   \\
  8\mathcal{F}^2_{\tau\tau^T\tau^T}&=&-\overline{\beta}^2-\overline{\gamma}^2-\overline{\mu}^2-\alpha_p^2-\alpha_q^2-\beta_p^2-\beta_q^2
      -3\left(\theta_{pq}^2+\gamma_p^2+\gamma_q^2\right)
   \\
  8\mathcal{F}^3_{\tau\tau^T\tau^T}&=&\overline{\beta}\overline{\gamma}\overline{\mu}+\overline{\beta}\left(\alpha_p\beta_p+\alpha_q\beta_q\right)
    +\overline{\gamma}\left(\alpha_p\gamma_p+\alpha_q\gamma_q\right)
     \\ &&
    +\overline{\mu}\left(\beta_p\gamma_p+\beta_q\gamma_q\right)
    +\theta_{pq}\left(3\alpha_p\alpha_q+3\beta_p\beta_q+5\gamma_p\gamma_q\right)
   \\
  8\mathcal{F}^4_{\tau\tau^T\tau^T}&=&\theta_{pq}^4+\theta_{pq}^2\left(3\overline{\beta}^2+\overline{\gamma}^2+\overline{\mu}^2\right)
    +2\overline{\beta}^2\left(\gamma_p^2+\gamma_q^2\right)
    -\overline{\beta}\left(\overline{\mu}\alpha_p\gamma_p+\overline{\gamma}\beta_q\gamma_q\right)
     \\ &&
    -\theta_{pq}\left(\overline{\beta}\left(\alpha_p\beta_q+2\alpha_q\beta_p\right)
      +2\overline{\gamma}\left(\alpha_p\gamma_q+\alpha_q\gamma_p\right)
      +2\overline{\mu}\left(\beta_p\gamma_q+\beta_q\gamma_p\right)\right)
     \\ &&
    +\theta_{pq}^2\left(\alpha_p^2+\alpha_q^2+\beta_p^2+\beta_q^2+\gamma_p^2+\gamma_q^2\right)
     \\ &&
    +2\left(\alpha_p^2+\alpha_q^2+\beta_p^2+\beta_q^2\right)\left(\gamma_p^2+\gamma_q^2\right)
    +\left(\alpha_p\alpha_q+\beta_p\beta_q\right)\gamma_p\gamma_q
   \\
  8\mathcal{F}^5_{\tau\tau^T\tau^T}&=&-2\overline{\beta}^2\theta_{pq}\gamma_p\gamma_q-\overline{\beta}\overline{\gamma}\overline{\mu}\theta_{pq}^2
    +2\overline{\beta}\overline{\gamma}\theta_{pq}\beta_p\gamma_q+2\overline{\beta}\overline{\mu}\theta_{pq}\alpha_q\gamma_p
    +\overline{\beta}\left(\alpha_p\beta_q-2\alpha_q\beta_p\right)\gamma_p\gamma_q
     \\ &&
    -\theta_{pq}^2\left(\overline{\beta}\left(\alpha_p\beta_p+\alpha_q\beta_q\right)
      -\overline{\gamma}\left(\alpha_p\gamma_p+\alpha_q\gamma_q\right)
      -\overline{\mu}\left(\beta_p\gamma_p+\beta_q\gamma_q\right)\right)
     \\ &&
    -2\overline{\beta}\left(\alpha_p\beta_p+\alpha_q\beta_q\right)\left(\gamma_p^2+\gamma_q^2\right)
    -\theta_{pq}^3\left(\alpha_p\alpha_q+\beta_p\beta_q+\gamma_p\gamma_q\right)
     \\ &&
    -2\theta_{pq}\left(\left(\alpha_p^2+\alpha_q^2+\beta_p^2+\beta_q^2\right)\gamma_p\gamma_q
      +\left(\alpha_p\alpha_q+\beta_p\beta_q\right)\left(\gamma_p^2+\gamma_q^2\right)\right)
   \\
  8\mathcal{F}^6_{\tau\tau^T\tau^T}&=&\overline{\beta}\theta_{pq}^2\left(\theta_{pq}\alpha_p\beta_q-\overline{\gamma}\beta_q\gamma_q
      -\overline{\mu}\alpha_p\gamma_p\right)
    +2\overline{\beta}\theta_{pq}\left(\alpha_p\beta_p+\alpha_q\beta_q\right)\gamma_p\gamma_q
     \\ &&
    +2\overline{\beta}\theta_{pq}\alpha_p\beta_q\left(\gamma_p^2+\gamma_q^2\right)
    +\theta_{pq}^2\left(\alpha_p\alpha_q+\beta_p\beta_q\right)\gamma_p\gamma_q
   \\
  8\mathcal{F}^7_{\tau\tau^T\tau^T}&=&-\overline{\beta}\theta_{pq}^2\alpha_p\beta_q\gamma_p\gamma_q
\eeas
and
\be
  \mathcal{F}_{\tau\tau^T\tau^T}=\sum_{n=0}^7\mathcal{F}_{\tau\tau^T\tau^T}^{n} .
\ee

The correlation between the traceless scalar and the tensors is even more complex; applying $\mathcal{A}^{ijklmn}=-Q^{ij}(\mathbf{k})\mathcal{P}^{Tklab}(\mathbf{p})\mathcal{P}^{Tmn}_{ab}(\mathbf{q})$ to $\mathcal{B}_{ijklmn}$ leads to an angular integrand which we express as
\be
  \mathcal{F}_{\tau_S\tau^T\tau^T}=\sum_{n=0}^8\mathcal{F}_{\tau_S\tau^T\tau^T}^n .
\ee
We present the forms of $\mathcal{F}^n_{\tau_S\tau^T\tau^T}$ in appendix \ref{Appendix-SourceStats-STT}.

The final rotationally-invariant cross-correlation is that between the vectors and the tensors, $\langle\tau^V_a\tau^{Ta}_b\tau^{Vb}\rangle$, found by applying $\mathcal{A}^{ijklmn}=\mathcal{P}^{Vij}_a(\mathbf{k})\mathcal{P}^{Tkla}_b(\mathbf{p})\mathcal{P}^{Vmnb}(\mathbf{q})$. This ultimately produces
\be
  \mathcal{F}_{\tau^V\tau^T\tau^V}=\sum_{n=2}^8\mathcal{F}_{\tau^V\tau^T\tau^V}^n
\ee
with $\mathcal{F}_{\tau^V\tau^T\tau^V}^n$ presented in appendix \ref{Appendix-SourceStats-VTV}.


\subsection{Tensor Auto-Correlation}
There is no rotationally-invariant vector auto-correlation possible at the three-point level; one has a residual vector freedom. Thus the last bispectra we consider is the full tensor auto-correlation, $\langle\tau^T_{ab}\tau^{bT}_c\tau^{acT}\rangle$, which can be found by the application of
\beas
  \mathcal{A}^{ijklmn}&=&\mathcal{P}^{Tij}_{ab}(\mathbf{k})\mathcal{P}^{Tklb}_c(\mathbf{p})
    \mathcal{P}^{mnac}_T(\mathbf{q}) .
\eeas
The result is far from pretty; it simplifies considerably for large-scales and colinear bispectra (for which, indeed, the result is vanishing). We present the full generality in appendix \ref{Appendix-SourceStats-TTT}; it has the form
\be
  \mathcal{F}_{\tau^T\tau^T\tau^T}=\sum_{n=0}^9\mathcal{F}_{\tau^T\tau^T\tau^T}^{n} .
\ee

\subsection{Results}
As was the case for the power spectra, there are two very different spectral r\'egimes for the bispectra. For ultra-violet spectra with $n>-1$, the integrals are dominated by the highest $k$ modes around the cutoff scale. For these spectral indices, the bispectra become independent of $k$ when $k\ll k_c$ and the analytic expressions are straight forward to integrate. Indeed, we can do them exactly in the limit $k\ll k_c$; we present these results in appendix \ref{Appendix-SourceStats-Results} for a generic geometry and here only the colinear case. We see that
\bea
  \langle\tau(\mathbf{k})\tau(\mathbf{p})\tau(\mathbf{q})\rangle&=&B\pi\delta(\mathbf{k}+\mathbf{p}+\mathbf{q})
    \nonumber \\
  \langle\tau(\mathbf{k})\tau(\mathbf{p})\tau_S(\mathbf{q})\rangle&=&0
    \nonumber \\
  \langle\tau(\mathbf{k})\tau_S(\mathbf{p})\tau_S(\mathbf{q})\rangle&=&
    \frac{7}{5}B\pi\delta(\mathbf{k}+\mathbf{p}+\mathbf{q})
    \nonumber \\
\label{LargeScaleColinearResults}
  \langle\tau_S(\mathbf{k})\tau_S(\mathbf{p})\tau_S(\mathbf{q})\rangle&=&
   -\frac{34}{35}B\pi\delta(\mathbf{k}+\mathbf{p}+\mathbf{q})
   \\
  \langle\tau(\mathbf{k})\tau_a^V(\mathbf{p})\tau^a_V(\mathbf{q})\rangle&=&
   -\frac{14}{15}B\pi\delta(\mathbf{k}+\mathbf{p}+\mathbf{q})
     \nonumber \\
  \langle\tau_S(\mathbf{k})\tau_a^V(\mathbf{p})\tau^a_V(\mathbf{q})\rangle&=&
   \frac{34}{105}B\pi\delta(\mathbf{k}+\mathbf{p}+\mathbf{q})
    \nonumber \\
  \langle\tau(\mathbf{k})\tau^{aT}_b(\mathbf{p})\tau^{bT}_a(\mathbf{q})\rangle&=&
   \frac{28}{15}B\pi\delta(\mathbf{k}+\mathbf{p}+\mathbf{q})
    \nonumber \\
  \langle\tau_S(\mathbf{k})\tau^{aT}_b(\mathbf{p})\tau^{bT}_a(\mathbf{q})\rangle&=&
   -\frac{136}{105}B\pi\delta(\mathbf{k}+\mathbf{p}+\mathbf{q})
    \nonumber \\
  \langle\tau^V_a(\mathbf{k})\tau^{aT}_b(\mathbf{p})\tau^b_V(\mathbf{q})\rangle&=&
   \frac{68}{105}B\pi\delta(\mathbf{k}+\mathbf{p}+\mathbf{q})
    \nonumber \\
  \langle\tau^{aT}_b(\mathbf{k})\tau^{cT}_a(\mathbf{p})\tau^{bT}_c(\mathbf{q})\rangle&=&0
    \nonumber
\eea
in excellent agreement with the results, both simulated and numerically integrated, shown in the figures below.

\subsubsection{Flat Spectrum Results}
\begin{figure}
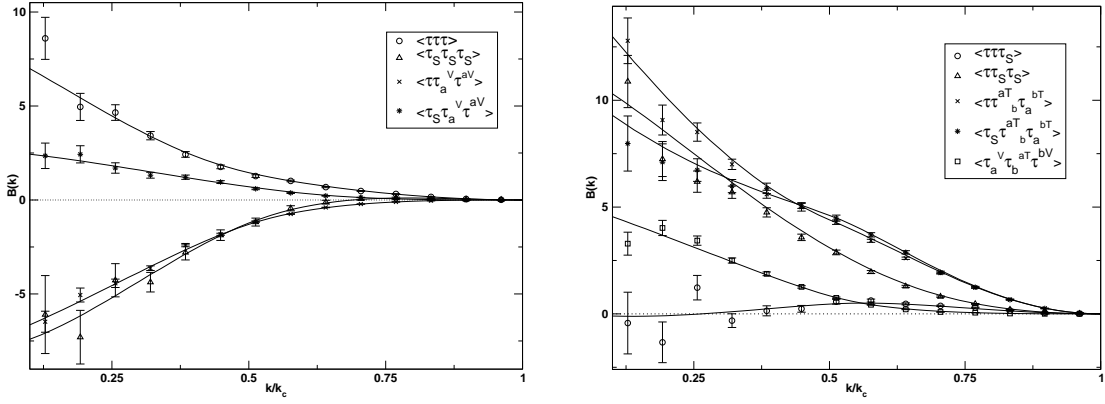
\begin{center}
\includegraphics[width=0.46\textwidth]{3Points_n0_1.eps}\qquad\includegraphics[width=0.46\textwidth]{3Points_n0_2.eps}
\caption{Here we show the various colinear bispectra for the flat spectrum $n=0$ generated from a collection of realisations. Analytic results agree well with the numerical simulations.}
\label{3Point_n0}\end{center}\end{figure}

The bispectra that are derived from the simulated fields are heavily compromised by the grid-size; unlike the two-point case the three-point moments use only a restricted number of the modes, selected by the geometry chosen for the wavevectors. The result from a single realisation is in most cases noise-dominated. To overcome this difficulty we have chosen to simulate a large number of different realisations, taking the mean signal and using their variance to provide an estimate for the errors involved. The results, for a flat power spectrum, a grid-size of $l_{\mathrm{dim}}=192$ (and a damping scale of $k_c=l_{\mathrm{dim}}/4.1$) and $1500$ combined realisations, are plotted rebinned into $64$ bins in figures \ref{3Point_n0} with the numerically-integrated predictions overlaid. For simplicity we have concentrated on the colinear case, wherein $\mathbf{p}=\mathbf{k}$ and so $\mathbf{q}=-\left(\mathbf{k}+\mathbf{p}\right)$. We plot $k/k_c$ against $B(k)$ where $B(k)$ represents the colinear bispectra.

\subsubsection{Steep spectrum results}
\begin{figure}
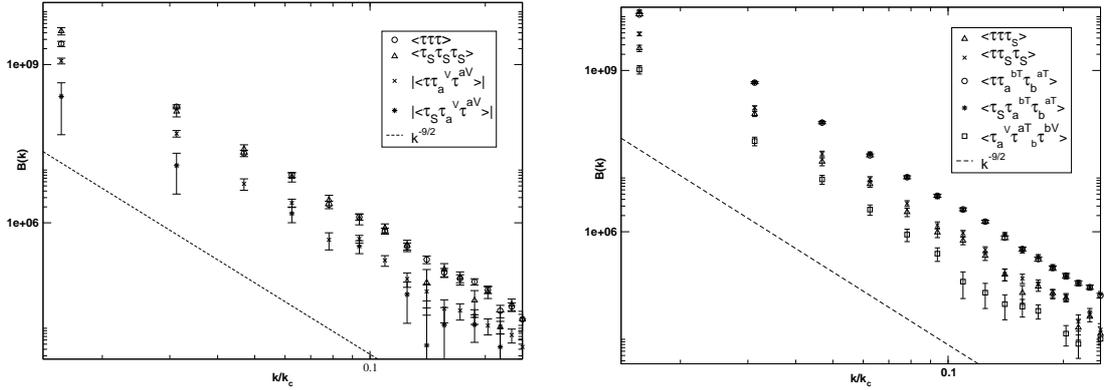
\begin{center}
\includegraphics[width=0.46\textwidth]{3Points_n-2.5_1.eps}\qquad\includegraphics[width=0.46\textwidth]{3Points_n-2.5_2.eps}
\caption{Here we show the colinear bispectra for a steep spectrum ($n=-2.5$.) The scaling is as expected naively, $\propto k^{3n+3} = k^{-9/2}$ (dotted line).}
\label{3Point_n-2.5}\end{center}\end{figure}

In the r\'egime $n<-1$ matters are, as with the two-point case, complicated by the presence of numerous poles; the integrals are dominated by the volume lying between the poles. Rather than attempt a solution, we use our realisations to calculate the bispectra. We present the results for $n=-2.5$ in figures \ref{3Point_n-2.5}. There is a dependence on the grid-size for this spectrum due to the paucity of modes of appreciable power given the strongly red spectrum. We tested this by running three realisations with differing grid-sizes, keeping the total number of modes constant in each case constant; specifically we took $1,500$ simulations at $l_{\mathrm{dim}}=192$, $5000$ simulations at $l_{\mathrm{dim}}=128$, and $40,000$ simulations at $l_{\mathrm{dim}}=64$. As might be expected we found a suppression for the case wherein $l_{\mathrm{dim}}=64$ -- due to scarcity of modes at low $k$ -- but there was good agreement between the other two cases. We have plotted the results from the $l_{\mathrm{dim}}=192$ run for greatest dynamic range, again rebinning into $64$ bins.

With this highly-tilted spectrum we see, as with the two-point case, that the features of the magnetic spectrum at the cut-off scale apparent in the flat case are washed out by the spectral tilt. The predominant shape again is the $k$-dependence we expect from the radial component of the integral; in this case $B(k)\propto k^{3(n+1)}$; these are plotted for comparison. Again, as with the two-point case there is an infra-red suppression. The magnitudes of the bispectra fall into three close bands; the strongest are the correlations between the scalars and the tensors, while the middle-band is composed of the scalar auto- and cross-correlations. The weakest bispectra are those involving correlations with the vectors and, in some cases, the $1-\sigma$ error bars are greater than the mean value; such points have been removed from these plots for aesthetic purposes. The $\langle\tau^V_a\tau^{aT}_b\tau^b_V\rangle$ correlation is particularly weak.

\section{The Lorentz Forces}
\begin{figure}
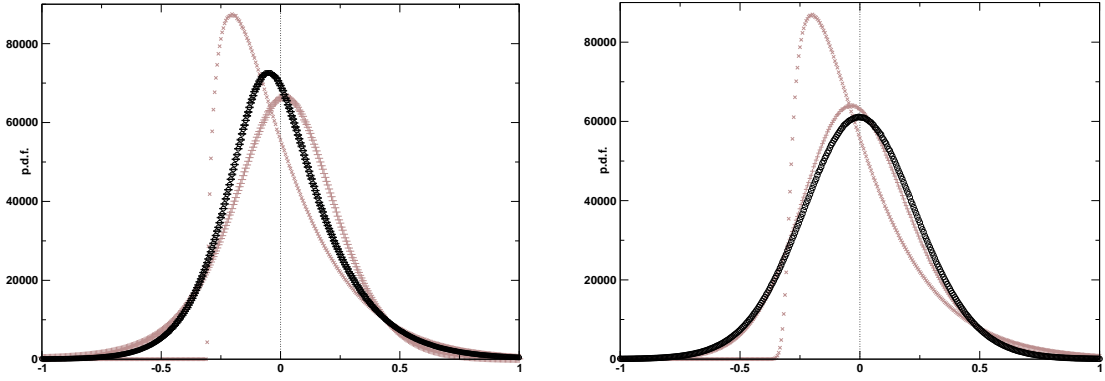
\begin{center}\includegraphics[width=0.46\textwidth]{PDF-Lorentz0.eps}\qquad\includegraphics[width=0.46\textwidth]{PDF-Lorentz2.5.eps}
\caption{Probability distribution function of the scalar Lorentz force for $n=0$ (left) and $n=-2.5$ (right); plotted alongside in grey are those of the scalar pressures.}
\label{LorentzPDF}\end{center}\end{figure}

For completeness we briefly present the results for the Lorentz forces. Recall (\ref{ScalarLorentzForce},\ref{VectorLorentzForce}) that these were
\bdm
  L_a(\mathbf{k})=ik_aL_S(\mathbf{k})+L^V_a(\mathbf{k}), \quad
  L_S(\mathbf{k})=\frac{2k}{3}P^{ij}(\mathbf{k})\tau_{ij}(\mathbf{k}),
   \quad L_a(\mathbf{k})=ik\tau^V_a(\mathbf{k}) .
\edm
Constructing these from our realised fields and evaluating the one-point moments of the scalar component, we find a skewness $\gamma_1=0.615\pm0.003$ and a kurtosis $\gamma_2=1.83\pm0.02$ for a flat magnetic spectrum, and $\gamma_1=0.034\pm0.001$, $\gamma_2=0.337\pm0.004$ for an index $n=-2.5$. The (unnormalised) probability density functions are plotted in figures \ref{LorentzPDF}, alongside those of the scalar pressures. From both the figures and the moments it is readily clear that the scalar Lorentz force is very nearly Gaussianly distributed; the difference between this and the energy density arises because combining the non-Gaussian energy density with the almost-Gaussian anisotropic pressure naturally reduces the level of non-Gaussianity.

\begin{figure}
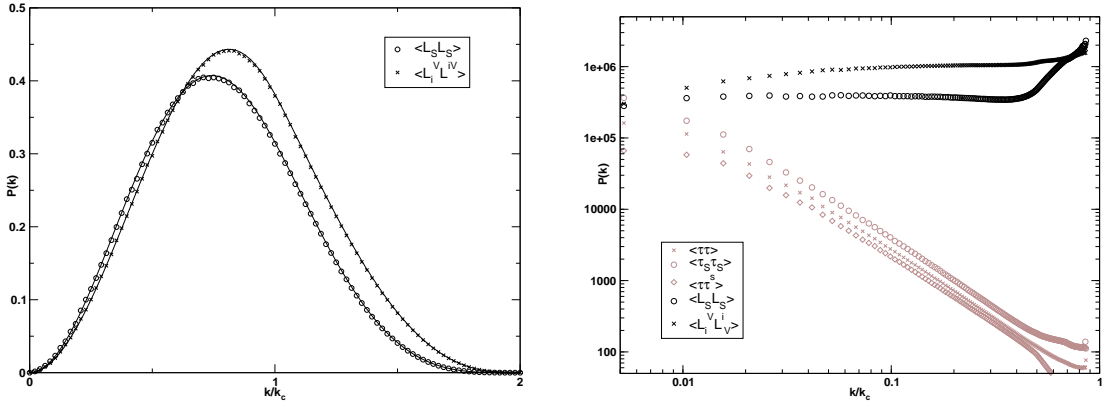
\begin{center}\includegraphics[width=0.46\textwidth]{Spectra-Lorentz0.eps}\qquad\includegraphics[width=0.46\textwidth]{Spectra-Lorentz2.5.eps}
\caption{Scalar and isotropic vector power spectra of the Lorentz force for a magnetic field with index $n=0$ (left) and $n=-2.5$ (right); plotted alongside in grey are those of the scalar pressures.}
\label{LorentzPower}\end{center}\end{figure}

The power spectra are plotted in figure \ref{LorentzPower}, alongside the pressures where helpful. The differences in the power spectra between the Lorentz forces and the corresponding stresses comes entirely from the multiplicative $k$. We have for the scalar case that
\bea
  \mathcal{F}_{L_S}&=&\frac{4k^2}{9}\left(\mathcal{F}_{\tau\tau}+\mathcal{F}_{\tau_S\tau_S}-2\mathcal{F}_{\tau\tau_S}\right) \nonumber \\
   &=&2k^2\left(1-\gamma^2\right)\left(1-\beta^2\right) .
\eea
(This can, of course, also be found by application of $\mathcal{A}^{ijkl}(k)=(2k/3)^2\kh^i\kh^j\kh^k\kh^l$ to $\mathcal{B}_{ijkl}$.) For the vectors,
\bdm
  L^i_V(\mathbf{k})=ik\tau^i_V(\mathbf{k})
\edm
which more simply implies
\be
  \mathcal{F}_{L^V}=k^2\mathcal{F}_{\tau^V\tau^V}=k^2\left(1-2\gamma^2\beta^2+\mu\gamma\beta\right) .
\ee
$\mathcal{F}$ is not now purely angular and the radial integration is modified. The resulting spectra then naturally display a suppression at low-$k$. This suppression also induces a length scale at which the Lorentz force has a maximum effect. For a flat magnetic spectrum and the scalar Lorentz force this length-scale is at approximately three-quarters of the damping scale, while for the vectors it is a little more, at about $0.8k_c$. Forming the power spectra of the Lorentz forces for a steeply-titled magnetic field ($n=-2.5$), we see that as before the features are washed out somewhat. However, due to the $k^2$ dependence above we also see that, for most of the dynamic range, the Lorentz force spectra go as $k^{2n+5}$ which in this specific case is obviously flat. On larger scales this breaks down somewhat and the correlations increase. It is interesting to note that the length-scales apparent in the flat case are still evident in the large-scale tail as a final inflection. The increase in power begins at roughly $k=0.4k_c$ for both the scalar and vector Lorentz forces. This corresponds to the points on the large-scale side of the flat-spectrum curve at which they curve inwards.

\begin{figure}
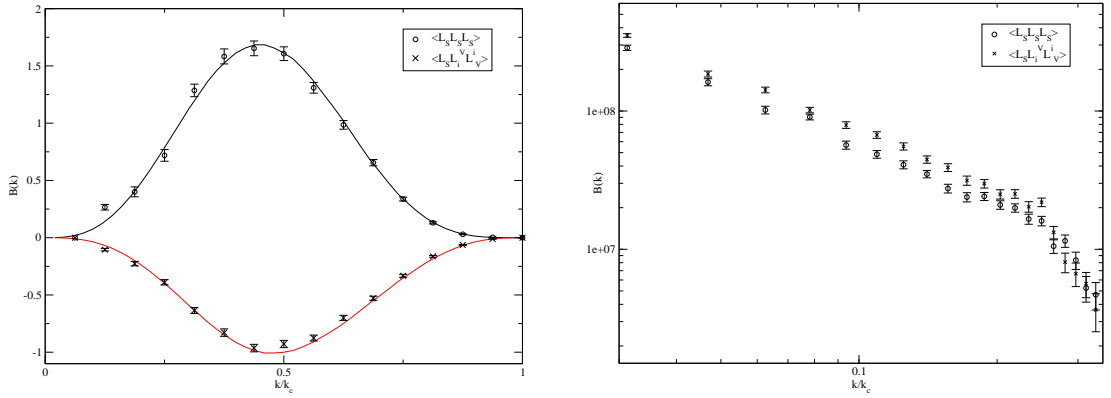
\begin{center}
\includegraphics[width=0.46\textwidth]{Bispectra-Lorentz0.eps}\qquad\includegraphics[width=0.46\textwidth]{Bispectra-Lorentz2.5.eps}
\caption{The Lorentz force bispectra for an $n=0$ field (left) and and $n=-2.5$ field (right).}
\label{LorentzBispectra}\end{center}\end{figure}

Finally, the bispectra we can construct from the Lorentz forces are the scalar auto-correlation $\langle L_SL_SL_S\rangle$ and the cross-correlation $\langle L_SL_a^VL^a_V\rangle$. These are plotted in figure \ref{LorentzBispectra} for $l_{\mathrm{dim}}=192$ and 1500 realisations.

\section{A Damped Causal Field}
\label{Section-DampedCausal}
Here we briefly consider the results of a field with a power spectrum
\be
  \mathcal{P}(k)=Ak^2\exp\left(-\left(\frac{7k}{4k_c}\right)^2\right) .
\ee
At the one-point level, as one might expect, the statistics of the scalar pressure are similar to those for the flat case; here we find $\gamma_1=1.63\pm0.01$ and $\gamma_2=3.96\pm0.07$. For the anisotropic pressure we have $\gamma_1=-0.25\pm0.01$ and $\gamma_2=1.11\pm 0.03$; we then see that for $n>0$ the one-point statistics of the pressures are insensitive to the spectral index.

\begin{figure}\begin{center}
\includegraphics{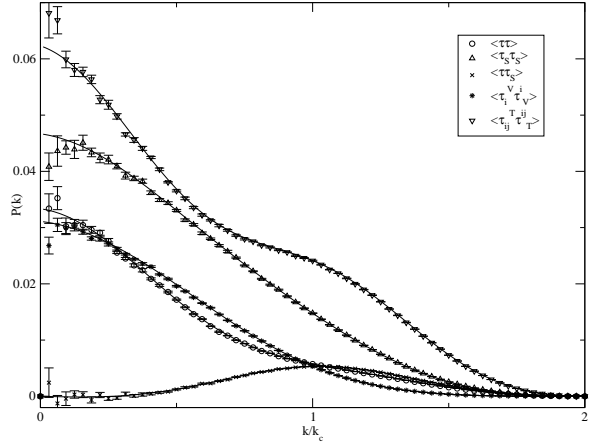}
\caption{The power spectra for the stresses of a damped causal magnetic field.}
\label{Spectra_DampedCausal}\end{center}\end{figure}

The power spectra are shown in figure \ref{Spectra_DampedCausal}. They are very similar to those produced by an undamped causal field, itself very similar to the flat field. For this field we content ourselves with the stresses and do not present the Lorentz forces which will also be very similar to the flat case. We retain the (1-$\sigma$) error bars due to increased large-scale errors coming from a large-scale suppression seen also in the analytical integration. Qualitatively, it is obvious that the features of a damped causal field are identical to a flat case; this is unsurprising since both are in the region $n>-3/2$. The details, however, do differ, most noticeably in the increased hump on the tensor spectrum.

\begin{figure}
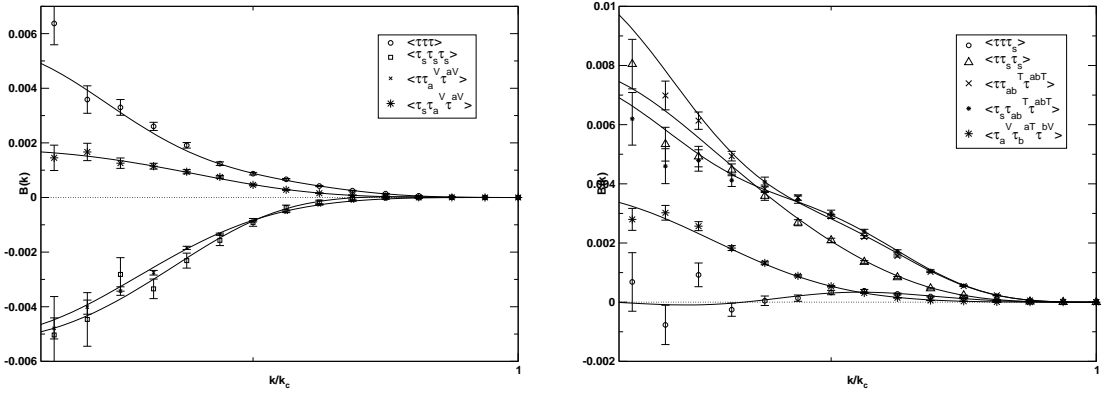
\begin{center}
\includegraphics[width=0.46\textwidth]{Bispectra-DampedCausal-1.eps}\qquad\includegraphics[width=0.46\textwidth]{Bispectra-DampedCausal-2.eps}
\caption{The bispectra of the stresses of a damped causal magnetic field.}
\label{CausalBispectra}\end{center}\end{figure}

The bispectra are in figure \ref{CausalBispectra}.

In all qualitative respects, the damped causal field closely resembles the case $n=0$.

\section{Discussion and Conclusions}
We have studied, analytically and via realisations, some of the higher point correlations for tangled large-scale magnetic fields.  The analysis is obviously highly model-dependent, and extending the analytic results to other models could be very difficult.  However, it would be a simple matter to generalise the numerical realisations to perform the same analysis for a wide variety of time-independent models; all we require is the statistical distribution of the underlying field, the power-spectrum and the form of the stress-energy tensor. For example, while we have assumed the simplest -- and perhaps most instructive -- case of a magnetic field with Gaussian statistics, it would be a simple matter to employ a $\chi^2$ probability distribution for a magnetic field, which is physically motivated by the creation mechanism considered by Matarrese \emph{et. al.} \cite{MatarreseEtAl04}. (Since our code is time-independent the analysis would be relatively complex.)  The realisations will however be limited by the narrow dynamic range allowed in the computation.

In our particular case of a tangled primordial magnetic field with a power-law spectrum, we have demonstrated that we can recover the 1-, 2- and 3-point statistics from simulations and with an excellent agreement to our analytic predictions. At the one-point level we have not only verified that the magnetic energy density follows a $\chi^2$ probability distribution function (as expected given that it is directly the square of the underlying Gaussian magnetic field), but that the anisotropic pressure is also non-Gaussian with a significantly more complex relation to the fields. There is also a spectral dependence on its probability distribution function affecting the skewness, which swaps sign as one passes through $n=-3/2$. The kurtosis, while exhibiting a spectral dependence, remains positive.

At the two-point level we have calculated the scalar, vector and tensor auto-correlations, as well as the scalar-cross correlation. For the scale-invariant spectrum we confirm the power-spectra with the expected ratios for scales longer than the cutoff; we also see that the scalar cross-correlation vanishes on large scales but is not in general entirely zero for modes close to the cutoff scale. For a highly-tilted spectrum, the cutoff scale is less important and the spectra behave with power law behaviour and with the relative ratios approximately constant. The $k$-dependence is the power-law that would be expected from a na\"{\i}ve point of view. The surprise is that in this r\'egime the scalar cross-correlation no longer vanishes on large-scales; rather, the correlation remains roughly constant at $\langle\tau\tau_S\rangle/\sqrt{\langle\tau\tau\rangle\langle\tau_S\tau_S\rangle}\approx 0.7.$ Thus, while the isotropic and anisotropic pressures are indeed correlated, they are not perfectly correlated.

At the three-point level we considered a number of rotationally-invariant bispectra, concentrating for simplicity on the colinear case. We find significant non-Gaussianities -- in excellent agreement with the analytic predictions.  For the flat spectrum, these approach fixed ratios on large scales.  These can be both positive or negative, or even zero.  As in the two-point case, some qualitative aspects change when we consider a strongly-tilted magnetic power spectrum; features arising from the cutoff scale tend to disappear, leaving instead a simple power-law drop off.  The relative ratios of the bispectra also change, and even their relative signs differ.  The correlation between the isotropic pressure squared and the anisotropic pressure, which vanishes for flat spectra, becomes non-zero.

It is interesting to note that the changeover between infra-red and ultraviolet behaviour changes as one considers higher-order moments. For an $a$-point function, the changeover will be at
\be
  n=-\frac{3}{a}
\ee
which implies that if one could take a very high-order moment, the stresses would exhibit an ultraviolet behaviour for the spectral index $n\rightarrow 0$ -- that is, the behaviour of sources would tend towards the behaviour of the field.

It remains for future work \cite{BrownCrittenden2} to fold these results in with the transfer functions for magnetised cosmologies and calculate the non-Gaussianities expected to be imprinted upon the cosmic microwave sky. We can then consider how such signals might be used to constrain the properties of a magnetic field of this type. It would also obviously be straight-forward to consider different magnetic power spectra and statistics.  While it is too early to speculate what these will discover, the higher order correlations in the sources will certainly lead to similar higher order correlations in the CMB observables, including perhaps cross correlations between the polarisation modes such as $\langle E^2 B^2 \rangle$. The foundations of this process form the basis of the next chapter of this thesis.

The techniques used in this chapter could be applied to a broad variety of models.
Moreover, we have here to our knowledge presented the first calculations of cross-correlations between, for example, scalars and tensors, and demonstrated that they can be of an equal magnitude to scalar auto-correlations.  This has great potential relevance to the wider field of sources with non-linear stress-energy tensors, or sources with non-Gaussian initial conditions. For example, it would be interesting to compare our results to the same moments evaluated for defect models, particularly given the current resurgence of interest into networks of cosmic strings. For non-coherent cases -- such as defect models, in which the sources evolve once they have entered the horizon, or very possibly magnetic models in which the intrinsic statistics are dependent on the evolving damping scale $k_D(\eta)$ -- a single static simulation will obviously not be sufficient. It is, however, straightforward in principle to construct the necessary time-dependent matrices.

\chapter{Observable Impacts of a Magnetic Field}
\label{Chapter-CMB}

\section{Overview}
The earliest probe into the nature of a magnetic field that we possess comes from the era of nucleosynthesis. A magnetic field alters nucleosynthesis directly and indirectly. The direct impact comes both from a shift in nuclear energy levels and from the magnetic energy density, decaying as radiation, altering the universe's expansion rate. The indirect impact arises from the gravitational waves that the magnetic field will generate. Strong nucleosynthesis constraints on the energy density in gravitational waves are already known and these can provide equally strong constraints on the field strength and spectral index of a power-law magnetic field. Likewise, constraints are known on the expansion rate of the universe during recombination and the possible energy density in radiation; given the strong bounds on current photon and neutrino densities, these constraints can then be passed on to the strength of the magnetic field.

However, the chief diagnostic for early universe magnetic fields -- and other primordial, non-linear sources -- is the CMB. For the case of a magnetic field, it has been shown that the temperature angular spectrum $C_{T,l}$ from magnetically-induced vector and scalar perturbations increases slightly across all angles, but that the magnetic signal remains subdominant to that from standard scalar perturbations until around $l\approx 2000$, depending on the field strength and spectral index. The field also sources tensor modes, but they are relatively low-amplitude; their signal is similar to the inflationary gravitational wave signature but likely to be rather weaker \cite{Lewis04-Mag}. The direct impact of a magnetic field onto the CMB $\langle TT\rangle$ correlation does not, then, appear to be an ideal probe of a magnetic field. Unfortunately this is the component of the CMB map that is known to the highest resolution; WMAP is limited only by cosmic variance up to $l\approx 350$ \cite{WMAP-Bennett} and in the smaller-scale r\'egime observations from CBI and VSA \cite{RajguruEtAl05}, ACBAR \cite{ACBAR} and the forthcoming SPT \cite{SPT04} missions will determine the spectrum to ever-greater accuracies. The Planck satellite \cite{Planck} is expected to extend the region limited only by cosmic variance to $l\approx 1500$. Even so, the high-$l$ r\'egime is littered with foreground sources and contamination from these will be high. The CBI mission observed a weak increase of power observed on small scales, as compared to the concordance model \cite{RajguruEtAl05}; should this be real and not a statistical or systematic artifact, it could possibly be explained in part by employing a primordial magnetic field (e.g. \cite{YamazakiIchikiKajino04,YamazakiEtAl06}). However, nucleosynthesis bounds imply that a primordial field is unlikely to account for all of the increase.

In addition to the contributions to the CMB temperature fluctuations, a magnetic field also sources polarisation $E$-modes; these will likely be subdominant to those from the standard model until the very small scales. Due to the presence of both vector and tensor perturbations, the field also induces $B$-mode polarisation. In the standard picture, these are produced only from lensed $E$-modes and from inflationary gravitational waves. One could conclude that, in principle, a $B$-mode power spectrum would give the clearest signature for primordial magnetic fields. Unfortunately, the CMB polarisation maps are poorly known compared to the temperature maps. While we currently possess a power spectrum $C_{TE,l}$, this is by no means cosmic-variance limited on any scale. The 2-, 3- or 4-year WMAP results, when they emerge, are expected to improve this situation somewhat, but WMAP was not designed to probe the CMB polarisation with this accuracy. The $\langle EE\rangle$ correlation has been detected to some accuracy (e.g. DASI \cite{LeitchEtAl04}, CAPMAP \cite{BarkatsEtAl04}, CBI \cite{ReadheadEtAl04} and Boomerang \cite{MontroyEtAl05}, the latter two of which also detected a rough power spectrum). The observations of the $B$-modes yield bounds consistent with zero (e.g. \cite{LeitchEtAl04,ReadheadEtAl04}). These observations are on relatively small scales, directly observing the region at which magnetic effects may come to dominate; however, we are some distance from the required accuracy, particularly for the $B$-modes. See \cite{EnsslinEtAl05} for a recent review of the prospects for detecting magnetic-induced polarisation with the Planck satellite.

Given the limitations of power spectra, the non-Gaussianity of the temperature map is a reasonable place to look to further the constraints on primordial magnetic fields and other non-linear sources. While thus far the observations are entirely consistent with Gaussian initial conditions, there are non-Gaussian features in the WMAP maps (e.g. \cite{LandMagueijo05-CubicAnomaly}) and the number of non-Gaussian features could well increase with the next generation of CMB experiments. There are many ways to characterise non-Gaussianity, and a search for non-Gaussian signatures from primordial magnetic fields will be eased by predictions of their form. The primordial bispectra we derived are generally strongest on very large scales; while this feature will naturally be smeared out in the 2-D multipole space we should still expect the most significant magnetic non-Gaussianities to lie on the large scales, at which WMAP is generally cosmic-variance limited. This chapter is concerned with planning how such predictions would be made, along with confirming and extending the range of two-point signals.

In this chapter we approach the magnetised CMB with a slightly different philosophy to previous studies in the literature, and demonstrate that our approach is not only valid but also highly extensible. While we concentrate on the CMB angular power spectrum for Gaussian power-law fields, these provide a material proof that one may take pre-computed power- or bi-spectra and predict from these the statistical nature of the CMB sky for higher-orders and more general fields. This then immediately implies that mapping the magnetic non-Gaussianities onto the CMB is realistically possible to a good accuracy, and far more generally than would be possible attempting to derive their forms analytically in the approach generally applied to magnetic fields (e.g. \cite{MackKahniashviliKosowsky02}).


We first briefly review the constraints placed on a magnetic field from nucleosynthesis before considering the impacts of a magnetic field upon large-scale cosmology. After justifying the approach we shall take, we derive approximate large-scale solutions to the vector and tensor photon transfer functions. We briefly provide an overview of the previous research into the magnetised CMB angular power spectrum and highlight the areas in which it is deficient, before deriving the two-point moments employing our own approach. This is used as a ``proof-of-concept'' of our methods and as demonstration that it can be applied with little extension to characterising CMB non-Gaussianities arising from a primordial magnetic field. We finish by introducing the CMB angular bispectrum and demonstrating that our approach is easily adapted to its study, and by outlining a procedure by which we could add magnetic support to the CMBFast code of Zaldarriaga and Seljak \cite{ZaldarriagaSeljak97}.

\section{Nucleosynthesis}
\label{Nucleosynthesis}
\subsection{Direct Bounds}
There are two major ways in which a magnetic field could directly influence nucleosynthesis. The first, and perhaps most obvious, is that electrons tend to circulate around a magnetic field, which leads \cite{Widrow02} to a quantisation of their energy to
\be
  E^2=p^2+m_e^2+2eB\hbar n_s
\ee
where $n_s$ is the quantum number characterising the induced Landau levels. With this altered energy the electron energy density will also naturally be altered, in turn altering the neutron decay rates. A stronger magnetic field will tend to increase neutron decay and hence lower the amount of (for example) He$_4$ generated during nucleosynthesis. The magnetic alteration becomes significant for field strengths above a critical
\be
  B_c\approx\frac{m_e^2}{e\hbar}
\ee
and assuming that our field is at or below this level provides us with a loose bound on the magnetic energy density during nucleosynthesis, which is approximately
\be
  B_{|\mathrm{nuc}}<B_c\approx 4.4\times 10^{13}\mathrm{G}.
\ee

The other direct impact a magnetic field has on nucleosynthesis comes from its energy density altering the expansion of the universe and thus the evolution of temperature. The presence of another radiative field will act to increase the universe's expansion rate and thus decrease the amount of time during which nucleosynthesis is possible -- and during which neutrons can decay. With fewer neutrons decaying, we thus have an \emph{increase} in He$_4$ due to the presence of a magnetic field. Since we know the helium abundance and the energy density in radiation during nucleosynthesis, we can then derive a bound on the magnetic field, which turns out to be tighter than that we can place from energy considerations. A magnetic field of
\be
  B\approx 6\times 10^{12}\mathrm{G}
\ee
present during recombination would provide an energy density equal to that of the universe and is thus rather too high. Taking a primordial magnetic field to be smaller than this (and scaling with $B\propto a^{-2}$ to the current epoch), direct nucleosynthesis bounds can constrain a magnetic field to be approximately
\be
  B\lesssim \mu\mathrm{G} .
\ee
For further details, see \cite{Widrow02} and its references.

\subsection{Indirect Bounds}
\label{Nucleosynthesis-Indirect}
In this section we summarise the arguments of Caprini and Durrer \cite{CapriniDurrer02,CapriniDurrer05} that impose stringent limits on tangled magnetic fields that arise from nucleosynthesis bounds. The tensor Einstein evolution equation in the presence of a magnetic field is
\be
  \ddot{h}_{ij}+\hub\dot{h}_{ij}+k^2h_{ij}=\frac{k\tau^T(\mathbf{k})}{a^2} .
\ee
Caprini and Durrer solve this equation approximately for a stochastic magnetic field with a power-law spectrum and characterise the resulting gravitational waves by their amplitude and their power spectra, which is bluer than the magnetic spectrum if $n<0$ and redder than the magnetic spectrum if $n>0$. They then use these gravitational waves, and the existing limits on gravitational waves present during nucleosynthesis, to constrain the magnetic fields. These limits are heavily dependent on the time at which the field was produced, the primordial damping scale at that time -- as we showed in \S\ref{Sec-DampingScaleMHD}, the damping scale is time-dependent -- and the time at which the viscous damping scale dominated over the primordial damping scale. They are also extraordinarily strong; for a field generated at a conformal time $\eta_{\mathrm{in}}$ and smoothed on a scale $\lambda$, they present the approximate limits
\be
  B_\lambda<700h\left(\frac{\eta_{\mathrm{in}}}{\lambda}\right)^{(n+3)/2}\sqrt{2^{(n+5)/2}\Gamma\left(\frac{n+5}{2}\right)}\mathrm{nG} .
\ee
For fields produced during inflation with $\eta_{\mathrm{in}}\approx 8\times10^{-9}s$, this provides limits down to $10^{-30}\mathrm{G}$ for $n>0$ and leaves weak constraints only for fields with $n\approx -3$. It should be noted, however, that these limits are derived with a smoothing scale $\lambda=0.1\mathrm{Mpc}$. The constraints for a more typical scale $\lambda=1\mathrm{Mpc}$ are correspondingly weaker.

The approach Caprini and Durrer take is not without criticism; Kosowsky \emph{et. al.} \cite{KosowskyEtAl04} comment that the conversion of energy from the magnetic fields into the gravitational waves has no impact on the energy density of the universe or its evolution, since both gravitational waves and the magnetic energy density decay as radiation, implying that the limits are on the total energy density of the magnetic field. This returns us to a limit of between nano- and micro-Gauss. Caprini and Durrer present a defence in \cite{CapriniDurrer05}. Caprini and Durrer's argument is that, since the damping scale evolves with time, the magnetic energy density does \emph{not} decay as radiation but faster. A bound placed on the magnetic energy density at nucleosynthesis will then yield a tighter bound than would otherwise be the case. So far as this goes, we will tend to view Kosowsky \emph{et. al.}'s bounds as somewhat conservative and consider the Caprini/Durrer bounds as perhaps slightly stronger than is warrented.

\section{Large-Scale Magnetised Cosmology}
\label{Sec-ApproxDeltaTl}
\subsection{General Approach}
The primordial magnetic field in our model is frozen into the plasma on large-scales and decays (\ref{MaxwellCosmological}) as
\bdm
  \mathbf{b}\propto\frac{1}{a^2} \Rightarrow \dot{\mathbf{B}}=0 .
\edm
The field is in a tangled configuration and possesses no directionality on scales larger than some length $\lambda$, of the order of megaparsecs, on which length scale the magnetic energy is $B_\lambda$. We characterise the statistics of this field with a power spectrum
\bdm
  \langle B(\mathbf{k})B(\mathbf{k}')\rangle=\mathcal{P}_B(k)P_{ab}(\mathbf{k})\delta(\mathbf{k-k}')
\edm
and in the simplest case of a Gaussian field the power spectrum contains all the information about the field. This spectrum is generally taken to be a power law
\bdm
  \mathcal{P}_B(k)=Ak^n .
\edm

In the later universe, the magnetic field damps on an evolving scale $k_D$ (\ref{FreeStreamDampingScale}). This scale is relevant back to neutrino decoupling; at the earliest times, photon viscosity dominates neutrino viscosity until $\eta\approx 10^5\mathrm{s}$, with neutrino viscosity dominating until decoupling when it again becomes subdominant to photon viscosity \cite{CapriniDurrer02}. The magnetic field was generated with a primordial damping scale $k_c^{\mathrm{prim}}$, and the effective magnetic damping scale will be
\be
  k_c=\mathrm{min}\left(k_c^{\mathrm{prim}},k_D(\eta)\right) .
\ee
At early times, then, the magnetic field is effectively static since damping processes occur on very small scales -- that is, at early times, $k_c^{\mathrm{prim}}<k_D$. However, since $k_D$ decays, at later times the situation is reversed and the damping scale of the magnetic field evolves. We return to this point in \S\ref{Sec-Correlations}.

The magnetic field possesses a stress-energy tensor (\ref{ScaledStressEnergy})
\be
  \tau^a_b(\mathbf{k})=\frac{1}{2}\tilde{\tau}^i_i(\mathbf{k})\delta^a_b-\tilde{\tau}^a_b(\mathbf{k}),
   \quad \tilde{\tau}^a_b(\mathbf{k})=\int B^a(\mathbf{k}')B_b(\mathbf{k-k}')\frac{d^3\mathbf{k}}{(2\pi)^3}
\ee
which can be separated into its scalar trace, traceless scale, vector and tensor components. The intrinsic two- and three-point correlations between these components were considered in detail in the last chapter. These correlations will form the underlying statistics for the magnetic sources and will be imparted onto the perturbations they produce. From the results of chapter \ref{Chapter-SourceStats} we can see that there are two r\'egimes -- the ultraviolet, in which the damping scale dominates the sources, and the infra-red in which the sources are generally independent of the damping scale. The boundary between these r\'egimes differs with the statistic that one is considering: for a 2-point moment it is at $n=-3/2$, while for a 3-point moment it is at $n=-1$.

We modelled the 2- and 3-point statistics of the magnetic sources by generating simulated fields on a grid, concentrating on those from power-law Gaussian fields for which we can also calculate the sources through a mixture of analysis and numerical integration. For the simulated fields, our use of a grid imposes an unphysical but unavoidable infra-red cut-off which we will have to deal with before employing the results for CMB study. It also results in a limited dynamic range from memory restrictions on the size of the grid. However, it also allows a great flexibility in the nature of the fields that we might choose to model. The results from the simulations, for both 2- and 3-point moments, were entirely consistent with the analysis.

The magnetic stresses source both plasma and gravitational perturbations over time. In the scalar case we have density and velocity perturbations in the plasmas and the two scalar metric perturbations. In the vector case there are the two velocity perturbations and two metric perturbations, while in the tensor case there are the two tensor perturbations. At each time, the statistics of the sources are also generated in the perturbations. However, we should note that coherence is not guaranteed -- perturbations generated relatively late may not be simply related to those sourced relatively early, due to the evolution of the magnetic damping scale. If the primordial field is infra-red, then this is not an issue and the simple outline that we present in this chapter is entirely valid, since the fields evolve purely as $a^{-2}$, and this time-dependence is easily factored out. For ultra-violet fields, however, the source statistics are evolving with the damping scale -- this causes the heavy dependence on cut-off scale noted by Koh and Lee \cite{KohLee00}, for example. It is unclear at the moment to what extent this is an issue; we briefly comment in more detail in \S\ref{Sec-Correlations}. As we mentioned before, it is also interesting that the boundary between infra-red and ultra-violet behaviour occurs at a different spectral index for 2- and 3-point moments; a bispectrum can possess a bluer index than a power spectrum while remaining infra-red.

Assuming that the imparted statistics remain coherent, we can then approach the induced CMB perturbations by evaluating the transfer functions generated by a magnetised cosmology, for which we can employ different levels of approximation or the results from a Boltzmann code. This process can be decoupled entirely from the statistics; with one set of transfer functions we can consider a wide variety of CMB statistics. The usual methods for predicting magnetised signatures on the sky (e.g. \cite{SubramanianBarrow02,MackKahniashviliKosowsky02,YamazakiEtAl06}) involve attempting to find closed analytical approximations for the properties of the magnetic field and then evaluating the statistics of the perturbations as they stand at recombination. These approaches are very hard to extend beyond simple Gaussian, power-law magnetic fields and will be difficult to extend to higher-order statistics. Our approach -- while obviously physically identical -- does not rely on the form of the magnetic field. We can thus with little effort model the intrinsic statistics of non power-law or non-Gaussian fields and, using the same transfer functions, generate the corresponding CMB statistics.

With coherent sources, we can find the CMB statistical imprints from static realisations; speaking heuristically, the imprint at one point in the sky will be given in Fourier space by a transfer function, embodying the evolution, multiplying a factor embodying the phase information,
\be
  \Delta(\mathbf{k},\mathbf{n})\sim \sum_l\Delta_{Tl}(k)S(\mathbf{k})P_l(\mu)
\ee
with
\bdm
  \langle S(\mathbf{k})S^*(\mathbf{k}')\rangle=\delta(\mathbf{k}-\mathbf{k}') .
\edm
The CMB statistics can then be built up by considering correlations of this and integrating across wavemodes. Since the transfer functions do not contain any statistics they can be taken outside the correlation and we can heuristically consider correlations of the form
\beas
  \langle\Delta\Delta\rangle&\propto&\int\left|\Delta_{Tl}(k)\right|^2\langle S(\mathbf{k})S(\mathbf{k'})\rangle d^3\mathbf{k}, \\
  \langle\Delta\Delta\Delta\rangle&\propto&\iiint\Delta_{Tl}(k)\Delta_{Tl}(p)\Delta_{Tl}(q)\langle S(\mathbf{k})S(\mathbf{p})S(\mathbf{q})\rangle d^3\mathbf{k}d^3\mathbf{p}d^3\mathbf{q} .
\eeas
We can employ any combination of approximations for the transfer functions and for the source statistics, which means that for each particular case we consider we can employ a spectrum evaluated analytically, or employ an approximation to the analytical spectrum, or interpolate between points on a realised grid. We can then fold these across transfer functions best suited for our aims, whether approximate or calculated from a Boltzmann routine.

In the remainder of this section we consider the magnetic damping scales and the parameters of the primordial magnetic field $B_\lambda$, $k_\lambda$ and the Alfv\'en velocity $v_A$, before briefly considering the resulting potential issue of the non-coherence of our sources. We then focus on approximations for the large-scale vector and tensor transfer functions, neglecting the induced scalar perturbations as subdominant to signals from the standard mode. We assume a flat universe evolving today as $a\propto\left(\eta/\eta_0\right)^2$ implying that the current conformal time is given by $\eta_0\approx 2/H_0$. This will be a reasonable approximation since the universe has only recently begun accelerating. Given $H_0$ and the current densities of baryons, matter, photons and neutrinos we then have a full system. We shall take as a first approximation the 1-year WMAP parameters, that is that
\be
  \Omega_{0b}h^2=0.0224, \quad \Omega_{0\gamma} h^2=2.48\times 10^{-5}h^{-2}, \quad \Omega_{0\nu}=1.69\times 10^{-5}h^{-2}, \quad h=0.71 .
\ee

\subsection{Damping Scales}
\label{Sec-DampingScales}
The primordial magnetic field is assumed to have been generated at some (brief) redshift $z_{\mathrm{in}}$ corresponding to some generation epoch $\eta_{\mathrm{in}}$. At generation, the field will have some primordial cut-off scale $k_c^{\mathrm{prim}}$. However, viscous processes operate in the interim period between magnetogenesis and recombination, both from neutrinos before decoupling and from photons at very early times and after neutrino decoupling, and this generates a second damping scale. While the viscous damping scales remain below the primordial damping scales -- that is, while $k_D>k_c^{\mathrm{prim}}$ -- the fields are static. However, the viscous damping scale evolves and eventually will pass the intrinsic damping scale and at this point the magnetic field will begin to damp on ever smaller scales. Caprini and Durrer \cite{CapriniDurrer02} present an overview of the history of the magnetic damping scale.

After neutrino decoupling, the damping scale is tied to the Alfv\'en scale, which we assume to be smaller than the intrinsic damping scale, so
\be
\label{AlfvenScale}
  k_c\approx\left(\sqrt{\frac{3}{5}}v_Al_\gamma\right)^{-1}
\ee
where $l_\gamma$ is the photon diffusion length (see \S\ref{Sec-DampingScaleMHD}) and $v_A$ the Alfv\'en velocity. This scale will govern the damping of the magnetic field; the two-point correlations of the sources are then damped at a scale $2k_c$.

The scalar and vector modes evolve up to the epoch of recombination; at recombination the Silk length is given by
\be
  \lambda_S\approx 2.7\left(\Omega_0\Omega_{0b}^2h^6\right)^{-\frac{1}{4}}\mathrm{Mpc}
\ee
(see \cite{Peacock}). For scalar and vector perturbations we then find the damping scale by substituting this for $l_\gamma$ in the Alfv\'en scale (\ref{AlfvenScale}).

In \S\ref{Sec-TensorCl} we demonstrate that magnetically-sourced tensor perturbations are effectively negligible following matter-radiation equality; as such we can take the Alfv\'en damping scale at equality to determine the damping scale for tensor perturbations. The photon diffusion length at equality is given \cite{MackKahniashviliKosowsky02} by
\be
  l_\gamma\approx 19.5\left(\frac{\Theta_{\mathrm{eq}}}{0.25\mathrm{eV}}\right)^{-5/4}\left(\frac{\Omega_{0b}h^2}{0.0125}\right)^{-1/2}h^{-1/2}\mathrm{Mpc}
\ee
with $\Theta_{\mathrm{eq}}$ the photon temperature at equality, and thus the tensor damping scale by
\be
  k_D^T\approx\left(\sqrt{\frac{3}{5}}v_Al_\gamma\right)^{-1} .
\ee

These damping scales were derived in a formalism that assumed a large homogeneous component to the magnetic field. This is not strictly the case here and so to evaluate them we employ an effective field strength from $B_\lambda$. The Alfv\'en velocity is given by
\be
\label{AlfvenVelocity}
  v_A^2=\frac{\overline{B}_{\mathrm{eff}}^2}{4\pi\left(\overline{\rho}_\gamma+\overline{p}_\gamma\right)} .
\ee
Equation (\ref{SpectrumNormalisation}) related the magnetic field strength when smoothed on some scale $\lambda$ to the power spectrum;
\bdm
  B_\lambda^2=\int\mathcal{P}(k)e^{-\lambda^2k^2}d^3\mathbf{k}=4\pi A\int k^2\mathcal{Q}(k)e^{-\lambda^2k^2}dk
\edm
with $\mathcal{P}(k)=A\mathcal{Q}(k)$ (see (\ref{AmplitudePowerSpectrum})). This implies that we can relate a field smoothed to a scale $k_\lambda$ to an effective field smoothed on a scale $k_c$. Taking the ratio between the field $B_\lambda$ smoothed on the scale $k_\lambda$ and the effective field $\overline{B}_{\mathrm{eff}}$ cut-off on the scale $k_c$ gives
\be
\label{BB}
  \frac{B_\lambda^2}{\int k^2\mathcal{P}(k)\exp{(-k^2/k_\lambda^2)}dk}=
   \frac{\overline{B}_{\mathrm{eff}}^2}{\int k^2\mathcal{P}(k)\exp{(-k^2/k_c^2)}dk} .
\ee
Using the Alfv\'en velocity (\ref{AlfvenVelocity}) and magnetic damping scale (\ref{AlfvenScale}) we can also see that
\be
\label{kC}
  k_c=\sqrt{\frac{10\Omega_{\gamma 0}}{3G}}\frac{a^2H_0}{\lambda_S\overline{B}_{\mathrm{eff}}}\mathrm{Mpc}^{-1} .
\ee
We then have two equations for our two unknowns $\overline{B}_{\mathrm{eff}}$ and $k_c$; $a$ and $\lambda_S$ are evaluated at recombination. A closed form of this system is not possible. We can, though, employ a simple iterative procedure wherein we make a rough approximation for either the smoothed field or the cut-off wavenumber and use (\ref{kC}) to determine the other. We then adjust our parameters until the integration of (\ref{BB}) is consistent. Once $k_c$ is known we can then employ (\ref{SpectrumNormalisation}) to determine the normalisation of the magnetic spectrum. While cumbersome, this approach allows us to retain generality. In the tensor case we substitute $l_\gamma$ evaluated at matter-radiation equality for the Silk length and employ $a_{\mathrm{eq}}$.

For a power-law field, an approximate solution to the normalisation of the magnetic spectrum (\ref{SpectrumNormalisation}) is \cite{MackKahniashviliKosowsky02}
\be
  A\approx \frac{2^{n+4}\pi^{n+5}}{\Gamma{\left(\frac{n+3}{2}\right)}}\frac{B_\lambda^2}{k_\lambda^{n+3}} .
\ee
We can then determine an approximate effective $\overline{B}_{\mathrm{eff}}$, damping on a scale $k_c$, by
\be
  \overline{B}_{\mathrm{eff}}\approx B_\lambda\left(\frac{k_c}{k_\lambda}\right)^{\frac{n+3}{2}} .
\ee
Substituting the effective background field into the Alfv\'en velocity gives us
\be
  v_A=\sqrt{\frac{G}{2H_0^2a^4\Omega_{\gamma 0}}}\left(\frac{k_c}{k_\lambda}\right)^{(n+3)/2}B_{\lambda},
\ee
leading to a magnetic damping scale at recombination of
\be
  k_c=\left(\frac{10}{3G}\right)^{1/(n+5)}\left(\frac{H_0^2\Omega_{\gamma 0}a_{\mathrm{rec}}^4}{B_\lambda^2\lambda_{S\mathrm{rec}}^2}\right)^{1/(n+5)}k_\lambda^{(n+3)/(n+5)} .
\ee
The tensor damping scale takes this form with $l_\gamma$ substituted for $\lambda_S$ and $a$ evaluated at matter-radiation equality rather than recombination.

\subsection{Non-Equal Time Correlations}
\label{Sec-Correlations}
In \S\ref{Line-of-Sight-CMB} we derived the equations by which the intrinsic statistics of a primordial source are wrapped on to the CMB across the transfer functions. However, in our derivation we implicitly assumed that the evolution of each wavemode remains coherent -- that is, that each $k$-mode evolves independently and the primordial statistics are easily transferred onto the CMB. To firm this statement up, consider the multipole moments of the scalar CMB,
\be
  a_{T,lm}=\int_\mathbf{k}\left(\frac{1}{4}\int_{\Omega_\mathbf{n}}Y^*_{lm}(\mathbf{n})\int_\eta S_T^S(k,\eta)e^{-ix\mu}d\eta d\Omega_\mathbf{n}\right)\frac{d^3\mathbf{k}}{(2\pi)^3} .
\ee
The general two-point correlation of this is
\bea
\label{FullCorrelation}
  \langle a_{T,lm}a_{T,l'm'}\rangle&=&\int_\mathbf{k}\int_{\mathbf{k}'}\Bigg(\frac{1}{16}\int_{\Omega_\mathbf{n}}\int_{\Omega_\mathbf{n'}}
   Y^*_{lm}(\mathbf{n})Y_{l'm'}(\mathbf{n'}) \\ &&\; \times \int_\eta\int_{\eta'}
   \langle S_T^S(\mathbf{k},\eta)S_T^{*S}(\mathbf{k'},\eta')\rangle e^{i(x'\mu'-x\mu)}d\eta d\eta'd\Omega_\mathbf{n}d\Omega_\mathbf{n'}\Bigg)
   \frac{d^3\mathbf{k}}{(2\pi)^3}\frac{d^3\mathbf{k'}}{(2\pi)^3} . \nonumber
\eea
The quantity $\langle S_T^S(\mathbf{k},\eta)S_T^{*S}(\mathbf{k'},\eta')\rangle$ is known as a non-equal time correlator; such correlators appear frequently in studies of defect models (e.g. \cite{VincentEtAl96,TurokPenSeljak97,Durrer97,DurrerKunzMelchiorri01,BevisEtAl04}) and should in general be calculated in their entirety. In the standard picture the perturbations are coherent -- the statistics are imprinted at the end of inflation and each mode then evolves independently. In this case, we can then decouple the non-equal time correlator into
\be
  \langle S_T^S(\mathbf{k},\eta)S_T^{*S}(\mathbf{k'},\eta')\rangle=\delta(\mathbf{k}-\mathbf{k'})\sqrt{\langle|S_T^S(k,\eta)|^2\rangle}\sqrt{\langle|S_T^S(k,\eta')|^2\rangle} .
\ee
Substitution in (\ref{FullCorrelation}) then ultimately reproduces the previous result, (\ref{StandardApproachCl}).

On large scales, the magnetic field is frozen in to the plasma and this simple evolution would lead one to believe that we can employ such an approach. For infra-red fields, with statistics that do not depend on the cutoff scale, this is still the case. However, for ultraviolet fields the statistics are dominated by the cutoff scale, which evolves with time. An active, evolving source causes an amount of decoherence; different wavemodes will evolve in a different manner, and we must employ the non-equal time correlators. In general, these have to be computed, which can be very intensive, albeit entirely realistic, especially when one works with field realisations.

This point does not appear to have been previously addressed in the literature, and correspondingly the results presented for ultraviolet fields should be treated with some caution until we better understand the microphysics around the cutoff scale. While we will work with the assumption that we can assume a simple, coherent approach for all types of field, it should be remembered that our approach is strictly applicable only to the infra-red r\'egime.

We also comment that, since an $a$-point moment leaves the ultraviolet r\'egime at a spectral index
\be
  n=-\frac{3}{a},
\ee
it is possible to consider fields with $n\in(-3/2,0)$ as independent of the cutoff scale so long as one considers a correlation of high enough order.

\subsection{Vector Transfer Functions}
In this section we produce large-scale approximations for the vector photon transfer functions, which will enable us to predict the nature of the low-$l$ CMB power spectra from magnetic vorticity and anisotropic stresses. We neglect the scalar mode; the scalar CMB signature is likely to be insensitive to a realistically observable magnetic field. It is, moreover, rather intractable. In contrast, the tensor mode might be substantially enhanced as compared to a standard inflationary model, and in our scenario the only source for vector perturbations comes from the magnetic field. Finding an approximation for the vector modes will require us to construct a tight-coupling limit similar to that we built for the scalars. The tensors are rather simpler. This section resembles the analysis of Mack \emph{et. al.} \cite{MackKahniashviliKosowsky02} but employs our own formalism.

The transfer functions we derive here should be trusted only on relatively large scales. Mack \emph{et. al.} derive similar transfer functions and apply them to multipoles as high as $l=500$; however, we feel this enormously overstates their validity due to the approximations involved. Comparison with the numerical results of Lewis \cite{Lewis04-Mag} suggests that these should be trusted for $l\lesssim 300$ at the most. For smaller-scale studies we will require the results of a numerical Boltzmann code. However, the approximations for the vector and tensor transfer functions will prove useful in demonstrating a ``proof-of-concept'' for our approach to the problem.

\label{Sec-VectorCl}
The evolution of the baryon vector velocity is given (\ref{VectorEulerMHD}) by
\be
  \dot{v}_b^V+\hub v_b^V+\frac{4}{3}\frac{\overline{\rho}_{\gamma a}}{\bkrO}an_e\sigma_T\left(v_b^V-v_\gamma^V\right)
   =\frac{ik}{\bkrO}\tau^V(\mathbf{k})
\ee
where we are considering only one mode and suppressing the index. We shall neglect the contribution from the neutrinos; while for a strictly accurate analysis we must include these, and Lewis has commented that much of the large-scale power is in fact controlled by a cancelling neutrino stress rather than regularity of the vorticity \cite{Lewis04-Mag}, our approach should be sufficient for a first order calculation.

This Euler equation is supplemented by the vector Einstein equations (\ref{EinsteinEqs})
\bea
\label{VectorEinsteinEq}
 \dot{\tilde{V}}+2\hub\tilde{V}&=&
  16\pi G\frac{a^2}{k}\left(\frac{1}{5}\overline{\rho}_{\gamma a}\left(\Delta^V_{T3}-\Delta^V_{T1}\right)
   +\frac{1}{a^2}\tau^V(\mathbf{k})\right), \\
  \tilde{V}&=&-16i\pi G\frac{a^2}{k^2}\left(\bkrO v_b^V+\overline{\rho}_{\gamma a} v_\gamma^V\right)
\eea
where we are employing the gauge-invariant metric perturbation $\tilde{V}=(i/k)\dot{h}^V$ (\ref{BardeenVariables}). We also have the vector Boltzmann hierarchies for the photons, (\ref{PhotonsVectorHierarchyT}-\ref{PhotonsVectorHierarchyP}). Adapting the procedure we employed for scalar perturbations in section \S\ref{Sec-TightlyCoupled} we write this hierarchy separated to zeroth and higher orders in $t_c=(an_e\sigma_T)^{-1}$,
\bea
  \Delta^V_{T0}&=&4iv_b^V+t_c\left(k\Delta^V_{T1}-\dot{\Delta}^V_{T0}\right), \nonumber \\
  \Delta^V_{T1}&=&-\frac{1}{3}\Phi^V+t_c\left(\frac{2}{3}k\Delta^V_{T2}-\frac{1}{3}k\Delta^V_{T0}
    -\frac{4}{3}ik\tilde{V}-\dot{\Delta}^V_{T1}\right), \nonumber \\
  \Delta^V_{Tl}&=&t_c\left(\frac{k}{2l+1}\left((l+1)\Delta^V_{Tl+1}-l\Delta^V_{Tl-1}\right)-\dot{\Delta}_{Tl}^V\right),
    \quad l\geq 2; \\
  \Delta^V_{P0}&=&\Phi^V+t_c\left(k\Delta^V_{P1}-\dot{\Delta}^V_{P0}\right), \nonumber \\
  \Delta^V_{Pl}&=&t_c\left(\frac{k}{2l+1}\left((l+1)\Delta^V_{Pl+1}-l\Delta^V_{Pl-1}\right)-\dot{\Delta}_{Pl}^V\right),
    \quad l\geq 1. \nonumber
\eea
Here
\bdm
 \Phi^V=\frac{3}{5}\Delta_{P0}^V+\frac{3}{7}\Delta_{P2}^V-\frac{6}{35}\Delta_{P4}^V
   -\frac{3}{10}\Delta_{T1}^V-\frac{3}{10}\Delta_{T3}^V .
\edm
Recursively substituting through the hierarchies down from $l=4$ for the temperature and $l=5$ for the polarisation and back up, we can readily see that to the first order in $t_c$,
\bea
  \Delta_{T0}^V&=&4iv_b^V-4it_c\dot{v}_b^V+\mathcal{O}(t_c^2), \nonumber \\
  \Delta_{T1}^V&=&-\frac{4}{15}ikt_c\left(v_b^V+\tilde{V}\right)+\mathcal{O}(t_c^2), \nonumber \\
  \Delta_{Tl}^V&=&\mathcal{O}(t_c^2), \quad l\geq 2; \\
  \Delta_{P0}^V&=&\frac{4}{10}ikt_c\left(v_b^V+\tilde{V}\right)+\mathcal{O}(t_c^2), \nonumber \\
  \Delta_{Pl}^V&=&\mathcal{O}(t_c^2), \quad l\geq 1, \nonumber \\
  \Phi^V&=&-\frac{4ik}{3}t_c\left(v_b^V+\tilde{V}\right) . \nonumber
\eea
Along with the photon velocity (\ref{PhotonVelocities})
\bdm
  v_\gamma^V=\frac{i}{4}\left(\Delta^V_{T2}-\Delta^V_{T0}\right)
\edm
we then have $v_b^V=v_\gamma^V$ and the tightly-coupled Euler equation
\be
  \left(\bkr_{ba}+\frac{4}{3}\bkr_{\gamma a}\right)\dot{v}_b^V+\bkr_{ba}\hub v_b^V=ik\tau^V .
\ee

Working with the Einstein equations (\ref{VectorEinsteinEq}) in radiation domination and neglecting quantities of first-order in the tight-coupling parameter we can quickly find the approximate solution
\be
  \tilde{V}\approx\frac{16\pi G\tau^V\eta}{ka^2}
\ee
where we have neglected a complementary function that grows in the past. This implies the velocity
\be
  v_b^V\approx\frac{ik\tau^V\eta}{a^4\left(\overline{\rho}_{\gamma a}+\bkrO\right)}=\frac{ik\eta\tau^V}{\bkr_{\gamma 0}(1+R_b)}
\ee
with $R_b=\bkrO/\overline{\rho}_{\gamma a}$. (Compare equations (4.10) and (4.13) in \cite{MackKahniashviliKosowsky02}.)

In matter domination we instead find
\be
  \tilde{V}=\frac{16i\pi G\tau^V}{k}\left(\frac{\eta}{a^4}-\frac{\eta_{\mathrm{eq}}}{a_{\mathrm{eq}}^4}\right)
\ee
where we retained the complementary function and matched the two solutions at the epoch of matter-radiation equality. This implies the velocity
\be
  v_b^V=\frac{ik\eta\tau^V(\mathbf{k})}{a^2\bkr_{\gamma 0}(1+R_b)}
  \left(1-\frac{\eta_{\mathrm{eq}}}{\eta}\left(\frac{a}{a_{\mathrm{eq}}}\right)^4\right) .
\ee

These forms are valid when the wavelength is longer than that of the Silk wavelength $\lambda_S$. To model the wavemodes on a smaller scale we add a photon shear viscosity $\xi$ to the system, giving the tightly-coupled Euler equation
\be
  \left(\bkrO+\frac{4}{3}\overline{\rho}_{\gamma a}\right)\dot{v}_b^V+\left(\bkrO\hub+k^2\frac{\xi}{a}\right)v_b^V=\frac{ik}{a^4}\tau^V
\ee
where we have rescaled the densities to their proper values. Here the photon viscosity is \cite{MackKahniashviliKosowsky02}
\be
  \xi=\frac{4}{15}\overline{\rho}_{\gamma a} al_\gamma
\ee
again with $l_\gamma=(n_e\sigma_T)^{-1}$. Since the modes we are interested in are those which are severely overdamped we can employ a terminal-velocity approximation \cite{SubramanianBarrow98-MHD} by assuming the damping is severe enough that the modes no longer accelerate; doing so we find
\be
  v_b^V=\frac{ik\eta}{a}\tau^V\left(2\bkr_{b0}+\frac{4}{15}(k\eta)(kl_\gamma)\bkr_{\gamma 0}\right)^{-1} .
\ee
(In radiation domination the $2$ in front of the baryon density becomes a $1$, but this term is generally subdominant regardless.) This applies when $k>k_S$. $\tilde{V}$ remains unchanged; we assume that the magnetic vector source dominates over the shear viscosity.

The instantaneous approximation for $\Delta^V_{Tl}$ (\ref{DeltaTV-Approx}) is
\bdm
  \Delta^V_{Tl}\approx\sqrt{\frac{(l+1)!}{(l-1)!}}
   \left(\left(4iv^v+\frac{1}{k}\dot{\Phi}^V\right|_{\eta_{\mathrm{rec}}}
   \frac{j_l(x_{\mathrm{rec}})}{x_{\mathrm{rec}}}
   +\frac{4}{k}\int_{\eta_{\mathrm{rec}}}^{\eta_0}\ddot{h}^V\frac{j_l(x)}{x}d\eta\right)
\edm
recalling that $x=k(\eta-\eta_0)$. $\Phi^V$ is first-order in the tight-coupling parameter and so is neglected; moreover, on large scales, the velocity will dominate over the metric perturbation and so we also neglect the integrated Sachs-Wolfe term. The vector transfer function sourced by a primordial magnetic field is then approximately
\be
\label{DeltaTV-MagApprox}
  \Delta_{Tl}^V\approx4i\sqrt{\frac{(l+1)!}{(l-1)!}}
   \frac{\tau^V(\mathbf{k})}{\bkr_{\gamma 0}}\frac{j_l(x_{\mathrm{rec}})}{x_{\mathrm{rec}}}
   \left\{\begin{array}{r}
    \frac{k\eta_{\mathrm{rec}}}{a^2(1+\left.R_b\right|_{\mathrm{rec}})}
     \left(1-\frac{\eta_{\mathrm{eq}}}{\eta_{\mathrm{rec}}}\right)\left(\frac{a_{\mathrm{rec}}}{a_{\mathrm{eq}}}\right)^4,
      \quad k<k_S, \\
    15/4kl_\gamma, \quad k>k_S
   \end{array}\right. .
\ee

\subsection{Tensor Transfer Functions}
\label{Sec-TensorCl}
Neglecting the anisotropic stresses of the neutrinos and photons (or considering the situation before neutrino decoupling; see \cite{Lewis04-Mag} for a detailed discussion), the tensor mode evolves as
\be
\label{MagTensorEv}
  \ddot{h}^T+2\hub\dot{h}^T+k^2h^T=16\pi G\frac{\tau^T(\mathbf{k})}{a^2} .
\ee
The complementary functions are simple to find; in radiation domination we quickly have
\be
  h^T_R=h_0j_0(k\eta)+h_1y_0(k\eta) ,
\ee
and in matter domination we have
\be
  h^T_M=h_2\frac{j_1(k\eta)}{k\eta}+h_3\frac{y_1(k\eta)}{k\eta}
\ee
where $y_l(x)$ is a spherical Bessel function of the second kind.

We find the full solutions with a Wronskian Green's function method; given an equation
\be
  a(t)\frac{d^2y(t)}{dt^2}+b(t)\frac{dy(t)}{dt}+c(t)y(t)=S(t)
\ee
with the homogeneous solutions $u_1(t)$ and $u_2(t)$, a general solution \cite{Jeffrey} is
\be
  y(t)=c_1u_1(t)+c_2u_2(t)+\int_1^2G(t,\overline{t})S(\overline{t})d\overline{t}
\ee
where
\be
  G(t,\overline{t})=-\frac{1}{A(t)}
   \frac{\left|\begin{array}{cc}u_1(t) & u_2(t) \\ u_1(\overline{t}) & u_2(\overline{t})\end{array}\right|}
    {\left|\begin{array}{cc}u_1(\overline{t}) & u_2(\overline{t}) \\ u'_1(\overline{t}) & u'_2(\overline{t})\end{array}\right|}
\ee
is the Green's function. This implies that we have a particular solution to (\ref{MagTensorEv})
\be
  h^T(y)=16\pi G\tau^T(\mathbf{k})\frac{\eta_{\mathrm{eq}}^2}{a_{\mathrm{eq}}^2}k^2\int_{y_{\mathrm{in}}}^y\frac{\sin(y-y')}{yy'}dy'
\ee
where we have assumed that the magnetic field was generated relatively rapidly at some time $\eta_{\mathrm{in}}$ and are employing the co-ordinate $y=k\eta$. We are interested merely in $\dot{h}^T(k,\eta)$ and large-scales. Since the magnetic source decays as $\eta^4$ in matter domination we also neglect the particular integral beyond matter-radiation equality; this means we may match the waves sourced by the stresses before equality with the homogeneous solutions after equality. Taking $y\ll 1$, retaining only the dominant contribution and matching the derivatives, we find
\be
  \dot{h}(k,\eta)\approx 4\pi G\eta_0^2z_{\mathrm{eq}}\ln\left(\frac{z_{\mathrm{in}}}{z_{\mathrm{eq}}}\right)
   k\tau^T(\mathbf{k})\frac{j_2(k\eta)}{k\eta} .
\ee

The tensor transfer function can then be recovered from the instantaneous-recombination approximation (\ref{DeltaTT-Approx})
\bdm
  \Delta^T_{Tl}\approx\sqrt{\frac{(l+2)!}{(l-2)!}}\left(\left.\Phi^T\right|_{\eta_\mathrm{rec}}
  \frac{j_l(x_{\mathrm{rec}})}{x_{\mathrm{rec}}}^2
   +2\int_{\eta_{\mathrm{rec}}}^{\eta_0}\dot{h}^T\frac{j_l(x)}{x^2}d\eta\right) .
\edm
We neglect the intrinsic term as it will be rapidly dominated by the integrated Sachs-Wolfe term, which we integrate from $\eta=0$ since we shall consider large-scale modes that were superhorizon at decoupling. We then have the approximate tensor transfer function
\be
\label{DeltaTT-MagApprox}
  \Delta^T_{Tl}=\sqrt{\frac{(l+2)!}{(l-2)!}}8\pi G\eta_0^2
   z_{\mathrm{eq}}\ln\left(\frac{z_{\mathrm{in}}}{z_{\mathrm{eq}}}\right)\tau^T(\mathbf{k})
   \int_0^{\eta_0}k\frac{j_2(k\eta)}{k\eta}\frac{j_l(x)}{x^2}d\eta .
\ee
As in the vector case we assume this to be valid up to $l\lesssim 300$.

\section{CMB Power Spectra from Tangled Magnetic Fields}
In this section we consider the CMB angular power spectrum induced by a primordial magnetic field, employing the approximations for the vector and tensor transfer functions (\ref{DeltaTV-MagApprox}, \ref{DeltaTT-MagApprox}) derived above. We begin by summarising the previous literature, both analytical and numerical. In the past, the effects of a magnetic field on the CMB have been derived with predominantly analytical or at most semi-analytical methods. Our approach instead employs the power spectra generated in chapter \ref{Chapter-SourceStats}, wrapping these directly across the approximate transfer functions. The main advantage of our approach is its extensibility -- while the analytical spectra we generated are constrained in the same manner as previous work to Gaussian, power-law fields, realised spectra can be produced for any field configuration. We also have a lessened reliance on approximate analytical results that assume power-law spectra. For the angular power spectrum we confirm and reproduce the previous results and it is merely a technical (albeit non-trivial) exercise to adapt our methods to the CMB bispectra.


\subsection{Analytical Estimates from the Literature}
\label{Sec-CMB-AnalyticLiterature}
Scalar modes have been much neglected in the literature, partly due to their complexity and the number of terms that appear in the estimates, but also because the scalar modes induced by magnetic fields are generally subdominant (e.g. \cite{MackKahniashviliKosowsky02,Lewis04-Mag}); however, numerical models of the scalar modes indicate that their effect can begin to become important on small scales. Analytical and semi-analytical efforts have concentrated primarily on the effects of the vector and tensor modes. While there have been many studies of these, we shall concentrate on very briefly reviewing studies in the two main r\'egimes, $l<500$ \cite{MackKahniashviliKosowsky02} and $l>1000$ \cite{SubramanianBarrow98,SubramanianBarrow02} which takes a somewhat separate approach. We shall concentrate on the temperature power spectrum although we shall mention the polarisation results these authors present.

Subramanian and Barrow \cite{SubramanianBarrow98,SubramanianBarrow02} concentrate on the vector perturbations induced by a magnetic field. With Seshadri \cite{SeshadriSubramanian00,SubramanianSeshadriBarrow03} they also consider the impacts of these modes on the small-scale polarisation. Working in two r\'egimes, that in which photon viscosity is important and that in which it may be neglected, they derive analytical approximations for the two-point auto-correlation function, one of which grows with $l$ and the other of which decays. For a concordance-like cosmology, the peak extrapolated from these two behaviours lies just before the magnetic signal begins to dominate over the standard signal, which happens at $l\approx 2000$ if $B_\lambda\approx 3\mathrm{nG}$ and $n=-2.9$. This signal is from the Alfv\'en (and slow magnetosonic) waves that survive Silk damping through extreme overdamping, and damp only slowly on small scales. The power increases with increasing spectral index and, correspondingly, the signals dominate the concordance signal for a lower $l$. For an early-universe field, though, one should ensure that nucleosynthesis bounds (\S\ref{Nucleosynthesis}) are not broken; if one trusts the gravitational wave bounds \cite{CapriniDurrer02}, nano-Gauss fields with indices much different from $n\approx-3$ are heavily constrained.

Modelling the polarisation signals from the vorticity in much the same manner, they find a $B$-mode beginning to dominate at $l\approx 1000-2000$ depending on the cosmological parameters and magnetic field. The case $B_\lambda\approx 3\mathrm{nG}$, $n=-2.9$ dominates over the standard signals at $l\approx 1700$. Again, shallower spectral indices give larger results on small scales They also found that the $E$-mode is much smaller than the $B$, not least because they are neglecting the scalar modes, and that the standard scalar modes dominate both the temperature and polarisation maps for all larger scales, making the detection of the magnetic signals difficult.

Mack \emph{et. al.} \cite{MackKahniashviliKosowsky02} provide a unifying review of the previous estimates \cite{SubramanianBarrow98, KohLee00,DurrerFerreiraKahniashvili00,  SubramanianBarrow02, SeshadriSubramanian00, MackKahniashviliKosowsky02, CapriniDurrerKahniashvili03, SubramanianSeshadriBarrow03, Lewis04-Mag, YamazakiIchikiKajino04, YamazakiEtAl06} of semi-analytical vector and tensor anisotropies from primordial magnetic fields for larger scales $l<500$. In deriving the statistical properties for their fields they provide various approximate results that have become standard in considering primordial magnetic fields, even for numerical studies.
%
Although widely used, these rely sensitively on the power-law form of the power spectrum and on the assumed Gaussianity of the fields. It would be much preferable to remove the reliance on such approximations, even at the cost of employing numerical integrations to produce the source statistics.
By considering a thin last-scattering surface and employing approximations similar to ours, Mack \emph{et. al.} find the vorticity at recombination in two limits, one in which photon damping may be neglected and one in which it cannot. They proceed to approximate the vector power spectra for a range of field strengths and spectral indices and find forms asymptotically rising to a limit for $l<500$, for all correlations between temperature and polarisation. As was found in the large-$l$ limit, the strengths of the effect rise with spectral index. Repeating the approximations for tensor modes they find similar qualitative results, excepting that the power spectra from gravitational waves are scale-invariant (and dominate over the vector modes) for $n\approx -3$.\footnote{This explains the use of the phrase ``scale-invariant'' when referring to the highly-red magnetic spectrum $n=-3$; this spectral index itself causes irremovable singularities but indices nearing scale-invariant are permitted and, indeed, are the most weakly constrained by big-bang nucleosynthesis \cite{CapriniDurrer02}.}

\subsection{Numerical Estimates from the Literature}
\label{Sec-CMB-NumericalLiterature}
Koh and Lee \cite{KohLee00} performed a limited numerical study of the impact of magnetic fields onto the CMB, considering only the scalar mode. They employed CMBFast, modified to incorporate a tangled magnetic field of the type we considered in chapter \ref{Chapter-SourceStats}. However, the damping scale they employ is not calculated from the photon viscosity but rather a phenomenological argument related to the cluster scale; they recognise this issue and comment that the resulting CMB power spectrum is very sensitive to $k_D$. This strong dependence on the cutoff scale arises from their potentially unwarrented use of ultraviolet fields. They consider both the temperature and polarisation auto-correlations, plus the cross-correlation. In each case they find, as the analytical studies suggested, that the power is boosted with increasing field strength and spectral index. The fields they consider ($n>0$ and $B_\lambda\approx\mathrm{nG}$) are ruled out by the Caprini/Durrer nucleosynthesis bounds.

Lewis incorporated a more rigorous model of a primordial magnetic field into the vector and tensor components of the CAMB code \cite{Lewis04-Mag}, although he neglected the scalar modes. He considered the temperature and both the polarisation spectra although his concentration was on the $B$-mode which gives the cleanest signal for magnetic fields.
For the tensor modes, he found that the gravitational waves sourced by magnetic fields after neutrino decoupling are negligible, the magnetic anisotropic stress being countered to a large degree by the presence of neutrinos. Those waves sourced before neutrino decoupling provide the dominant contribution; this closely resembles the signal from inflation-produced gravitational waves with a modified spectral index. The non-Gaussianity of the signal would discriminate between the two, as would the corresponding impacts on the high-$l$ temperature spectrum and the non-Gaussianities on larger scales. Lewis found that the magnetic signals from vector modes could begin to dominate at a high $l\approx 2000$ for a field $B_{\lambda}\approx 8\mathrm{nG}$ and $n\approx -2.9$. Lewis concluded that fields with $B_\lambda\gtrsim 0.1\mathrm{nG}$ can yield observable impacts on the CMB, albeit requiring good power spectra for the $E$- and $B$-modes.

Perhaps the greatest benefit of Lewis' study is his detailed analysis of the initial conditions of the problem, albeit in the GIC formalism; this area had been somewhat neglected previously. Although he did not perform the calculations, Giovannini \cite{Giovannini04-Mag} provided the scalar initial conditions, in both conformal Newtonian and synchronous gauges, that Lewis neglected. However, while his work presents full temperature and polarisation power spectra from the magnetised vector and tensor perturbations, and transfer functions including the impact of a magnetic field across all scales, Lewis is heavily restricted in the type of field he can consider by his reliance on approximations based on Mack \emph{et. al.}'s formalism. A particular issue is the approximation for $\mathcal{P}^{S,V}(k)$ which for ultraviolet fields is only a marginal fit and depends on the field being both Gaussian and power law. He is also reliant on power-law approximations for the normalisation of the power spectrum $A$, the effective magnetic field $\overline{B}_{\mathrm{eff}}(\eta)$ and the damping scale $k_c(\eta)$. The approximations are regularly good only to the $10\%$ level and this lessens the accuracy of his study somewhat.

Yamazaki \emph{et. al.} \cite{YamazakiIchikiKajino04,YamazakiEtAl06} performed a parameter estimation employing a modified CMBFast including support for scalar and vector modes. In the earlier study \cite{YamazakiIchikiKajino04}, they confirmed for high $n$ the previous results that the CMB begins to deviate from the concordance model at $l\gtrsim 1000$ from the impact of scalar modes, and that the vector mode easily dominates the scalar at larger scales. Unlike Koh and Lee, Yamazaki \emph{et. al.} employed a viscous damping scale rather than roughly determining it from the galactic scale, although it is unclear whether this is taken into account in the evolution equations. They, too, consider only power law spectra and employ the common approximations summarised in, e.g. \cite{MackKahniashviliKosowsky02,Lewis04-Mag}. They find a $2\sigma$ constraint of $B_\lambda\approx 3.9\mathrm{nG}$ for $n\approx 1.1$ where $\lambda=1\mathrm{Mpc}$. They also comment that there is a strong degeneracy between these parameters. If reliable, the Caprini/Durrer nucleosynthesis bounds rule out the existence of such a field in reality. In the more recent study \cite{YamazakiEtAl06} they find a ``concordance region'' on the magnetic parameters of $B_\lambda\in(1,4.7)\mathrm{nG}$ at $1\mathrm{Mpc}$, and $n\in(-3,-2.4)$. This is at the $1\sigma$ level. As with Lewis, the work of Yamazaki and his collaborators is compromised by their reliance on the formalism reviewed in Mack \emph{et. al.} and the approximations they employ -- that is, they are restricted to purely power-law magnetic fields with Gaussian statistics, and rely upon approximate solutions for $A$, $\overline{B}_{\mathrm{eff}}$ and $\mathcal{P}^{S,V}(k)$ that are not necessarily accurate.

\subsection{A New Approach to the Magnetised CMB Power Spectrum}
\label{Sec-ClApprox}
Assuming that the time-evolution of the damping scale has a negligible impact and that we can approximately transfer the statistics employing an equal-time correlation, the integral producing the CMB angular power spectrum is
\be
\label{ClApprox}
  C^A_l=\frac{2}{\pi}\int_0^\infty\mathcal{P}_A(k)\left|\Delta^A_{Tl}(k,\eta)\right|^2k^2dk .
\ee
As a demonstration that we may realistically recover CMB properties by a direct integration of this equation employing the results of chapter \ref{Chapter-SourceStats} we can employ the approximate vector and tensor transfer functions (\ref{DeltaTV-MagApprox}, \ref{DeltaTT-MagApprox}), and either use the power spectra from analysis or from our realisations. The more general method will be to employ the power spectra derived from our realisations; the realised fields, while of a limited dynamic range, can be entirely arbitrary in nature. In addition to the limited range realisations are also hampered by an unphysical infra-red cutoff; we deal with this issue by excising the unphysical region from the realisation and extrapolating its behaviour back to low-$k$. The theoretical spectra, while lacking the errors that somewhat hamper the large-scale realisations, are constrained to be Gaussian in nature and, currently, of a power-law spectrum. While employing analytical results would remove the need to interpolate a pre-existing look-up table, the generality allowed by employing the realisations -- and the saving in time to produce the CMB angular power spectra, even with different approximations for $\Delta_T$ -- can outweigh this advantage.

The form of the vector transfer function (\ref{DeltaTV-MagApprox}) implies that the angular power spectrum will take the form
\be
  C_l^V\approx\alpha_0\int_0^{k_S}k^2\mathcal{P}^V(k)j^2_l(k\eta_{\mathrm{rec}})dk+\alpha_1\int_{k_S}^\infty\frac{j_l(k\eta_{\mathrm{rec}})}{l_\gamma}\mathcal{P}^V(k)dk
\ee
where $\alpha_0$ and $\alpha_1$ are functions of $l$ and the cosmology. On numerical evaluation, the integration on the interval $[k_S,\infty)$ is seen to be subdominant to that on the interval $(0,k_S]$, and so, as did Mack \emph{et. al.}, we neglect the vector transfer function in the damped r\'egime. This can be understood by noting that while the first term will contain a factor $k_S^2$, the second term will contain extremal factors $j_l(k\eta_{\mathrm{rec}})$ and $j_l(\infty)$; these are much smaller than the Silk wavenumber. We can thus take the vector transfer function to be
\be
  \Delta_{Tl}^V\approx4i\sqrt{\frac{(l+1)!}{(l-1)!}}
   \frac{\tau^V(\mathbf{k})}{\bkr_{\gamma 0}}\frac{j_l(x_{\mathrm{rec}})}{x_{\mathrm{rec}}}
    \frac{k\eta_{\mathrm{rec}}}{a_{\mathrm{rec}}^2(1+\left.R_b\right|_{\mathrm{rec}})}
     \left(1-\frac{\eta_{\mathrm{eq}}}{\eta_{\mathrm{rec}}}\right)\left(\frac{a_{\mathrm{rec}}}{a_{\mathrm{eq}}}\right)^4,
      \quad k<k_S .
\ee
We have tested this assumption by integrating across the entire range of $k$; the dominant contribution indeed arises from the $k<k_S$ wavemodes.

The tensor transfer function
\bdm
  \Delta_{Tl}^T(k,\eta_0)\approx2\sqrt{\frac{(l+2)!}{(l-2)!}}\left(4\pi G\eta_0^2z_{\mathrm{eq}}\ln\left(\frac{z_{\mathrm{in}}}{z_{\mathrm{eq}}}\right)
   \tau^T(\mathbf{k})\right)k\int_0^{\eta_0}\frac{j_2(k\eta)}{k\eta}\frac{j_l(x)}{x^2}d\eta
\edm
contains an ISW integration which it would be preferable to pre-evaluate to save on computation time. Changing variables to $y=k\eta$ renders the integrated Sachs-Wolfe term as
\be
  I=\frac{1}{k}\int_0^{y_0}\frac{j_2(y)}{y}\frac{j_l(y-y_0)}{(y-y_0)^2}dy
  =\frac{\pi}{2k}\int_0^{y_0}\frac{J_{5/2}(y)}{y^{3/2}}\frac{J_{l+1/2}(y-y_0)}{(y-y_0)^{5/2}}dy
\ee
where we have converted the spherical Bessel functions into Bessel functions. This can be approximated \cite{MackKahniashviliKosowsky02} by
\be
  I\approx \frac{7\pi}{20k}\sqrt{l}\int_0^{y_0}\frac{J_{5/2}(y)}{y}\frac{J_{l+1/2}(y-y_0)}{(y-y_0)^3}dy
\ee
which is a standard integral evaluating \cite{AbramowitzStegun} to
\be
  I\approx\frac{7\pi}{20k}\sqrt{\frac{3l}{2}}\frac{J_{l+3}(k\eta_0)}{k^3\eta_0^3} .
\ee
This leaves us the transfer function
\be
  \Delta^T_{Tl}=\sqrt{\frac{3l}{2}}\sqrt{\frac{(l+2)!}{(l-2)!}}\frac{56\pi^2 G}{20\eta_0}
   z_{\mathrm{eq}}\ln\left(\frac{z_{\mathrm{in}}}{z_{\mathrm{eq}}}\right)\tau^T(\mathbf{k})
   \frac{J_{l+3}(k\eta_0)}{k^3} .
\ee

It remains to consider the source power spectra. There are three main ways in which we can employ these; the most extensible is to employ the realisations, while the easier is to employ the analytical integrations or the large-scale approximate results. We can also employ analytical approximations to these, as in Mack \emph{et. al.} \cite{MackKahniashviliKosowsky02}, which are useful in certain regions. We consider the Mack \emph{et. al.} approxiations, the analytical integrations, and the realisations in order.  

\subsubsection{$C_l$ from Analytic Approximations}
\begin{figure}\begin{center}
\includegraphics{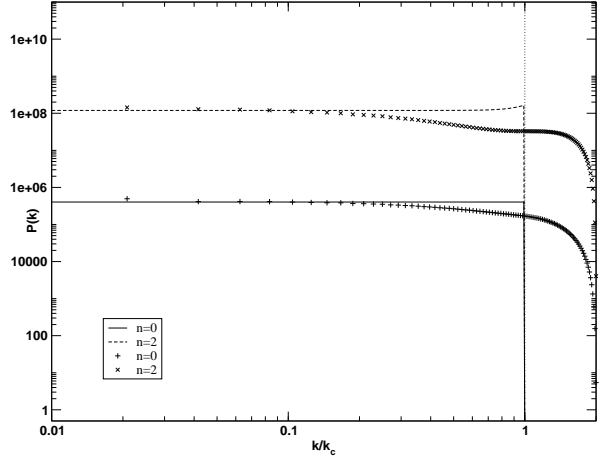}
\caption{A comparison of the approximate tensor spectra from \cite{DurrerFerreiraKahniashvili00,MackKahniashviliKosowsky02} (lines) with integrations of equations (\ref{SpectrumTt}, \ref{TensorTwoPoint}) (points). We have factored out the non-$n$ dependent parts of the amplitude and the scale is arbitrary.}
\label{MKKSpectraApprox}\end{center}\end{figure}

Mack \emph{et. al.} present the approximation for the source power spectrum
\be
\label{MKKSource}
  \mathcal{P}^T(k)=\frac{\left(2\pi\right)^9}{16}A^2\left(k_c^{2n+3}+\frac{n}{n+3}k^{2n+3}\right), \quad \mathcal{P}^V(k)=2\mathcal{P}^T(k) .
\ee
Factoring out the amplitude, we plot this approximation in figure \ref{MKKSpectraApprox} for $n=0$ and $n=2$, compared to the accurate spectra we derived in chapter \ref{Chapter-SourceStats}. It is readily apparent that this approximation is good only on the largest scales; for $k\ll k_c$ it reduces to our large-scale approximations (\ref{LargestScaleTwoPoint}), while for $k<k_c$ it is little better than reasonable. The discrepancies for $k>k_c$ are unimportant since the transfer functions vanish at the damping scale. We then do not expect the CMB signatures from ultraviolet fields to necessarily agree perfectly with those we generate from our own spectra. However, this expression is a very good approximation for fields lying in the infra-red r\'egime and is, indeed, more useful than employing our realisations due to the infra-red suppression in that r\'egime.

\begin{figure}
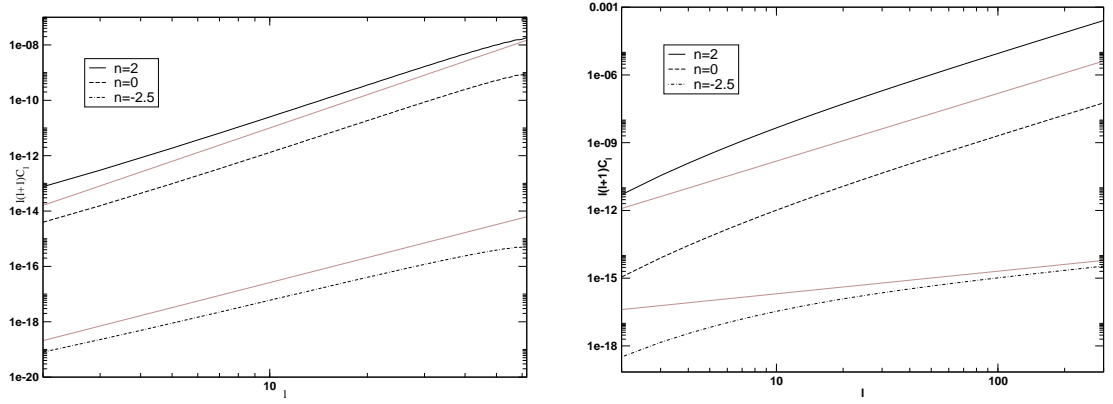
\begin{center}
\includegraphics[width=0.46\textwidth]{Cl-MKK-Vector.eps}\qquad\includegraphics[width=0.46\textwidth]{Cl-MKK-Tensor.eps}
\caption{CMB temperature auto-correlations from power-law stochastic Gaussian magnetic fields employing the source spectrum approximation (\ref{MKKSource}). Vector signals are plotted on the left and tensor on the right, for spectral indices $n=2$, $n=0$ and $n=-2.5$ from top to bottom. The estimates from \cite{MackKahniashviliKosowsky02} are plotted in grey. The variation in amplitude arises from varying $n$, on which $k_c$ sensitively depends.}
\label{Cl-MKK}\end{center}\end{figure}

In figure \ref{Cl-MKK} we present the results of integrating equation (\ref{ClApprox}) with the above power source spectrum and approximate large-scale transfer functions. As one can see, these agree with those presented by Mack \emph{et. al.}; the ultraviolet fields -- which should, perhaps, not be entirely trusted -- cause CMB tensor perturbations that follow an $l(l+1)C_l\propto l^3$ power law. The infra-red field with $n=-2.5$ creates tensor perturbations following the power law $l(l+1)C_l\propto l$, agreeing with the prediction $l(l+1)C_l\propto l^{2n+6}$. We plot the vector results only on the largest scale due to an unresolved numerical instability associated with the spherical Bessel function in the vector transfer function. We also observe that the powers predicted by Mack \emph{et. al.}, $l(l+1)C_l\propto l^4$ for $n>-3/2$ and $l(l+1)C_l\propto l^{2n+8}$ for $n\in(-3,-2.3)$, are not quite those we find with our model. It is likely that this slight difference is again caused by the numerical instability observed with the spherical Bessel functions rather than inaccurate analysis on the parts of Mack \emph{et. al.}. The relative amplitudes between signals are real, arising because we are employing $B_\lambda$, $k_\lambda$ and $n$ as input parameters to fix $k_c$. The absolute amplitudes, however, should be taken as indicative rather than necessarily exact.

\subsubsection{$C_l$ from Analytic Spectra}
\begin{figure}
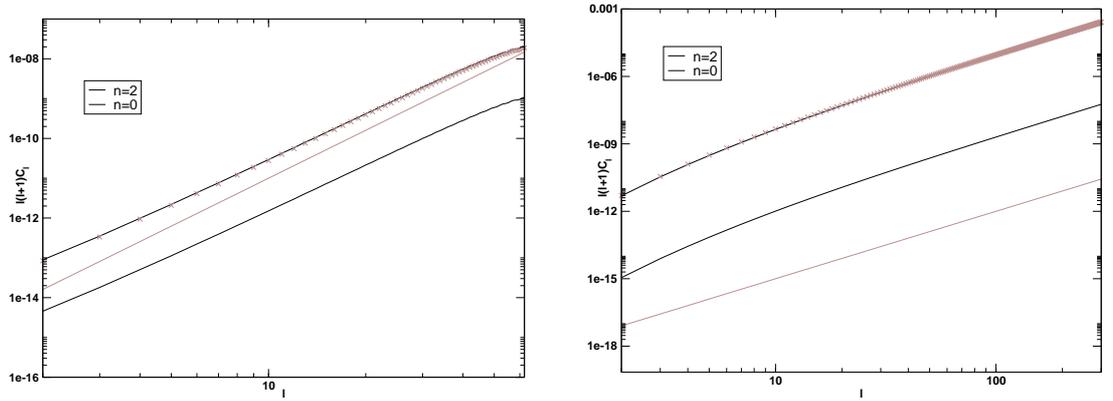
\begin{center}
\includegraphics[width=0.46\textwidth]{Cl-Th-Vector.eps}\qquad\includegraphics[width=0.46\textwidth]{Cl-Th-Tensor.eps}
\caption{Vector (left) and tensor (right) CMB temperature auto-correlations from power-law stochastic Gaussian magnetic fields employing the theoretical predictions (\ref{MagneticSourcePowerSpectra}) by interpolating a table. Agreement with those plotted in figure \ref{Cl-MKK} is good. Also plotted as crosses is the signal from a damped causal field which as expected is very close to that from a standard causal field.}
\label{Cl-Th}\end{center}\end{figure}

In this case we can use equation (\ref{MagneticSourcePowerSpectra}) to generate a power spectrum as we perform the integration over $k$. Such a process is computationally intensive, and it might be preferable to employ a pre-computed look-up table where the integration has already been performed. To do so we require a reliable interpolator and extrapolator; since the functions are in general very smooth a linear interpolator is likely to be more than sufficient for our purposes. In figure \ref{Cl-Th} we present the results for the case $B_\lambda=1\mathrm{nG}$, $k_\lambda=1\mathrm{Mpc}$ and $n=0$, $n=2$ and the damped causal field considered in \S\ref{Section-DampedCausal}. We also confirmed the accuracy of the interpolation by directly integrating equation (\ref{MagneticSourcePowerSpectra}).

As one can see, we have replicated for $n=0$ and $n=2$ the behaviours demonstrated previously; the angular power spectra rise as power-laws in the region $l<300$, with a stronger signal arising from stronger fields with higher spectral indices. The signal from the damped causal spectrum, which has not been considered before, behaves very similarly to the pure causal field. While not necessarily a surprising result -- we have, after all, merely replaced a hard, sudden cut-off at small scales for a softer decay -- this demonstrates that we can consider a variety of non-power law spectra without modification of our formalism.

\subsubsection{$C_l$ from Realised Spectra}
\label{ClRealisation}
If one wishes to employ realisations, one must take into account the infra-red cutoff. Both the vector and the tensor cases are compromised by a low mode-coverage on very large scales; this is naturally more severe for the vector case as the integration is dominated by the scales below the Silk scale, which is typically much smaller than the damping scale. The damping scale at recombination for a flat field is typically $k_c\approx 3.4\mathrm{Mpc}^{-1}$, and $k_c\approx20\mathrm{Mpc}^{-1}$ for an index $n\approx-3$. The Silk scale at recombination, above which the vector transfer function is negligable, is typically of the order of $k_S\approx 0.3\mathrm{Mpc}^{-1}$. For a flat field, then, the vector integration is dominated by wavenumbers below $k/k_c\approx 0.11$, while for the steeply-tilted fields it is dominated by wavenumbers below $k/k_c\approx 0.02$. With a simulation of side-length $l_{\mathrm{dim}}=192$ and a cut-off scale of $k_c=l_{\mathrm{dim}}/R$, our smallest mode is at $k/k_c\approx 0.005R$. For the flat field (taking $R=4$ to avoid aliasing of power from small scales to large), we will then only pick up modes above $k_{\mathrm{min}}/k_c\approx 0.02$ and thus a relatively small number of modes below the Silk scale. For a steeply-tilted field, taking $R=1$ to maximise the available large-scale modes, we have $k_{\mathrm{min}}/k_c\approx 0.005$. With an even spacing of large-scale modes this gives us a maximum of four modes contributing to the vector integrals. We will thus need to extrapolate the source power spectrum back towards $k=0$, correspondingly increasing the error.

\begin{figure}
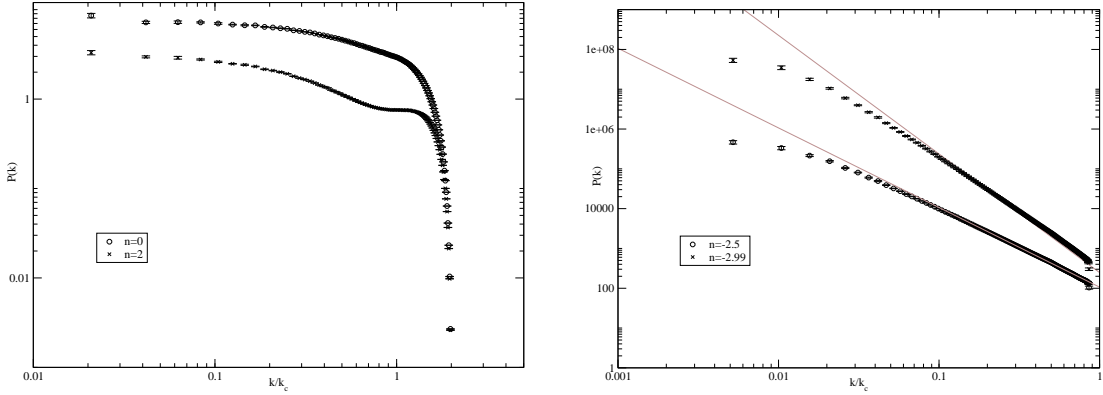
\begin{center}
\includegraphics[width=0.46\textwidth]{TTSpectra02.eps}\qquad\includegraphics[width=0.46\textwidth]{TTSpectrum25.eps}
\caption{Tensor-Tensor source statistics, demonstrating infra-red cut-off for indices $n<-3/2$.}
\label{RealisedSpectrumTt}\end{center}\end{figure}

The easiest way of achieving this is to consider each spectrum we wish to employ individually. If we consider the tensor case for $n>-3/2$ (figure \ref{RealisedSpectrumTt}, left), we can see that the infra-red cut-off is not a major issue; a smooth interpolation will extend these spectra to the very large-scales without issue. However, if we consider infra-red spectra (figure \ref{RealisedSpectrumTt}, right), the infra-red cut-off becomes highly significant. In this case, we choose to excise the region $k/k_c<0.1$ -- which, coincidentally, corresponds approximately to the Silk scale at recombination -- and replace it with a line fitted to the source spectrum. Doing so in this case effectively takes us back to the approximation in equation (\ref{MKKSource}), with $\mathcal{P}\propto k^{2n+3}$. However, for a more general field this approach is still workable, while the approximation is not valid. Since a perfect fit of our realisations for the infra-red region is provided by the approximation (\ref{MKKSource}) we do not repeat this procedure here.


\subsection{An Outline for Generating Statistics from CMBFast}
A more accurate approach for generating spectra from magnetic fields will be to employ a Boltzmann code; our formalism has been designed in the synchronous gauge to ensure it is consistent with the widely-employed CMBFast \cite{SeljakZaldarriaga96}. To construct our power spectra and bispectra we shall first need to build a vector component to the code; purely magnetised initial conditions can be readily adapted from Giovannini \cite{Giovannini04-Mag} and Lewis \cite{Lewis04-Mag}. Employing the viscous damping scales derived in \S\ref{Sec-DampingScaleMHD} we can then apply separate cut-offs to the scalar, vector and tensor evolution equations to account for photon damping. To model the Alfv\'en wave oscillations we shall follow Durrer \emph{et. al.} \cite{DurrerFerreiraKahniashvili00} and take $v_A\rightarrow v_A\cos (k\eta)$.

The sources in the evolution equations will be presented without their $k$-dependence as this will arise when we later fold across the power spectra or bispectra. The sources are then given purely by $A^2$ which can be related to the smoothed field strength $B_\lambda$ on a scale $\lambda$ again found from
\be
  A=\frac{B_\lambda^2}{4\pi\int k^2\mathcal{Q}(k)e^{-\lambda^2k^2}dk}
\ee
with $\mathcal{P}(k)=A\mathcal{Q}(k)$; see (\ref{SpectrumNormalisation}). This will be performed numerically at the beginning of the run.

CMBFast does not by default generate the transfer functions for a particular run but rather the $C_l$s directly. However it is not difficult to modify the code such that the usual calls to integrate up the angular power spectrum instead merely return the transfer functions. We outline the process we shall follow to generate our two-point spectra below.

A wrapper routine generates the magnetic sources from an input smoothed field and damping scale (of the order of $B_\lambda=10^{-9}\mathrm{G}$ at $\lambda=1\mathrm{Mpc}$) and then calls CMBFast with these sources. Once CMBFast has completed its run, instead of building the angular power spectrum it will return the scalar, vector and tensor transfer functions to the wrapper. Once we have the transfer functions, generating the CMB angular power spectra is a simple matter of performing the integration
\be
  C_l=\int_0^\infty\mathcal{P}_{AB}(k)\Delta^A_{Tl}(k)\Delta^{*B}_{Tl}(k)k^2dk .
\ee
The code to perform this integration is that employed in the previous section. This integration can again obviously be performed over power spectra resulting from the analytic integrations or from realisations; as before the more extensible approach is to employ the realised spectra. $A$ and $B$ can take any of the values permitted in chapter \ref{Chapter-SourceStats} -- that is, $AB\in\left\{\tau\tau,\tau\tau_S,\tau_S\tau_S,\tau^V\tau^V,\tau^T\tau^T\right\}$.

\section{The CMB Bispectrum}
\label{CMB-Bispectrum}
The CMB angular power spectrum is a two-point statistic and the integration over wavemodes used to generate it includes a contribution from two transfer functions. Perhaps the most instructive case is that of the temperature/polarisation cross-correlation sourced by scalar perturbations:
\be
  C_l^{TE}=\frac{2}{\pi}\int k^2\mathcal{P}(k)\Delta^S_{Tl}(k,\eta_0)\Delta^{*S}_{El}(k,\eta_0)dk .
\ee
In this integration we have the relevant power spectrum -- here the initial spectrum of scalar perturbations -- and one transfer function corresponding to the correlated quantities at two points on the sky. A sensible suggestion for the form of the three-point CMB angular correlation, the \emph{bispectrum}, is then that we integrate a primordial bispectrum across three transfer functions. For the vector-tensor-vector correlation, for example, we might suggest that the CMB angular bispectrum is
\be
  B_{ll'l''}^{VTV}=\frac{8}{\pi^3}\iiint k^2p^2q^2\mathcal{B}_{VTV}(k,p,q)\Delta^V_{Tl}(k,\eta_0)\Delta^T_{Tl'}(p,\eta_0)\Delta^V_{Tl''}(q,\eta_0)
   dkdpdq
\ee
where the integral is across three wavenumbers because the three wavemodes form a closed triangle rather than being degenerate.

This expression is not entirely accurate; we have not, for example, enforced the triangle conditions on the wavenumbers $k$, $p$ and $q$. We should also be suspicious about whether this expression has accurately mapped the primordial bispectrum onto a sphere; while in the two-point case the spherical Bessel functions that one would imagine should arise are cancelled by the delta function this is not necessarily the case when the three wavemodes form a closed triangle. If one works accurately through the calculation then one sees that, indeed, we are missing important terms. The integrand includes a term
\be
  J_{ll'l''}(k,p,q)=\int_Xj_l(kX)j_{l'}(pX)j_{l''}(qX)X^2dX
\ee
where $X$ is an arbitrary real-space variable. This term accounts for an accurate wrapping of the bispectrum onto the 2-dimensional CMB sky and can be calculated recursively (see \S\ref{JlllAlgorithm}) with relative ease. We also have a Wigner 3-$j$ symbol premultiplying the integration; the conditions of this symbol enforce the triangle inequality 
\be
  |l-l'|\leq l''\leq|l+l'|
\ee
on the multipole moments, and ensures that they sum to an even integer. The CMB angular bispectrum generated by a primordial bispectrum $\mathcal{B}_{ABC}$ can then be written as
\bea
  B_{ll'l''}&=&
  \frac{8}{\pi^3}\sqrt{\frac{(2l+1)(2l'+1)(2l''+1)}{4\pi}}
   \left(\begin{array}{ccc}l&l'&l''\\0&0&0\end{array}\right)
   \iiint
   k^2p^2q^2\mathcal{B}_{ABC}(k,p,q)
  \nonumber \\ && \times
\label{CMBBispectrum}
   \Delta^A_{T,l}(k,\eta_0)\Delta^B_{T,l'}(p,\eta_0)\Delta^C_{T,l''}(q,\eta_0)
   J_{ll'l''}(k,p,q)dkdpdq .
\eea
We derive this in the line-of-sight approach in appendix \ref{Appendix-Bispectra}; see also Wang and Kamionkowski \cite{WangKamionkowski00} and Ferreira, Magueijo and G\'orski \cite{FerreiraMagueijoGorski98}. Here $\left\{k,p,q\right\}$ are the amplitudes of the wavevectors forming a closed triangle on the sky; these are, of course, entirely equivalent to the variables employed in chapter \ref{Chapter-SourceStats} and are related by $r=p/k$ and $\cos\phi=(q^2-k^2-p^2)/kp$.

It is common to follow \cite{FerreiraMagueijoGorski98} and define a more limited measure of the bispectrum
\be
\label{CMBReducedBispectrum}
  \hat{B}_l=\left(\frac{2}{\pi}\right)^3\iiint k^2p^2q^2\mathcal{B}_{ABC}(k,p,q)J_{lll}(k,p,q)
    \Delta^A_{T,l}(k,\eta_0)\Delta^B_{T,l}(p,\eta_0)\Delta^C_{T,l}(q,\eta_0)
    dkdpdq
\ee
which more intuitively resembles to the angular power spectrum and will contain much of the information contained within the bispectrum. Whether or not we employ the angular bispectrum or this limited bispectrum is immaterial; it is clear from their form that, given pre-computed primordial bispectra, it is simple to predict the form of the corresponding CMB angular bispectra. Solving this equation is clearly very similar to solving the two-point case; given a pre-computed primordial bispectra -- such as those from chapter \ref{Chapter-SourceStats} -- we merely require the relevant transfer functions to calculate the CMB angular bispectrum.

The main problem generating estimates for the CMB bispectra is the sheer size of the primordial bispectra -- even for $\hat{B}_l$ we require a full primordial bispectrum. In principle the integration is similar to that in the two-point case, merely needing an extra call to the function $J_{lll}(k,p,q)$. However, in the three-point case we must integrate over a 3-D function rather than the effective 1-D function we had for the spectra. Again, if we wished we could call a primordial bispectrum generation routine for each $\left\{k,p,q\right\}$; this, however, would be ruinously slow. Instead we shall employ look-up tables, generated either from analysis or from realisations. Even this has its drawbacks; while the look-up tables produced for a power spectrum are relatively small, being of length $l_{\mathrm{dim}}=192$, those for a bispectrum are anything but, being of size $l_{\mathrm{dim}}^3>7\times10^6$. Moreover, a full three-dimensional primordial bispectrum takes a long time to compute if one wants to run enough simulations with a large enough dynamic range to reduce the errors in the final CMB plots. Nonetheless, for reasons of extensibility it will again be preferable to employ bispectra and use analytical plots only to verify simple, Gaussian power-law cases. To overcome the infra-red cut-off we can again excise the low-$k$ region and fit a three-dimensional curve in a case-by-case basis to employ in the integration.


The main advantage of the method we are proposing is not that the physics is in any way different -- obviously, that is not the case -- but rather that we have exploited the simple time-evolution of the magnetic sources to decouple the integration forming the angular (bi-)spectra to enable us to generate intrinsic bispectra separately from the transfer functions. We are also relying on numerical methods to produce our results and this gives us great freedom in the choice of our magnetic field. While we have chosen to consider primarily Gaussian, power-law magnetic fields the formalism is general enough to consider \emph{any} underlying statistical nature and any power spectrum. As we have shown in this brief section, extending our method from two-point to three-point angular spectra can be achieved with relative ease -- it is merely a question of implementation, and the problems are technical and concern the size and lengthy computation of the intrinsic bispectra rather than questions of the physics. In the approach employed thus far in the literature (\cite{DurrerFerreiraKahniashvili00,SubramanianBarrow02,MackKahniashviliKosowsky02} for example), on the other hand, the authors are concerned with finding closed-form, analytic solutions. This necessitates them making repeated approximations, many with relatively limited validity, and deriving even the two-point spectra is relatively complex. Extending the formalism to consider CMB bispectra would be extremely non-trivial.

One should not overstate the advantages of our methods. The intrinsic bispectra are heavily compromised by our limited dynamic range. Requiring a selection of three wavevectors, forming a closed triangle, severely restricts the number of contributory modes at each point. We must then repeat the simulations -- around 1,500 times for a gridsize of $l_{\mathrm{dim}}=192$ -- to reduce the errors to workable amounts. This compounds the previous issue with a limited dynamic range for the realisations causing inaccuracies in, for example, integrations across the vector transfer functions. The time required to na\"{\i}vely produce an intrinsic bispectrum is large and increases linearly with repeated simulations, and the resulting files are correspondingly huge. For infra-red fields, we will also again have significant divergences at low-$k$, which will need to be removed by examination. It is likely that this will be most easily achieved by examining the colinear case to determine a large-scale cut-off, excising the bispectrum for all larger-scale modes, and then extrapolating onto the large scales from the more trustworthy points. In the infra-red r\'egime, bispectra are relatively simple, and a three-dimensional interpolater should have little trouble fitting a curve we can extrapolate onto the large scales. However, we feel that despite these drawbacks this approach will most realistically allow us to make accurate predictions concerning the various three-point correlations induced by magnetic fields onto the microwave background. Moreover, it is again merely a matter of implementation to extend our results to correlations between temperature and polarisation, or to four-, five- and $n$-point correlations -- this involves calculating the intrinsic statistics and folding them in a consistent manner onto the microwave sky by use of the transfer functions, something that is relatively simple using the realised fields but rather complicated analytically.



\section{Conclusion}
We have demonstrated in this chapter that our formalism, developed in chapters \ref{Chapter-MagnetisedCosmology} and \ref{Chapter-SourceStats}, is sufficient to model a linear cosmology including a stochastic magnetic field and produce accurate photon transfer functions. After reviewing the current literature considering the impacts of a magnetic field -- first briefly on nucleosynthesis but then concentrating on the CMB sky -- we then demonstrated that one can numerically integrate the expression for $C_l$ employing pre-computed primordial power spectra and reproduce the correlations induced by a magnetic field. While the resulting CMB angular power spectra we presented in this chapter are generally not new, they should serve as a demonstration that our approach is valid. Rather than restrict ourselves to very simple field configurations, we have developed an approach that is highly extensible and based upon flexible models of magnetic fields, capable of dealing with non-power law spectra and arbitrary statistics. Our presentation of the statistics of the simply damped causal magnetic field demonstrates this. The flexibility allowed by the presented formalism will enable us to consider the impact on the CMB temperature and polarisation angular power spectra of more realistic primordial scenarios than has hitherto been possible. This could be significant, particularly in the light of the CBI excess observed in the temperature auto-correlation. Ongoing and upcoming missions -- such as CBI and VSA \cite{RajguruEtAl05}, ACBAR \cite{ACBAR}, SPT \cite{SPT04}, CAPMAP \cite{BarkatsEtAl04}, Boomerang \cite{MontroyEtAl05} and Planck \cite{Planck}, in particular -- will extend our knowledge of the small-scale temperature and polarisation power spectra enormously, and the two-point measures of these will allow us to constrain the possible properties of a primordial magnetic field with much more accuracy than is currently possible. With the formalism presented in this thesis, we are no longer restricted to considering only simple Gaussian power-law fields and can instead directly test field configurations arising from particular magnetogenesis mechanisms. Simulations are, however, limited in dynamic range -- the mechanism of Matarrese \emph{et. al.} \cite{MatarreseEtAl04}, for example, is difficult to model with the range allowed by today's computers. We must also add that our approach, and those that preceded it, may not necessarily be valid for ultraviolet fields, due to their strong dependence on the time-evolving damping scale. This would conceivably introduce decoherence in the perturbations, requiring us to employ full non-equal time correlators to determine the imprint on the CMB. This will be considered in more detail in the future.

A more realistic probe for primordial magnetic fields that the CMB provides is likely to be the non-Gaussianity that a magnetic field will induce. The field strength of the magnetic field is necessarily small to avoid violation of nucleosynthesis bounds; however, a magnetic field is also likely to be one of the few primordial sources capable of imprinting non-Gaussianity on the CMB sky at all scales. In the standard approach, predictions of CMB non-Gaussianity arising from such a magnetic field would be extremely complex, even assuming purely power-law power spectra and Gaussian statistics. In our approach, generating maps of CMB non-Gaussianity is simply a matter of integrating equation (\ref{CMBBispectrum}) or (\ref{CMBReducedBispectrum}) employing $\mathcal{B}_{ABC}$ and $\Delta_{TEB,l}$ calculated previously, which is a more realistic proposition. It will then be a relatively simple matter to predict the forms of the various possible bispectra arising from primordial magnetic fields both of the standard Gaussian power-law type -- for which we might employ analytical as well as purely numerical methods to derive the intrinsic statistics -- and of more general forms. Again, we are limited by the dynamic range, and more severely than is the case for the two-point moments, due to the mode-selection necessary in forming closed triangles. There is also a severe infra-red divergence in the primordial bispectra and we will necessarily have to excise the large-scale region and replace it with a surface fitted to the smaller-scale bispectra. We are also limited by matters of practicality -- intrinsic bispectra take a long time to generate and are relatively large. However, they need only be computed the once and can be employed many times for different scenarios; a continuously-sourced field, for example, will generate different transfer functions than a field that has been in existence since before neutrino decoupling. Care would need to be taken in such a case that the sources remained coherent. The non-Gaussianity of the CMB is currently only weakly constrained by WMAP, even on the large scales; this is partly due to the method by which we assess it. Knowing the signatures to test for given a particular scenario will help target future searches. Forthcoming missions are expected to improve detection significantly. A usual measure of non-Gaussianity is known as $f_{\mathrm{NL}}$; this characterises the strength of a non-Gaussian component to the initial curvature perturbation \cite{BartoloEtAl04}. While it is not particularly meaningful at present to relate this parameter to the non-Gaussianities from magnetic fields, it is indicative that while the first year of WMAP data can only constrain $f_{\mathrm{NL}}$ to $f_{\mathrm{NL}}\in(-58,134)$ \cite{WMAP-Komatsu}, the full eight-year WMAP data might constrain it to $|f_{\mathrm{NL}}|\leq 20$ \cite{KomatsuSpergel01}. The Planck satellite is likely to constrain it to the order of unity, $|f_{\mathrm{NL}}|<5$ \cite{KomatsuSpergel01}. It would then seem plausible that Planck will measure the CMB non-Gaussianity sensitively enough to allow us to search for the imprint of primordial magnetic fields, and tighten the constraints derived from the two-point statistics.

%
\chapter{Discussion and Conclusions}
In this thesis we have considered various aspects of the cosmological magnetic field and the impact it has on the pre-recombination universe and on the microwave background. In chapter \ref{Chapter-PerturbationTheory} we constructed a unified formalism for an unmagnetised, linear cosmology in a synchronous-gauge FLRW universe, considering the vector perturbations on an equal footing to the scalar and the tensor perturbations. While this does not involve much modification for fluid matter, we have presented the formalism for effective fluids governed by the Boltzmann equation, adapting and extending the results of Landriau and Shellard \cite{LandriauShellard03}. We derived the form of the CMB transfer functions for temperature in a traditional and in a line-of-sight formalism for scalar, vector and tensor modes. Polarisation is considered in appendix \ref{Appendix-EandB}.

In chapter \ref{Chapter-MagnetisedCosmology} we then incorporated a magnetic field into this perturbation theory. Some results were presented for the first time in a greater detail than has been previously, leaving the perturbation character of the magnetic fields entirely unspecified and leaving open the simple adaptation of the equations to a two-parameter system wherein the geometry is linearised with an explicit parameter $\varepsilon$ while the magnetic field is linearised with an independent parameter $\varepsilon_B$, and combinations of the two are not neglected. We considered the effects of the magnetic field on the Lorentz forces and the stresses in some detail and reviewed the results of Subramanian and Barrow \cite{SubramanianBarrow98-MHD} and Jedamzik, Katalini\'c and Olinto \cite{JedamzikKatalinicOlinto98}, in which they derived the damping of magnetic fields from photon viscosity. We returned to this matter in chapter \ref{Chapter-CMB}.

Chapter \ref{Chapter-SourceStats} considered in detail the full statistics of the magnetic sources that would be included in a numerical model of the magnetised cosmology. Exploiting the large-scale time-independence of a scaled field $\mathbf{B}=a^2\mathbf{b}$, we presented analytical expressions for the two- and three-point moments of the static magnetic stresses, separated into a scalar trace, traceless scalar, two vector and two tensor degrees of freedom. We then employed these analytical expressions, derived for the very specific (and perhaps unrealistic) case of a Gaussian-distributed magnetic field with a power-law spectrum, to confirm the accuracy of numerically simulated fields generated on a finite grid. These numerical models can be constructed with arbitrary power spectra and statistical character and are vastly more flexible than the analytical approach; the drawback is the limited dynamical range and computational time required. Excellent agreement was found between the analytical approach and the simulated fields, leaving open the study of more generic power spectra and statistical natures.

At the two-point level we constructed the power spectra of the different components of the stress-energy tensor, confirming and greatly extending the previous results. By the nature of the separation, the cross-correlations vanish. Differing behaviours for the ultraviolet r\'egime $n>-3/2$ and the infra-red r\'egime $n<-3/2$ were confirmed, with the bluer spectra being dominated on large-scales by the intrinsic damping scale, and the infra-red spectra losing their features and behaving as $k^{-(2n+3)}$.

Even for the simple power-law and Gaussian case significant non-Gaussianities were found at both the one-point level for the scalar pressures, and the three-point level for a wide variety of cross-correlations. At the one-point level the isotropic pressure (and thus energy density) of the field behaves, as expected, as a $\chi^2$-distributed field and this is insensitive to spectral index. The anisotropic pressure, on the other hand, is relatively Gaussian, due to its non-local dependence on the magnetic fields diluting the non-Gaussianity. The statistics of the anisotropic stress are dependent on the spectral index.

At the three-point level we modelled non-Gaussianity with a bispectrum method, considering rotationally-invariant combinations such as $\langle\tau\tau_i^V\tau^i_V\rangle$. This is, to our knowledge, the first time that cross-correlations between different components of the scalar-vector-tensor split have been studied with a bispectrum method and we presented the full analytical results. Although severely limited by a sparse mode-selection, averaging a large number of random simulations generated a superb agreement with the predictions. The intrinsic bispectra for the infra-red power-law case with $n>-1$ contain many features and are dependent on the damping scale. These spectra are tightly constrained to unobservable levels by the Caprini/Durrer nucleosynthesis bounds. For the more realistically observable case with an infra-red spectrum $n\approx -3$ the features observed are washed out by the extreme tilt; more interestingly the nature of the $\langle\tau\tau\tau_S\rangle$ correlation on large scales changes drastically -- for ultraviolet fields this cross-correlation vanishes on large scales while for infra-red fields, along with the other bispectra, it behaves as $k^{-(3n+3)}$.

We also briefly presented the one-, two- and three-point results for the Lorentz forces. The Lorentz forces are directly proportional to components of the stress-energy tensor; the proportionality does, however, include the wavenumber and this modifies the angular integrations. At the one point level, the scalar Lorentz force inherits the nature of the anisotropic pressure and is relatively Gaussian compared to the isotropic pressure. It is also sensitive to the spectral index. At the two-point level, the dependence on the wavenumber drives the auto-correlations to zero on large-scales and a length-scale is introduced at which the power spectra turn over. The three-point correlations have a similar nature; only two rotationally-invariant combinations are possible, the scalar auto-correlation and the cross-correlation between the scalar and the vector isotropic spectrum. Both agree to good accuracy between simulated fields and analysis. These bispectra are again vanishing on large-scales for a flat spectrum. While they do also vanish on large-scales for a strongly-tilted spectrum (by construction) the behaviour close to $k=0$ is very similar to that of the stresses.

As an illustrative example we then briefly considered a modified power spectrum, taking a causally generated field with $n=2$ and introducing an exponential damping rather than a sudden cutoff. This is a crude approximation for the type of field that would be generated by a process other than inflation (or some other acausal early-universe source) such as plasma processes before and at recombination. These precise magnetogenesis models are unlikely to generate Gaussian fields -- indeed, fields sourced at second-order by the first-order density perturbation are $\chi^2$ distributed -- and are, moreover, likely to be too weak to have an observable impact upon the CMB. However, it is not inconceivable that some process existed at some time before or after neutrino decoupling that generated a causal magnetic field with a greater strength. The gross features of a damped causal magnetic field are very similar to the $n=0$ case considered previously. We again confirmed the simulated fields with an analytical integration.

The techniques developed and employed in this chapter, and particularly the simulations of the fields, have a greater validity than just the toy model of a primordial magnetic field that we have considered. There are various possible non-linear sources, and each such source will contribute non-Gaussian signatures. Cosmic strings are a natural example of such a source. Modification to our codes would naturally be necessary; however the basic techniques could be highly adaptable if care was taken to consider the decoherence introduced by an evolving source.

In chapter \ref{Chapter-CMB} we turned to the observational impacts of a simple primordial magnetic field, first reviewing limits from nucleosynthesis. A conservative estimate \cite{Widrow02} shows that the current magnetic field strength is constrained from nucleosynthesis to be of the order of $\mu\mathrm{G}$, while stringent -- albeit controversial -- limits can be found by considering the gravitational waves produced from a magnetic field \cite{CapriniDurrer02}; for a field with a flat spectral index, smoothed on a sub-megaparsec scale, these can be as strict as $B\lesssim 10^{-30}\mathrm{G}$.

However, it is with the CMB that we chiefly occupy ourselves. We described how one can evaluate the damping scales without recourse to power-law approximations before outlining the basic procedure by which the intrinsic statistics derived in the previous chapter could be wrapped onto the CMB. To do so requires the photon brightness functions, and as a first approximation, and to act as illustrative examples, we derive the large-scale approximations of these for the vector and tensor perturbations. These reproduce from our own formalism those in Mack \emph{et. al.} \cite{MackKahniashviliKosowsky02} and include the vector tight-coupling approximation. The vector transfer functions are found to differ dramatically between different damping r\'egimes for perturbations below and above the Silk scale, while the tensor transfer functions are found to consist predominantly of an integrated Sachs-Wolfe term.

We considered the two-point statistics of the CMB, first briefly reviewing the results of previous analytical studies. The small-scale studies suggest that for a steeply-tilted spectral index a magnetic field of approximately nano-Gauss strength would begin to dominate the standard $\Lambda$CDM at around $l\approx 2000$, with stronger signals for shallower spectra. The large-scale studies consider the magnetic field when its impact is generally subdominant for the temperature auto-correlation and thus concentrate on the polarisation spectra. We then reviewed the more accurate numerical results which employed CAMB \cite{Lewis04-Mag} or CMBFast \cite{KohLee00,YamazakiIchikiKajino04,YamazakiEtAl06} with magnetic sources. These studies, each limited in their own way, confirmed and extended the previous analytical estimates. Magnetic signals on the temperature auto-correlation spectrum dominate for highly-red nano-Gauss fields at around $l\approx 2000$ and at much lower levels for shallower indices. A causal field might be expected to contribute a large amount towards the possible excess of small-scale power observed by, for example, CBI \cite{RajguruEtAl05} and ACBAR \cite{ACBAR}, and this suggests a constraint on the amplitude of such a field. Lewis performed far the most complex study and his contributions to the initial conditions required, in particular, are invaluable. Despite being magnetised Boltzmann codes, and generating in principle fully accurate transfer functions, these studies still rely on the previous analytical approximations and are thus restricted solely to Gaussian power law fields, and reliant on approximations for the normalisation of the power spectrum and the source spectra that are not necessarily particularly accurate.

Using our own formalism, we demonstrated that our approach is valid, reproducing the forms of Mack \emph{et. al.} \cite{MackKahniashviliKosowsky02} for the angular power spectrum. We also presented the angular power spectrum imprinted by a Gaussian field with a damped causal spectrum; this is not significantly different from a causal field with a sharp cut-off, as should be expected. However, it serves as a demonstration that not only can we model the CMB imprints of fields that have been considered before, but also that we can consider more general fields that would be intractable analytically. By generating their intrinsic power spectra through realisations and folding them across the usual magnetised transfer functions, we can then predict the observable impact of an entirely generic magnetic field. A motivation for this comes from the field produced by second-order mixing of scalar and vector perturbations \cite{MatarreseEtAl04,GopalSethi04,TakahashiEtAl05}. Regardless of the details of these studies, such a magnetic field will be naturally $\chi^2$-distributed for Gaussian initial conditions, and will possess a damped causal power spectrum of some form. Unfortunately our formalism is limited by dynamic range and it is not possible to use the form of power spectrum from, for example, \cite{MatarreseEtAl04} with any great accuracy. Nonetheless with a new generation of small-scale ($l\gtrsim 1000$) CMB results expected from CBI and VSA \cite{RajguruEtAl05}, ACBAR \cite{ACBAR}, Boomerang \cite{MontroyEtAl05},  CAPMAP \cite{BarkatsEtAl04}, SPT \cite{SPT04} and Planck \cite{Planck}, we might expect to constrain the possible properties of a primordial magnetic field with much more accuracy than is currently possible, and our approach allows us to consider a wide variety of fields from a wide variety of magnetogenesis mechanisms.

We finally turned to briefly considering the non-Gaussianity of the CMB. While magnetic fields contribute fluctuations to the CMB angular power spectra, their effect is relatively minor compared to that from standard cosmology, until one considers the very small scales. While the upcoming CMB observations will greatly improve observations on a small scale and the potential is there to constrain magnetic fields -- Yamazaki \emph{et. al.} have derived impressive constraints on a Gaussian power law field in \cite{YamazakiEtAl06} and these might be tightened as observations improve -- they will still be compromised by foregrounds. The non-Gaussianity imprinted by even a Gaussian magnetic field, however, will be on large-scales as well as small scales, and a primordial magnetic field is one of relatively few sources that is expected to imprint a significant primordial non-Gaussian signature. The eight-year WMAP data and results from the Planck satellite are expected to constrain non-Gaussianities well -- to the level of $|f_{\mathrm{NL}}|<20$ for optimal WMAP results, and to $|f_{\mathrm{NL}}|<5$ for Planck. While there is no direct correspondance between $f_{\mathrm{NL}}$ and the non-Gaussian signatures of a magnetic field, this does suggest that the sensitivity might be there to detect magnetic non-Gaussianities. We choose to characterise the CMB non-Gaussianity with an angular bispectrum and demonstrate that this can be evaluated straightforwardly using an extension of our two-point methods. The situation is complicated by the unwieldy nature of the intrinsic magnetic bispectra, but this is merely a technical issue rather than a fundamental flaw. Given the transfer functions for a magnetised cosmology, which would ideally be generated by a modified Boltzmann code such as CAMB \cite{CAMB}, CMBFast \cite{SeljakZaldarriaga96} or CMBEasy \cite{CMBEasy}, it is a matter of integrating equation (\ref{CMBReducedBispectrum}) for a chosen pre-generated intrinsic bispectrum. This process is, however, lengthy and the bispectra are severely limited by the dynamic range of the realisations. However, this method will be both more accurate and more extensible than an analytic approximation would be. It seems plausible that the next generation of CMB observations will be capable of strongly constraining the CMB non-Gaussianity; our formalism will then allow researchers to search for the imprint of a primordial magnetic field and perhaps constrain magnetogenesis models more tightly than is possible with the two-point spectra.

Magnetic fields are likely to be present in the universe from the earliest times and fields strong enough to affect the CMB sky are entirely likely. In this thesis we have presented a magnetised cosmology in a formalism easily incorporated into the Boltzmann code CMBFast and examined in detail the statistical nature of the magnetic sources. We have then demonstrated that we can numerically produce the CMB anisotropies in an extensible manner that is not restricted by the nature of the fields. This approach is easily adapted for considering the non-Gaussianity such a field would induce onto the microwave sky. With the improvement in CMB data within the next decade expected to tightly constrain the CMB non-Gaussianity, we can then have a solid test of a primordial magnetic field.

\appendix
\chapter{Further Issues in Perturbation Theory}
\label{Appendix-Perturbations}
In this appendix we shall briefly cover some fundamental issues in relativistic cosmological perturbation theory that we did not mention in the main text. We begin with a derivation of the Robertson-Walker metric and proceed to a discussion of gauge issues in a linearly-perturbed FLRW universe.

\section{Derivation of the Robertson-Walker Metric}
We arrive at our assumptions for the metric by extending the Copernican principle out into the universe. In its bare form, the Copernican principle states that the Earth is in no special location in either the solar system or the universe. It seems ridiculous to think that our solar system -- in an unremarkable part of the galaxy -- is at a preferred location in the universe and the Copernican principle is precisely that it is not. When married to observations of the CMB demonstrating that it is isotropic to at least one part in ten thousand the Copernican principle leads us to model the current universe as a maximally-symmetric space of some sort. It is maybe worth commenting that merely because the current observations strongly support a universe isotropic about the Earth there is no good reason to believe that this always had to be the case and models of cosmology that tend asymptotically to the isotropic are allowed by the observations so long as they satisfy nucleosynthesis and CMB bounds. Our derivation of the Robertson-Walker metric follows that of Carroll \cite{Carroll}.

\subsection{Maximally-Symmetric Spaces}
Consider a four-dimensional spacetime with the greatest possible number of isometries -- transformations on the metric that leave it unchanged, equivalent to Killing vectors. In this spacetime there are obviously four translation symmetries, corresponding to the four axes of some co-ordinate system. If we set up a co-ordinate system about a point $P$ we can also see that there are six independant rotations after we have removed those that mirror another. This leaves us with the expected ten isometries in a four-dimensional space. This argument is strictly applicable directly only in flat space; however, it only concerned the nature of the space local to the point $P$ -- that is, in a Riemannian co-ordinate system. The co-ordinate system chosen will not affect the maximum number of independant isometries, and the point $P$ was chosen arbitrarily, and so this conclusion can be applied in any situation regardless of the curvature of the space.

In a maximally-symmetric space, the curvature obviously has to be the same at each point (by the translation symmetries); this restricts the number of maximally-symmetric spaces -- one merely needs the Ricci scalar and we can surmise that we can take a prefactor to be negative, zero or positive. That is, we expect only three maximally-symmetric spaces.

To quantify this statement, let us construct the Riemann-Christoffel tensor. Consider a Riemannian co-ordinate system about some arbitrary point in the maximally-symmetric spacetime with metric $\eta_{\mu\nu}$. This metric is invariant under a Lorentz transformation to another Riemannian co-ordinate system and we would wish for the associated Riemann-Christoffel tensor to behave in the same way. It must then be constructed from the metric and from the few other tensors that remain invariant under Lorentz transformations -- the Kronecker delta and the Levi-Civita tensor density. If we make combinations of these tensors we can then match them against the required -- and stringent -- symmetries for the Riemann-Christoffel tensor. The only combination that matches the symmetries is
\bdm
  R_{\mu\nu\rho\sigma}\propto\eta_{\mu\rho}\eta_{\nu\sigma}-\eta_{\mu\sigma}\eta_{\nu\rho}.
\edm
This is a tensorial relation and so is applicable in all reference frames; moreover, our choice of $P$ was arbitrary, and so the equation holds across the whole spacetime. We then express the curvature tensor of a maximally-symmetric spacetime as
\bdm
  R_{\mu\nu\rho\sigma}=\mathrm{const}\times\left(g_{\mu\rho}g_{\nu\sigma}-g_{\mu\sigma}g_{\nu\rho}\right) .
\edm
Contracting across the first and fourth indices leaves
\bdm
  R_{\nu\rho}=\mathrm{const}\times \left(g_{\nu\rho}-n g_{\nu\rho}\right)
\edm
and contracting this again leaves us with
\bdm
  R=\mathrm{const}\times n(n-1)
\edm
which fixes the proportionality constant, finally yielding
\be
\label{MaximalRiemannTensor}
  R_{\mu\nu\rho\sigma}=-\frac{R}{n(n-1)}\left(g_{\mu\rho}g_{\nu\sigma}-g_{\mu\sigma}g_{\nu\rho}\right) .
\ee

So if
\bdm
  \mathcal{K}=\frac{R}{n(n-1)}
\edm
is the measure of curvature of the space -- which can be either positively curved, negatively curved, or zero and is usually normalised to $\mathcal{K}\in\{-1,0,1\}$ -- then in four dimensions we have the Ricci tensor and scalar
\bdm
  R_{\mu\nu}=3\mathcal{K} g_{\mu\nu}, \quad R=12\mathcal{K}
\edm
and so the Einstein tensor
\bdm
  G_{\mu\nu}=-3\mathcal{K}g_{\mu\nu}.
\edm
This then implies (by the Einstein equations) a matter that obeys
\bdm
  T_{\mu\nu}=\frac{3\mathcal{K}}{8\pi G}g_{\mu\nu}=\Lambda g_{\mu\nu};
\edm
being directly proportional to the metric tensor, this is a cosmological constant.

We have thus shown that there are only three spaces of maximal symmetry, which are called de Sitter space if $\Lambda>0$, Minkowski if $\Lambda=0$ and anti-de Sitter if $\Lambda<0$. The physical interpretation is that the de Sitter universes contain only a vacuum energy and are in a state of either eternal expansion or eternal collapse (see equation \ref{deSitter}), while Minkowski space is non-gravitating. (Quasi-)de Sitter spaces are vital in inflationary and dark energetic theories (see for example \S\ref{SecScalarFields}), corresponding as they do to a universe filled with a cosmological constant (a vacuum energy density) or a scalar field mimicking such, while anti-de Sitter spaces are frequently employed in string and string-inspired theories, such as braneworld models of cosmology \cite{Langlois03, Maartens04, Durrer05}.

\subsection{The Robertson-Walker Metric}
The (Friedmann-LeMa\^{\i}tre-)Robertson-Walker metric is the basis of modern cosmology, being the metric that results when one imposes isotropy and homogeneity upon a spacelike slicing of the universe. One could attempt to use a fully maximally-symmetric metric but, as we have shown, there are only three of these, de-Sitter space, anti de-Sitter space, and Minkowski space, and these are not acceptable models of the current universe; even though this would hold only on the largest scales, the exponential expansion or collapse of the de Sitter models  is simply not observed, even taking into account the recent supernova results implying an acceleration in the expansion of the universe \cite{RiessEtAl04,WMAP-Bennett}. Instead, we wish to find a maximally-symmetric 3-space but allow for a generic time evolution.

We thus impose a milder implementation of homogeneity and isotropy, for a spacelike slicing with metric $\gamma_{ij}$, by taking the spacetime line element to be
\bdm
  ds^2=-dt^2+R^2(t)\gamma_{ij}(u)du^idu^j
\edm
where $u^i$ are some (comoving) co-ordinates that ensure the time-space cross-terms vanish. This metric models a maximally-symmetric spacelike surface evolving with a time parameter $t$.

The Riemann-Christoffel tensor on this 3-dimensional spacelike hypersurface is
\bdm
  {}^{(3)}R_{ijkl}=\frac{{}^{(3)}R}{6}\left(\gamma_{ik}\gamma_{jl}-\gamma_{il}\gamma_{jk}\right),
\edm
which has the Ricci tensor
\bdm
  {}^{(3)}R_{ij}=2\frac{{}^{(3)}R}{6}\gamma_{ij} .
\edm
As a convenient choice of co-ordinates, we shall employ spherically symmetric co-ordinates -- a space of maximum symmetry will certainly be spherically symmetric. Appropriating the spacial part of the Schwarzchild solution we express
\bdm
  \gamma_{ij}du^idu^j=e^{2\beta(\overline{r})}d\overline{r}^2+\overline{r}^2d\Omega^2 .
\edm
which has the Ricci tensor
\bdm
  {}^{(3)}R_{11}=\frac{2}{\overline{r}}\partial_1\beta, \quad
  {}^{(3)}R_{22}=e^{-2\beta}\left(\overline{r}\partial_1\beta-1\right)+1, \quad
  {}^{(3)}R_{33}=\left(e^{-2\beta}\left(\overline{r}\partial_1\beta-1\right)+1\right)
\edm
with all other components zero. We can then use the relationship between the Ricci tensor and the metric to read off the components (defining $\mathcal{K}={}^{(3)}R/6$) and solve for $\beta$; that is use the equations that result from
\bdm
  {}^{(3)}R_{ij}=2\mathcal{K}\gamma_{ij}
\edm
to determine $\beta$ and the Robertson-Walker metric. This expression contains only two independant equations,
\be
  \frac{d\beta(\overline{r})}{d\overline{r}}-\mathcal{K}\overline{r}e^{2\beta(\overline{r})}=0, \quad
  \overline{r}\frac{d\beta(d\overline{r})}{\overline{r}}-1=\left(2\mathcal{K}\overline{r}^2-1\right)e^{2\beta(\overline{r})} .
\ee
The solution of the first of these is found by setting $\beta=\ln\alpha$ and integrating, whence
\bdm
  \beta(\overline{r})=-\frac{1}{2}\ln\left(c-\mathcal{K}\overline{r}^2\right)
\edm
with $c$ the constant of integration. Substitution into the second equation quickly gives us
\be
  \beta(\overline{r})=-\frac{1}{2}\ln\left(1-\mathcal{K}\overline{r}^2\right) .
\ee

We then have that the line element on the maximally-symmetric hypersurfaces is
\be
  {}^{(3)}ds^2=\frac{d\overline{r}^2}{1-\mathcal{K}\overline{r}^2}+\overline{r}^2d\Omega^2 .
\ee
A common alternative is to employ co-ordinates
\be
  d\chi=\frac{d\overline{r}}{\sqrt{1-\mathcal{K}\overline{r}^2}}
\ee
implying
\be
  \overline{r}=\left\{\begin{array}{rl}\sin\chi, &\mathcal{K}=1\\ \chi, &\mathcal{K}=0\\
   \sinh\chi, &\mathcal{K}=-1 \end{array}\right.=S_\mathcal{K}(\chi)
\ee
and
\be
  {}^{(3)}ds^2=d\chi^2+S_\mathcal{K}^2(\chi)d\Omega^2 .
\ee
We then find Euclidean space for $\mathcal{K}=0$, a sphere for $\mathcal{K}=1$ and a hyperboloid for $\mathcal{K}=-1$.

Going to our full spacetime we then have the FLRW metric,
\be
  ds^2=-dt^2+R^2(t)\left(\frac{d\overline{r}^2}{1-\mathcal{K}\overline{r}^2}+\overline{r}^2d\Omega^2\right)
\ee
where $\overline{r}$ is dimensionless and $R(t)$ is the ``scale factor'' giving the physical size of the spacelike slices. In this work we have taken the more usual approach of setting $a(t)$ as the dimensionless scale factor and a radial co-ordinate $r$ with the units of distance,
\bea
  ds^2&=&-dt^2+a^2(t)\left(\frac{dr^2}{1-\mathcal{K}r^2}+r^2d\Omega^2\right) \nonumber \\
      &=&a^2(\eta)\left(-d\eta^2+\frac{dr^2}{1-\mathcal{K}r^2}+r^2d\Omega^2\right)
\eea
where in the second equation we have converted to ``conformal time'' $d\eta=a^{-1}(t)dt$ which puts us into a co-ordinate system that is conformally Minkowski if $\mathcal{K}=0$. For simplicity, we shall take the metric to be conformally-Minkowski throughout, which reduces the harmonic functions -- the eigenvectors of
\bdm
  \nabla^2_\nu f\left(x^\mu\right)=-k^2f\left(x^\mu\right),
\edm
where $\nabla^2_\nu$ is the covariant derivative with respect to $x^\nu$ -- to the Fourier modes $\exp(\pm i\mathbf{k}.\mathbf{x})$.

\section{Perturbation Theory and Gauge Issues}
The (Friedmann-LeMa\^itre-)Robertson-Walker metric derived in the last section forms the large-scale ``background'' model for our cosmology; to begin to approach a realistic model of the universe, we must perturb this metric to some order in an implicit perturbation parameter $\varepsilon$. Doing so immediately raises the issue of gauge variance; general relativity is a gauge theory and, while the metric has ten degrees of freedom, there are only six independently contained within the perturbations -- two of scalar form under arbitrary co-ordinate changes, two of vector (gradient and divergenceless modes) and two of tensor ($+$- and $\times$-type polarisations). We thus have four degrees of freedom in the metric perturbed to first-order in the geometry which must be removed to ensure the physicality of the results.

There are two main approaches to perturbation theory that have been employed in cosmology; the first is the metric-based approach pioneered by Lifshitz as long ago as 1947 \cite{Lifshitz46,LifshitzKhalatnikov63,LandauLifshitz-ClassicalTheoryFields}; see \cite{Efstathiou90,MukhanovFeldmanBrandenberger92,MaBertschinger95,Durrer01} for an inexhaustive list of a few more modern reviews. The basis of the metric based approach is familiar in style from perturbation theory in other areas of physics and particularly in fluid mechanics; we assume a ``background'' geometry that we know to be fictional, and then add in perturbations to some order to gain a better approximation of the physical situation. In effect, we are making a map from a fictional, smooth manifold onto a perturbed, ``physical'' manifold. The problems in this approach arise from the gauge-dependence of general relativity and the complications from assuming from the outset a purely fictional background metric. The older approach to metric-based perturbation theory was to select a gauge to work in throughout, the most common choices being the conformal Newtonian (or longitudinal) gauge -- or the related Poisson gauge when vector (vortical) perturbations are also considered -- and the synchronous gauge initially used by Lifshitz, in which the universe is foliated around either the cosmic or conformal time with comoving hypersurfaces. The synchronous gauge has the benefit of physical lucidity -- the line element of the flat Robertson-Walker geometry closely resembles that of Minkowski space, as opposed to that of the conformal Newtonian gauge which, as its name might suggest, more closely resembles linearised gravity -- but has a distinct drawback: as we show in the next section, synchronous gauge is \emph{not} a uniquely defined gauge and possesses spurious ``gauge modes'' which one has to be careful to remove. Ma and Bertschinger provide a useful review of linearised perturbation theory -- for scalar perturbations -- in both the synchronous and conformal Newtonian gauges, though one should be aware that their definitions of perturbed quantities differ significantly from those we employ in this thesis. Beginning with Bardeen (for example, \cite{Bardeen80,Bardeen88}) a ``gauge-invariant'' approach to metric-based cosmological perturbation theory has grown in popularity; this involves removing gauge ambiguities by combining variables (defined in some gauge) in such a way that they are invariant under some gauge transformation. While admittedly contrived, this approach has significant advantages over the older approach of fixing a single gauge and working within it. This approach is now widely employed and the review of Mukhanov, Feldman and Brandenberger \cite{MukhanovFeldmanBrandenberger92} is an excellent introduction to this formalism. The other main approach that has been employed is the gauge-invariant and covariant (GIC) approach to cosmology, introduced by Ellis and Bruni \cite{EllisBruni89,EllisHwangBruni89}; see also the comprehensive review \cite{EllisVanElst98} and \cite{TsagasBarrow97,TsagasMaartens99-Perts,Tsagas04} for selected examples of its application to magnetised spacetimes. The GIC approach is a 3+1 split superficially similar to the ADM formalism \cite{MisnerWheelerThorne}, in which spacetime is separated around the congruence of ``fundamental'' observers' four-velocities -- where ``fundamental'' observers would be roughly equivalent to comoving observers in the standard metric-based approach. Using the four-velocity as the timelike parameter, spacetime is then foliated with planes instantaneously orthogonal to the four-velocity.
The main benefit of the GIC comes from the Stewart-Walker lemma \cite{StewartWalker74} which states that any quantity that vanishes on a background manifold is automatically gauge-invariant on the perturbed manifold.\footnote{It is worth pointing out that the definition of ``gauge-invariant'' between the standard and the GIC approaches is thus somewhat different.} Applying this to cosmology, we may define a Robertson-Walker manifold as an implicit ``background'' and, by forming quantities that vanish in a Robertson-Walker metric, we may automatically reduce the full non-linear equations to those for a zeroth- and first-order perturbed Robertson-Walker metric with the assurance that gauge issues have been avoided.

\subsection{Metric-Based Perturbation Theory}
In this section we shall briefly discuss first-order metric perturbations of the flat FLRW metric and the gauge issues that arise. We employ a flat metric partly for the pure simplicity it introduces to the equations; however, it should be noted that the current observations strongly support a universe that is, at the most, only mildly differing from a flat case \cite{WMAP-Bennett}. With the simplifying assumption $\mathcal{K}=0$ we can expand our spatial variables across Fourier modes $\exp{(-i\mathbf{k}.\mathbf{x})}$, rather than constructing harmonic eigenvectors in a closed or an open geometry (see for example \cite{Durrer01} for a generalised discussion).

Mukhanov, Feldman and Brandenberger \cite{MukhanovFeldmanBrandenberger92} present a lucid introduction to gauge transformations and perturbed geometries; further details may be found in, for example, Weinberg \cite{Weinberg}, Wald \cite{Wald} or Carroll \cite{Carroll}. We present the first-order FLRW metric here as
\bea
  \lefteqn{ds^2=a^2(\eta)\left(-d\eta^2+\gamma_{ij}dx^idx^j\right)} \nonumber \\&&
  +a^2(\eta)\left(-\delta g_{00}d\eta^2+2\delta g_{oi}d\eta dx^i+\delta g_{ij}dx^idx^j\right) .
\eea
The perturbed section of the metric can be made up of quantities derived from scalar, vector and tensor quantities.

Consider first the scalars: $\delta g_{00}$ is a scalar quantity and so we set $\delta g_{00}=-2\phi$, while $\delta g_{0i}$ is a vector and can be constructed as $\delta g_{0i}=B_{|i}$ where $|$ denotes a covariant derivative with respect to the background metric -- while for a scalar this is obviously identical to a partial derivative, for other objects this specification will be  necessary since the perturbed metric is as yet undefined. Similarly, $\delta g_{ij}$ is a rank-two tensor and can be constructed from scalars in two forms, one proportional to the background metric and one constructed from the second derivative of a scalar, that is, $\delta g_{ij}=2\left(E_{|ij}-\psi\gamma_{ij}\right)$. We cannot add a multiple of the totally anti-symmetric tensor density since the metric must be symmetric. The signs on $\phi$ and $\psi$ are chosen to resemble the Newtonian gravitational potential.

We do much the same for the vector perturbations; here, obviously, we do not have a contribution to $\delta g_{00}$ since any scalar formed from the divergence of a vector can be incorporated into the scalar $\phi$; for $\delta g_{0i}$ we can simply take $\delta g_{0i}=-S_i$, and for the tensor we can take $\delta g_{ij}=F_{i|j}+F_{j|i}$ which has been symmetrised to ensure the symmetry of the metric tensor. We impose $S^i_{\phantom{i}|i}=F^i_{\phantom{i}|i}=0$ to ensure an unambiguous split from the scalars. Raising and lowering of indices is performed with the background metric $\gamma_{ij}$ and its inverse $\gamma^{ij}$; see, for example, Wald \cite{Wald} for a justification. Acting on a first-order variable with the full metric obviously yields the same result as acting with the background metric.

In the tensor case we can only perturb the space-space part of the metric, and here we denote this by $\delta g_{ij}=G_{ij}$ with $G^i_{\phantom{i}i}=G^{ij}_{\phantom{ij}|j}=0$.

Thus we have the perturbed metric
\bea
  \lefteqn{ds^2=a^2(\eta)\bigg(-\left(1+2\phi\right)d\eta^2+2\left(B_{|i}-S_i\right)d\eta dx^i } \nonumber \\
  &&+\left(\left(1-2\psi\right)\gamma_{ij}+2E_{|ij}+2F_{(i|j)}+G_{ij}\right)dx^idx^j\bigg)
\eea
where $A_{(ij)}=(1/2)(A_{ij}+A_{ji})$ is the symmetrised part of a tensor $A_{ij}$. The time-time component, $\phi$, is known as the ``lapse'' function, while the space-time component, $B_{|i}-S_i$, is known as the ``shift'' for readily apparent intuitive reasons; for an event on the spacelike hypersurface the lapse function gives the change in the time co-ordinate -- the time ``elapsed'' -- while the shift function maps it to a corresponding event on the next hypersurface.

This line-element is redundant since we have not yet imposed a gauge. We turn briefly to considering transformations between gauges and comment on ``gauge-invariance'' in this formalism.

A gauge transformation swaps the ``reference'' manifold on which quantities are defined from one to another; these quantities, however, must be measured at the same event as viewed from the two manifolds. The map that uniquely determines the co-ordinates of the event on a manifold is called the gauge choice, and we can view the choice of a gauge as the choice of a co-ordinate system -- in the language of the 3+1 split of spacetime, the gauge choice is a choice of the threading and slicing of spacetime. Applied directly to our situation, the gauge choice is a choice of the map between an event on the ``physical'' (\emph{i.e.}, first-order) universe and an event on the fictional Robertson-Walker background.

Let a gauge transformation between two gauge specifications be given by
\be
  \overline{x}^\mu=x^\mu+\xi^\mu(x)
\ee
where $\xi^\mu(x)$ is infinitesimal and for the purposes of this section a bar denotes a different gauge rather than a ``background'' quantity. This then implies that
\be
  \frac{\partial\overline{x}^\mu}{\partial x^\nu}=\delta^\mu_\nu+\xi^\mu_{\phantom{\mu},\nu} \Rightarrow
  \frac{\partial x^\mu}{\partial\overline{x}^\nu}=\delta^\mu_\nu-\xi^\mu_{\phantom{\mu},\nu}+\mathcal{O}(\xi^2) .
\ee
With these we can consider how scalar, vector and tensor objects transform under a gauge shift.

Consider first a scalar $V(x^\mu)$. Then, by definition,
\bdm
  \overline{V}(\overline{x}^\mu)=V(x^\mu) .
\edm
But we can expand $\overline{x}^\mu$ to give
\bdm
  V(x^\mu)=\overline{V}(\overline{x}^\mu)=\overline{V}\left(x^\mu+\xi^\mu(x)\right)=\overline{V}(x^\mu)+\frac{\partial V}{\partial x^\mu}\xi^\mu+\mathcal{O}\left(\xi^2\right)
\edm
implying that the gauge shift in a scalar quantity is
\be
  \overline{V}(x^\mu)=V(x^\mu)-\frac{\partial V}{\partial x^\mu}\xi^\mu=V(x^\mu)-\xi^\mu V_{|\mu}
\ee
where the last step obviously follows because, for a scalar quantity, the partial and the covariant derivative are identical.

For a vector $A_\alpha(x^\mu)$, by the tensor transformation laws,
\beas
  \overline{A}_\alpha(\overline{x})&=&\frac{\partial x^\beta}{\partial\overline{x}^\alpha}A_\beta(x)
   =\left(\delta^\beta_\alpha-\frac{\partial\xi^\beta}{\partial x^\alpha}\right)A_\beta(x)+\mathcal{O}(\xi^2) \\
   &=&A_\alpha(x)-A_\beta\xi^\beta_{\phantom{\beta},\alpha}+\mathcal{O}(\xi^2)
\eeas
But
\bdm
  \overline{A}_\alpha(\overline{x}^\mu)=\overline{A}_\alpha(x^\mu+\xi^\mu)=\overline{A}_\alpha(x^\mu)+\frac{\partial A_\alpha}{\partial x^\nu}\xi^\nu+\mathcal{O}(\xi^2)
\edm
which then implies
\bdm
  \overline{A}_\mu(x)=A_\mu(x)-A_\alpha\xi^\alpha_{\phantom{\alpha},\mu}-\xi^\alpha A_{\mu,\alpha}
\edm
which, by converting the partial derivatives into covariant derivatives can be shown to be
\be
  \overline{A}_\mu(x)=A_\mu(x)-A_\alpha\xi^\alpha_{\phantom{\alpha}|\mu}-\xi^\alpha A_{\mu|\alpha} .
\ee
Repeating the calculation for a contravariant vector gives
\be
  \overline{A}^\mu(x)=A^\mu(x)+A_\alpha\xi^{\alpha|\mu}-\xi^\alpha A^\mu_{\phantom{\alpha}|\alpha} .
\ee

A similar calculation for a tensor $B^\mu_\nu(x)$ rapidly establishes that
\be
  \overline{B}_{\mu\nu}(x)=B_{\mu\nu}(x)+B_{\mu\alpha}\xi^\alpha_{\phantom{\alpha}|\nu}
   -B_{\alpha\nu}\xi^\alpha_{\phantom{\alpha}|\mu}-\xi^\alpha B_{\mu\nu|\alpha} .
\ee

We may generalise these considerations by defining the Lie derivative, which: for each covariant index $\mu$ subtracts a contraction of the tensor with $\xi^\alpha_{\phantom{\alpha}|\mu}$; for each contravariant index $\nu$ adds a contraction of the tensor with $\xi_\alpha^{\phantom{\alpha}|\nu}$; and subtracts off a final term $\xi^\mu\nabla_\mu$ where $\nabla$ denotes the covariant derivative. That is, the difference in some arbitrary tensor under a gauge transformation is
\bea
  \Delta T^{\mu\ldots}_{\nu\ldots}(x)&=&\overline{T}^{\mu\ldots}_{\nu\ldots}(x)-T^{\mu\ldots}_{\nu\ldots}(x)
   =\Delta_\xi T^{\mu\ldots}_{\nu\ldots}(x)
   \nonumber \\
   &=&T^{\alpha\ldots}_{\nu\ldots}\xi_{\alpha}^{\phantom{\alpha}|\mu}+\cdots-T^{\mu\ldots}_{\alpha\ldots}\xi^{\alpha}_{\phantom{\alpha}|\nu}-\cdots-\xi^\alpha T^{\mu\ldots}_{\nu\ldots|\alpha}
\eea
and $\Delta_\xi$ denotes the Lie derivative in the direction of $\xi$.

These are the general gauge transformations of scalar, vector and tensor quantities. We can now apply these transformations to the perturbations of the metric. First, let us specify the gauge transform as
\be
  \xi^\mu=\left(\xi^0,\xi^{|i}+\breve{\xi}^i\right)
\ee
where $\breve{\xi}^i_{\phantom{i}|i}=0$ -- \emph{i.e.} $\breve{\xi}^i$ is the transverse component of the spatial gauge shift. From this we can immediately show
\bdm
  \overline{\delta g}_{\mu\nu}(x)=\delta g_{\mu\nu}(x)-g_{0\nu}^{(0)}\xi^0_{\phantom{0},\mu}-g_{\mu0}^{(0)}\xi^0_{\phantom{0},\nu}
   -\xi^0g_{\mu\nu,0}^{(0)}-\gamma^{ik}\left(g_{\mu i}^{(0)}\xi_{\nu k}+g_{\nu i}^{(0)}\xi_{\mu k}\right)
   -\left(g_{\mu i}^{(0)}\breve{\xi}^i_{\phantom{i}\nu}+g_{\nu i}^{(0)}\breve{\xi}^i_{\phantom{i}\mu}\right)
\edm
and so
\bea
  \overline{\delta g}_{00}&=&\delta g_{00}-2g_{00}^{(0)}\xi^0_{\phantom{0},0}-g_{00,0}^{(0)}\xi^0 ,
   \nonumber \\
  \overline{\delta g}_{0i}&=&\delta g_{0i}-g_{00}^{(0)}\xi^0_{\phantom{0},i}-\gamma^{jk}g_{ij}^{(0)}\xi_{,0k}
   -g_{ij}^{(0)}\breve{\xi}^j_{\phantom{j},0}
     \\
  \overline{\delta g}_{ij}&=&\delta g_{ij}-\xi^0g_{ij,0}^{(0)}-\gamma^{ak}\left(g_{jk}^{(0)}\xi_{,ia}+g_{ik}^{(0)}\xi_{,ja}\right)
   -\left(g_{jk}^{(0)}\breve{\xi}^k_{\phantom{k},i}+g_{ik}^{(0)}\breve{\xi}^k_{\phantom{k},j}\right), \nonumber
\eea
where we have used the vanishing spatial derivatives and $g_{0i}$ components of the background metric. Inserting now the full annsatz for the background metric we see that
\bea
  \frac{\overline{\delta g}_{00}}{a^2(\eta)}&=&\frac{\delta g_{00}}{a^2(\eta)}+2\left(\dot{\xi}^0+\hub\xi^0\right)
   \nonumber \\
  \frac{\overline{\delta g}_{0i}}{a^2(\eta)}&=&\frac{\delta g_{0i}}{a^2(\eta)}+\xi^0_{\phantom{0},i}-\dot{\xi}_{,i}-\dot{\breve{\xi}}_i
   \nonumber \\
  \frac{\overline{\delta g}_{ij}}{a^2(\eta)}&=&\frac{\delta g_{ij}}{a^2(\eta)}-2\hub\gamma_{ij}\xi^0-2\xi_{,ij}-2\breve{\xi}_{(i,j)}
\eea
where an overdot denotes differentiation with respect to the conformal time.

With the definitions of our metric perturbations we can now rapidly derive a set of gauge transformations,
\bea
  &\overline{\phi}=\phi-\left(\dot{\xi}^0+\hub\xi^0\right), \quad
  \overline{\psi}=\psi-\hub\xi^0, \quad
  \overline{B}=B+\xi^0-\dot{\xi}, \quad
  \overline{E}=E-\xi,& \nonumber \\
  &\overline{S}_i=S_i+\dot{\breve{\xi}}_i, \quad
  \overline{F}_i=F_i-\breve{\xi}_i, \quad
  \overline{G}_{ij}=G_{ij}&
\eea
demonstrating that the tensor perturbations -- the gravitational waves -- are naturally gauge-invariant.

There are naturally many -- indeed, an infinity -- of possible gauges one may select. We shall concentrate on two often encountered within the literature: synchronous gauge, first examined in the cosmological context by Lifshitz, where we take a purely spatial slicing of spacetime foliated along the cosmic time; and Poisson gauge, which is a generalisation \cite{Bertschinger95} of the longitudinal, gauge employed by Mukhanov, Feldman and Brandenberger \cite{MukhanovFeldmanBrandenberger92} and Ma and Bertschinger \cite{MaBertschinger95}. Both gauges have their advantages and disadvantages. Other common choices of gauge include the uniform curvature gauge, wherein spacetime is sliced along surfaces of constant curvature, and uniform density gauge, wherein the slicing is performed along surfaces of constant matter density.

\subsection{Synchronous Gauge}
Throughout this thesis we shall work almost entirely within synchronous gauge, which is the gauge found by explicitly writing the FLRW metric as a simple foliation of spacetime. That is, we set the lapse and shift functions to zero and spacetime is foliated by conformally Euclidean spacelike hypersurfaces. As we shall shortly see, synchronous gauge is not well defined and contains spurious gauge modes. However, it is useful for two respects -- the metric manifestly resembles our usual Minkowski space with purely spatial perturbations, which can provide a useful aid to intuition, and there is an unambiguous time parameter which is simply either the cosmic time or the conformal time. Moreover, the CMBFast code \cite{SeljakZaldarriaga96} is written in synchronous gauge, partly to exploit this simple time parameter, and if we at any point should desire to incorporate our results into a numerical code this will prove a large advantage.

Synchronous gauge is found by setting
\be
  \phi_{\mathrm{Syn}}=B_{\mathrm{Syn}}=S_{\mathrm{Syn}}^i=0
\ee
and from an arbitrary gauge the transformation into synchronous gauge is then defined by
\be
  \phi_{\mathrm{Syn}}=0=\phi-\left(\dot{\xi}^0+\hub\xi^0\right), \quad B_{\mathrm{Syn}}=0=B+\xi^0-\dot{\xi}, \quad
  S^{\mathrm{Syn}}_i=0=S_i+\dot{\breve{\xi}}_i .
\ee
Solving this set of equations for the gauge transformation ultimately gives us
\bea
  a(\eta)\xi^0&=&c_1(\mathbf{x})+\int a(\tilde{\eta})\phi(\tilde{\eta},\mathbf{x})d\tilde{\eta}, \nonumber \\
  \xi&=&c_2(\mathbf{x})+\int B(\tilde{\eta},\mathbf{x})d\tilde{\eta}+\iint\frac{1}{a(\tilde{\eta})}a(\breve{\eta})\phi(\breve{\eta},\mathbf{x})d\breve{\eta}d\tilde{\eta}, \\
  \breve{\xi}_i&=&c^V_i(\mathbf{x})-\int S_i(\tilde{\eta},\mathbf{x})d\tilde{\eta}
\eea
where $c_1$, $c_2$ and $c^V_i$ are arbitrary functions of the spatial variables but constant in time. It is these functions that provide the ambiguity within synchronous gauge and results from within this gauge need to be analysed carefully to remove any spurious, unphysical modes. See for example \cite{PressVishniac80,Bardeen80} for more details on this point.

\subsection{Poisson Gauge}
We shall employ Poisson Gauge as employed by Bertschinger \cite{Bertschinger95}, introducing first the conformal Newtonian gauge (\cite{MukhanovFeldmanBrandenberger92,MaBertschinger95}). Conformal Newtonian gauge contains only scalar perturbations and so, by definition, $S^{\mathrm{Nwt}}_i=F^{\mathrm{Nwt}}_i=G^{\mathrm{Nwt}}_{ij}=0$. The gauge condition on the remaining scalar degrees of freedom is that the lapse vanishes and there are no scalar derivatives in the spatial perturbation -- that is,
\be
  B_{\mathrm{Nwt}}=E_{\mathrm{Nwt}}=0 .
\ee
Since the only perturbations that remain are on the time-time and isotropic space-space terms the resulting metric closely resembles a perturbed Minkowski (weak-field) metric and the perturbations $\phi$, $\psi$ the Newtonian potentials.
Following the same process as for the synchronous gauge, we rapidly find that the gauge-transform from an arbitrary (scalar) gauge into conformal Newtonian gauge is
\be
  \xi=E, \quad \xi^0=\dot{\xi}-B=\dot{E}-B
\ee
which, unlike the transformation into synchronous gauge, is uniquely defined.

Poisson gauge is little more complicated than conformal Newtonian gauge; it is defined by retaining a transverse vector part in the shift, which then renders the shift merely $S_i$, and eliminating the vectorial space perturbation. The additional transform needed from an arbitrary gauge into Poisson gauge is then
\be
  \breve{\xi}_i=F_i .
\ee

The tensor perturbations being gauge-invariant, Poisson gauge is then well suited for studying cosmological perturbations, even in the presence of active anisotropic sources, as synchronous gauge.

\subsection{Transferral between Synchronous and Poisson Gauge}
Transferral from synchronous into Poisson gauge is simple; using the above expressions for transferring into the Poisson gauge we rapidly see that
\be
  \xi^0=\dot{E}_{\mathrm{Syn}}-B_{\mathrm{Syn}}=\dot{E}_{\mathrm{Syn}}, \quad
  \xi=E_{\mathrm{Syn}}, \quad
  \breve{\xi}^i=F^i_{\mathrm{Syn}} .
\ee
Thus the Poisson gauge variables are given in terms of the synchronous variables as
\be
  \phi_{\mathrm{Poi}}=-\left(\ddot{E}_{\mathrm{Syn}}+\hub\dot{E}_{\mathrm{Syn}}\right), \quad
  \psi_{\mathrm{Poi}}=\psi_{\mathrm{Syn}}-\hub\dot{E}_{\mathrm{Syn}}, \quad
  S^i_{\mathrm{Poi}}=F^i_{\mathrm{Syn}} .
\ee

Due to the ambiguity in transferring into synchronous gauge, the reverse is far from pleasant. The gauge transformation turns out to be
\bea
  a(\eta)\xi^0&=&c_1(\mathbf{x})+\int a(\tilde{\eta})\phi_{\mathrm{Poi}}(\tilde{\eta},\mathbf{x})d\tilde{\eta}, \nonumber \\
  \xi&=&c_2(\mathbf{x})+\iint\frac{a(\breve{\eta})}{a(\tilde{\eta})}\phi_{\mathrm{Poi}}(\breve{\eta},\mathbf{x})d\breve{\eta}d\tilde{\eta}, \\
  \breve{\xi}^i&=&c^i_V(\mathbf{x})-\int S^i_{\mathrm{Poi}}(\tilde{\eta},\mathbf{x})d\tilde{\eta} .
\eea

The explicit forms for the metric perturbations are now simple to write down but not particularly illuminating.

\subsection{Gauge Invariance and the Bardeen Variables}
From the form of the generic gauge transformations it is easy to see that we can construct quantities that are invariant under gauge transformations. It should be emphasised that, in different gauges, these variables will retain different physical interpretations even though they have the same numerical value. Moreover, it is worth commenting that these combinations are in many ways arbitrary constructions designed purely to eliminate the difficulties introduced by gauge variance. A more natural form of gauge-invariance occurs in the GIC approach with quite a different interpretation.

First it is worth noticing that
\bdm
  \overline{B}-\dot{\overline{E}}=B-\dot{E}+\xi^0 .
\edm
The time derivative of this, multiplied by the scale factor, is then
\bdm
  \left(a\left(\overline{B}-\dot{\overline{E}}\right)\right)^\cdot=\left(a\left(B-\dot{E}\right)\right)^\cdot+a\left(\dot{\xi}^0+\hub\xi^0\right) .
\edm
From here it is immediately obvious that we can define
\be
  \Phi=\phi+\frac{1}{a}\left(a\left(B-\dot{E}\right)\right)^\cdot, \quad
  \Psi=\psi+\hub\left(B-\dot{E}\right)
\ee
as two gauge-invariant variables, known as the Bardeen potentials \cite{Bardeen80,Bardeen88}.

For the vectors it is readily apparent that
\be
  \tilde{V}^i=S^i+\dot{F}^i
\ee
is also gauge-invariant. Along with the gravitational waves $G_{ij}$ we thus have a complete set of gauge-invariant variables.

In Poisson gauge, these variables reduce to
\be
  \Phi=\phi_{\mathrm{Poi}}, \quad \Psi=\psi_{\mathrm{Poi}}, \quad \tilde{V}^i=S^i_{\mathrm{Poi}}
\ee
which explains the terminology ``Bardeen potentials''; the scalar gauge-invariant combinations reduce to the conformal Newtonian gravitational potentials. The vector gauge-invariant variable reduces to the shift function.

In synchronous gauge, on the other hand,
\be
\label{BardeenSynchronous}
  \Phi=-\frac{1}{a}\left(a\dot{E}_{\mathrm{Syn}}\right)^\cdot, \quad
  \Psi=\psi_{\mathrm{Syn}}-\hub\dot{E}_{\mathrm{Syn}}, \quad
  \tilde{V}^i=\dot{F}^i_{\mathrm{Syn}} .
\ee
For the scalar variables these do not possess as transparent a physical definition; the vector variable reduces to the time derivative of the spatial geometrical vorticity.

\chapter{Viscous Fluids}
\label{Appendix-ViscousFluids}
In this appendix we briefly derive the conservation laws omitted from \S\ref{Sec-Fluids} for a fluid containing bulk and shear viscosities and conducting heat. The full stress-energy tensor we employ \cite{Weinberg} is
\beas
  T^\mu_\nu&=&\rho u^\mu u_\nu+pH^\mu_\nu-\chi\left(H^{\mu\alpha}u_\nu+H^\alpha_\nu u^\mu\right)Q_\alpha
  -\xi H^{\mu\alpha}H_{\nu\beta}W_\alpha^\beta-\zeta H^\mu_\nu u^{\alpha}_{\phantom{\alpha};\alpha}
   \\ &=&\rho u^\mu u_\nu+pH^\mu_\nu+\tilde{\chi}^\mu_\nu+\tilde{\xi}^\mu_\nu+\tilde{\zeta}^\mu_\nu
\eeas
with $p=w(\rho)\rho$, the four-velocity
\be
  u^\mu=\frac{dx^\mu}{d\sqrt{-ds^2}},
\ee
the projection tensor onto a hypersurface orthogonal to $u^\mu$
\bdm
  H^\mu_\nu=\delta^\mu_\nu+u^\mu u_\nu,
\edm
the heat-flow vector
\be
  Q_\mu=\Theta_{;\mu}+\Theta u_{\mu;\nu}u^\nu,
\ee
and the shear tensor is defined as
\be
  W^\mu_\nu=u^\mu_{\phantom{\mu};\nu}+u_\nu^{\phantom{\nu};\mu}-\frac{2}{3}\delta^\mu_\nu u^\alpha_{\phantom{\alpha};\alpha} .
\ee

This reduces, with $\Theta=\overline{\Theta}(\eta)+\theta(\mathbf{x},\eta)$, to
\bea
  T^0_0&=&-\rho, \\
  T^i_0&=&-\left(\rho+p\right)v^i+\chi_a\overline{\Theta}\left(\dot{v}^i
    +\left(\hub+\frac{\dot{\overline{\Theta}}}{\overline{\Theta}}\right)v^i
    +\frac{\partial^i\theta}{\overline{\Theta}}\right)
   +3\zeta_a\hub v^i, \\
  T^i_j&=&p\delta^i_j-\zeta_a\left(3\hub+\nabla.\mathbf{v}+\frac{1}{2}\dot{h}\right)\delta^i_j
   -\xi_a\left(\partial^iv_j+\partial_jv^i+\dot{h}^i_j-\frac{1}{3}\delta^i_j\left(2\nabla.\mathbf{v}+\dot{h}\right)\right)
\eea
If we assume no further collisional processes then we can set the collision term $\mathcal{C}_\mu$ in the conservation equations to zero. As saw earlier, the ideal components reduce to a background contribution
\bea
  -\overline{T}^\mu_{0;\mu}&=&\dot{\bkr}+3\hub\bkr\left(1+w(\rho)\right), \nonumber \\
  \overline{T}^\mu_{i;\mu}&=&\partial_i\overline{p}
\eea
and a foreground contribution
\bea
  \dot{\delta}+3\hub\left(c_s^{\phantom{s}2}-w\right)\delta+\left(1+w\right)
  \left(\nabla.\fgv+\frac{1}{2}\dot{h}\right)&=&-\delta T^\mu_{0;\mu}, \\
  \dot{v}^i+\left(\frac{\dot{w}}{1+w}+\hub\left(1-3w\right)\right)v^i
  +\frac{c_s^{\phantom{s}2}}{1+w}\partial^i\delta&=&\gamma^{ij}\delta T^\mu_{j;\mu}. \nonumber
\eea

For the heat conduction we find that
\bea
  -\tilde{\chi}^\nu_{0;\nu}&=&-\chi_a\overline{\Theta}\left(
    \nabla.\dot{\mathbf{v}}+\left(\hub+\frac{\dot{\overline{\Theta}}}{\overline{\Theta}}\right)\nabla.\mathbf{v}
    +\frac{\nabla^2\theta}{\overline{\Theta}}
     \right. \nonumber \\ && \quad \left.
    +\left(\frac{\nabla\chi_a}{\chi_a}\right)\left(\dot{\mathbf{v}}
    +\left(\hub+\frac{\dot{\overline{\Theta}}}{\overline{\Theta}}\right)\mathbf{v}
    +\frac{\nabla\theta}{\overline{\Theta}}\right)\right), \\
  \tilde{\chi}^\nu_{i;\nu}&=&-\chi_a\overline{\Theta}\left\{
    \ddot{v}_i+\left(\hub+\frac{\dot{\overline{\Theta}}}{\overline{\Theta}}\right)\dot{v}_i
    +\left(\frac{\ddot{a}}{a}-\left(\hub\right)^2+\frac{\ddot{\overline{\Theta}}}{\overline{\Theta}}
    -\left(\frac{\dot{\overline{\Theta}}}{\overline{\Theta}}\right)^2\right)v_i
     \right. \\ && \quad \left. 
    +\frac{\partial_i\dot{\theta}}{\overline{\Theta}}
    -\frac{\dot{\overline{\Theta}}}{\overline{\Theta}}\frac{\partial_i\theta}{\overline{\Theta}}
    +\left(\frac{\dot{\chi}_a}{\chi_a}
    +\frac{\dot{\overline{\Theta}}}{\overline{\Theta}}+4\hub\right)\left(\dot{v}_i+\left(\hub
    +\frac{\dot{\overline{\Theta}}}{\overline{\Theta}}\right)v_i+\frac{\partial_i\theta}{\overline{\Theta}}\right)
    \right\} \nonumber
\eea
which are all first-order while, predictably, the shear viscous sector only contributes to the first-order Euler equation, with
\bea
  \lefteqn{\tilde{\xi}^\nu_{i;\nu}=-\xi_a\left\{\nabla^2v_i+\frac{1}{3}\partial_i\left(\nabla.\mathbf{v}
    -\dot{h}\right)+\partial_a\dot{h}^a_i
    \nonumber \right.} \\ && \left.
    +\left(\frac{\partial_j\xi_a}{\xi_a}\right)\left(\partial^jv_i+\partial_iv^j+\dot{h}^j_i
     -\frac{1}{3}\left(2\nabla.\mathbf{v}+\dot{h}\right)\right)\right\} .
\eea
The bulk viscous sector contributes to both the conservation and Euler equations with
\bea
  -\tilde{\zeta}^\nu_{0;\nu}&=&-3\hub\zeta_a\left(3\hub+2\nabla.\mathbf{v}+\dot{h}
    +\frac{\left(\nabla.\mathbf{v}\right)\zeta_a}{\zeta_a}\right), \\
  \tilde{\zeta}^\nu_{i;\nu}&=&-3\zeta_a\left(
    \frac{1}{3}\partial_i\left(\nabla.\mathbf{v}+\frac{1}{2}\dot{h}\right)+\hub\dot{v}_i
    +\left(\frac{\dot{\zeta}_a}{\zeta_a}\hub+\frac{\ddot{a}}{a}+3\left(\hub\right)^2\right)v_i
   \right. \nonumber \\ && \left.
    +\frac{1}{3}\left(\nabla.\mathbf{v}+\frac{1}{2}\dot{h}\right)\frac{\partial_i\zeta_a}{\zeta_a}
    +\hub\frac{\partial_i\zeta_a}{\zeta_a}\right) .
\eea
Assuming we can take $\zeta\approx\mathcal{O}(0)$ we then have a background contribution from the bulk viscosity
\be
  -\overline{\zeta}^\nu_{0;\nu}=-9\left(\hub\right)^2\zeta_a
\ee
with all other effects linear order only. Should the coefficiants of bulk viscosity be effectively first-order then this forms the only contribution and enters at linear order.

While we could certainly retain the generality of non-constant coefficiants of heat conductivity and viscosities, we shall instead assume that they are constant. That is,
\be
  \frac{\dot{\chi}_a}{\chi_a}=-\hub, \quad \partial_i\chi_a=0
\ee
and similar.

\subsubsection{Fourier Space}
In the background,
\bdm
  \overline{T}^0_0=-\bkr, \quad \overline{T}^i_0=0, \quad \overline{T}^i_j=c_s^2\bkr\delta^i_j-3\hub\zeta_a\delta^i_j
\edm
and in the foreground,
\bdm
  \delta T^0_0=-\bkr\delta, \quad \delta T^i_0=-\bkr\left(1+w\right)v_i, \quad \delta T^i_j=c_s^2\bkr\delta\delta^i_j
\edm
which, separating across the scalar, vector and tensor parts, go to
\bdm
  \overline{T}^0_0=-\bkr, \quad \overline{T}^i_0=0, \quad \overline{T}^i_i=3c_s^2\bkr, \quad \overline{\pi}^i_j=0
\edm
in the background and
\beas
  &\delta T^0_0=-\bkr\delta, \quad \delta T^{(i)}_{(0)S}=-\bkr\left(1+w\right)v_S,& \\
  &\delta T^i_{(0)V}=-\bkr\left(1+w\right)\V{v}_i, \quad \delta T=3c_s^2\bkr\delta+9\zeta_a\hub, \quad \delta\pi^i_j=0 .&
\eeas

After the scalar/vector/tensor separation, the foreground contributions from the heat conduction contribute only to the $0i$ components of the stress-energy -- that is, the energy flow -- and separating this we find
\be
  \tilde{\chi}^{(i)}_{(0)S}=\chi_a\overline{\Theta}\left(\dot{v}_S
   +\left(\hub+\frac{\dot{\overline{\Theta}}}{\overline{\Theta}}\right)v_S
   -\frac{k\theta}{\overline{\Theta}}\right), \;
  \tilde{\chi}^i_{(0)V}=\chi_a\overline{\Theta}\left(\V{\dot{v}}_i
   +\left(\hub+\frac{\dot{\overline{\Theta}}}{\overline{\Theta}}\right)\V{v}_i\right) .
\ee
The shear viscosity impacts purely on the space-space components of the stress-energy and, by construction, has a vanishing trace. It separates to
\be
  \tilde{\xi}^i_i=0, \;
  \tilde{\xi}_S=-\xi_a\left(2kv_S+\dot{h}_S\right), \;
  \V{\tilde{\xi}}_i=-\xi_a\left(\V{h}_i-ik\V{v}_i\right), \;
  \tilde{\xi}^{iT}_j=-\xi_ah^{iT}_j .
\ee
Finally, the first-order bulk viscosity contributes to an energy flow and, obviously and by construction, only to an isotropic stress:
\be
  \tilde{\zeta}^{(i)}_{(0)S}=3\zeta_a\hub v_s, \quad
  \tilde{\zeta}^{i}_{(0)V}=3\zeta_a\hub v^i_V, \quad
  \tilde{\zeta}^i_i=-3\zeta_a\left(kv_S+\frac{1}{2}\dot{h}_S\right).
\ee

Turning to the fluid dynamical equations, in the presence of non-ideal terms the background matter conservation equation is modified to
\be
  \dot{\bkr}+3\hub\bkr\left(1+w-3\hub\zeta_a\right)=0 .
\ee

The perturbed ideal components are
\bea
  \dot{\delta}+3\hub\left(c_s^{\phantom{s}2}-w\right)\delta+\left(1+w\right)
   \left(kv_S+\frac{1}{2}\dot{h}\right)&=&-\delta T^\mu_{0;\mu}, \\
  \dot{v}_S+\left(\frac{\dot{w}}{1+w}+\hub\left(1-3w\right)\right)v_S
   -k\frac{c_s^{\phantom{s}2}}{1+w}\delta&=&\left(\delta T^\mu_{i;\mu}\right)^S, \\
  \dot{v}^i_V+\left(\frac{\dot{w}}{1+w}+\hub\left(1-3w\right)\right)v^i_V&=&\gamma^{ij}\left(\delta T^\mu_{j;\mu}\right)^V
\eea
while the non-ideal contribution to the perturbed matter conservation equations comes from
\be
  -\left(\chi+\zeta\right)^\mu_{0;\mu}=-k\chi_a\overline{\Theta}\left(\dot{v}_S+\left(\hub
   +\frac{\dot{\overline{\Theta}}}{\overline{\Theta}}\right)v_S-\frac{k\theta}{\overline{\Theta}}\right)
   -3\zeta_a\hub\left(2kv_S+\dot{h}\right) .
\ee

The contribution to the scalar Euler equation from heat conduction is
\bea
  \lefteqn{\tilde{\chi}^\mu_{(i)S;\mu}=-\chi_a\overline{\Theta}\left(\ddot{v}S+2\left(2\hub
   +\frac{\dot{\overline{\Theta}}}{\overline{\Theta}}\right)\dot{v}_S+
   -\frac{k\dot{\theta}}{\overline{\Theta}}-3k\hub\frac{\theta}{\overline{\Theta}}
    \right. } \nonumber \\ && \left.
   \left(\frac{\ddot{a}}{a}+2\left(\hub\right)^2
   +\frac{\ddot{\overline{\Theta}}}{\overline{\Theta}}+4\hub\frac{\dot{\overline{\Theta}}}{\overline{\Theta}}
   \right)v_S\right) ,
\eea
from shear viscosity is
\be
  \tilde{\xi}^\mu_{(i)S;\mu}=\frac{2k}{3}\xi_a\left(2kv_s+\dot{h}_S\right)
\ee
and from bulk viscosity is
\be
  \tilde{\zeta}^\mu_{(i)S;\mu}=-3\zeta_a\left(\hub\dot{v}_S+\left(\frac{\ddot{a}}{a}+2\left(\hub\right)^2
   -\frac{1}{3}k^2\right)v_S-\frac{1}{6}k\dot{h}\right)
\ee

Non ideal effects contribute
\be
  \tilde{\chi}^\mu_{iV;\mu}=-\chi_a\overline{\Theta}\left(\ddot{v}_i^V
   +2\left(2\hub+\frac{\dot{\overline{\Theta}}}{\overline{\Theta}}\right)\dot{v}_i^V
   +\left(\frac{\ddot{a}}{a}+2\left(\hub\right)^2+\frac{\ddot{\overline{\Theta}}}{\overline{\Theta}}
   +4\hub\frac{\dot{\overline{\Theta}}}{\overline{\Theta}}\right)v_i^V\right) ,
\ee
\be
  \tilde{\xi}^\mu_{iV;\mu}=k\xi_a\left(kv_i^V-ik\dot{h}_i^V\right)
\ee
and
\be
  \tilde{\zeta}^\mu_{iV;\mu}=-3\zeta_a\left(\hub\dot{v}_i^V+\left(\frac{\ddot{a}}{a}+2\left(\hub\right)^2
   \right)v_i^V\right)
\ee
to the vector Euler equations, from heat conduction, shear and bulk viscosities respectively.

It is readily apparent that bulk viscosity is far the simplest viscous effect to include in a cosmological study; however, as we demonstrated in \S\ref{Sec-Fluids-SETensor} it is not particularly physical to include a bulk viscosity in the background. Shear viscosities are widely used for small-scale models of the photon drag, as in \S\ref{Sec-DampingScaleMHD}.

Many of the effects of heat conduction cancel if one takes the simplifying assumption
\be
  \dot{\overline{\Theta}}+2\hub\overline{\Theta}=0
\ee
which is true for a species of non-interacting particles and may be derived from the first law of thermodynamics.

\chapter{Vector and Tensor CMB Statistics}
In this appendix we derive in detail the vector and tensor CMB angular power spectra in the traditional approach.

\section{Vector Contributions to the CMB in the Traditional Approach}
\label{VectorCMBSpectraAppendix}
Here we derive the result stated in section (\ref{VectorCMBSpectra}). The vector contribution to the CMB temperature shift was
\bdm
  \V{\frac{\delta T}{T}}\left(\mathbf{x},\nhv,\eta\right)=\frac{1}{4}
  \sqrt{1-\mu^2}\sum_l(-i)^l(2l+1)P_l(\mu)\left(\alpha_l^{(1)}(\kv,\eta)\cos\phi
   +\alpha_l^{(2)}(\kv,\eta)\sin\phi\right)
\edm
and so the vector temperature auto-correlation function is the ugly
\beas
  \V{C}(q)=\frac{1}{16}\sum_l\sum_p(-i)^li^p(2l+1)(2p+1)\iint \sqrt{1-\mu^2}
  \sqrt{1-\overline{\mu}^2}P_l(\mu)P_p(\overline{\mu}) \qquad \qquad \\ \times \left(
   \left<\alpha_l^{(1)}(\kv)\alpha_p^{*(1)}(\kv')\right>\cos\overline{\phi}\cos\phi+
   \left<\alpha_l^{(1)}(\kv)\alpha_p^{*(2)}(\kv')\right>\sin\overline{\phi}\cos\phi+ \right.\quad \\ \left.
   \left<\alpha_l^{(2)}(\kv)\alpha_p^{*(1)}(\kv')\right>\cos\overline{\phi}\sin\phi+
   \left<\alpha_l^{(2)}(\kv)\alpha_p^{*(2)}(\kv')\right>\sin\overline{\phi}\sin\phi
  \right)e^{i(\kv'-\kv).\xv}\frac{d^3\kv}{\left(2\pi\right)^3}\frac{d^3\kv'}{\left(2\pi\right)^3}
\eeas
where $\overline{\mu}=\khv'.\nhv'$ and $\overline{\phi}$ and $\overline{\theta}$ are associated with $\nhv'$.

However, since the hierarchies are symmetric with respect to change of vector mode, we shall assume that the two are uncorrelated but similar -- that is, we shall assume that
\bea
  \left<\alpha_l^{(1)}(\kv)\alpha_p^{*(1)}(\kv')\right>&\approx&
  \left<\alpha_l^{(2)}(\kv)\alpha_p^{*(2)}(\kv')\right>, \nonumber \\
  \left<\alpha_l^{(1)}(\kv)\alpha_p^{*(2)}(\kv')\right>&=&
  \left<\alpha_l^{(2)}(\kv)\alpha_p^{*(1)}(\kv')\right>^*\approx0.
\eea
As in the scalar case, the independence of the evolution equations on the direction of $\kv$ allows us to define
\be
  \left<\alpha_l^{(1)}(\kv)\alpha_p^{*(1)}(\kv')\right>=\mathcal{P}_V(k)\alpha_l(k)\alpha^*_p(k')\delta
  (\kv'-\kv).
\ee
With these identifications, the temperature auto-correlation is
\beas
  \V{C}(q)=\frac{1}{16(2\pi)^6}\sum_l\sum_p(-i)^li^p(2l+1)(2p+1)\iint \sqrt{1-\mu^2}
  \sqrt{1-\overline{\mu}^2}P_l(\mu)P_p(\overline{\mu}) \quad \\
   \times \mathcal{P}_V(k) \alpha_l(k)\alpha^*_p(k')\delta(\kv'-\kv)\cos\left(\overline{\phi}-\phi\right)
   e^{i(\kv'-\kv).\xv}d^3\kv d^3\kv' .
\eeas

We can remove the terms dependent on the azimuthal angle by considering $\nhv'.\nhv$:
\beas
  \nhv'.\nhv&=&\sin\overline{\theta}\sin\theta\left(\cos\overline{\phi}\cos\phi+\sin\overline{\phi}
   \sin\phi\right)+\mu\overline{\mu}\\
   &=&\sqrt{1-\mu^2}\sqrt{1-\overline{\mu}^2}\cos\left(\overline{\phi}-\phi\right)+\mu\overline{\mu}
\eeas
which gives us
\beas
  \V{C}(q)=\frac{1}{16(2\pi)^6}\sum_l\sum_p(-i)^li^p(2l+1)(2p+1)\iint
  P_l(\mu)P_p(\overline{\mu})\mathcal{P}_V(k)\alpha_l(k)\alpha^*_p(k') \quad \\
   \times\left(\nhv'.\nhv-\mu\overline{\mu}\right)
   \delta(\kv'-\kv)e^{i(\kv'-\kv).\xv}d^3\kv d^3\kv' .
\eeas

Considering first the term proportional to $\nhv'.\nhv$, we can follow much the same process as for the scalars, expanding the Legendre polynomials across the spherical harmonics and integrating over $\kv'$ and $d\Omega_\kv$ to leave
\bdm
  \V{C}_1(q)=\frac{4\pi}{16(2\pi)^6}\int\sum_l(2l+1)\mathcal{P}_V(k)\left|\alpha_l(k)\right|^2qP_l(q)k^2dk .
\edm
We then employ the recursion relation on $P_l(q)$ to find
\bdm
  \V{C}_1(q)=\frac{4\pi}{16(2\pi)^6}\int\sum_l\mathcal{P}_V(k)\left|\alpha_l(k)\right|^2
    \left((l+1)P_{l+1}(q)+lP_{l-1}(q)\right)k^2dk .
\edm
Relabelling the summation indices on the two terms then allows us to write the contribution to the vector $C_l$:
\bdm
  \V{C}_{1l}=A\int\mathcal{P}_V(k)\frac{l\left|\alpha_{l-1}(k)\right|^2+(l+1)\left|\alpha_{l+1}(k)\right|^2}{2l+1}k^2dk.
\edm

The second term, proportional to $\mu\overline{\mu}$, yields a slightly more convoluted analysis. Firstly we employ the recurrence relation for Legendre polynomials to write
\beas
  \lefteqn{\V{C}_2(q)=-\frac{1}{16(2\pi)^6}\iint\sum_l\sum_p(-i)^li^p\mathcal{P}_V(k)\alpha_l\alpha^*_p} \\ &&
  \times \left\{ (l+1)(p+1)P_{l+1}(\mu)P_{p+1}(\overline{\mu})+lpP_{l-1}(\mu)P_{p-1}(\overline{\mu}) \right. \\&& \quad \left.
  +p(l+1)P_{l+1}(\mu)P_{p-1}(\overline{\mu})+l(p+1)P_{l-1}(\mu)P_{p+1}(\overline{\mu})
  \right\}\delta\left(\kv'-\kv\right)e^{i\left(\kv'-\kv\right)}d^3\kv d^3\kv'
\eeas
Expanding the Legendres across the spherical harmonics and integrating over both $d\kv'$ and $d\Omega_\kv$ gives us
\beas
  \V{C}_2(q)=-\frac{(4\pi)^2}{16(2\pi)^6}\sum_{l,m}\int\mathcal{P}_V(k)\left\{\left|\alpha_l\right|^2
   \frac{(l+1)^2}{(2l+3)^2}Y^*_{l+1m}(\nhv')Y_{l+1m}(\nhv) \right. \qquad \qquad \qquad \\
   +\left|\alpha_l\right|^2\frac{l^2}{(2l-1)^2}Y^*_{l-1m}(\nhv')Y_{l-1m}(\nhv)
   -\alpha_l\alpha^*_{l+2}\frac{(l+1)(l+2)}{(2l+3)^2}Y^*_{l+1m}(\nhv')Y_{l+1m}(\nhv) \; \\ \left.
   -\alpha_l\alpha^*_{l-2}\frac{(l-1)l}{(2l-1)^2}Y^*_{l-1m}(\nhv')Y_{l-1m}(\nhv) \right\}k^2dk
\eeas
whence we rapidly find
\bdm
  \V{C}_{2l}=-A\int\mathcal{P}_V(k)\left\{\frac{l^2\left|\alpha_{l-1}\right|^2+(l+1)^2\left|\alpha_{l+1}\right|^2
  -l(l+1)\left(\alpha_{l-1}\alpha^*_{l+1}+\alpha_{l+1}\alpha^*_{l-1}\right)}{(2l+1)^2}\right\}k^2dk.
\edm

Combining this with $\V{C}_{1l}$, we can reduce the contribution from vector perturbations to the angular power spectrum of the temperature auto-correlation function to finally reach
\be
  \V{C}_l=A\frac{l(l+1)}{(2l+1)^2}\int\mathcal{P}_V(k)\left|\alpha_{l-1}+\alpha_{l+1}\right|^2k^2dk
\ee
as stated.

\section{Tensor Contributions to the CMB in the Traditional Approach}
\label{TensorCMBSpectraAppendix}
The tensor contribution to the temperature auto-correlation was
\beas
  \T{C}(q)&=&\frac{1}{16}\sum_l\sum_p(-i)^li^p(2l+1)(2p+1)\iint
  (1-\mu^2)(1-\overline{\mu}^2)P_l(\mu)P_p(\overline{\mu}) \\ && \qquad \qquad \quad \times
  \left<\alpha_l^+(\kv)\alpha_p^{*+}(\kv')\right>\cos2\left(\overline{\phi}-\phi\right)
  e^{i(\kv'-\kv).\xv}\frac{d^3\kv}{(2\pi)^3}\frac{d^3\kv'}{(2\pi)^3} .
\eeas
We proceed in an analogous manner to our approach to the vectors. First we define
\bdm
  \left<\alpha_l(\kv)\alpha_p^*(\kv')\right>=\mathcal{P}_T(k)\alpha_l(k)\alpha^*_p(k')\delta(\kv'-\kv).
\edm
It is then necessary to remove the dependence on $\phi$; we do this by evaluating the square of $\nhv'.\nhv$, which leads to
\bdm
  (1-\mu^2)(1-\overline{\mu}^2)\cos2\left(\overline{\phi}-\phi\right)
  =2(\nhv'.\nhv)^2+\mu^2\overline{\mu}^2+\overline{\mu}^2+\mu^2-4\mu\overline{\mu}\nhv'.\nhv-1
\edm
and separates the correlation function into six distinct parts which we shall tackle in turn.

Firstly consider the contribution proportional to $(\nhv'.\nhv)^2$; the standard process employed for the scalars and vectors above rapidly leads to
\bdm
  C^{(1)}(q)=2\frac{4\pi}{16(2\pi)^6}\sum_{l,m}(2l+1)q^2\int\mathcal{P}_T(k)
  \left|\alpha_l\right|^2P_l(q)k^2dk .
\edm
We now employ the second recursion relation for $P_l(q)$ (which may readily be derived from the usual recursion relation),
\bdm
  (2l+1)q^2P_l(q)=\frac{(l+1)(l+2)}{2l+3}P_{l+2}(q)+\left(\frac{l^2}{2l-1}
  +\frac{(l+1)^2}{2l+3}\right)P_l(q)+\frac{(l-1)l}{2l-1}P_{l-2}(q)
\edm
and reduce this first contribution to
\beas
  \lefteqn{C^{(1)}_l=2A\int\mathcal{P}_T(k)\left\{\frac{(l-1)l}{(2l-1)(2l+1)}\left|\alpha_{l-2}\right|^2 \right.} \\ && \left.
   +\left(\frac{l^2}{(2l-1)(2l+1)}+\frac{(l+1)^2}{(2l+1)(2l+3)}\right)\left|\alpha_l\right|^2
   +\frac{(l+1)(l+2)}{(2l+1)(2l+3)}\left|\alpha_{l+2}\right|^2\right\}k^2dk .
\eeas

Dealing with the second contribution, proportional to $\nhv'.\nhv\mu\overline{\mu}$, is a lengthy procedure; first employing the recurrence relation for the Legendres, expanding them into spherical harmonics and integrating over both $\kv'$ and $\Omega_\kv$ gives us
\beas
  \lefteqn{C^{(2)}(q)=-4q\frac{4\pi}{16(2\pi^6)}\sum_l\int\mathcal{P}_T(k)\left\{
  \left(\frac{(l+1)^2}{2l+3}P_{l+1}(q)+\frac{l^2}{2l-1}P_{l-1}(q)\right)\left|\alpha_l\right|^2 \right.}
  \\&& \qquad \qquad \left.
  -\frac{(l-1)l}{2l-1}P_{l-1}(q)\alpha_l\alpha^*_{l-2}
  -\frac{(l+1)(l+2)}{2l+3}P_{l+1}(q)\alpha_l\alpha^*_{l+2}\right\}k^2dk
\eeas
We now again employ the recursion relation for $P_l(q)$ and, after a certain amount of algebra, can reduce this to
\beas
  \lefteqn{C^{(2)}_l(q)=-\frac{4A}{2l+1}\int\mathcal{P}_T(k)\left\{\frac{(l-1)^2l}{(2l-1)^2}\left|\alpha_{l-2}\right|^2
  -\frac{(l-1)l^2}{(2l-1)^2}\left(\alpha_{l-2}\alpha^*_l+\alpha_l\alpha^*_{l-2}\right) \right.} \\ &&
  +\left(\frac{l^3}{(2l-1)^2}+\frac{(l+1)^3}{(2l+3)^2}\right)\left|\alpha_l\right|^2
  -\frac{(l+1)^2(l+2)}{(2l+3)^2}\left(\alpha_{l+2}\alpha^*_l+\alpha^*_{l+2}\alpha_l\right) \\ && \quad \left.
  +\frac{(l+1)(l+2)^2}{(2l+3)^2}\left|\alpha_{l+2}\right|^2\right\}k^2dk
\eeas

The third contribution is that proportional to $\mu^2\overline{\mu}^2$; employing immediately the second recursion relation for the Legendres, expanding over the spherical harmonics and integrating over $\kv'$ and directions of $\kv$ leads, after much algebra, to
\beas
  \lefteqn{ C^{(3)}(q)=\frac{4\pi}{16(2\pi^6)}\sum_l\int\mathcal{P}_T(k)\left\{
  \left[
   \frac{(l-1)l(l-3)(l-2)}{(2l-1)(2l-3)(2l-5)}\alpha^*_{l-4} \right. \right.} \\&& \left.
   -\frac{(l-1)l}{2l-1}\left(\frac{(l-2)^2}{(2l-5)(2l-3)}+\frac{(l-1)^2}{(2l-3)(2l-1)}\right)\alpha^*_{l-2}
   +\frac{(l-1)^2l^2}{(2l-3)(2l-1^2)}\alpha_l
  \right]P_{l-2}(q) \\&&\
  -\left[
   \left(\frac{l^2}{(2l-1)(2l+1)}+\frac{(l+1)^2}{(2l+1)(2l+3)}\right)\frac{(l-1)l}{2l-1}\alpha^*_{l-2} \right.\\&& \ \
   -(2l+1)\left(\frac{l^2}{(2l-1)(2l+1)}+\frac{(l+1)^2}{(2l+1)(2l+3)}\right)^2\alpha_l\\&&\ \ \ \left.
   +\left(\frac{l^2}{(2l-1)(2l+1)}+\frac{(l+1)^2}{(2l+1)(2l+3)}\right)\frac{(l+1)(l+2)}{2l+3}\alpha^*_{l+2}
  \right]P_l(q)\\&&\ \ \ \
  +\left[
   \frac{(l+1)^2(l+2)^2}{(2l+3)^2(2l+5)}\alpha_l
   -\frac{(l+1)(l+2)}{(2l+3)(2l+5)}\left(\frac{(l+2)^2}{2l+3}+\frac{(l+3)^2}{2l+7}\right)\alpha^*_{l+2} \right.
    \\&&\ \ \ \ \ \left.\left.
   +\frac{(l+1)(l+2)}{(2l+3)(2l+5)}\frac{(l+3)(l+4)}{2l+7}\alpha^*_{l+4}
  \right]P_{l+2}(q)
  \right\}\alpha_lk^2dk
\eeas
whence
\beas
  \lefteqn{C_l^{(3)}=A\int\mathcal{P}_T(k)\left|\frac{(l-1)l}{(2l-1)(2l+1)}\alpha_{l-2} \right.} \\&&\left.
   -\left(\frac{l^2}{(2l-1)(2l+1)}+\frac{(l+1)^2}{(2l+1)(2l+3)}\right)\alpha_l
   +\frac{(l+1)(l+2)}{(2l+1)(2l+3)}\alpha_{l+2}\right|^2k^2dk .
\eeas

Hereon the contributions get simpler; for the fourth contribution, proportional to $\overline{\mu}^2$, we first employ the second recursion relation for $P_l(\overline{\mu})$ and then follow the usual procedure, expanding across the spherical harmonics and integrating twice; the result is
\beas
  \lefteqn{C_l^{(4)}=-A\int\mathcal{P}_T(k)\alpha_l\left\{\frac{(l-1)l}{(2l-1)(2l+1)}\alpha^*_{l-2} \right.}\\&&\left.
  -\left(\frac{l^2}{(2l-1)(2l+1)}+\frac{(l+1)^2}{(2l+1)(2l+3)}\right)\alpha^*_l
  +\frac{(l+1)(l+2)}{(2l+1)(2l+3)}\alpha^*_{l+2}\right\}k^2dk.
\eeas

Obviously enough, the fifth contribution, proportional to $\mu^2$, is the complex conjugate of this,
\beas
  \lefteqn{C_l^{(5)}=-A\int\mathcal{P}_T(k)\alpha^*_l\left\{\frac{(l-1)l}{(2l-1)(2l+1)}\alpha_{l-2} \right.}\\&&\left.
  -\left(\frac{l^2}{(2l-1)(2l+1)}+\frac{(l+1)^2}{(2l+1)(2l+3)}\right)\alpha_l
  +\frac{(l+1)(l+2)}{(2l+1)(2l+3)}\alpha_{l+2}\right\}k^2dk.
\eeas

The sixth and last contribution is identical in form to the negative of the scalar $C_l$, that is
\bdm
  C^{(6)}_l=-A\int\mathcal{P}_T(k)\left|\alpha_l\right|^2k^2dk.
\edm

Putting these six contributions together and performing much tedious algebra, we finally arrive at the contribution to the power spectrum of temperature anisotropies,
\be
  \T{C}_l=A(l-1)l(l+1)(l+2)\int\mathcal{P}_T(k)\left|\tilde{\alpha}_{l-2}
  +\tilde{\alpha}_l+\tilde{\alpha}_{l+2}\right|^2k^2dk
\ee
with
\be
  \tilde{\alpha}_{l-2}=\frac{\alpha_{l-2}}{(2l-1)(2l+1)}, \
  \tilde{\alpha}_l=\frac{2\alpha_l}{(2l-1)(2l+3)}, \
  \tilde{\alpha}_{l+2}=\frac{\alpha_{l+2}}{(2l+1)(2l+3)}
\ee
as we previously asserted.

\chapter{E and B Modes}
\label{Appendix-EandB}
\section{Spin-Weighted Polarisation -- E and B Modes}
An all-sky analysis of the statistical properties is considerably confused by the rotational variance of the Stokes parameters characterising polarisation; this caused earlier authors \cite{CrittendenDavisSteinhardt93,FrewinPolnarevColes93,CrittendenCoulsonTurok94,Seljak97,Kosowsky96,Polnarev85} to operate in a small-angle approximation in which the integration across wavevectors does not cause ambiguities to arise in the definitions of $\Delta_P$ and $U$. However, in the mid-late 1990s a new formalism arose that circumvented these problems, due to two independant teams, of Kamionkowski, Kosowsky and Stebbins \cite{KamionkowskiKosowskyStebbins97} and Zaldarriaga and Seljak \cite{ZaldarriagaSeljak97}. Our approach is based heavily on that of Zaldarriaga and Seljak \cite{ZaldarriagaSeljak97}.

Consider a rotation of the angular basis vectors with respect to which the Stokes parameters are defined -- that is, a rotation about the radial vector. Then the set of parameters $\left\{{}^RI, {}^LI, U\right\}$ varies on rotation about $\hat{\mathbf{r}}$ by an angle $\psi$ according to (see Chandrasekhar, \cite{Chandrasekhar60})
\bdm
  \left(\begin{array}{c}{}^RI'\\{}^LI'\\U'\end{array}\right)=
  \left(\begin{array}{ccc}\cos^2\psi & \sin^2\psi & -\frac{1}{2}\sin2\psi \\
   \sin^2\psi & \cos^2\psi & \frac{1}{2}\sin2\psi \\
   \sin2\psi & -\sin2\psi & \cos2\psi \end{array}\right)
  \left(\begin{array}{c}{}^RI\\{}^LI\\U\end{array}\right) .
\edm
Converting this to our standard set $\left\{\Delta_T,\Delta_P,U\right\}$ we see that
\bdm
  \Delta_T'=\Delta_T, \quad \Delta_P'=\Delta_P\cos2\psi+U\sin2\psi, \quad
  U'=-\Delta_P\sin2\psi+U\cos2\psi .
\edm
The temperature anisotropies are invariant under rotation, rendering the analysis on the CMB considerably simplified, but the polarisation parameters most definitely are not.\footnote{This is strictly not true; had we retained the circular polarisation $V$ -- and for a genuinely complete treatment we would have to, incorporating both Faraday rotation from $Q$ to $U$ and also Faraday ``conversion'' from $U$ to $V$; see for example Cooray \emph{et. al.} \cite{CoorayEtAl02} and Matsuyima and Ioka \cite{MatsumiyaIoka03} -- then we would have seen that $V$ is also invariant under co-ordinate rotations and is analysed as simply as the temperature perturbations are. Note however that the latter demonstrate that to produce circular polarisation we would require an ordered component to our field and while many studies of ordered cosmological fields exist, we are going to follow the currently conventional route of considering a field with a tangled component only.} Consider, however, some linear combination of $\Delta_P$ and $U$, and convert the trigonometric expressions to exponentials:
\beas
  \Delta_P'+\alpha U'&=&\Delta_P\left(\frac{e^{2i\psi}+e^{-2i\psi}}{2}
   -\frac{\alpha}{i}\left(\frac{e^{2i\psi}-e^{-2i\psi}}{2}\right)\right) \\ && \quad
   +U\left(\frac{e^{2i\psi}-e^{-2i\psi}}{2i}+\alpha\left(
   \frac{e^{2i\psi}+e^{-2i\psi}}{2}\right)\right) .
\eeas
Then we see that we can reduce this to a simple form by choosing $\alpha=i$, which gives us
\bdm
  \Delta_P'+iU'=e^{-2i\psi}\left(\Delta_P+iU\right);
\edm
similarly, we can show that
\bdm
  \Delta_P'-iU'=e^{2i\psi}\left(\Delta_P-iU\right) .
\edm

While this has not given us two objects invariant under rotation, it has given us objects with useful properties, if we employ the spin-weighted spherical harmonics introduced by Newman and Penrose \cite{NewmanPenrose66}. We shall leave a detailed consideration of spin-weighted functions and harmonics to a rainy-day appendix, and content ourselves with referring the reader to the article by Newman and Penrose, and the follow-up article by Goldberg \emph{et. al.} \cite{GoldbergEtAl67}, and the related article by Thorne \cite{Thorne80}. Useful information in this precise context may also be found in the appendices of Zaldarriaga and Seljak \cite{ZaldarriagaSeljak97} and Koh and Lee \cite{KohLee00}.

A function $f$ is called ``spin-$s$'' if, under a right-handed rotation $\psi$ about the radial vector it transforms as
\bdm
  f'=e^{-is\psi}f .
\edm
Note that, following Zaldarriaga and Seljak, we take the opposite spin-definition to that of Newman and Penrose or Goldberg \emph{et. al.}. A function $f$ with a spin-weighting $s$ is conveniently written $f_{\mathrm{s}}$.

We may define a differential operator $\eth$ on the surface of the sphere by
\be
  \eth f_{\mathrm{s}}=-\left(\sin\theta\right)^s\left(\frac{\partial}{\partial\theta}
  +i\csc\theta\frac{\partial}{\partial\phi}\right)\left(\sin\theta\right)^{-s}
  f_{\mathrm{s}}
\ee
which is related to the covariant derivative on the surface of the sphere. (Obviously no summation across the spin index in any of these formulae.) This has the related operator
\be
  \overline{\eth}f_{\mathrm{s}}=-\left(\sin\theta\right)^{-s}\left(\frac{\partial}{\partial\theta}
  -i\csc\theta\frac{\partial}{\partial\phi}\right)\left(\sin\theta\right)^s
  f_{\mathrm{s}} .
\ee
These operators (dubbed ``thop'' by Newman and Penrose) obey a commutator
\bdm
  \left[\overline{\eth},\eth\right]f_{\mathrm{s}}=2sf_{\mathrm{s}}
\edm
and, vitally, have the property under rotation through $\psi$ that
\bdm
  \left(\eth f_{\mathrm{s}}\right)'=e^{-i(s+1)\psi}\eth f_{\mathrm{s}}, \quad
  \left(\overline{\eth}{}_sf\right)'=e^{-i(s-1)\psi}\overline{\eth}f_{\mathrm{s}};
\edm
that is, they act as spin raising and lowering operators of some sort. This being so, one is immediately tempted to act these upon the familiar spherical harmonics, and obtain functions that we might dub ``spin-weighted spherical harmonics'' $Y_{\mathrm{s},lm}$ if they possess the correct properties:
\be
\label{SpinWeightedHarmonics}
  Y_{\mathrm{s},lm}=\left\{\begin{array}{ll}
   \sqrt{\frac{(l-s)!}{(l+s)!}}\eth^sY_{lm}, & s\in[0,l] \\
   \sqrt{\frac{(l+s)!}{(l-s)!}}(-1)^s\overline{\eth}^{-s}Y_{lm}, & s\in[-l,0]
\end{array} \right. \ee
and leaving them undefined for $|s|>l$. Note that if one substitutes $|s|$ for $s$ in the second definition, the form becomes identical to the first, other than a factor of $(-1)^{|s|}$.

It can then be shown that, as we hoped, these functions form, for each spin-weighting, a complete and orthonormal set, and so we can expand
\bdm
  f_{\mathrm{s}}=\sum_{l,m}a_{\mathrm{s},lm}Y_{\mathrm{s},lm} .
\edm

The orthonormality and completeness relations are, as in the standard spherical harmonics,
\bea
  \int_0^{2\pi}\int_{-1}^1Y_{\mathrm{s},lm}^*(\theta,\phi)
   Y_{\mathrm{s},l'm'}(\theta',\phi')d\mu d\phi &=& \gamma_{ll'}
   \gamma_{mm'}, \nonumber \\
  \sum_{l,m}Y_{\mathrm{s},lm}^*(\theta,\phi)Y_{\mathrm{s},lm}(\theta',\phi')
   &=&\delta(\phi'-\phi)\delta(\cos\theta'-\cos\theta) .
\eea

The spin-weighted harmonics also obey the useful relations
\bea
\label{SWHarmonicsRelations}
  &Y_{\mathrm{s},lm}^*=(-1)^{m+s}Y_{\mathrm{-s},lm},& \\
  &\eth Y_{\mathrm{s},lm}=\sqrt{(l-s)(l+s+1)}Y_{\mathrm{s+1},lm}, \quad
  \overline{\eth}Y_{\mathrm{s},lm}=-\sqrt{(l+s)(l-s+1)}Y_{\mathrm{s-1},lm},& \nonumber \\
  &\overline{\eth}\eth Y_{\mathrm{s},lm}=-(l-s)(l+s+1)Y_{\mathrm{s},lm} \Rightarrow
  \eth\overline{\eth}Y_{\mathrm{s},lm}=-(l+s)(l-s+1)Y_{\mathrm{s},lm} ,& \nonumber
\eea
by which one may see that the $Y_{\mathrm{s},lm}$ are the eigenvectors of $\overline{\eth}\eth$, and that $\eth$ and $\overline{\eth}$ are raising and lowering operators respectively.

It may also be shown that
\beas
  Y_{\mathrm{s},lm}&=&e^{im\phi}\sqrt{\frac{(l+m)!}{(l+s)!}\frac{(l-m)!}{(l-s)!}
   \frac{2l+1}{4\pi}}\left(\sin\frac{1}{2}\theta\right)^{2l} \\ && \quad \times
   \sum_p\left(\begin{array}{c}l-s\\p\end{array}\right)
   \left(\begin{array}{c}l+s\\p+s-m\end{array}\right)(-1)^{l-p-s+m}
   \left(\cot\frac{1}{2}\theta\right)^{2p+s-m}
\eeas
where
\bdm
  \left(\begin{array}{c}a\\p\end{array}\right)=\frac{a!}{(a-b)!b!}
\edm

To return to the matter of polarisation, from the transformation under rotation
\be
  \left(\Delta_P\pm iU\right)'=e^{\mp 2i\psi}
  \left(\Delta_P\pm iU\right) ,
\ee
we see that we are dealing with two objects of spin-weighting $\pm 2$. We may expand these across the relevant harmonics as
\bdm
  \left(\Delta_P+iU\right)\left(\nhv\right)=\sum_{l,m}
   a_{\mathrm{2},lm}Y_{\mathrm{2},lm}\left(\nhv\right), \quad
  \left(\Delta_P-iU\right)\left(\nhv\right)=\sum_{l,m}
   a_{\mathrm{-2},lm}Y_{\mathrm{-2},lm} .
\edm

To get rotationally-invariant measures of polarisation, then, we might act on these with $\eth^2$ and $\overline{\eth}^2$ as required. Doing so, and employing the definitions of the spin-weighted harmonics (\ref{SpinWeightedHarmonics}) and the effects of $\eth$ and $\overline{\eth}$ (\ref{SWHarmonicsRelations}), leaves our two quantities,
\bea
  \overline{\eth}^2\left(\Delta_P+iU\right)&=&\sum_{l,m}a_{\mathrm{2},lm}
   \sqrt{(l+2)(l+1)l(l-1)}Y_{lm} , \\
  \eth^2\left(\Delta_P-iU\right)&=&\sum_{l,m}a_{\mathrm{-2},lm}
   \sqrt{(l+2)(l+1)l(l-1)}Y_{lm} .
\eea
We can then find the coefficients from either integrating $\Delta\pm iU$ over the spin-$2$ spherical harmonics, or by integrating the transformed analogues over the standard, spin-$0$ harmonics:
\beas
  a_{\mathrm{2},lm}=\int\left(\Delta_P+iU\right)Y_{\mathrm{2},lm}^*\left(\nhv\right)d\Omega_{\nhv}
   &=&\sqrt{\frac{(l-2)!}{(l+2)!}}\int\overline{\eth}^2\left(\Delta_P+iU\right)Y_{lm}^*(\nhv)
   d\Omega_{\nhv}, \\
  a_{\mathrm{-2},lm}=\int\left(\Delta_P-iU\right)Y_{\mathrm{-2},lm}^*\left(\nhv\right)d\Omega_{\nhv}
   &=&\sqrt{\frac{(l-2)!}{(l+2)!}}\int\eth^2\left(\Delta_P-iU\right)Y_{lm}^*(\nhv)
   d\Omega_{\nhv} .
\eeas
From considerations of parity (see again Newman and Penrose \cite{NewmanPenrose66}), we can construct two objects with opposite parities,
\bea
  \tilde{E}(\nhv)&=&-\frac{1}{2}\left(\overline{\eth}^2\left(\Delta_P+iU\right)
   +\eth^2\left(\Delta_P-iU\right)\right) , \\
  \tilde{B}(\nhv)&=&-\frac{1}{2i}\left(\overline{\eth}^2\left(\Delta_P+iU\right)
   -\eth^2\left(\Delta_P-iU\right)\right) .
\eea
To simplify the expressions, Zaldarriaga and Seljak define
\bdm
  a_{E,lm}=-\frac{1}{2}\left(a_{\mathrm{2},lm}+a_{\mathrm{-2},lm}\right), \quad
  a_{B,lm}=-\frac{1}{2i}\left(a_{\mathrm{2},lm}-a_{\mathrm{-2},lm}\right)
\edm
which gives us the variables
\beas
  \tilde{E}=\sum_{l,m}\sqrt{\frac{(l+2)!}{(l-2)!}}a_{E,lm}Y_{lm}(\nhv),& &
  \tilde{B}=\sum_{l,m}\sqrt{\frac{(l+2)!}{(l-2)!}}a_{B,lm}Y_{lm}(\nhv); \\
  E=\sum_{l,m}a_{E,lm}Y_{lm}(\nhv),& &
  B=\sum_{l,m}a_{B,lm}Y_{lm}(\nhv) .
\eeas

The spin-$2$ harmonics are given explicitly by (see Hu and White, \cite{HuWhite97})
\be
\begin{array}{rrrrcl}
  m &=& 2, & \quad Y_{\mathrm{2},2m} &=& \frac{1}{8}\sqrt{\frac{5}{\pi}}\left(1-\mu\right)^2e^{2i\phi} \\
    && 1, &&& \frac{1}{4}\sqrt{\frac{5}{\pi}}\sqrt{1-\mu^2}\left(1-\mu\right)e^{i\phi} \\
    && 0, &&& \frac{3}{4}\sqrt{\frac{5}{6\pi}}\left(1-\mu^2\right) \\
    && -1, &&& \frac{1}{4}\sqrt{\frac{5}{\pi}}\sqrt{1-\mu^2}\left(1+\mu\right)e^{-i\phi} \\
    && -2, &&& \frac{1}{8}\sqrt{\frac{5}{\pi}}\left(1+\mu\right)^2e^{-2i\phi}
\end{array}
\ee
with the spin $-2$ harmonics related by relations (\ref{SWHarmonicsRelations}). We can also write the twice-applied spin-lowering operator acting on a spin $2$ function $f_{\mathrm{2}}$ with angular dependence $\partial_\phi f_{\mathrm{2}}=imf_{\mathrm{2}}$ as
\bea
  \overline{\eth}^2f_{\mathrm{2}}&=&-\frac{1}{\sin\theta}\left(\frac{\partial}{\partial\theta}-\frac{i}{\sin\theta}\frac{\partial}{\partial\phi}\right)\cdot\left(-\frac{1}{\sin\theta}\right)\left(\frac{\partial}{\partial\theta}-\frac{i}{\sin\theta}\frac{\partial}{\partial\phi}\right)\left(\sin^2\theta f_{\mathrm{2}}\right)
  \nonumber \\
  &=&\left(-\frac{\partial}{\partial\mu}+\frac{m}{1-\mu^2}\right)^2\left(\left(1-\mu^2\right)f_{\mathrm{2}}\right)
\eea
with an analogous expression for a double-raise on a spin $-2$ function
\be
  \overline{\eth}^2f_{-\mathrm{2}}=
  \left(-\frac{\partial}{\partial\mu}-\frac{m}{1-\mu^2}\right)^2\left(\left(1-\mu^2\right)f_{-\mathrm{2}}\right) .
\ee

While in principle we could expand the Boltzmann hierarchies across these spin-2 harmonics and construct the evolution equations for the polarisation parameters, it is easier to follow the method of Zaldarriaga and Seljak and instead consider a line-of-sight approach. Working in the na\"{\i}ve manner would yield a coupled system of thousands of differential equations; the line-of-sight approach renders this problem rather more tractable than attempting to manipulate the $a_{\pm\mathrm{2},lm}$s and would be quicker by far to implement in a Boltzmann code. We will still employ the hierarchies in their earlier Stokes parameter formalism to evaluate the sources of the perturbations, but shall reconstruct the microwave background sky -- including the $E$ and $B$ modes -- from the line-of-sight approach, to which we now turn.

\section{The Line-of-Sight Approach}
As with the temperature case considered in the body of the thesis we shall separate the polarisation Boltzmann equations into sections dependent purely on the sources and sections dependent purely on the geometry. As before, $\tau=\int_{\eta}^{\eta_0}\dot{\tau}(\overline{\eta})d\overline{\eta}$ is the optical depth at a time $\eta$, $g(\eta)=\dot{\tau}\exp(-\tau)$ is the visibility function and we sometimes use the notation $x=k(\eta-\eta_0)$.

\subsection{Scalar Evolution}
Consider first the scalar perturbations. Here
\bdm
  \dot{\Delta}_P-\left(ik\mu-\dot{\tau}\right)\Delta_P=\frac{3}{2}\dot{\tau}\left(1-\mu^2\right)\Phi_S
\edm
where
\bdm
  \Phi_S=\Delta_{T2}+\Delta_{P0}+\Delta_{P2}.
\edm
Formally integrating this between $\eta=0$ and $\eta=\eta_0$ gives
\be
  \Delta_P(\eta_0)=\frac{3}{4}\int_0^{\eta_0}e^{-ix\mu}g(\eta)\left(1-\mu^2\right)\Phi_Sd\eta.
\ee
We can thus say
\be
  \Delta_P(\eta_0)=\int_0^{\eta_0}d\eta e^{-ix\mu}S^{(S)}_P(k,\eta)
\ee
with
\be
  S^{(S)}_P(k,\eta)=g\frac{3}{4}\left(\left(\Phi_S+\frac{\ddot{\Phi}_S}{k^2}\right)
   +2\dot{g}\frac{\dot{\Phi}_S}{k^2}+\ddot{g}\frac{\Phi_S}{k^2}\right)=\frac{3}{4}\left(1-\mu^2\right)g(\eta)\Sc{\Phi} .
\ee

Now, here $U=0$ and so $\Delta_P=\Delta_P\pm iU$ implying that $a_{\mathrm{2},lm}=a_{-2,lm}$ which immediately shows that scalar perturbations do not source $B$-mode polarisation. We recover the $E$ and $B$ modes simply by applying $\eth^2$ and $\overline{\eth}^2$ to the line-of-sight results; in this case this means we apply either of the operators to $\Delta_P$. We can then see that
\be
  \tilde{E}^{(S)}=\frac{\partial^2}{\partial\mu^2}\Delta_P, \; \tilde{B}^{(S)}=0 .
\ee

\subsection{Vector Evolution}
Here the physical Stokes parameters for polarisation were
\bdm
  \Delta_P=\mu\sqrt{1-\mu^2}\left(\Delta_P^1\cos\phi+\Delta_P^2\sin\phi\right), \quad
  U=\sqrt{1-\mu^2}\left(\Delta_P^1\sin\phi-\Delta_P^2\cos\phi\right)
\edm
and these obeyed the evolution equations
\be
  \dot{\Delta}_P^\epsilon-\left(ik\mu-\dot{\tau}\right)\Delta_P^\epsilon=\dot{\tau}\Phi_\epsilon^{(V)}
\ee
with
\bdm
  \V{\Phi}_\epsilon=\frac{3}{5}\Delta_{P0}^\epsilon+\frac{3}{7}\Delta_{P2}^\epsilon
  -\frac{6}{35}\Delta_{P4}^\epsilon-\frac{3}{10}\Delta_{T1}^\epsilon-\frac{3}{10}\Delta_{T3}^\epsilon .
\edm
We then rotate our basis to
\be
  2\tilde{\Delta}_{P,U}=\Delta_{P,U}^1-i\Delta_{P,U}^2, 2\breve{\Delta}_{P,U}=\Delta_{P,U}^1+i\Delta_{P,U}^2
\ee
as with the temperature case. $\tilde{\zeta}$ and $\breve{\zeta}$ with the properties
\be
  \langle\tilde{\zeta}(\mathbf{k})\tilde{\zeta}(\mathbf{k}')\rangle=\langle\breve{\zeta}(\mathbf{k})\breve{\zeta}(\mathbf{k}')\rangle=\frac{1}{2}\mathcal{P}_V(k)\delta\left(\mathbf{k}-\mathbf{k}'\right), \quad \langle\tilde{\zeta}(\mathbf{k})\breve{\zeta}(\mathbf{k}')\rangle=0
\ee
again characterise the vector perturbations. Assuming that our modes are thus uncorrelated but similar, we can employ the source term generated from, for example, the $1$ mode, and write our line-of-sight integrals as
\bea
  \Delta_P+iU&=&\sqrt{1-\mu^2}\left(\left(\mu+1\right)\tilde{\zeta}(\mathbf{k})e^{i\phi}+\left(\mu-1\right)\breve{\zeta}(\mathbf{k})e^{-i\phi}\right)\int_0^{\eta_0}e^{-ix\mu}\V{S}_P(k,\eta)d\eta, \nonumber \\
  \Delta_P-iU&=&\sqrt{1-\mu^2}\left(\left(\mu-1\right)\tilde{\zeta}(\mathbf{k})e^{i\phi}+\left(\mu+1\right)\breve{\zeta}(\mathbf{k})e^{-i\phi}\right)\int_0^{\eta_0}e^{-ix\mu}\V{S}_P(k,\eta)d\eta, \nonumber
\eea
with
\be
  \V{S}_P(k\eta)=g(\eta)\V{\Phi}_1 .
\ee
%

\subsection{Tensor Evolution}
With the tensors,
\bdm
  \Delta_P=\left(1+\mu^2\right)\left(\beta^+\cos2\phi+\beta^\times\sin2\phi\right), \quad
  U=2\mu\left(\beta^+\sin2\phi-\beta^\times\cos2\phi\right)
\edm
with the evolution equations
\beas
  \dot{\beta}^*-\left(ik\mu-\dot{\tau}\right)\beta^*&=&-\dot{\tau}\T{\Phi}_*
\eeas
and
\bdm
  \T{\Phi}_*=\frac{1}{10}\alpha_0^*-\frac{3}{5}\beta_0^*+\frac{1}{7}\left(\alpha_2^*+6\beta_2^*\right)+\frac{3}{70}\left(\alpha_4^*-\beta_4^*\right) .
\edm
From here we rapidly find the source terms
\be
  S_\alpha^*(k,\eta)=-2\T{\dot{h}}_*e^{-2\tau}+g\T{\Phi}_*, \quad S_{\beta}^*=-g\T{\Phi}_* .
\ee

The rotation we employed for the temperature puts our Stokes parameters into the forms
\bea
  \Delta_P+iU&=&\left(1+\mu\right)^2\beta^1e^{2i\phi}+\left(1-\mu\right)^2\beta^2e^{-2i\phi}, \\
  \Delta_P-iU&=&\left(1-\mu\right)^2\beta^1e^{2i\phi}+\left(1+\mu\right)^2\beta^2e^{-2i\phi} \nonumber .
\eea
Characterising the statistics of the gravity waves with variables $\xi^1$ and $\xi^2$, uncorrelated but similar, we have the solutions
\bea
  \Delta_P+iU&=&\left(\left(1+\mu^2\right)\xi^1(\mathbf{k})e^{2i\phi}+\left(1-\mu^2\right)\xi^2(\mathbf{k})e^{-2i\phi}\right)\int_0^{\eta_0}e^{-ix\mu}S^{(T)}_P(k,\eta)d\eta, \\
  \Delta_P-iU&=&\left(\left(1-\mu^2\right)\xi^1(\mathbf{k})e^{2i\phi}+\left(1+\mu^2\right)\xi^2(\mathbf{k})e^{-2i\phi}\right)\int_0^{\eta_0}e^{-ix\mu}S^{(T)}_P(k,\eta)d\eta, \nonumber
\eea
with
\be
  \langle\xi^1(\mathbf{k})\xi^{1*}(\mathbf{k}')\rangle=\langle\xi^2(\mathbf{k})\xi^{2*}(\mathbf{k}')\rangle=\frac{1}{2}\mathcal{P}_T(k)\delta\left(\mathbf{k}-\mathbf{k}'\right), \quad \langle\xi^1(\mathbf{k})\xi^{2*}(\mathbf{k})\rangle=0
\ee
and
\be
  \T{S}_P=-g\T{\Phi}_+ .
\ee
%

\subsection{Scalar CMB Angular Power Spectra}
Here we found that
\bea
  \Delta_T\left(\mathbf{k},\eta_0\right)&=&\int_\eta d\eta\Sc{S}_T\left(k,\eta\right)e^{-ix\mu}, \\
  \Delta_P\left(\mathbf{k},\eta_0\right)&=&\left(\Delta_P\pm iU\right)\left(\mathbf{k},\eta_0\right)=\frac{3}{4}\int_\eta d\eta\left(1-\mu^2\right)g(\eta)\Phi_S(k,\eta)e^{-ix\mu} .
\eea

\subsubsection{Polarisation Auto-Correlation}
Since there is no $U$ polarisation for the scalar modes we can write
\be
  \tilde{E}(\mathbf{k},\mathbf{n})=\frac{\partial^2}{\partial\mu^2}\Delta_P(\mathbf{k},\mathbf{n})
   =-\frac{3}{4}\frac{\partial^2}{\partial\mu^2}\int_\eta g(\eta)\Phi_S(k,\eta)\left(1-\mu^2\right)^2e^{-i\mu x}d\eta
\ee
Rewriting the $\mu$ in the integrand as a derivative of the exponential with respect to $x$ we then have
\be
  \tilde{E}(\mathbf{k},\mathbf{n})=-\frac{3}{4}\int_\eta g(\eta)\Phi_S(k,\eta)\left(1+\partial_x^{2}\right)^2
   \left(x^2e^{-i\mu x}\right)d\eta
\ee
Taking this into real space and calculating the $a_{\tilde{E},lm}$s we have
\be
  a_{\tilde{E},lm}=-\frac{3}{4}\int_{\mathbf{k}}\int_{\Omega_\mathbf{n}}Y^*_{lm}(\mathbf{n})\int_\eta
  g(\eta)\Phi_S(k,\eta)\left(1+\partial_x^{2}\right)^2\left(x^2e^{-i\mu x}\right)d\eta d\Omega_\mathbf{n}
  \frac{d^3\mathbf{k}}{(2\pi)^3} .
\ee
In a manner entirely analogous to the temperature case we can then find
\bdm
  \Delta_{E,l}=\frac{3}{4}\sqrt{\frac{(l-2)!}{(l+2)!}}\int_\eta g(\eta)\Phi_S(k,\eta)\left(1+\partial_x^{2}\right)^2
   \left(x^2j_l(x)\right)d\eta .
\edm
As in the tensor temperature correlation, we expand the derivative out and simplifying using the spherical Bessel equation (\ref{SphericalBesselEquation}) to find
\bdm
  \left(1+\partial_x^{2}\right)^2\left(x^2j_l(x)\right)=\frac{(l+2)!}{(l-2)!}\frac{j_l(x)}{x^2} ,
\edm
and with this the $E$-mode transfer function reduces to
\be
  \Delta_{E,l}=\sqrt{\frac{(l+2)!}{(l-2)!}}\int_\eta S^{(S)}_E(k,\eta)j_l(x)d\eta
\ee
with a source term
\be
  \Sc{S}_E(k,\eta)=\frac{3}{4x^2}g(\eta)\Phi_S(k,\eta) .
\ee

\subsubsection{Temperature-Polarisation Cross-Correlation}
From the form of $a_{T,lm}$ and $a_{\tilde{E},lm}$, and noting that for two scalar functions $f=f(\mathbf{k})$ and $g=g(\mathbf{k})$,
\be
  \langle f(\mathbf{k})g^*(\mathbf{k}')\rangle=\mathcal{P}_S(k)f(k)g^*(k)\delta(\mathbf{k}-\mathbf{k}')
\ee
we can write
\beas
  \lefteqn{\langle a_{T,lm}a^*_{\tilde{E},lm}\rangle=}
  \\ &&
  -\frac{3}{4}\int_\mathbf{k}\mathcal{P}_S(k)
  \int_{\Omega_{\mathbf{n}}}\int_{\Omega_{\mathbf{n}'}}
  Y^*_{lm}(\mathbf{n})Y_{lm}(\mathbf{n}')
  \int_{\eta}\int_{\eta'}S_T^S(k,\eta)g^*(\eta')\Phi_S^*(k,\eta')
  \\ && \times
  e^{-ix\mu}\left(1+\partial_{x'}^{\phantom{x'}2}\right)^2\left(x'^2e^{ix'\mu'}\right)
  d\eta'd\eta d\Omega_{\mathbf{n}'}d\Omega_{\mathbf{n}}\frac{d^3\mathbf{k}}{(2\pi)^3}
\eeas
with $x'=k(\eta'-\eta_0)$. Again integrating over the directions of $\mathbf{k}$, expanding the exponentials in $x\mu$ across the Legendre polynomials and spherical Bessel functions, and expanding the spherical harmonics into associated Legendre polynomials and exponentials, and using (\ref{x^2j_l(x)}), we get
\beas
  \lefteqn{\langle a_{T,lm}a^*_{\tilde{E},lm}\rangle=}
  \\ &&
  -\frac{3}{4}\frac{4\pi}{(2\pi)^3}\left(\frac{2l+1}{4\pi}\frac{(l-m)!}{(l+m)!}\right)\frac{(l+2)!}{(l-2)!}
  \int_k\mathcal{P}_S(k)\sum_p\sum_{p'}(-i)^pi^{p'}(2p+1)(2p'+1)
  \\ && \times
  \int_{\Omega_{\mathbf{n}}}\int_{\Omega_{\mathbf{n}'}}\int_\eta\int_{\eta'}
  P^m_l(\mu)P_p(\mu)P^m_l(\mu')P_{p'}(\mu')e^{-im\phi}e^{im\phi'}
  \\ && \times
  S^S_T(k,\eta)g^*(\eta')\Phi_S^*(k,\eta')j_p(x)\frac{j_{p'}(x')}{x'^2}
  d\eta'd\eta d\Omega_{\mathbf{n}'}d\Omega_{\mathbf{n}}k^2dk
\eeas
We can now integrate over $\phi$ and $\phi'$ -- which both yield a term $2\pi\delta^0_m$ -- and over $\mu$ and $\mu'$ which give terms $2/(2p+1)\delta^p_l$ and similar. Summing over $m$, then, gives
\bdm
  C_{T\tilde{E},l}=
  -\frac{3}{4}\frac{(l+2)!}{(l-2)!}\frac{2}{\pi}
  \int_k\mathcal{P}_S(k)
  \int_\eta S^S_T(k,\eta)j_p(x)d\eta
  \int_{\eta'}g^*(\eta')\Phi_S^*(k,\eta')\frac{j_{p'}(x')}{x'^2}d\eta'
  k^2dk
\edm
or, converting $\tilde{E}$ to $E$ and substituting in the form of the transfer functions,
\be
  C_{TE,l}=\frac{2}{\pi}\int_k\mathcal{P}_S(k)\Delta_{T,l}(k,\eta_0)\Delta^*_{E,l}(k,\eta_0)k^2dk
\ee
in agreement with that found in the standard approach.

\subsection{Vector and Tensor CMB Angular Power Spectra}
The vector and tensor angular power spectra can be evaluated in broadly equivalent ways; for details of the tensor case, the reader is referred to \cite{ZaldarriagaSeljak97}.

\chapter{Supplementary Results concerning Source Statistics}
\label{Appendix-SourceStats}
In this appendix we present results that were excluded from the main body of chapter \ref{Chapter-SourceStats} due to their complexity.

\section{Magnetic Bispectra}
\subsection{Traceless Scalar-Tensor-Tensor Correlation}
\label{Appendix-SourceStats-STT}
The correlation between the traceless scalar part and the tensors is recovered by applying $\mathcal{A}^{ijklmn}=-Q^{ij}(\mathbf{k})\mathcal{P}^{Tklab}(\mathbf{p})\mathcal{P}^{Tmn}_{ab}(\mathbf{q})$ to $\mathcal{B}_{ijklmn}$ (\ref{SixFieldCorrelation}); the full result for the angular integrand is

\beas
  \mathcal{F}_{\tau_S\tau^T\tau^T}&=&\sum_{n=0}^8\mathcal{F}_{\tau_S\tau^T\tau^T}^{n}, \\
  \mathcal{F}^0_{\tau_S\tau^T\tau^T}&=&5
   \\
  \mathcal{F}^1_{\tau_S\tau^T\tau^T}&=&0
   \\
  -\mathcal{F}^2_{\tau_S\tau^T\tau^T}&=&
    \overline{\beta}^2+\overline{\gamma}^2+\overline{\mu}^2
    +3\left(\theta_{kp}^2+\theta_{kq}^2+4\theta_{pq}^2\right)+3\left(\alpha_k^2+\beta_k^2\right)
    +\alpha_p^2+\beta_p^2
     \\ &&
    +\alpha_q^2+\beta_q^2+3\left(\gamma_k^2+3\gamma_p^2+3\gamma_q^2\right)
   \\
  \mathcal{F}^3_{\tau_S\tau^T\tau^T}&=&
    \overline{\beta}\overline{\gamma}\overline{\mu}
    +3\left(\overline{\beta}\alpha_k\beta_k+\overline{\gamma}\alpha_k\gamma_k+\overline{\mu}\beta_k\gamma_k\right)
    +\overline{\beta}\left(\alpha_p\beta_p+\alpha_q\beta_q\right)
     \\ &&
    +\overline{\gamma}\left(\alpha_p\gamma_p+\alpha_q\gamma_q\right)
    +\overline{\mu}\left(\beta_p\gamma_p+\beta_q\gamma_q\right)
    +3\theta_{kp}\left(\theta_{kq}\theta_{pq}+\alpha_k\alpha_p+\beta_k\beta_p+\gamma_k\gamma_p\right)
     \\ &&
    +3\theta_{kq}\left(\theta_{kp}\theta_{pq}+\alpha_k\alpha_q+\beta_k\beta_q+\gamma_k\gamma_q\right)
    +3\theta_{pq}\left(\theta_{kp}\theta_{kq}+\alpha_p\alpha_q+\beta_p\beta_q\right)
    +11\theta_{pq}\gamma_p\gamma_q
   \\
  \mathcal{F}^4_{\tau_S\tau^T\tau^T}&=&
    \left(3\overline{\beta}^2+\overline{\gamma}^2+\overline{\mu}^2\right)\theta_{pq}^2
    +2\overline{\beta}^2\left(\gamma_p^2+\gamma_q^2\right)
    -3\overline{\gamma}\overline{\mu}\alpha_k\beta_k+\overline{\beta}\overline{\gamma}\beta_q\gamma_q+\overline{\beta}\overline{\mu}\alpha_p\gamma_p
     \\ &&
    -3\overline{\mu}\theta_{kp}\beta_k\gamma_p+3\overline{\gamma}\theta_{kq}\alpha_k\gamma_q+\overline{\beta}\theta_{pq}\alpha_p\beta_q
     \\ &&
    -2\theta_{pq}\left(\overline{\gamma}\left(\alpha_p\gamma_q+\alpha_q\gamma_p\right)+\overline{\beta}\alpha_q\beta_p
      +\overline{\mu}\left(\beta_p\gamma_q+\beta_q\gamma_p\right)\right)
    +\theta_{pq}^4
     \\ &&
    +3\left(\theta_{kp}^2+\theta_{kq}^2\right)\theta_{pq}^2
    -3\left(\theta_{kp}\theta_{pq}\beta_k\beta_q-\theta_{kp}\theta_{kq}\gamma_p\gamma_q
      +\theta_{kq}\theta_{pq}\alpha_k\alpha_p\right)
     \\ &&
    +6\left(\theta_{kp}^2+\theta_{kq}^2\right)\left(\gamma_p^2+\gamma_q^2\right)
    -6\theta_{pq}\left(\theta_{kp}\left(\alpha_k\alpha_q+\gamma_k\gamma_q\right)
      +\theta_{kq}\left(\beta_k\beta_p+\gamma_k\gamma_p\right)\right)
     \\ &&
    +\theta_{pq}^2\left(9\alpha_k^2+9\theta_{pq}^2\beta_k^2+3\theta_{pq}^2\gamma_k^2
      +\alpha_p^2+\beta_p^2+\gamma_p^2+\alpha_q^2+\beta_q^2+\gamma_q^2\right)
     \\ &&
    +2\left(3\alpha_k^2+\alpha_p^2+\alpha_q^2+3\beta_k^2+\beta_p^2+\beta_q^2\right)\left(\gamma_p^2+\gamma_q^2\right)
    -3\alpha_k\left(\alpha_p\beta_p+\alpha_q\beta_q\right)\beta_k
     \\ &&
    -3\left(\alpha_k\alpha_p\gamma_p+\beta_k\beta_q\gamma_q\right)\gamma_k
    +\left(\alpha_p\alpha_q+\beta_p\beta_q\right)\gamma_p\gamma_q
\eeas
\newpage
\beas
  -\mathcal{F}^5_{\tau_S\tau^T\tau^T}&=&
    \overline{\beta}\overline{\gamma}\overline{\mu}\theta_{pq}^2
    -6\theta_{pq}\left(\overline{\gamma}\theta_{kp}\alpha_k\gamma_q+\overline{\mu}\theta_{kq}\beta_k\gamma_p\right)
    +\overline{\beta}\theta_{pq}^2\left(9\alpha_k\beta_k+\alpha_p\beta_p+\alpha_q\beta_q\right)
     \\ &&
    +\overline{\gamma}\theta_{pq}^2\left(3\alpha_k\gamma_k-\alpha_p\gamma_p-\alpha_q\gamma_q\right)
    +\overline{\mu}\theta_{pq}^2\left(3\beta_k\gamma_k-\beta_p\gamma_p-\beta_q\gamma_q\right)
     \\ &&
    +2\overline{\beta}\theta_{pq}\left(\overline{\beta}\gamma_p\gamma_q-\overline{\gamma}\beta_p\gamma_q-\overline{\mu}\alpha_q\gamma_p\right)
    +2\overline{\beta}\left(3\alpha_k\beta_k+\alpha_p\beta_p+\alpha_q\beta_q\right)\left(\gamma_p^2+\gamma_q^2\right)
     \\ &&
    +\overline{\beta}\left(2\alpha_q\beta_p-\alpha_p\beta_q\right)\gamma_p\gamma_q
    -3\left(\overline{\mu}\alpha_k\alpha_p\beta_k\gamma_p+\overline{\gamma}\alpha_k\beta_k\beta_q\gamma_q\right)
     \\ &&
    +3\theta_{kp}\theta_{kq}\theta_{pq}^3
    +6\theta_{pq}\left(\theta_{kp}^2+\theta_{kq}^2\right)\gamma_p\gamma_q
    +6\left(\theta_{kp}\alpha_k\alpha_q+\theta_{kq}\beta_k\beta_p\right)\gamma_p\gamma_q
     \\ &&
    +6\left(\theta_{kp}\theta_{kq}\theta_{pq}+\theta_{kp}\left(\alpha_k\alpha_p+\beta_k\beta_p\right)
      +\theta_{kq}\left(\alpha_k\alpha_q+\beta_k\beta_q\right)\right)\left(\gamma_p^2+\gamma_q^2\right)
     \\ &&
    +3\theta_{kp}\theta_{pq}^2\left(\alpha_k\alpha_p+\beta_k\beta_p-\gamma_k\gamma_p\right)
    +3\theta_{kq}\theta_{pq}^2\left(\alpha_k\alpha_q+\beta_k\beta_q-\gamma_k\gamma_q\right)
     \\ &&
    -3\left(\theta_{pq}\alpha_k\alpha_p\beta_k\beta_q+\theta_{kq}\alpha_k\alpha_p\gamma_p\gamma_q
      +\theta_{kp}\beta_k\beta_q\gamma_p\gamma_q\right)
     \\ &&
    +\theta_{pq}^3\left(\alpha_p\alpha_q+\beta_p\beta_q+\gamma_p\gamma_q\right)
    +2\theta_{pq}\left(3\alpha_k^2+\alpha_p^2+\alpha_q^2+3\beta_k^2+\beta_p^2+\beta_q^2\right)\gamma_p\gamma_q
     \\ &&
    -6\theta_{pq}\left(\alpha_k\alpha_q\beta_k\beta_p+\alpha_k\alpha_q\gamma_k\gamma_p
      +\beta_k\beta_p\gamma_k\gamma_q\right)
    +2\theta_{pq}\left(\alpha_p\alpha_q+\beta_p\beta_q\right)\left(\gamma_p^2+\gamma_q^2\right)
   \\
  \mathcal{F}^6_{\tau_S\tau^T\tau^T}&=&
    +\overline{\beta}\theta_{pq}^3\alpha_p\beta_q
    -\theta_{pq}^2\left(\overline{\beta}\overline{\gamma}\beta_q\gamma_q+\overline{\beta}\overline{\mu}\alpha_p\gamma_p
      -3\overline{\gamma}\overline{\mu}\alpha_k\beta_k\right)
     \\ &&
    +2\overline{\beta}\theta_{pq}\alpha_p\beta_q\left(\gamma_p^2+\gamma_q^2\right)
    +2\overline{\beta}\theta_{pq}\left(3\alpha_k\beta_k+\alpha_p\beta_p+\alpha_q\beta_q\right)\gamma_p\gamma_q
     \\ &&
    -3\theta_{pq}^2\left(\overline{\gamma}\theta_{kq}\alpha_k\gamma_q+\overline{\mu}\theta_{kp}\beta_k\gamma_p\right)
    -6\theta_{pq}\alpha_k\beta_k\left(\overline{\gamma}\beta_p\gamma_q+\overline{\mu}\alpha_q\gamma_p\right)
     \\ &&
    +3\theta_{pq}^2\left(\theta_{kp}\theta_{pq}\beta_k\beta_q+\theta_{kq}\theta_{pq}\alpha_k\alpha_p
      +\theta_{kp}\theta_{kq}\gamma_p\gamma_q\right)
     \\ &&
    +6\theta_{pq}\left(\theta_{kp}\alpha_k\alpha_p+\theta_{kp}\beta_k\beta_p+\theta_{kq}\alpha_k\alpha_q
      +\theta_{kq}\beta_k\beta_q\right)\gamma_p\gamma_q
     \\ &&
    +6\theta_{pq}\left(\theta_{kq}\alpha_k\alpha_p+\theta_{kp}\beta_k\beta_q\right)\left(\gamma_p^2+\gamma_q^2\right)
     \\ &&
    -3\theta_{pq}^2\left(\left(\alpha_k\alpha_p+\beta_k\beta_q\right)\gamma_k\gamma_q
      +\alpha_k\beta_k\left(\alpha_p\beta_p+\alpha_q\beta_k\right)\right)
     \\ &&
    +\theta_{pq}^2\left(\alpha_p\alpha_q+\beta_p\beta_q\right)\gamma_p\gamma_q
    +6\alpha_k\beta_k\left(\alpha_p\beta_p+\alpha_q\beta_q\right)\left(\gamma_p^2+\gamma_q^2\right)
     \\ &&
    -3\alpha_k\beta_k\left(\alpha_p\beta_q-2\alpha_q\beta_p\right)\gamma_p\gamma_q
   \\
  -\mathcal{F}^7_{\tau_S\tau^T\tau^T}&=&
    3\theta_{pq}^3\alpha_k\alpha_p\beta_k\beta_q
    +\theta_{pq}^2\left(\overline{\beta}\alpha_p\beta_q\gamma_p\gamma_q-3\overline{\gamma}\alpha_k\beta_k\beta_q\gamma_q
      -3\overline{\mu}\alpha_k\alpha_p\beta_k\gamma_p\right)
     \\ &&
    +3\theta_{pq}^2\left(\theta_{kq}\alpha_k\alpha_p+\theta_{kp}\beta_k\beta_q\right)\gamma_p\gamma_q
    +6\theta_{pq}\alpha_k\left(\alpha_p\beta_k\beta_p+\alpha_q\beta_k\beta_q\right)\gamma_p\gamma_q
     \\ &&
    +6\theta_{pq}\alpha_k\alpha_p\beta_k\beta_q\left(\gamma_p^2+\gamma_q^2\right)
   \\
  \mathcal{F}^8_{\tau_S\tau^T\tau^T}&=&3\theta_{pq}^2\alpha_k\alpha_p\beta_k\beta_q\gamma_p\gamma_q
\eeas

\subsection{Vector-Tensor-Vector Correlation}
\label{Appendix-SourceStats-VTV}
The cross-correlation between the vectors and the tensors, $\langle\tau^V_a\tau^{Ta}_b\tau^{Vb}\rangle$, is found by applying $\mathcal{A}^{ijklmn}=\mathcal{P}^{Vij}_a(\mathbf{k})\mathcal{P}^{Tkla}_b(\mathbf{p})\mathcal{P}^{Vmnb}(\mathbf{q})$ to $\mathcal{B}_{ijklmn}$. This ultimately produces
\be
  \mathcal{F}_{\tau^V\tau^T\tau^V}=\sum_{n=2}^8\mathcal{F}_{\tau^V\tau^T\tau^V}^n
\ee
with
\beas
  8\mathcal{F}^2_{\tau^V\tau^T\tau^V}&=&
    \theta_{kp}\theta_{kq}-2\alpha_p\alpha_q-2\beta_p\beta_q-\gamma_p\gamma_q
   \\
  8\mathcal{F}^3_{\tau^V\tau^T\tau^V}&=&
    5\overline{\beta}\alpha_p\beta_q+\overline{\gamma}\alpha_p\gamma_q+\overline{\mu}\beta_q\gamma_p
    +\theta_{kp}\left(2\beta_k\beta_q+\gamma_k\gamma_q\right)
     \\ &&
    +\theta_{kq}\left(2\alpha_k\alpha_p+\gamma_k\gamma_p\right)
    +\theta_{pq}\left(2\alpha_p^2+2\beta_q^2-\gamma_k^2\right)
   \\
  8\mathcal{F}^4_{\tau^V\tau^T\tau^V}&=&
    -2\overline{\gamma}\overline{\mu}\alpha_p\beta_q
    -2\left(\overline{\gamma}\theta_{kq}\alpha_p+\overline{\mu}\theta_{kp}\beta_q\gamma_k\right)
    -2\theta_{pq}\left(\overline{\gamma}\theta_{pq}\alpha_p\gamma_p+\overline{\mu}\beta_q\gamma_q\right)
     \\ &&
    -2\theta_{kp}\theta_{kq}\left(\theta_{kp}^2+\theta_{kq}^2\right)
    +2\left(\theta_{kp}^2-\theta_{kq}^2\right)\left(\alpha_p\alpha_q-\beta_p\beta_q\right)
    -2\theta_{pq}\left(\theta_{kp}\alpha_k\alpha_p+\theta_{kq}\beta_k\beta_q\right)
     \\ &&
    -\theta_{kp}\theta_{kq}\left(\alpha_k^2+4\alpha_p^2+\beta_k^2+4\beta_q^2+\gamma_k^2\right)
    -\left(\alpha_k^2+4\alpha_p^2\right)\beta_p\beta_q
    -5\alpha_k\alpha_p\beta_k\beta_q
     \\ &&
    -\left(\alpha_k\alpha_p\gamma_q+\beta_k\beta_q\gamma_p\right)\gamma_k
    -\alpha_p\alpha_q\left(\beta_k^2+4\beta_q^2\right)
    +\left(\alpha_p\alpha_q+\beta_p\beta_q\right)\left(\gamma_k^2+2\gamma_p^2\right)
\eeas
\beas
 8\mathcal{F}^5_{\tau^V\tau^T\tau^V}&=&
    -2\overline{\beta}\left(\theta_{kp}^2+\theta_{kq}^2+2\theta_{pq}^2\right)\alpha_p\beta_q
    -\overline{\beta}\alpha_p\beta_q\left(\gamma_k^2+2\gamma_p^2+2\gamma_q^2\right)
     \\ &&
    +2\theta_{kp}\theta_{kq}\left(\overline{\gamma}\alpha_p\gamma_p+\overline{\mu}\beta_q\gamma_q\right)
    +2\left(\overline{\gamma}\theta_{kq}^2\alpha_p\gamma_q+\overline{\mu}\theta_{kp}^2\beta_q\gamma_p\right)
     \\ &&
    +2\left(\overline{\gamma}\beta_k\gamma_k+\overline{\gamma}\beta_p\gamma_p+\overline{\mu}\alpha_k\gamma_k\right)
    +\overline{\gamma}\alpha_p\left(\beta_k^2+2\beta_q^2\right)\gamma_q
    +\overline{\mu}\left(\alpha_k^2+2\alpha_p^2\right)\beta_q\gamma_p
     \\ &&
    +4\theta_{kp}^2\theta_{kq}^2\theta_{pq}
    +2\theta_{kp}\bigg(\theta_{kp}\theta_{kq}\alpha_k\alpha_p
      +\theta_{kp}^2\beta_k\beta_q+\theta_{kq}^2\beta_k\beta_q
      +\theta_{kp}\theta_{kq}\gamma_k\gamma_p
      \\ && \quad
      +\theta_{kq}^2\gamma_k\gamma_q+2\theta_{kp}\theta_{pq}\left(\beta_k^2+\beta_q^2\right)\bigg)
     \\ &&
    +2\theta_{kq}\bigg(\theta_{kq}^2\alpha_k\alpha_p
      +\theta_{kq}\theta_{pq}\alpha_k^2+2\theta_{kq}\theta_{pq}\alpha_p^2
      +2\alpha_k\alpha_p\beta_q^2
       \\ && \quad
      -\alpha_k\alpha_p\gamma_p^2+\alpha_p^2\gamma_k\gamma_p
      +2\alpha_p\alpha_q\beta_k\beta_q-\beta_p\beta_q\gamma_k\gamma_q\bigg)
    +\theta_{kp}\beta_k\beta_q\left(\gamma_k^2-2\gamma_q^2\right)
     \\ &&
    +\theta_{kp}\left(\left(\alpha_k^2+4\alpha_p^2\right)\beta_k\beta_q
      +\left(\beta_k^2+2\beta_q^2\right)\gamma_k\gamma_q\right)
    +2\theta_{kp}\left(2\alpha_k\alpha_p\beta_p\beta_q-\alpha_p\alpha_q\gamma_k\gamma_p\right)
     \\ &&
    +\theta_{kq}\left(\alpha_k^2\gamma_k\gamma_p+\alpha_k\alpha_p\beta_k^2
      +\frac{1}{2}\alpha_k\alpha_p\gamma_k^2\right)
     \\ &&
    +\theta_{pq}\left(\alpha_k^2\beta_k^2+2\alpha_k^2\beta_q^2+2\alpha_p^2\beta_k^2+2\alpha_p^2\beta_q^2
      +2\alpha_k\alpha_p\gamma_k\gamma_p+2\alpha_p\alpha_q\beta_p\beta_q
      +2\beta_k\beta_q\gamma_k\gamma_q\right)
   \\
 8\mathcal{F}^6_{\tau^V\tau^T\tau^V}&=&
    +4\overline{\beta}\theta_{kp}\theta_{kq}\theta_{pq}\alpha_p\beta_q
    +2\overline{\beta}\theta_{kp}\alpha_p\beta_q\gamma_k\gamma_p
    +2\overline{\beta}\theta_{kq}\alpha_p\beta_q\gamma_k\gamma_q
     \\ &&
    +4\overline{\beta}\theta_{pq}\alpha_p\beta_q\gamma_p\gamma_q
    -2\left(\left(\theta_{kp}\gamma_p+\theta_{kq}\gamma_q\right)
      \left(\overline{\mu}\alpha_k+\overline{\gamma}\beta_k\right)\right)\alpha_p\beta_q
     \\ &&
    -4\theta_{kp}\theta_{kq}\left(\theta_{kq}\theta_{pq}\alpha_k\alpha_p
      +\theta_{kp}\theta_{pq}\beta_k\beta_q+\theta_{kp}\theta_{kq}\gamma_p\gamma_q\right)
     \\ &&
    -2\left(\theta_{kq}^2\left(\alpha_k^2+2\alpha_p^2\right)
      +\theta_{kp}^2\left(\beta_k^2+2\beta_q^2\right)\right)\gamma_p\gamma_q
     \\ &&
    -2\left(\theta_{kq}\alpha_k\alpha_p+\theta_{kq}\beta_k\beta_q\right)
      \left(\theta_{kp}\gamma_p+\theta_{kq}\gamma_q\right)\gamma_k
    -4\theta_{pq}\left(\theta_{kq}\alpha_k\beta_p+\theta_{kp}\alpha_q\beta_k\right)\alpha_p\beta_q
     \\ &&
    -2\theta_{kq}\theta_{pq}\left(\alpha_k^2+2\alpha_p^2\right)\beta_k\beta_q
    -2\theta_{kp}\theta_{pq}\alpha_k\alpha_p\left(\beta_k^2+2\beta_q^2\right)
     \\ &&
    -2\left(\theta_{kp}^2+\theta_{kq}^2-2\theta_{pq}^2\right)\alpha_k\alpha_p\beta_k\beta_q
    -\left(\alpha_k^2+2\alpha_p^2\right)\left(\beta_k^2+2\beta_q^2\right)\gamma_p\gamma_q
\\
    &&-\left(\alpha_k^2+2\alpha_p^2\right)\beta_k\beta_q\gamma_k\gamma_p
    -\alpha_k\alpha_p\left(\beta_k^2+2\beta_q^2\right)\gamma_k\gamma_q
    -\alpha_k\alpha_p\beta_k\beta_q\left(\gamma_k^2-2\gamma_p^2-2\gamma_q^2\right)
     \\ &&
    -2\alpha_k\alpha_p\beta_p\beta_q\gamma_k\gamma_p
    -2\alpha_p\alpha_q\beta_k\beta_q\gamma_k\gamma_q
    -4\alpha_p\alpha_q\beta_p\beta_q\gamma_p\gamma_q
   \\
 8\mathcal{F}^7_{\tau^V\tau^T\tau^V}&=&
    -4\overline{\beta}\theta_{kp}\theta_{kq}\alpha_p\beta_q\gamma_p\gamma_q
    +4\theta_{kp}\theta_{kq}\left(\theta_{pq}\alpha_k\alpha_p\beta_k\beta_q
      +\theta_{kp}\beta_k\beta_q\gamma_p\gamma_q+\theta_{kq}\alpha_k\alpha_p\gamma_p\gamma_q\right)
     \\ &&
    +2\theta_{kq}\left(\alpha_k^2+2\alpha_p^2\right)\beta_k\beta_q\gamma_p\gamma_q
    +2\theta_{kp}\alpha_k\alpha_p\left(\beta_k^2+\beta_q^2\right)\gamma_p\gamma_q
    +2\theta_{kp}\alpha_k\alpha_p\beta_k\beta_q\gamma_k\gamma_p
     \\ &&
    +2\theta_{kq}\alpha_k\alpha_p\beta_k\beta_q\gamma_k\gamma_q
    +4\left(\theta_{kp}\alpha_q\beta_k+\theta_{kq}\alpha_k\beta_p
      -\theta_{pq}\alpha_k\beta_k\right)\alpha_p\beta_q\gamma_p\gamma_q
   \\
 8\mathcal{F}^8_{\tau^V\tau^T\tau^V}&=&
    -4\theta_{kp}\theta_{kq}\alpha_k\alpha_p\beta_k\beta_q\gamma_p\gamma_q .
\eeas

\subsection{Tensor Auto-Correlation}
\label{Appendix-SourceStats-TTT}
The full tensor auto-correlation, $\langle\tau^T_{ab}\tau^{bT}_c\tau^{acT}\rangle$, is found by the application of
\beas
  \mathcal{A}^{ijklmn}&=&\mathcal{P}^{Tij}_{ab}(\mathbf{k})\mathcal{P}^{Tklb}_c(\mathbf{p})
    \mathcal{P}^{mnac}_T(\mathbf{q})
\eeas
to $\mathcal{B}_{ijklmn}$. This leads ultimately to
\be
  \mathcal{F}_{\tau^T\tau^T\tau^T}=\sum_{n=0}^9\mathcal{F}_{\tau^T\tau^T\tau^T}^{n} .
\ee
and
\beas
  8F^0_{\tau^T\tau^T\tau^T}&=&
    -18
   \\
  8F^1_{\tau^T\tau^T\tau^T}&=&
    0
   \\
  8F^2_{\tau^T\tau^T\tau^T}&=&
    \alpha_k^2+\alpha_p^2+\alpha_q^2+\beta_k^2+\beta_p^2+\beta_q^2+\gamma_k^2+\gamma_p^2+\gamma_q^2
     \\ &&
    5\left(\overline{\beta}^2+\overline{\gamma}^2+\overline{\mu}^2\right)
    +12\left(\theta_{kp}^2+\theta_{kq}^2+\theta_{pq}^2\right)
\eeas
\beas
  8F^3_{\tau^T\tau^T\tau^T}&=&
    -4\overline{\beta}\overline{\gamma}\overline{\mu}
    -\overline{\beta}\left(5\alpha_k\beta_k+7\alpha_p\beta_p+7\alpha_q\beta_q\right)
    -\overline{\gamma}\left(5\alpha_p\gamma_p+7\alpha_k\gamma_k+7\alpha_q\gamma_q\right)
      \\ &&
    -\overline{\mu}\left(5\beta_q\gamma_q+7\beta_k\gamma_k+7\beta_p\gamma_p\right)
    -20\theta_{kp}\theta_{kq}\theta_{pq}
    -12\theta_{kp}\left(\alpha_k\alpha_p+\beta_k\beta_p+\gamma_k\gamma_p\right)
      \\ &&
    -12\theta_{kq}\left(\alpha_k\alpha_q+\beta_k\beta_q+\gamma_k\gamma_q\right)
    -12\theta_{pq}\left(\alpha_p\alpha_q+\beta_p\beta_q+\gamma_p\gamma_q\right)
\\
  8F^4_{\tau^T\tau^T\tau^T}&=&
    -2\left(\overline{\beta}^2+\overline{\gamma}^2+\overline{\mu}^2\right)
      \left(\theta_{kp}^2+\theta_{kq}^2+\theta_{pq}^2\right)
    -\overline{\mu}^2\left(\alpha_k^2+\alpha_p^2+\alpha_q^2\right)
    -\overline{\gamma}^2\left(\beta_k^2+\beta_p^2+\beta_q^2\right)
      \\ &&
    -\overline{\beta}^2\left(\gamma_k^2+\gamma_p^2+\gamma_q^2\right)
    +2\overline{\beta}\overline{\mu}\left(\alpha_k\gamma_k+\alpha_p\gamma_p+\alpha_q\gamma_q\right)
    +2\overline{\gamma}\overline{\mu}\left(\alpha_k\beta_k+\alpha_p\beta_p+\alpha_q\beta_q\right)
      \\ &&
    +2\overline{\beta}\overline{\gamma}\left(\beta_k\gamma_k+\beta_p\gamma_p+\beta_q\gamma_q\right)
    +\overline{\gamma}\theta_{kq}\left(5\alpha_k\gamma_q+4\alpha_q\gamma_k\right)
    +\overline{\mu}\theta_{kp}\left(5\beta_k\gamma_p+4\beta_p\gamma_k\right)
      \\ &&
    +\overline{\beta}\theta_{pq}\left(5\alpha_p\beta_q+4\alpha_q\beta_p\right)
    +4\overline{\beta}\theta_{kp}\left(\alpha_k\beta_p+\alpha_p\beta_k\right)
    +4\overline{\beta}\theta_{kq}\left(\alpha_k\beta_q+\alpha_q\beta_k\right)
      \\ &&
    +4\overline{\gamma}\theta_{kp}\left(\alpha_k\gamma_p+\alpha_p\gamma_k\right)
    +4\overline{\gamma}\theta_{pq}\left(\alpha_p\gamma_q+\alpha_q\gamma_p\right)
    +4\overline{\mu}\theta_{kq}\left(\beta_k\gamma_q+\beta_q\gamma_k\right)
      \\ &&
    +4\overline{\mu}\theta_{pq}\left(\beta_p\gamma_q+\beta_q\gamma_p\right)
    -4\theta_{pq}^2\left(\alpha_k^2+\beta_k^2+\gamma_k^2\right)
    -4\theta_{kq}^2\left(\alpha_p^2+\beta_p^2+\gamma_p^2\right)
      \\ &&
    -4\theta_{kp}^2\left(\alpha_q^2+\beta_q^2+\gamma_q^2\right)
    +\theta_{kp}\theta_{kq}\left(\alpha_p\alpha_q+\beta_p\beta_q+\gamma_p\gamma_q\right)
      \\ &&
    +\theta_{kq}\theta_{pq}\left(\alpha_k\alpha_p+\beta_k\beta_p+\gamma_k\gamma_p\right)
    +\theta_{kp}\theta_{pq}\left(\alpha_k\alpha_q+\beta_k\beta_q+\gamma_k\gamma_q\right)
      \\ &&
    -2\alpha_k^2\left(\beta_k^2+\beta_p^2+\beta_q^2+\gamma_p^2+\gamma_q^2\right)
    -2\alpha_p^2\left(\beta_k^2+\beta_q^2+\gamma_k^2+\gamma_p^2+\gamma_q^2\right)
      \\ &&
    -2\alpha_q^2\left(\beta_k^2+\beta_p^2+\gamma_k^2+\gamma_p^2\right)
    -2\beta_k^2\left(\gamma_p^2+\gamma_q^2\right)
    -2\beta_p^2\left(\gamma_k^2+\gamma_q^2\right)
      \\ &&
    -2\beta_q^2\left(\gamma_k^2+\gamma_p^2+\gamma_q^2\right)
    +3\alpha_k\alpha_p\left(\beta_k\beta_p+\gamma_k\gamma_p\right)
    +\alpha_k\alpha_q\left(3\beta_k\beta_q+4\gamma_k\gamma_q\right)
      \\ &&
    +\alpha_p\alpha_q\left(4\beta_p\beta_q+3\gamma_p\gamma_q\right)
    +4\beta_k\beta_p\gamma_k\gamma_p+3\beta_k\beta_q\gamma_k\gamma_q+3\beta_p\beta_q\gamma_p\gamma_q
   \\
  8F^5_{\tau^T\tau^T\tau^T}&=&
    \overline{\beta}\overline{\gamma}\overline{\mu}\left(\theta_{kp}^2+\theta_{kq}^2+\theta_{pq}^2\right)
    +2\left(\overline{\beta}^2+\overline{\gamma}^2+\overline{\mu}^2\right)\theta_{kp}\theta_{kq}\theta_{pq}
      \\ &&
    +\overline{\beta}^2\left(\theta_{kp}\gamma_k\gamma_p+\theta_{kq}\gamma_k\gamma_q
      +\theta_{pq}\gamma_p\gamma_q\right)
    +\overline{\gamma}^2\left(\theta_{kp}\beta_k\beta_p+\theta_{kq}\beta_k\beta_q+\theta_{pq}\beta_p\beta_q\right)
      \\ &&
    -\overline{\beta}\overline{\gamma}\bigg(\theta_{kp}\left(\beta_k\gamma_p+\beta_p\gamma_k\right)
      +\theta_{kq}\left(\beta_k\gamma_q+\beta_q\gamma_k\right)
      +\theta_{pq}\left(\beta_p\gamma_q+\beta_q\gamma_p\right)\bigg)
      \\ &&
    -\overline{\beta}\overline{\mu}\left(\theta_{kp}\left(\alpha_k\gamma_p+\theta_{kp}\alpha_p\gamma_k\right)
      +\theta_{kq}\left(\alpha_k\gamma_q+\alpha_q\gamma_k\right)
      +\theta_{pq}\left(\alpha_p\gamma_q+\alpha_q\gamma_p\right)\right)
      \\ &&
    -\overline{\gamma}\overline{\mu}\left(\theta_{kp}\left(\alpha_k\beta_p+\alpha_p\beta_k\right)
      +\theta_{kq}\left(\alpha_k\beta_q+\alpha_q\beta_k\right)
      +\theta_{pq}\left(\alpha_p\beta_q+\alpha_q\beta_p\right)\right)
      \\ &&
    +2\theta_{pq}^2\left(\overline{\gamma}\left(\alpha_k\gamma_k-\alpha_p\gamma_p\right)
       +\overline{\mu}\left(\beta_k\gamma_k-\beta_q\gamma_q\right)\right)
      \\ &&
    -\theta_{kp}^2\left(\overline{\beta}\left(\alpha_k\beta_k-\alpha_q\beta_q\right)
       +\overline{\gamma}\left(\alpha_p\gamma_p-\alpha_q\gamma_q\right)\right)
      \\ &&
     -\theta_{kq}^2\left(\overline{\beta}\left(\alpha_k\beta_k-\alpha_p\beta_p\right)
      -\overline{\mu}\left(\beta_p\gamma_p-\beta_q\gamma_q\right)\right)\bigg)
      \\ &&
    +\overline{\mu}^2\left(\theta_{kp}\alpha_k\alpha_p+\theta_{kq}\alpha_k\alpha_q+\theta_{pq}\alpha_p\alpha_q\right)
      \\ &&
    -2\theta_{kp}\theta_{pq}\bigg(\overline{\beta}\left(\alpha_k\beta_q+\alpha_q\beta_k\right)
      +\overline{\gamma}\left(\alpha_k\gamma_q+\alpha_q\gamma_k\right)
      +\overline{\mu}\left(\beta_k\gamma_q+\beta_q\gamma_k\right)\bigg)
      \\ &&
    +2\left(\overline{\beta}\theta_{pq}^2\alpha_k\beta_k+\overline{\gamma}\theta_{kq}^2\alpha_p\gamma_p
      +\overline{\mu}\theta_{kp}^2\beta_q\gamma_q\right)
      \\ &&
    -2\theta_{kp}\theta_{kq}\left(\overline{\beta}\left(\alpha_p\beta_q+\alpha_q\beta_p\right)
      +\overline{\gamma}\left(\alpha_p\gamma_q+\alpha_q\gamma_p\right)
      +\overline{\mu}\left(\beta_p\gamma_q+\beta_q\gamma_p\right)\right)
      \\ &&
    -2\theta_{kq}\theta_{pq}\left(\overline{\beta}\left(\alpha_k\beta_p+\alpha_p\beta_k\right)
      +\overline{\gamma}\left(\alpha_k\gamma_p+\alpha_p\gamma_k\right)
      +\overline{\mu}\left(\beta_k\gamma_p+\beta_p\gamma_k\right)\right)
      \\ &&
    +\overline{\beta}\left(\alpha_q\beta_q+\alpha_p\beta_p\right)\left(\gamma_k^2+\gamma_p^2+\gamma_q^2\right)
    -\overline{\beta}\alpha_k\beta_k\left(\gamma_k^2-\gamma_p^2-\gamma_q^2\right)
      \\ &&
    +\overline{\gamma}\left(\alpha_k\gamma_k+\alpha_q\gamma_q\right)\left(\beta_k^2+\beta_p^2+\beta_q^2\right)
    -\overline{\gamma}\alpha_p\left(\beta_p^2-\beta_q^2-\beta_k^2\right)\gamma_p
      \\ &&
    +\overline{\mu}\left(\alpha_k^2+\alpha_p^2+\alpha_q^2\right)\left(\beta_k\gamma_k+\beta_p\gamma_p\right)
    -\overline{\mu}\left(\alpha_q^2-\alpha_k^2-\alpha_p^2\right)\beta_q\gamma_q
      \\ &&
    -\overline{\beta}\alpha_k\gamma_k\left(\beta_p\gamma_p+\beta_q\gamma_q\right)
    -\overline{\gamma}\alpha_k\beta_k\left(\beta_p\gamma_p+\beta_q\gamma_q\right)
    -\overline{\beta}\beta_k\gamma_k\left(\alpha_p\gamma_p+\alpha_q\gamma_q\right)
      \\ &&
    -\overline{\beta}\left(\alpha_p\beta_q+\alpha_q\beta_p\right)\gamma_p\gamma_q
    -\overline{\gamma}\beta_k\gamma_k\left(\alpha_p\beta_p+\alpha_q\beta_q\right)
    -\overline{\gamma}\left(\alpha_p\gamma_q+\alpha_q\gamma_p\right)\beta_p\beta_q
      \\ &&
    -\overline{\mu}\alpha_k\alpha_p\left(\beta_k\gamma_p+\beta_p\gamma_k\right)
    -\overline{\mu}\alpha_k\alpha_q\left(\beta_k\gamma_q+\beta_q\gamma_k\right)
    -\overline{\mu}\alpha_p\alpha_q\left(\beta_p\gamma_q+\beta_q\gamma_p\right)
\\
    &&+2\theta_{kp}\left(\alpha_k^2+\alpha_q^2\right)\beta_k\beta_p
    +2\theta_{kp}\left(\alpha_p^2+\alpha_q^2\right)\gamma_k\gamma_p
    +2\theta_{kq}\left(\alpha_k^2+\alpha_p^2\right)\beta_k\beta_q
      \\ &&
    +2\theta_{pq}\left(\alpha_k^2+\alpha_p^2\right)\gamma_p\gamma_q
    +2\theta_{kp}\alpha_k\alpha_p\left(\beta_k^2+\beta_q^2\right)
    +2\theta_{kq}\alpha_k\alpha_q\left(\beta_k^2+\beta_p^2\right)
      \\ &&
    +2\theta_{kq}\left(\beta_p^2+\beta_q^2\right)\gamma_k\gamma_q
    +2\theta_{pq}\left(\beta_k^2+\beta_q^2\right)\gamma_p\gamma_q
    +2\theta_{kp}\alpha_k\alpha_p\left(\gamma_p^2+\gamma_q^2\right)
      \\ &&
    +2\theta_{kq}\beta_k\beta_q\left(\gamma_p^2+\gamma_q^2\right)
    +2\theta_{pq}\alpha_p\alpha_q\left(\gamma_k^2+\gamma_p^2\right)
    +2\theta_{pq}\beta_p\beta_q\left(\gamma_k^2+\gamma_q^2\right)
      \\ &&
    +2\left(\theta_{kq}\alpha_p^2\gamma_k\gamma_q+\theta_{pq}\alpha_k^2\beta_p\beta_q
      +\theta_{kp}\beta_q^2\gamma_k\gamma_p
     +\theta_{pq}\alpha_p\alpha_q\beta_k^2+\theta_{kq}\alpha_k\alpha_q\gamma_p^2
      +\theta_{kp}\beta_k\beta_p\gamma_q^2\right)
      \\ &&
    -2\theta_{kp}\alpha_k\alpha_q\left(\beta_p\beta_q+\gamma_p\gamma_q\right)
    -2\theta_{kp}\alpha_p\alpha_q\left(\beta_k\beta_q+\gamma_k\gamma_q\right)
    -3\left(\theta_{kq}\alpha_k\alpha_p+\theta_{kp}\beta_k\beta_q\right)\gamma_p\gamma_q
\eeas
\beas
    && -2\left(\theta_{kq}\alpha_k\alpha_p+\theta_{pq}\alpha_k\alpha_p\right)\gamma_k\gamma_q
    -2\left(\theta_{kq}\beta_p\beta_q+\theta_{pq}\alpha_k\alpha_q\right)\gamma_k\gamma_p
      \\ &&
    -2\left(\theta_{kp}\beta_p\beta_q+\theta_{pq}\alpha_k\alpha_p\right)\gamma_k\gamma_q
    -2\theta_{kq}\alpha_p\alpha_q\left(\beta_k\beta_p+\gamma_k\gamma_p\right)
      \\ &&
    -\theta_{pq}\alpha_k\beta_k\left(3\alpha_p\beta_q+2\alpha_q\beta_p\right)
    -2\beta_k\gamma_p\left(\theta_{kq}\beta_p\gamma_q+\theta_{pq}\beta_q\gamma_k\right)
   \\
  8F^6_{\tau^T\tau^T\tau^T}&=&
    -\overline{\beta}\overline{\gamma}\overline{\mu}\theta_{kp}\theta_{kq}\theta_{pq}
    +\overline{\beta}\overline{\gamma}\theta_{kp}\left(\theta_{kq}\beta_p\gamma_q+\theta_{pq}\beta_q\gamma_k\right)
    +\overline{\beta}\overline{\mu}\theta_{kq}\left(\theta_{kp}\alpha_q\gamma_p+\theta_{pq}\alpha_p\gamma_k\right)
      \\ &&
    +\overline{\gamma}\overline{\mu}\theta_{pq}\left(\theta_{kp}\alpha_q\beta_k+\theta_{kq}\alpha_k\beta_p\right)
    -\overline{\beta}\overline{\gamma}\left(\theta_{kp}^2-\theta_{kq}^2-\theta_{pq}^2\right)\beta_q\gamma_q
      \\ &&
    +\overline{\beta}\overline{\mu}\left(\theta_{kp}^2-\theta_{kq}^2+\theta_{pq}^2\right)\alpha_p\gamma_p
    +\overline{\gamma}\overline{\mu}\left(\theta_{kp}^2+\theta_{kq}^2-\theta_{pq}^2\right)\alpha_k\beta_k
      \\ &&
    -\overline{\mu}^2\left(\theta_{kq}\theta_{pq}\alpha_k\alpha_p
     +\overline{\gamma}^2\theta_{kp}\theta_{pq}\beta_k\beta_q
     +\overline{\beta}^2\theta_{kp}\theta_{kq}\gamma_p\gamma_q\right)
      \\ &&
    -\left(\overline{\gamma}\theta_{kq}^3\alpha_k\gamma_q+\overline{\beta}\theta_{pq}^3\alpha_p\beta_q
     +\overline{\mu}\theta_{kp}^3\beta_k\gamma_p\right)
    +2\theta_{kp}\theta_{kq}\theta_{pq}\left(\overline{\beta}\alpha_k\beta_k
     +\overline{\gamma}\alpha_p\gamma_p+\overline{\mu}\beta_q\gamma_q\right)
      \\ &&
    -\overline{\beta}\theta_{pq}\alpha_p\beta_q\left(\gamma_k^2+2\gamma_p^2+2\gamma_q^2\right)
    -\overline{\gamma}\theta_{kq}\alpha_k\left(\beta_p^2+\beta_q^2+\beta_k^2\right)\gamma_q
      \\ &&
    -\overline{\mu}\theta_{kp}\left(\alpha_q^2+\alpha_k^2+\alpha_p^2\right)\beta_k\gamma_p
    +\overline{\beta}\theta_{kp}\left(\alpha_k\beta_k+\alpha_p\beta_p-\alpha_q\beta_q\right)\gamma_k\gamma_p
      \\ &&
    +\overline{\beta}\theta_{kq}\left(\alpha_k\beta_k-\alpha_p\beta_p-\alpha_q\beta_q\right)\gamma_k\gamma_q
    -\overline{\beta}\theta_{pq}\left(\alpha_k\beta_k+\alpha_p\beta_p+\alpha_q\beta_q\right)\gamma_p\gamma_q
      \\ &&
    +\overline{\beta}\theta_{kp}\left(\alpha_k\beta_q+\alpha_q\beta_k\right)\gamma_p\gamma_q
    +\overline{\beta}\theta_{kq}\left(\alpha_k\beta_p+\alpha_p\beta_k\right)\gamma_p\gamma_q
      \\ &&
    -\overline{\gamma}\theta_{kp}\left(\alpha_k\gamma_k-\alpha_p\gamma_p+\alpha_q\gamma_q\right)
    +\overline{\gamma}\theta_{kp}\left(\alpha_k\beta_p+\alpha_p\beta_k\right)\beta_q\gamma_q
      \\ &&
    -\overline{\gamma}\theta_{kq}\left(\alpha_k\gamma_k+\alpha_p\gamma_p+\alpha_q\gamma_q\right)\beta_k\beta_q
    +\overline{\gamma}\theta_{pq}\left(\alpha_k\gamma_p+\alpha_p\gamma_k\right)\beta_k\beta_q
      \\ &&
    +\overline{\gamma}\theta_{pq}\left(\alpha_p\gamma_p-\alpha_q\gamma_q-\alpha_k\gamma_k\right)\beta_p\beta_q
    -\overline{\mu}\theta_{kp}\alpha_k\alpha_p\left(\beta_k\gamma_k+\beta_p\gamma_p+\beta_q\gamma_q\right)
      \\ &&
    +\overline{\mu}\theta_{kq}\alpha_k\alpha_p\left(\beta_p\gamma_q+\beta_q\gamma_p\right)
    -\overline{\mu}\theta_{kq}\alpha_k\alpha_q\left(\beta_k\gamma_k+\beta_p\gamma_p-\beta_q\gamma_q\right)
      \\ &&
    +\overline{\mu}\theta_{pq}\alpha_k\alpha_p\left(\beta_k\gamma_q+\beta_q\gamma_k\right)
    -\overline{\mu}\theta_{pq}\alpha_p\alpha_q\left(\beta_k\gamma_k+\beta_p\gamma_p-\beta_q\gamma_q\right)
      \\ &&
    +\theta_{kq}\alpha_p\left(\overline{\mu}\alpha_q\beta_k+\overline{\beta}\beta_q\gamma_k\right)\gamma_p
    +\theta_{kp}\left(\overline{\beta}\alpha_p\gamma_k\gamma_q+\overline{\gamma}\alpha_q\beta_k\gamma_p\right)\beta_q
      \\ &&
    +\theta_{pq}\alpha_k\beta_k\left(\overline{\gamma}\beta_p\gamma_q+\overline{\mu}\alpha_q\gamma_p\right)
    +\theta_{kp}^2\theta_{kq}^2\theta_{pq}^2
    +\theta_{kq}^2\theta_{pq}^2\left(\alpha_k^2+\alpha_p^2\right)
    +\theta_{kp}^2\theta_{pq}^2\left(\beta_k^2+\beta_q^2\right)
      \\ &&
    +\theta_{kp}^2\theta_{kq}^2\left(\gamma_p^2+\gamma_q^2\right)
    +\theta_{kq}^2\left(\alpha_k^2+\alpha_p^2\right)\left(\gamma_p^2+\gamma_q^2\right)
    +\theta_{pq}^2\left(\alpha_k^2+\alpha_p^2\right)\left(\beta_k^2+\beta_q^2\right)
      \\ &&
    +\theta_{kp}^2\left(\beta_k^2+\beta_q^2\right)\left(\gamma_p^2+\gamma_q^2\right)
    +\theta_{kp}^2\left(2\alpha_k\alpha_q\beta_k\beta_q+2\alpha_p\alpha_q\gamma_p\gamma_q
      +\beta_k\beta_p\gamma_k\gamma_p\right)
      \\ &&
    +\theta_{kq}^2\left(2\alpha_k\alpha_p\beta_k\beta_p+\alpha_k\alpha_q\gamma_k\gamma_q
      +2\beta_p\beta_q\gamma_p\gamma_q\right)
      \\ &&
    +\theta_{pq}^2\left(2\alpha_k\alpha_p\gamma_k\gamma_p+\alpha_p\alpha_q\beta_p\beta_q
      +2\beta_k\beta_q\gamma_k\gamma_q\right)
      \\ &&
    -2\theta_{kp}\theta_{kq}\alpha_k\left(\alpha_p\beta_q+\alpha_q\beta_p\right)\beta_k
    -2\theta_{kp}\theta_{pq}\left(\alpha_k^2\beta_k+\beta_k\gamma_q^2\right)\beta_q
      \\ &&
    -2\theta_{kp}\theta_{pq}\left(\alpha_k\gamma_q+\alpha_q\gamma_k\right)\alpha_p\gamma_p
    -2\theta_{kp}\theta_{kq}\left(\alpha_p^2+\beta_q^2\right)\gamma_p\gamma_q
      \\ &&
    -2\theta_{kq}\theta_{pq}\left(\beta_k\gamma_p+\beta_p\gamma_k\right)\beta_q\gamma_q
    -2\theta_{kq}\theta_{pq}\alpha_k\alpha_p\left(\beta_k^2+\gamma_p^2\right)
      \\ &&
    +\left(\alpha_k^2+\alpha_p^2\right)\left(\beta_k^2+\beta_q^2\right)\left(\gamma_p^2+\gamma_q^2\right)
    +\left(\alpha_k^2+\alpha_p^2\right)\left(\beta_k\beta_p\gamma_k\gamma_p-\beta_q\gamma_k\gamma_q
     -\beta_p\beta_q\gamma_p\gamma_q\right)
     \\ &&
    +\alpha_k\beta_k\left(\alpha_p\beta_p+\alpha_q\beta_q\right)\left(\gamma_k^2-\gamma_p^2-\gamma_q^2\right)
    -\alpha_p\gamma_p\left(\beta_k^2-\beta_p^2+\beta_q^2\right)\left(\alpha_k\gamma_k+\alpha_q\gamma_q\right)
      \\ &&
    +\alpha_q^2\left(\beta_k\gamma_k+\beta_p\gamma_p\right)\beta_q\gamma_q
    +\alpha_k\alpha_q\left(\beta_k^2+\beta_q^2\right)\gamma_k\gamma_q
      \\ &&
    +\alpha_p\alpha_q\beta_p\beta_q\left(\gamma_p^2+\gamma_q^2\right)
    +\alpha_q\beta_p\left(\alpha_k\beta_q\gamma_p+\alpha_p\beta_k\gamma_q\right)\gamma_k
      \\
  8F^7_{\tau^T\tau^T\tau^T}&=&
    \theta_{kp}\theta_{kq}\theta_{pq}\left(\overline{\beta}\theta_{pq}\alpha_p\beta_q
      +\overline{\gamma}\theta_{kq}\alpha_k\gamma_q+\overline{\mu}\theta_{kp}\beta_k\gamma_p\right)
      \\ &&
    -\left(\overline{\beta}\overline{\gamma}\beta_q\gamma_q+\overline{\beta}\overline{\mu}\alpha_p\gamma_p
      +\overline{\gamma}\overline{\mu}\alpha_k\beta_k\right)\theta_{kp}\theta_{kq}\theta_{pq}
    -\overline{\beta}\left(\theta_{kp}^2+\theta_{kq}^2-\theta_{pq}^2\right)\alpha_p\beta_q\gamma_p\gamma_q
      \\ &&
    -\overline{\gamma}\left(\theta_{kp}^2-\theta_{kq}^2+\theta_{pq}^2\right)\alpha_k\beta_k\beta_q\gamma_q
    +\overline{\mu}\left(\theta_{kp}^2-\theta_{kq}^2-\theta_{pq}^2\right)\alpha_k\alpha_p\beta_k\gamma_p
      \\ &&
    +\overline{\mu}\theta_{kq}\theta_{pq}\left(\alpha_k^2+\alpha_p^2\right)\beta_k\gamma_p
    +\overline{\gamma}\theta_{kp}\theta_{pq}\alpha_k\left(\beta_k^2+\beta_q^2\right)\gamma_q
    +\overline{\beta}\theta_{kp}\theta_{kq}\alpha_p\beta_q\left(\gamma_p^2+\gamma_q^2\right)
      \\ &&
    -\overline{\beta}\theta_{kp}\theta_{kq}\left(\alpha_k\beta_k-\alpha_p\beta_p-\alpha_q\beta_q\right)\gamma_p\gamma_q
    +\overline{\beta}\left(\theta_{kp}\gamma_p+\theta_{kq}\gamma_q\right)\theta_{pq}\alpha_p\beta_q\gamma_k
      \\ &&
    +\overline{\gamma}\gamma_q\left(\theta_{kp}\theta_{kq}\alpha_k\beta_k\beta_p
     +\theta_{kp}\theta_{pq}\alpha_q\beta_k\beta_q+\theta_{kq}\theta_{pq}\alpha_k\beta_p\beta_q\right)
      \\ &&
    +\overline{\gamma}\theta_{kp}\theta_{pq}\left(\alpha_k\gamma_k-\alpha_p\gamma_p\right)\beta_k\beta_q\gamma_p
    +\overline{\mu}\theta_{kq}\theta_{pq}\alpha_k\alpha_p\left(\beta_k\gamma_k+\beta_p\gamma_p-\beta_q\gamma_q\right)
      \\ &&
    +\overline{\mu}\theta_{kp}\left(\theta_{kq}\alpha_k+\theta_{pq}\alpha_p\right)\alpha_q\beta_k\gamma_p
    +\theta_{pq}^3\alpha_k\alpha_p\beta_k\beta_q+\theta_{kq}^3\alpha_k\alpha_p\gamma_p\gamma_q
    +\theta_{kp}^3\beta_k\beta_q\gamma_p\gamma_q
      \\ &&
    -\theta_{kp}\theta_{kq}\theta_{pq}\left(\theta_{kp}\theta_{kq}\gamma_p\gamma_q
     +\theta_{kp}\theta_{pq}\beta_k\beta_q+\theta_{kq}\theta_{pq}\alpha_k\alpha_p\right)
      \\ &&
    -\theta_{kq}\theta_{pq}\left(\theta_{kq}\left(\alpha_k^2+\alpha_q^2\right)\gamma_p\gamma_q
     +\theta_{pq}\left(\alpha_k^2+\alpha_p^2\right)\beta_k\beta_q\right)
      \\ &&
    -\theta_{kp}\theta_{pq}\left(\theta_{kp}\left(\beta_k^2+\beta_q^2\right)\gamma_p\gamma_q
     +\theta_{pq}\alpha_k\alpha_p\left(\beta_k^2+\beta_q^2\right)\right)
      \\ &&
    -\theta_{kp}\theta_{kq}\left(\theta_{kp}\beta_k\beta_q\left(\gamma_p^2+\gamma_q^2\right)
     +\theta_{kq}\alpha_k\alpha_p\left(\gamma_p^2+\gamma_q^2\right)\right)
\eeas
\beas &&
    -\theta_{kp}\theta_{kq}\left(\theta_{kp}\beta_k\beta_p+\theta_{kq}\alpha_k\alpha_q\right)\gamma_p\gamma_q
    -\theta_{kp}\theta_{pq}\left(\theta_{kp}\gamma_k\gamma_p+\theta_{pq}\alpha_p\alpha_q\right)\beta_k\beta_q
      \\ &&
    -\theta_{kq}\theta_{pq}\left(\theta_{kq}\gamma_k\gamma_q+\theta_{pq}\beta_p\beta_q\right)\alpha_k\alpha_p
    -\theta_{pq}\left(\alpha_k^2+\alpha_p^2\right)\left(\beta_k^2+\beta_q^2\right)\gamma_p\gamma_q
      \\ &&
    -\theta_{kq}\left(\alpha_k^2+\alpha_p^2\right)\beta_k\beta_q\left(\gamma_p^2+\gamma_q^2\right)
    -\theta_{kp}\alpha_k\alpha_p\left(\beta_k^2+\beta_q^2\right)\left(\gamma_p^2+\gamma_q^2\right)
      \\ &&
    +2\theta_{kp}\left(\alpha_k^2+\alpha_p^2-\alpha_q^2\right)\beta_k\beta_q\gamma_p\gamma_q
    -\left(\theta_{kq}\left(\alpha_k^2+\alpha_p^2\right)\beta_p\gamma_q
     +\theta_{pq}\left(\alpha_k^2+\alpha_p^2\right)\beta_q\gamma_k\right)\beta_k\gamma_p
      \\ &&
    +2\theta_{kq}\alpha_k\alpha_p\left(\beta_k^2+\beta_q^2\right)\gamma_p\gamma_q
    -\alpha_k\left(\theta_{kp}\alpha_q\beta_k^2+\theta_{kq}\alpha_p\beta_p^2
     +\theta_{kp}\alpha_q\beta_q^2\right)\gamma_p\gamma_q
      \\ &&
    -\theta_{pq}\alpha_k\alpha_p\left(\beta_k^2+\beta_q^2\right)\gamma_k\gamma_q
    -\alpha_p\beta_q\left(\theta_{kp}\alpha_q\beta_k\left(\gamma_p^2+\gamma_q^2\right)
     +\theta_{kq}\alpha_k\beta_p\left(\gamma_p^2+\gamma_q^2\right)\right)
      \\ &&
    -\theta_{pq}\alpha_k\alpha_p\beta_k\beta_q\left(\gamma_k^2-2\gamma_p^2-2\gamma_q^2\right)
    -\alpha_k\alpha_p\beta_k\left(\theta_{kp}\left(\beta_p+\beta_q\right)
      +\theta_{kq}\left(\beta_p+\beta_q\right)\right)\gamma_k\gamma_q
      \\ &&
    +\theta_{kq}\alpha_q\beta_q\left(\alpha_p\beta_k-\alpha_k\beta_p\right)\gamma_p\gamma_q
    +\theta_{kq}\alpha_k\alpha_p\beta_k\beta_q\gamma_k\gamma_p
      \\ &&
    -\theta_{pq}\alpha_p\alpha_q\left(\beta_k+\beta_p\right)\beta_q\gamma_p\gamma_q
    +\theta_{kp}\alpha_p\left(\alpha_k\beta_p\beta_q-\alpha_q\beta_k\beta_p\right)\gamma_p\gamma_q
      \\ &&
    -\theta_{kp}\alpha_k\alpha_q\beta_k\beta_q\gamma_k\gamma_p
    +\theta_{pq}\alpha_k\left(\alpha_p\beta_k\beta_p\gamma_p\gamma_q-\alpha_p\beta_p\beta_q\gamma_k\gamma_p
     +\alpha_q\beta_k\beta_q\gamma_p\gamma_q\right)
   \\
  8F^8_{\tau^T\tau^T\tau^T}&=&
    -\theta_{kp}\theta_{kq}\theta_{pq}\left(\overline{\beta}\alpha_p\beta_q\gamma_p\gamma_q
      +\overline{\gamma}\alpha_k\beta_k\beta_q\gamma_q+\overline{\mu}\alpha_k\alpha_p\beta_k\gamma_p\right)
      \\ &&
    +\theta_{kp}\theta_{kq}\theta_{pq}\left(\theta_{kp}\beta_k\beta_q\gamma_p\gamma_q
     +\theta_{kq}\alpha_k\alpha_p\gamma_p\gamma_q+\theta_{pq}\alpha_k\alpha_p\beta_k\beta_q\right)
      \\ &&
    -\left(\theta_{kp}^2+\theta_{kq}^2+\theta_{pq}^2\right)\alpha_k\alpha_p\beta_k\beta_q\gamma_p\gamma_q
    +\theta_{kq}\theta_{pq}\left(\alpha_k^2+\alpha_p^2\right)\beta_k\beta_q\gamma_p\gamma_q
      \\ &&
    +\theta_{kp}\theta_{pq}\alpha_k\alpha_p\left(\beta_k^2+\beta_q^2\right)\gamma_p\gamma_q
    +\theta_{kp}\theta_{kq}\alpha_k\alpha_p\beta_k\beta_q\left(\gamma_p^2+\gamma_q^2\right)
      \\ &&
    +\theta_{kp}\theta_{kq}\alpha_k\beta_k\left(\alpha_p\beta_p+\alpha_q\beta_q\right)\gamma_p\gamma_q
    +\theta_{kp}\theta_{pq}\left(\alpha_k\gamma_k+\alpha_q\gamma_q\right)\alpha_p\beta_k\beta_q\gamma_p
      \\ &&
    +\theta_{kq}\theta_{pq}\alpha_k\alpha_p\left(\beta_k\gamma_k+\beta_p\gamma_p\right)\beta_q\gamma_q
   \\
  8F^9_{\tau^T\tau^T\tau^T}&=&-\theta_{kp}\theta_{kq}\theta_{pq}\alpha_k\alpha_p\beta_k\beta_q\gamma_p\gamma_q
\eeas

\section{Geometry-Independent Large-Scale Results}
\label{Appendix-SourceStats-Results}
Here we present the full generality behind equations (\ref{LargeScaleColinearResults}). Taking $n>-1$ and $k\ll k_c$, and for convenience defining the generic geometry by the two angle cosines $\alpha_k=\cos(\bar{\theta})$ and $\theta_{kq}=-\cos(\xi_{kq})$ (see Figure \ref{BispectraGeometry}) we have
\beas
  \lefteqn{\langle\tau(\mathbf{k})\tau(\mathbf{p})\tau(\mathbf{q})\rangle=
    B\pi\delta(\mathbf{k}+\mathbf{p}+\mathbf{q})} \nonumber \\ 
  \lefteqn{\langle\tau(\mathbf{k})\tau(\mathbf{p})\tau_S(\mathbf{q})\rangle=0} \nonumber \\
  \lefteqn{\langle\tau(\mathbf{k})\tau_S(\mathbf{p})\tau_S(\mathbf{q})\rangle=} \nonumber \\ &&
   \frac{21}{30}B\pi\bigg(2-3\cos^2(\xi_{kq})-3\cos^2(\phi)+6\cos^2(\phi)\cos^2(\xi_{kq})
   +6\cos(\phi)\cos(\xi_{kq})
    \nonumber \\ &&
   -6\cos(\phi)\cos^3(\xi_{kq})-6\cos^3(\phi)\cos(\xi_{kq})
   +6\cos^3(\phi)\cos^3(\xi_{kq})\bigg)\delta(\mathbf{k}+\mathbf{p}+\mathbf{q})
    \nonumber \\
  \lefteqn{\langle\tau_S(\mathbf{k})\tau_S(\mathbf{p})\tau_S(\mathbf{q})\rangle=} \nonumber \\ &&
   \frac{17}{35}B\pi\bigg(1-3\cos^2(\phi)\cos^2(\xi_{kq})
   -3\cos(\phi)\cos(\xi_{kq})\sin^2(\phi)\sin^2(\xi_{kq})\bigg)\delta(\mathbf{k}+\mathbf{p}+\mathbf{q})
    \\
  \lefteqn{\langle\tau(\mathbf{k})\tau_a^V(\mathbf{p})\tau^a_V(\mathbf{q})\rangle=} \nonumber \\ &&
   -\frac{14}{15}B\pi\bigg(3\cos(\phi)\cos(\xi_{kq})+\sin^2(\phi)\sin^2(\xi_{kq})
   -3\cos^3(\phi)\cos(\xi_{kq})
    \nonumber \\ &&
   -3\cos(\phi)\cos^3(\xi_{kq})+4\cos^3(\phi)\cos^3(\xi_{kq})
    \nonumber \\ &&
   -\sin^2(\phi)\cos^2(\phi)\sin^2(\xi_{kq})
   -\sin^2(\phi)\sin^2(\xi_{kq})\cos^2(\xi_{kq})
    \nonumber \\ &&
   +4\cos^2(\phi)\cos^2(\xi_{kq})\sin^2(\phi)\sin^2(\xi_{kq})\bigg)\delta(\mathbf{k}+\mathbf{p}+\mathbf{q})
\eeas
\beas &&
  \lefteqn{\langle\tau_S(\mathbf{k})\tau_a^V(\mathbf{p})\tau^a_V(\mathbf{q})\rangle=} \nonumber \\ &&
   \frac{17}{105}B\pi\bigg(6\cos(\phi)\cos(\xi_{kq})-6\cos^3(\phi)\cos(\xi_{kq})-6\cos(\phi)\cos^3(\xi_{kq})
    \nonumber \\ &&
   -\sin^2(\phi)\sin^2(\xi_{kq})+8\cos^3(\phi)\cos^3(\xi_{kq})
    \nonumber \\ &&
   -2\sin^2(\phi)\sin^2(\xi_{kq})\cos^2(\xi_{kq})
   -2\sin^2(\phi)\sin^2(\xi_{kq})\cos^2(\phi)
    \nonumber \\ &&
   +8\cos^2(\phi)\cos^2(\xi_{kq})\sin^2(\phi)\sin^2(\xi_{kq})\bigg)\delta(\mathbf{k}+\mathbf{p}+\mathbf{q})
    \nonumber \\
  \lefteqn{\langle\tau(\mathbf{k})\tau^{aT}_b(\mathbf{p})\tau^{bT}_a(\mathbf{q})\rangle=} \nonumber \\ &&
   \frac{1}{30}B\pi\bigg(30\cos^3(\phi)\cos^3(\xi_q)\sin^2(\phi)\sin^2(\xi_q)
    \nonumber \\ &&
   +70\cos(\phi)\cos(\xi_q)\sin^2(\phi)\sin^2(\xi_q)+40\sin^4(\phi)\sin^4(\xi_q)+8\cos^4(\phi)\cos^4(\xi_q)
    \nonumber \\ &&
   +8\sin^8(\phi)\sin^8(\xi_q)-5\sin^4(\xi_q)-5\sin^4(\phi)+\cos^2(\phi)\cos^4(\xi_q)\sin^4(\phi)+6\sin^8(\phi)
    \nonumber \\ &&
   +6\sin^8(\xi_q)+3\sin^4(\phi)\sin^8(\xi_q)+5\cos^2(\phi)\sin^4(\xi_q)+12\cos^2(\phi)\sin^4(\phi)
    \nonumber \\ &&
   +3\sin^8(\phi)\sin^4(\xi_q)+12\cos^2(\xi_q)\sin^4(\xi_q)+5\sin^4(\phi)\cos^2(\xi_q)
    \nonumber \\ &&
   +2\cos^3(\phi)\cos(\xi_q)\sin^2(\phi)\sin^6(\xi_q)+\cos^4(\phi)\cos^2(\xi_q)\sin^4(\xi_q)
    \nonumber \\ &&
   +\sin^8(\phi)\sin^4(\xi_q)\cos^2(\xi_q)+3\cos^2(\phi)\cos^2(\xi_q)\sin^4(\xi_q)
    \nonumber \\ &&
   +\sin^4(\phi)\sin^8(\xi_q)\cos^2(\phi)+6\cos(\phi)\cos(\xi_q)\sin^6(\phi)\sin^2(\xi_q)
    \nonumber \\ &&
   +6\cos(\phi)\cos(\xi_q)\sin^2(\phi)\sin^6(\xi_q)+2\cos(\phi)\cos^3(\xi_q)\sin^6(\phi)\sin^2(\xi_q)
    \nonumber \\ &&
   +3\cos^2(\phi)\cos^2(\xi_q)\sin^4(\phi)+6\cos^4(\xi_q)-5\cos^2(\xi_q)+6\cos^4(\phi)
    \nonumber \\ &&
   +3\sin^4(\phi)\sin^4(\xi_q)\cos^2(\phi)+3\sin^4(\phi)\sin^4(\xi_q)\cos^2(\xi_q)+3\cos^4(\phi)\cos^2(\xi_q)
    \nonumber \\ &&
   +6\sin^2(\phi)\sin^2(\xi_q)\cos^3(\xi_q)\cos(\phi)+40\cos^2(\phi)\cos^2(\xi_q)
    \nonumber \\ &&
   +6\cos^3(\phi)\cos(\xi_q)\sin^2(\phi)\sin^2(\xi_q)+3\cos^2(\phi)\cos^4(\xi_q)-5\cos^2(phi)^2
    \nonumber \\ &&
   +44\cos^2(\phi)\cos^2(\xi_q)\sin^4(\phi)\sin^4(\xi_q)+30\cos(\phi)\cos(\xi_q)\sin^6(\phi)\sin^6(\xi_q)\bigg)
   \delta(\mathbf{k}+\mathbf{p}+\mathbf{q})
    \nonumber \\
  \lefteqn{\langle\tau_S(\mathbf{k})\tau^{aT}_b(\mathbf{p})\tau^{bT}_a(\mathbf{q})\rangle=} \nonumber \\ &&
   -\frac{1}{210}B\pi\bigg(210-210\cos^3(\phi)\cos^3(\xi_q)\sin^2(\phi)\sin^2(\xi_q)^2
    \nonumber \\ &&
   -86\cos(\phi)\cos(\xi_q)\sin^2(\phi)\sin^2(\xi_q)-74\sin^4(\phi)\sin^4(\xi_q)
    \nonumber \\ &&
   -88\cos^4(\phi)\cos^4(\xi_q)+44\sin^8(\phi)\sin^8(\xi_q)-224\sin^4(\xi_q)-224\sin^4(\phi)
    \nonumber \\ &&
   -5\cos^2(\phi)\cos^4(\xi_q)\sin^4(\phi)+60\sin^8(\phi)+60\sin^8(\xi_q)+6\sin^4(\phi)\sin^8(\xi_q)
    \nonumber \\ &&
   +185\cos^2(\phi)\sin^4(\xi_q)+150\cos^2(\phi)\sin^4(\phi)+6\sin^8(\phi)\sin^4(\xi_q)
    \nonumber \\ &&
   +150\cos^2(\xi_q)\sin^4(\xi_q)+185\sin^4(\phi)\cos^2(\xi_q)
    \nonumber \\ &&
   -10\cos^3(\phi)\cos(\xi_q)\sin^2(\phi)\sin^6(\xi_q)-5\cos^4(\phi)\cos^2(\xi_q)\sin^4(\xi_q)
    \nonumber \\ &&
   -5\sin^8(\phi)\sin^4(\xi_q)\cos^2(\xi_q)-24\cos^2(\phi)\cos^2(\xi_q)\sin^4(\xi_q)
    \nonumber \\ &&
   -5\sin^4(\phi)\sin^8(\xi_q)\cos^2(\phi)-18\cos(\phi)\cos(\xi_q)\sin^6(\phi)\sin^2(\xi_q)
    \nonumber \\ &&
   -18\cos(\phi)\cos(\xi_q)\sin^2(\phi)\sin^6(\xi_q)-10\cos(\phi)\cos^3(\xi_q)\sin^6(\phi)\sin^2(\xi_q)
    \nonumber \\ &&
   -24\cos^2(\phi)\cos^2(\xi_q)\sin^4(\phi)+90\cos^4(\xi_q)-287\cos^2(\xi_q)+90\cos^4(phi)
    \nonumber \\ &&
   +123\sin^4(\phi)\sin^4(\xi_q)\cos^2(\phi)+123\sin^4(\phi)\sin^4(\xi_q)\cos^2(\xi_q)
    \nonumber \\ &&
   +93\cos^4(\phi)\cos^2(\xi_q)+216\sin^2(\phi)\sin^2(\xi_q)\cos^3(\xi_q)\cos(\phi)+358\cos^2(\phi)\cos^2(\xi_q)
    \nonumber \\ &&
   +216\cos^3(\phi)\cos(\xi_q)\sin^2(\phi)\sin^2(\xi_q)+93\cos^2(\phi)\cos^4(\xi_q)-287\cos^2(\phi)
    \nonumber \\ &&
   -112\cos^2(\phi)\cos^2(\xi_q)\sin^4(\phi)\sin^4(\xi_q)+54\cos(\phi)\cos(\xi_q)\sin^6(\phi)\sin^6(\xi_q)
   \bigg)\delta(\mathbf{k}+\mathbf{p}+\mathbf{q})
\eeas
\beas
  \lefteqn{\langle\tau^V_a(\mathbf{k})\tau^{aT}_b(\mathbf{p})\tau^b_V(\mathbf{q})\rangle=} \nonumber \\ &&
   \frac{2}{105}B\pi\bigg(11\sin^2(\phi)\sin^2(\xi_q)\cos^2(\xi_q)-13\cos(\xi_q)\cos(\phi)\sin^4(\xi_q)
    \nonumber \\ &&
   +56\cos^2(\phi)\cos^2(\xi_q)\sin^2(\phi)\sin^2(\xi_q)+21\cos^3(\xi_q)\cos(\phi)\sin^4(\phi)
    \nonumber \\ &&
   -28\cos(\phi)\cos^3(\xi_q)-28\cos^3(\phi)\cos(\xi_q)-10\sin^2(\phi)\sin^2(\xi_q)+70\cos^3(\phi)\cos^3(\xi_q)
    \nonumber \\ &&
   +20\cos(\phi)\cos(\xi_q)+4\sin^6(\xi_q)\sin^2(\phi)+4\sin^6(\phi)\sin^2(\xi_q)+2\sin^6(\phi)\sin^6(\xi_q)
    \nonumber \\ &&
   +12\cos^2(\phi)\sin^2(\phi)\sin^6(\xi_q)-13\cos(\phi)\sin^4(\phi)\cos(\xi_q)
    \nonumber \\ &&
   +6\cos(\phi)\sin^4(\phi)\sin^4(\xi_q)\cos(\xi_q)+12\cos^2(\xi_q)\sin^6(\phi)\sin^2(\xi_q)
    \nonumber \\ &&
   +21\cos^3(\phi)\cos(\xi_q)\sin^4(\xi_q)+11\sin^2(\phi)\sin^2(\xi_q)\cos^2(\phi)
   \bigg)\delta(\mathbf{k}+\mathbf{p}+\mathbf{q})
    \nonumber \\
  \lefteqn{\langle\tau^{aT}_b(\mathbf{k})\tau^{cT}_a(\mathbf{p})\tau^{bT}_c(\mathbf{q})\rangle=}
 \nonumber \\ &&
    \frac{1}{210}B\pi\bigg(-476+24\cos^4(\phi)\sin^4(\xi_q)+24\cos^4(\xi_q)\sin^4(\phi)+15\cos^2(\xi_q)\sin^8(\phi)
    \nonumber \\ &&
   +15\cos^2(\phi)\sin^8(\xi_q)+126\cos^3(\phi)\cos^3(\xi_q)\sin^2(\phi)\sin^2(\xi_q)
    \nonumber \\ &&
   -250\cos(\phi)\cos(\xi_q)\sin^2(\phi)\sin^2(\xi_q)-32\sin^4(\phi)\sin^4(\xi_q)+70\cos^4(\phi)\cos^4(\xi_q)
    \nonumber \\ &&
   +2\sin^8(\phi)\sin^8(\xi_q)+272\sin^4(\xi_q)+272\sin^4(\phi)+21\cos^2(\phi)\cos^4(\xi_q)\sin^4(\phi)
    \nonumber \\ &&
   -15\sin^8(\phi)-15\sin^8(\xi_q)+30\sin^4(\phi)\sin^8(\xi_q)-244\cos^2(\phi)\sin^4(\xi_q)
    \nonumber \\ &&
   -17\cos^2(\phi)\sin^4(\phi)+30\sin^8(\phi)\sin^4(\xi_q)-17\cos^2(\xi_q)\sin^4(\xi_q)
    \nonumber \\ &&
   -244\sin^4(\phi)\cos^2(\xi_q)+33\cos^3(\phi)\cos(\xi_q)\sin^2(\phi)\sin^6(\xi_q)
    \nonumber \\ &&
   +21\cos^4(\phi)\cos^2(\xi_q)\sin^4(\xi_q)+12\sin^8(\phi)\sin^4(\xi_q)\cos^2(\xi_q)
    \nonumber \\ &&
   +48\cos^2(\phi)\cos^2(\xi_q)\sin^4(\xi_q)+12\sin^4(\phi)\sin^8(\xi_q)\cos^2(\phi)
    \nonumber \\ &&
   +19\cos(\phi)\cos(\xi_q)\sin^6(\phi)\sin^2(\xi_q)+19\cos(\phi)\cos(\xi_q)\sin^2(\phi)\sin^6(\xi_q)
    \nonumber \\ &&
   +33\cos(\phi)\cos^3(\xi_q)\sin^6(\phi)\sin^2(\xi_q)+48\cos^2(\phi)\cos^2(\xi_q)\sin^4(\phi)-7\cos^4(\xi_q)
    \nonumber \\ &&
   +499\cos^2(\xi_q)-7\cos^4(\phi)+73\sin^4(\phi)\sin^4(\xi_q)\cos^2(\phi)+73\sin^4(\phi)\sin^4(\xi_q)\cos^2(\xi_q)
    \nonumber \\ &&
   +57\cos^4(\phi)\cos^2(\xi_q)+62\sin^2(\phi)\sin^2(\xi_q)\cos^3(\xi_q)\cos(\phi)-692\cos^2(\phi)\cos^2(\xi_q)
    \nonumber \\ &&
   +62\cos^3(\phi)\cos(\xi_q)\sin^2(\phi)\sin^2(\xi_q)+57\cos^2(\phi)\cos^4(\xi_q)+499\cos^2(\phi)
    \nonumber \\ &&
   +62\cos^2(\phi)\cos^2(\xi_q)\sin^4(\phi)\sin^4(\xi_q)+8\cos(\phi)\cos(\xi_q)\sin^6(\phi)\sin^6(\xi_q)
   \bigg)\delta(\mathbf{k}+\mathbf{p}+\mathbf{q})
    \nonumber
\eeas
where
\be
  B=A^2k_c^{3(n+1)}/3(n+1)
\ee
with $A$ the amplitude of the magnetic field power spectrum.

Taking the colinear case ($\phi=\xi_{kq}=0$) then recovers the expressions (\ref{LargeScaleColinearResults}).

\chapter{The CMB Bispectrum}
\label{Appendix-Bispectra}
In this appendix we derive the CMB bispectra, chiefly following Wang and Kamionkowski \cite{WangKamionkowski00}, itself based on work by Ferreira, Magueijo and G\'orski \cite{FerreiraMagueijoGorski98}, although neither employed the line-of-sight approach and our derivation thus follows a different route. We construct the form for the bispectrum from the scalar modes, these being far the simplest; for vector and tensor modes the relevant transfer functions should naturally be employed.

\section{The CMB Angular Bispectrum}
Let us first consider a primordial bispectrum analogous to the primordial power spectrum; letting $\left\{A,B,C\right\}$ denote independently trace or traceless components and assuming statistical homogeneity, we can define
\be
  \langle A(\mathbf{k})B(\mathbf{p})C(\mathbf{q})\rangle=
   \mathcal{B}_{ABC}(k,p,q)A(\mathbf{k})B(\mathbf{p})C(\mathbf{q})\delta(\mathbf{k}+\mathbf{p}+\mathbf{q}) .
\ee
Considering for simplicity the scalars and using equation (\ref{Scalar_aTlm}) we can then express the scalar three-point correlation of the $a_{T,lm}$s in the line-of-sight approach as
\beas
  \lefteqn{\langle a^A_{T,lm}a^B_{T,l'm'}a^C_{T,l''m''}\rangle=
   \int_{\mathbf{k}}\int_{\mathbf{p}}\int_{\mathbf{q}}
   \int_{\Omega_{\mathbf{n}}}\int_{\Omega_{\mathbf{n}'}}\int_{\Omega_{\mathbf{n}''}}
   \int_{\eta}\int_{\eta'}\int_{\eta''}}
  \\ && \; \times
   Y^*_{lm}(\mathbf{n})Y^*_{l'm'}(\mathbf{n})Y^*_{l'm'}(\mathbf{n}')Y^*_{l''m''}(\mathbf{n}'')
   \mathcal{B}_{ABC}(k,p,q)
   S^S_T(k,\eta)S^S_T(p,\eta')S^S_T(q,\eta'')
  \\ && \qquad\times
   e^{-ix\mu}e^{-ix'\mu'}e^{-ix''\mu''}
   \delta(\mathbf{k}+\mathbf{p}+\mathbf{q})
   d\eta''d\eta'd\eta
   d\Omega_{\mathbf{n}''}d\Omega_{\mathbf{n}'}d\Omega_{\mathbf{n}}
   \frac{d^3\mathbf{q}}{(2\pi)^3}\frac{d^3\mathbf{p}}{(2\pi)^3}\frac{d^3\mathbf{k}}{(2\pi)^3}
\eeas
where we have defined $x'=p(\eta'-\eta_0)$ and $x''=q(\eta''-\eta_0)$.

Following the method of Wang and Kamionkowski, we now expand the Dirac delta function as an integral across some real-space variable $\mathbf{X}$,
\bdm
  \delta(\mathbf{k}+\mathbf{p}+\mathbf{q})=\int e^{i(\mathbf{k}+\mathbf{p}+\mathbf{q}).\mathbf{X}}\frac{d^3\mathbf{X}}{(2\pi)^3}
\edm
and then expand the exponentials with equation (\ref{ExponentialSphericalHarmonics}) to give
\beas
  \lefteqn{\langle a^A_{T,lm}a^B_{T,l'm'}a^C_{T,l''m''}\rangle=
   \int_{\mathbf{k}}\int_{\mathbf{p}}\int_{\mathbf{q}}
   \int_{\Omega_{\mathbf{n}}}\int_{\Omega_{\mathbf{n}'}}\int_{\Omega_{\mathbf{n}''}}
   \int_{\eta}\int_{\eta'}\int_{\eta''}
   Y^*_{lm}(\mathbf{n})Y^*_{l'm'}(\mathbf{n})Y^*_{l'm'}(\mathbf{n}')Y^*_{l''m''}(\mathbf{n}'')}
  \\ && \times
   \mathcal{B}_{ABC}(k,p,q)
   S^S_T(k,\eta)S^S_T(p,\eta')S^S_T(q,\eta'')
   \sum_{ab}4\pi(-i)^a(2a+1)j_a(x)Y^*_{ab}(\mathbf{k})Y_{ab}(\mathbf{n})
  \\ && \times
   \sum_{a'b'}4\pi(-i)^{a'}(2a'+1)j_a'(x')Y^*_{a'b'}(\mathbf{p})Y_{a'b'}(\mathbf{n}')
   \sum_{a''b''}4\pi(-i)^{a''}(2a''+1)j_a''(x'')Y^*_{a''b''}(\mathbf{q})
  \\ && \times
   Y_{a''b''}(\mathbf{n}'')
   \int_{\mathbf{X}}
   \sum_{cd}4\pi i^c(2c+1)j_c(kX)Y_{cd}(\mathbf{k})Y^*_{cd}(\mathbf{X})
   \sum_{c'd'}4\pi i^{c'}(2c'+1)j_c'(pX)Y_{c'd'}(\mathbf{p})
  \\ && \times 
   Y^*_{c'd'}(\mathbf{X})
   \sum_{c''d''}4\pi i^{c''}(2c''+1)j_c''(qX)Y_{c''d''}(\mathbf{q})Y^*_{c''d''}(\mathbf{X})
   \frac{d^3\mathbf{X}}{(2\pi)^3}
  \\ && \times
   d\eta''d\eta'd\eta
   d\Omega_{\mathbf{n}''}d\Omega_{\mathbf{n}'}d\Omega_{\mathbf{n}}
   \frac{d^3\mathbf{q}}{(2\pi)^3}\frac{d^3\mathbf{p}}{(2\pi)^3}\frac{d^3\mathbf{k}}{(2\pi)^3} .
\eeas
Immediately integrating over the directions of $\mathbf{n}$, $\mathbf{n}'$ and $\mathbf{n}''$ we find a series of Kronecker deltas $\delta^a_l\delta^b_m$ and similar, and performing the sum over $a,a',a''$ we are left with
\beas
  \lefteqn{\langle a^A_{T,lm}a^B_{T,l'm'}a^C_{T,l''m''}\rangle=
   \frac{1}{(2\pi)^9}\int_{k}\int_{p}\int_{q}
   \int_{\Omega_{\mathbf{k}}}\int_{\Omega_{\mathbf{p}}}\int_{\Omega_{\mathbf{q}}}
   \int_{\eta}\int_{\eta'}\int_{\eta''}
   \mathcal{B}_{ABC}(k,p,q)}
  \\ && \times
   S^S_T(k,\eta)S^S_T(p,\eta')S^S_T(q,\eta'')
   (4\pi)^3\sum_{cd}\sum{c'd'}\sum{c''d''}
   i^{c+c'+c''}(-i)^{l+l'+l''}
  \\ && \times
   (2c+1)(2c'+1)(2c''+1)
   (2l+1)(2l'+1)(2l''+1)
   (4\pi)^3
   j_l(x)j_l'(x')j_l''(x'')
  \\ && \times
   Y^*_{lm}(\mathbf{k})Y^*_{l'm'}(\mathbf{p})Y^*_{l''m''}(\mathbf{q})
   Y_{cd}(\mathbf{k})Y_{c'd'}(\mathbf{p})Y_{c''d''}(\mathbf{q})
   \frac{1}{(2\pi)^3}\int_{X}j_c(kX)j_c'(pX)j_c''(qX)X^2dX
  \\ && \times
   \int_{\Omega_{\mathbf{X}}}Y^*_{cd}(\mathbf{X})Y^*_{c'd'}(\mathbf{X})Y^*_{c''d''}(\mathbf{X})d\Omega_{\mathbf{X}}
   d\eta''d\eta'd\eta
   d\Omega_{\mathbf{k}}d\Omega_{\mathbf{p}}d\Omega_{\mathbf{q}}
   q^2dqp^2dpk^2dk
\eeas
which we can then immediately integrate over the directions of $\mathbf{k}$, $\mathbf{p}$ and $\mathbf{q}$ which gives us Kronecker deltas $\delta^c_l\delta^d_m$ and similar. After summation, then, we have found
\beas
  \lefteqn{\langle a^A_{T,lm}a^B_{T,l'm'}a^C_{T,l''m''}\rangle=
   \frac{(4\pi)^6}{(2\pi)^9}
   (2l+1)^2(2l'+1)^2(2l''+1)^2}
  \\ && \times
   \int_{k}\int_{p}\int_{q}
   \mathcal{B}_{ABC}(k,p,q)
   \left(\int_{\eta}S^S_T(k,\eta)j_l(x)d\eta\right)
   \left(\int_{\eta'}S^S_T(p,\eta')j_l'(x')d\eta'\right)
  \\ && \times
   \left(\int_{\eta''}S^S_T(q,\eta'')j_l''(x'')d\eta''\right)
   \left(\int_{X}j_c(kX)j_c'(pX)j_c''(qX)X^2dX\right)
  \\ && \times
   \left(\int_{\Omega_{\mathbf{X}}}Y^*_{lm}(\mathbf{X})Y^*_{l'm'}(\mathbf{X})Y^*_{l''m''}(\mathbf{X})d\Omega_{\mathbf{X}}\right)
   q^2dqp^2dpk^2dk
\eeas

The integral of three spherical harmonics is called the Gaunt integral and it evaluates \cite{WangKamionkowski00} to
\bea
  \lefteqn{\int_{\Omega_{\mathbf{n}}}Y^*_{lm}(\mathbf{n})Y^*_{l'm'}(\mathbf{n})Y^*_{l''m''}(\mathbf{n})d\Omega_{\mathbf{n}}=}
  \nonumber \\ &&
  \sqrt{\frac{(2l+1)(2l'+1)(2l''+1)}{4\pi}}\left(\begin{array}{ccc}l&l'&l''\\0&0&0\end{array}\right)
   \left(\begin{array}{ccc}l&l'&l''\\m&m'&m''\end{array}\right)
\eea
where the objects in brackets are Wigner $3j$ symbols, closely related to the Clebsch-Gordan coefficients.

If we also denote the integral over three spherical Bessel functions by
\be
  J_{ll'l''}(k,p,q)=\int_{X}j_l(kX)j_{l'}(pX)j_{l''}(qX)X^2dX
\ee
and recall the definition of the scalar transfer functions (\ref{ScalarTemperatureTransferFunctions}) then we can finally see that
\bea
  \lefteqn{\langle a^A_{T,lm}a^B_{T,l'm'}a^C_{T,l''m''}\rangle=
   \frac{8}{\pi^3}
   \sqrt{\frac{(2l+1)(2l'+1)(2l''+1)}{4\pi}}
   \left(\begin{array}{ccc}l&l'&l''\\0&0&0\end{array}\right)}
     \nonumber \\ && \times
   \left(\begin{array}{ccc}l&l'&l''\\m&m'&m''\end{array}\right)
   \int_{k}\int_{p}\int_{q}
   \mathcal{B}_{ABC}(k,p,q)
   (2l+1)^2(2l'+1)^2(2l''+1)^2
     \nonumber \\ && \qquad\times
   \Delta^S_{T,l}(k,\eta_0)\Delta^S_{T,l'}(p,\eta_0)\Delta^S_{T,l''}(q,\eta_0)
   J_{ll'l''}(k,p,q)
   q^2dqp^2dpk^2dk \nonumber \\
   &&=\left(\begin{array}{ccc}l&l'&l''\\m&m'&m''\end{array}\right)B_{ll'l''}(k,p,q)
\eea
where
\bea
  B_{ll'l''}&=&\frac{1}{(2l+1)^2(2l'+1)^2(2l''+1)^2}\sum_{mm'}\langle a^A_{T,lm}a^B_{T,l'm'}a^C_{T,l''m''}\rangle
  \nonumber \\ &=&
  \frac{8}{\pi^3}\sqrt{\frac{(2l+1)(2l'+1)(2l''+1)}{4\pi}}
   \left(\begin{array}{ccc}l&l'&l''\\0&0&0\end{array}\right)
   \int_{k}\int_{p}\int_{q}
   \mathcal{B}_{ABC}(k,p,q)
  \nonumber \\ && \times
   (2l+1)^2(2l'+1)^2(2l''+1)^2
   \Delta^S_{T,l}(k,\eta_0)\Delta^S_{T,l'}(p,\eta_0)\Delta^S_{T,l''}(q,\eta_0)
  \nonumber \\ && \times
\label{ScalarBispectrum}
   J_{ll'l''}(k,p,q)q^2dqp^2dpk^2dk
\eea
is called the CMB angular bispectrum in clear analogy with the CMB angular power spectrum. Other than the inclusion of the Wigner $3j$ symbol and the integration across the Bessel functions this clearly has a familiar form, with the transfer functions merely wrapping the primordial bispectrum onto the CMB sky.

Note that $m''$ is not summed across due to the restrictions on $l,l',l''$ and $m,m',m''$ placed by the Wigner $3j$ symbol,
\be
  l+l'+l''\in\mathcal{I}, \quad m+m'+m''=0, \quad \left|l-l'\right|\leq l''\leq \left|l+l'\right|
\ee
\emph{i.e.}, the sum of $l,l',l''$ is an integer, $m''$ is determined from $m$ and $m'$, and the triangle inequality must be obeyed. If these conditions are not satisfied then the symbol is vanishing.

From Abramowitz and Stegun \cite{AbramowitzStegun} we see that
\be
  \left(\begin{array}{ccc}l&l'&l''\\0&0&0\end{array}\right)=
  (-1)^g\sqrt{\frac{(2g-2l)!(2g-2l')!(2g-2l'')!}{(2g+1)!}}\frac{g!}{(g-l)!(g-l')!(g-l'')!}
\ee
for even $l+l'+l''$ and where $2g=l+l'+l''$. For odd $l+l'+l''$ the symbol vanishes.

As with the two-point case, this form for the bispectra holds regardless of the nature of variable one is employing; $\left\{A,B,C\right\}$ can ultimately denote scalar trace, traceless scalar, vector or tensor pieces.

Rather than work with the full bispectrum (\ref{ScalarBispectrum}) we follow Ferreira, Magueijo and G\'orski \cite{FerreiraMagueijoGorski98} and Wang and Kamionkowski \cite{WangKamionkowski00} in defining a reduced bispectrum
\bea
  \hat{B}^{ABC}_l&=&\frac{4\pi}{\sqrt{(2l+1)^3}}\left(\begin{array}{ccc}l&l'&l''\\0&0&0\end{array}\right)^{-1}B^{ABC}_{lll}
    \\
   &=&\left(\frac{2}{\pi}\right)^3\int_k\int_p\int_q\mathcal{B}_{ABC}(k,p,q)J_{lll}(k,p,q)
    \Delta^S_{T,l}(k,\eta_0)\Delta^S_{T,l}(p,\eta_0)\Delta^S_{T,l}(q,\eta_0)
    q^2dqp^2dpk^2dk . \nonumber
\eea
Note that our definition is a factor of $\sqrt{4\pi}$ different to those in \cite{FerreiraMagueijoGorski98,WangKamionkowski00}. We have sacrificed a large amount of the information from the full bispectrum in the interests of a quantity that more closely resembles the familiar angular power spectrum for the CMB.

\section{An Algorithm for Evaluating $J_{ll'l''}$}
\label{JlllAlgorithm}
The function $J_{lll}(k,p,q)$ could be calculated recursively, as outlined in the appendix of Wang and Kamionkowski, or by direct integration. Due to the number of calls that a direct integration approach would make we choose to evaluate it recursively and here outline the approach. First we derive a recursion relation for the full integral $J_{l_1,l_2,l_3}(k,p,q)$ by employing the standard spherical Bessel relation
\be
  j_{l-1}(x)+j_{l+1}(x)=\frac{2l+1}{x}j_l(x),
\ee
from which we can see that
\beas
  \int_0^\infty j_{l_1}(kx)j_{l_2}(px)j_{l_3}(qx)xdx
   &=&\int_0^\infty \frac{k}{2l_1+1}\left(j_{l_1-1}(kx)+j_{l1+1}(kx)\right)j_{l_2}(px)j_{l_3}(qx)x^2dx \\
    &&=\frac{k}{2l_1+1}\left(J_{l1-1,l2,l3}+J_{l1+1,l2,l3}\right) \\
   &=&\int_0^\infty \frac{p}{2l_2+1}j_{l_1}(kx)\left(j_{l_2-1}(px)+j_{l2+1}(px)\right)j_{l_3}(qx)x^2dx \\
    &&=\frac{p}{2l_2+1}\left(J_{l1,l2-1,l3}+J_{l1,l2+1,l3}\right) \\
   &=&\int_0^\infty \frac{q}{2l_3+1}j_{l_1}(kx)j_{l_2}(px)\left(j_{l_3-1}(qx)+j_{l3+1}(qx)\right)x^2dx \\
    &&=\frac{q}{2l_3+1}\left(J_{l1,l2,l3-1}+J_{l1,l2,l3+1}\right)
\eeas
whence the two recursion relations
\beas
  J_{l_1,l_2+1,l_3}(k,p,q)&=&\frac{k}{p}\frac{2l_2+1}{2l_1+1}\left(J_{l_1-1,l_2,l_3}(k,p,q)
   +J_{l_1+1,l_2,l_3}(k,p,q)\right)-J_{l_1,l_2-1,l_3}(k,p,q), \\
  J_{l_1,l_2,l_3+1}(k,p,q)&=&\frac{k}{p}\frac{2l_3+1}{2l_1+1}\left(J_{l_1-1,l_2,l_3}(k,p,q)
   +J_{l_1+1,l_2,l_3}(k,p,q)\right)-J_{l_1,l_2,l_3-1}(k,p,q) .
\eeas
Obviously there is a third recurrence relation between $p$ and $q$ but due to the Dirac delta this is not needed.

To construct $J_{l_1,l_2,l_3}(k,p,q)$ for any combination of $\left\{l_1,l_2,l_3\right\}$ and $\left\{k,p,q\right\}$, then, we follow the algorithm below:
\begin{itemize}
\item Rearrange $J_{l_1,l_2,l_3}(k,p,q)$ such that $l_1>l_2>l_3$; for example, $J_{1,3,2}(k,p,q)\rightarrow J_{3,2,1}(p,q,k)$;
\item For $a\in [-1,l_1+1]$ generate $J_{a,0,0}$, $J_{a,-1,0}$, $J_{a,0,-1}$ and $J_{a,-1,-1}$;
\item For $a\in [0,l_1+1]$ and $b\in [1,l_2]$ generate $J_{a,b,-1}$ from $J_{a,0,-1}$, producing $J_{l_1,l_2,0}$ and $J_{l_1+1,l_2,0}$;
\item For $c\in [0,l_1+1]$ generate $J_{l_1,l_2,c}$, ultimately producing $J_{l_1,l_2,l_3}(k,p,q)$.
\end{itemize}

We then require the four bases from which we will produce any required $J_{l_1,l_2,l_3}$, $J_{a,0,0}$, $J_{a,-1,0}$, $J_{a,0,-1}$ and $J_{a,-1,-1}$. These can each be evaluated analytically, as shown in Wang and Kamionkowski, by applying the stringent conditions on $\left\{l_1,l_2,l_3\right\}$ and $\left\{k,p,q\right\}$ enforced by the Clebsch-Gordan coefficient and statistical isotropy. That is, the wavenumbers must obey the triangle relation and $l_1+l_2+l_3$ must be even.

Now, employing that
\be
  j_0(x)=\mathrm{sinc}(x), \quad j_l(x)=\int_{-1}^{1}\frac{1}{2i^l}P_l(y)e^{ixy}dy
\ee
we can write
\bdm
  J_{l,0,0}(k,p,q)=
   \int_0^\infty\frac{1}{2i^l}\int_{-1}^{1}P_l(y)e^{ikxy}dy\frac{\sin(px)}{px}\frac{\sin(qx)}{qx}dx .
\edm
Imposing that $l$ is even ensures that the integrand is an even function; we can convert the $\sin(px)$ and $\sin(qx)$ into exponentials and integrate over $x$ to leave the integral over $y$ and four delta functions. Enforcing the triangle relation on the wavemodes then leaves a tractable integration that evaluates to
\bdm
  J_{l,0,0}(k,p,q)=\frac{\pi}{4i^l}\frac{1}{kpq}P_l\left(\frac{p-q}{k}\right) .
\edm
We can follow much the same procedure for $J_{l,-1,0}$ (which contains an odd function and requires $l$ to be odd, implying that the overall function is even) and $J_{l,-1,-1}$and find that
\beas
  J_{l,0,0}(k,p,q)&=&\frac{\pi}{4i^l}\frac{1}{kpq}P_l\left(\frac{p-q}{k}\right), \\
  J_{l,-1,0}(k,p,q)&=&\frac{\pi}{4i^{l+1}}\frac{1}{kpq}P_l\left(\frac{p-q}{k}\right), \\
  J_{l,0,-1}(k,p,q)&=&\frac{\pi}{4i^{l+1}}\frac{1}{kpq}P_l\left(\frac{q-p}{k}\right), \\
  J_{l,-1,-1}(k,p,q)&=&\frac{\pi}{4i^l}\frac{1}{kpq}P_l\left(\frac{p-q}{k}\right) .
\eeas
This gives us the basis from which all possible $J_{l_1,l_2,l_3}$ can be evaluated. This algorithm is easy to implement in Fortran 90; we can build it as a recursive function and rapidly and accurately generate $J_{l_1,l_2,l_3}$ on the fly.

\chapter{Mathematical Relations}
We list here a brief selection of definitions, relations and identities for some of the functions we have been employing throughout this thesis. For further detail see, for example, Abramowitz and Stegun \cite{AbramowitzStegun}.

\section{Legendre Polynomials}
The Legendre polynomials satisfy the equation
\be
\label{LegendreEquation}
  \left\{\left(1-\mu^2\right)\frac{d^2}{d\mu^2}-2\mu\frac{d}{d\mu}+\left(l(l+1)-\frac{m^2}{1-\mu^2}\right)
   \right\}P_l^m(\mu)
\ee
where $P^m_l(\mu)$ is an associated Legendre polynomial. The Legendre polynomials are given by $m=0$ and the associated Legendre polynomials are recovered from
\be
\label{AssociatedLegendreDefinition}
  P^m_l(\mu)=(-1)^m\left(1-\mu^2\right)^{m/2}\frac{\partial^mP_l(\mu)}{\partial\mu^2} .
\ee
Legendre polynomials obey the recursion relation
\be
\label{LegendreRecursion}
  \left(2l+1\right)\mu P^m_l(\mu)=\left(l-m+1\right)P^m_{l+1}(\mu)+(l+m)P^m_{l-1}(\mu)
\ee
and are orthogonal over $x\in[-1,1]$,
\be
\label{LegendreOrgogonality}
  \int_{-1}^{1}P^m_l(\mu)P^m_a(\mu)d\mu = \frac{2}{2l+1}\frac{(l+m)!}{(l-m)!}\gamma_{la} .
\ee
The first few Legendre polynomials are
\beas
  P_0(\mu)&=&1 \\
  P_1(\mu)&=&\mu \\
  P_2(\mu)&=&\frac{1}{2}\left(3\mu^2-1\right) \\
  P_3(\mu)&=&\frac{1}{2}\mu\left(5\mu^2-3\right) \\
  P_4(\mu)&=&\frac{1}{8}\left(35\mu^4-20\mu^2+3\right) \\
  P_5(\mu)&=&\frac{1}{8}\mu\left(63\mu^4-70\mu^2+15\right)
\eeas
and so the first few powers of $\mu$ can be expressed as
\beas
  1&=&P_0(\mu) \\
  \mu&=&P_1(\mu) \\
  \mu^2&=&\frac{1}{3}\left(P_0(\mu)+2P_2(\mu)\right) \\
  \mu^3&=&\frac{1}{5}\left(3P_1(\mu)+2P_3(\mu)\right) \\
  \mu^4&=&\frac{1}{35}\left(7P_0(\mu)+20P_2(\mu)+8P_4(\mu)\right) \\
  \mu^5&=&\frac{1}{63}\left(27P_1(\mu)+28P_3(\mu)+8P_5(\mu)\right) .
\eeas

\section{Spherical Harmonics}
The spherical harmonics are defined in terms of the associated Legendre polynomials as
\be
\label{SphericalHarmonicsDefinition}
  Y_{lm}(\mathbf{n})=\sqrt{\frac{2l+1}{4\pi}\frac{(l-m)!}{(l+m)!}}P^m_l(\mu)e^{im\phi} .
\ee
They are orthogonal over the directions of $\mathbf{n}$:
\be
\label{SphericalHarmonicsOrthogonality}
  \int_\mathbf{n}Y^*_{lm}(\mathbf{n})Y_{pn}(\mathbf{n})d\Omega_\mathbf{n}=\gamma_{lp}\gamma_{mn} .
\ee
The Legendre polynomials can be recovered from
\be
\label{LegendreSphericalHarmonicRelation}
  \left(2l+1\right)P_l\left(\khv.\nhv\right)=4\pi\sum_mY^{}_{lm}(\khv)
  Y^*_{lm}(\nhv) .
\ee

\section{Bessel Functions}
The Bessel functions are a solution of the equation
\be
\label{BesselEquation}
  x^2\frac{d^2J_l(x)}{dx^2}+x\frac{dJ_l(x)}{dx}+\left(x^2-l^2\right)J_l(x)=0 .
\ee
While there are various solutions to this equation we are interested solely in the Bessel functions of the first kind. Bessel functions of the first kind satisfy the derivative relation
\be
  2\frac{dJ_l(x)}{dx}=J_{l-1}(x)-J_{l+1}(x)
\ee
and the recursion relation
\be
\label{BesselRecursion}
  2l\frac{J_l(x)}{x}=J_{l+1}(x)+J_{l-1}(x) .
\ee

The spherical Bessel function is defined by
\be
\label{SphericalBesselDefinition}
  j_l(x)=\sqrt{\frac{\pi}{2x}}J_{l+1/2}(x) .
\ee
Using this we may transfer the relations for the Bessel functions to the spherical Bessel functions, again concentrating on those of the first kind. Firstly we transfer the fundamental equation,
\be
\label{SphericalBesselEquation}
  x^2\frac{d^2j_l(x)}{dx^2}+2x\frac{dj_l(x)}{dx}+\left(x^2-l(l+1)\right)j_l(x)=0
\ee
The recursion relation (\ref{BesselRecursion}) becomes
\be
\label{SphericalBesselRecursion}
  \left(2l+1\right)\frac{j_l(x)}{x}=j_{l+1}(x)+j_{l-1}(x) .
\ee

\section{Exponential Expansions}
The complex exponential is related to the Legendre polynomials and spherical Bessel functions by
\be
\label{ExponentialLegendre}
  \exp(ix\mu)=\sum_li^l(2l+1)j_l(x)P_l(\mu)
\ee
and thus to the spherical harmonics by
\be
\label{ExponentialSphericalHarmonics}
  \exp(ix\mu)=4\pi\sum_{lm}i^lj_l(x)Y^*_{lm}(\hat{\mathbf{k}})Y_{lm}(\hat{\mathbf{n}}) .
\ee

\section{Miscellaneous}
We sometimes abbreviate
\be
  \mathrm{sinc}(x)=\frac{\sin(x)}{x} .
\ee

\chapter{Conventions and Notation}
\section{Conventions}
\subsection{Fundamental}
Perhaps our most fundamental assumption is that we assume General Relativity to provide an accurate picture of the universe from the smallest scales to beyond the Hubble distance. For a derivation of general relativity we refer the reader to, for example, Carroll \cite{Carroll}, Weinberg \cite{Weinberg}, Wald \cite{Wald}, Rindler \cite{Rindler} or Misner, Wheeler and Thorne \cite{MisnerWheelerThorne}. Misner, Wheeler and Thorne \cite{MisnerWheelerThorne} characterise the sign conventions possible in GR into
\bea
  \eta_{\mu\nu}&=&[\mathrm{S1}]\times\mathrm{diag}\left(-1,+1,+1,+1\right), \nonumber \\
  R^\mu_{\phantom{\mu}\alpha\beta\gamma}&=&[\mathrm{S2}]\times\left(\Gamma^\mu_{\alpha\gamma,\beta}
   -\Gamma^\mu_{\alpha\beta,\gamma}+\Gamma^\mu_{\sigma\beta}\Gamma^\sigma_{\alpha\gamma}
   -\Gamma^\mu_{\sigma\gamma}\Gamma^\sigma_{\alpha\beta}\right), \\
  G_{\mu\nu}&=&[\mathrm{S3}]\times 8\pi GT_{\mu\nu} .
\eea
where $\eta_{\mu\nu}$ is the Minkowski metric; (see also Peacock \cite{Peacock} \S1.5.) The third sign $[\mathrm{S3}]$ can also be fixed by
\be
  R_{\mu\nu}=[\mathrm{S2}]\times[\mathrm{S3}]\times R^\alpha_{\phantom{\alpha}\mu\alpha\nu} .
\ee
$[\mathrm{S1}]$ is commonly known as the ``signature'' of the spacetime. Under these traditional definitions, some of the main general relativity and cosmology textbooks can be classified; Misner, Wheeler and Thorne \cite{MisnerWheelerThorne} and Carroll \cite{Carroll} are $(+++)$, Weinberg \cite{Weinberg} is $(+ - -)$, Peebles \cite{Peebles-LargeScaleStructure,Peebles-PrinciplesPhysicalCosmology} is $(- + +)$, and Rindler \cite{Rindler} and Peacock \cite{Peacock} are $(- + -)$. We assume Misner, Wheeler and Thorne's conventions $(+ + +)$.

Our co-ordinates shall be labelled $x^\mu=(x^0,x^1,x^2,x^3)$ where $x^0$ is a timelike co-ordinate and $x^1,x^2,x^3$ are spacelike co-ordinates. We use Greek indices $\{\mu,\nu,\ldots\}$ to denote spacetime co-ordinates and lower-case Latin $\{i,j,\ldots\}$ to denote spatial co-ordinates. Summation over repeated indices is implied unless otherwise noted.

\subsection{Perturbation Theory}
An overdot represents differentiation with respect to the ``conformal time'' while a prime generally denotes differentiation with respect to the co-ordinate time. Variables lie in Fourier space unless an explicit dependence on the space variables $\mathbf{x}$ is declared or the situation is unambiguous. The transformation character of a variable will, in cases of ambiguity, be denoted with a capital Roman superscript (${}^{S,V,T}$) or subscript $({}_{S,V,T})$. Unless otherwise specified, we operate in natural units -- \emph{i.e.} $c=k_B=\hbar=1$. Averaged -- background -- variables are denoted with an overline (e.g. $\overline{p})$ and perturbed variables with a preceding $\delta$ (e.g. $\delta p$) unless otherwise noted. Unit vectors are denoted with an overhat, e.g. $\khv=\mathbf{k}/k$. Unless otherwise noted a subscript ${}_0$ refers to the current (observed) value of a quantity.

\subsection{Magnetised Plasmas}
When we come to consider magnetised plasmas, we must take great care with our conventions. There are many different conventions within electromagnetism and one has to be careful with the units one is working with; an appendix in Jackson \cite{Jackson} summarises these. In cosmology it is still standard to work within Heaviside-Lorentz units, a centimetre-gram-second system.

We denote real-space electromagnetic fields and the three-current with lower case Roman letters $\mathbf{e}$, $\mathbf{b}$, $\mathbf{j}$; although technically $\mathbf{b}$ is the magnetic induction we habitually refer to it as the magnetic field. We shall later scale our variables with powers of the scale factor, and these will be denoted by capital Roman letters $\mathbf{E}$, $\mathbf{B}$, $\mathbf{J}$. We will not work with unscaled variables while in Fourier space. Other scaled magnetohydrodynamic variables are denoted with a subscript $a$, e.g. $\rho_{ba}$. The electromagnetic charge density and conductivity are denoted by $\varrho$ and $\sigma$ and the baryon density by $\rho_b$, and the scaled cases by $\varrho_a$, $\sigma_a$ and $\rho_{ba}$ respectively.

\section{Notation}
Due to the large number of variables we consider, we present a table summarising our notation. Some few symbols are used twice (the entropy and action, for example). These cases are cross-referenced one to another; in the thesis the context should make it clear which property is meant.

\begin{longtable}{r|p{0.75\textwidth}}
  \bf{Symbol} & \bf{Physical Quantity; First Appearance} \\ \hline \hline
  $B$ & Magnetic induction strength; \S\ref{Chapter-Introduction}  \\
  $\mathcal{P}(k)$ & Power spectrum -- two-point correlation in Fourier space; \S\ref{Chapter-Introduction} \\
  $k$ & Wavenumber; \S\ref{Chapter-Introduction} \\
  $n$ & Spectral index; \S\ref{Chapter-Introduction} \\
  $l$ & CMB angular power spectrum multipole number; \S\ref{Chapter-Introduction} \\
  $ds$ & Spacetime interval; \S\ref{Sec-Metric} \\
  $g_{\mu\nu}$ & Spacetime metric; \S\ref{Sec-Metric}  \\
  $t$ & Co-ordinate time; \S\ref{Sec-Metric} \\
  $a$ & Scale factor; \S\ref{Sec-Metric} \\
  $\gamma_{ij}$ & Metric on a spacelike hypersurface; \S\ref{Sec-Metric} \\
  $h_{ij}$ & Metric perturbation on a spacelike hypersurface; \S\ref{Sec-Metric} \\
  $\eta$ & Conformal time; \S\ref{Sec-Metric} \\
  $\varepsilon$ & (Usually implicit) perturbation parameter; \S\ref{Sec-Metric} \\
  $\partial_aA$ or $A_{,a}$ & Partial derivative of a (tensor-valued) function $A$ with respect to co-ordinate $x^a$; \S\ref{Sec-Metric} \\
  $\nabla_aA$ or $A_{;a}$ &
    Covariant derivative of a (tensor-valued) function $A$ with respect to co-ordinate $x^a$; \S\ref{Sec-Metric} \\
  $\Gamma^i_{jk}$ & Affine connections -- Christoffel symbols; \S\ref{Sec-Metric} \\
  $\nabla$ & $\left(\partial/\partial x^1,\partial/\partial x^2,\partial/\partial x^3\right)$
    ; \S\ref{Sec-Metric} \\
  $R^\alpha_{\phantom{\alpha}\beta\gamma\mu}$ & Riemann-Christoffel curvature tensor; \S\ref{Sec-Metric} \\
  $R_{\mu\nu}$ & Ricci curvature tensor; \S\ref{Sec-Metric} \\
  $R$ & Ricci curvature scalar; \S\ref{Sec-Metric} \\
  $h$ & Trace of metric perturbation; \S\ref{Sec-Metric} \\
  $T^\mu_\nu$ & Generic stress-energy tensor; \S\ref{Sec-Metric} \\
  $\rho$ & Matter mass/energy density; \S\ref{Sec-Metric} \\
  $\mathcal{E}^i$ & Matter energy flux; \S\ref{Sec-Metric} \\
  $\mathcal{P}^i_j$ & Matter flux density; \S\ref{Sec-Metric} \\
  $p$ & Matter pressure; \S\ref{Sec-Metric} \\
  $u^\mu$ & Matter four-velocity; \S\ref{Sec-Metric} \\
  $\delta^\mu_\nu$ & Kronecker delta; \S\ref{Sec-Metric} \\
  $\Pi^i_j$ & Traceless component of the momentum flux density, the anisotropic stresses; \S\ref{Sec-Metric} \\
  $G^\mu_\nu$ & Einstein tensor; \S\ref{Sec-Metric} \\
  $G$ & Newtonian gravitational constant; \S\ref{Sec-Metric} \\
  $\Lambda$ & Cosmological constant; \S\ref{Sec-Metric} \\
  $\mathcal{H}$ & Hubble parameter in conformal time; \S\ref{Sec-Metric} \\
  $H$ & Observed Hubble parameter in co-ordinate time; \S\ref{Sec-Metric} \\
  $w$ & Fluid equation of state $w=\rho/p$; \S\ref{Sec-Metric} \\
  $k^i$ & Wavemode; co-ordinate in Fourier space; \S\ref{Sec-Metric} \\
  $P^i_j(\mathbf{k})$ & Projection operator $\delta^i_j-\kh^i\kh^j$ onto a hypersurface orthogonal to $\khv$; \S\ref{Sec-Metric} \\
  $Q^i_j(\mathbf{k})$ & Projection operator $(3/2)(\delta^i_j-P^i_j(\mathbf{k}))$; \S\ref{Sec-Metric} \\
  ${}_{(\ldots)}$ & Symmetrisation on enclosed indices; \S\ref{Sec-Metric} \\
  $\mathcal{P}^k_{ij}(\mathbf{k})$ & Vector-valued projection operator $\kh_iP^k_j(\mathbf{k})$;
    \S\ref{Sec-Metric} \\
  $\mathcal{P}^{ib}_{ja}(\mathbf{k})$ & Tensor-valued projection operator
     $P^i_a(\mathbf{k})P^b_j(\mathbf{k})-(1/2)P^i_j(\mathbf{k})P^b_a(\mathbf{k})$; \S\ref{Sec-Metric} \\
  $\Phi,\Psi$ & Scalar Bardeen variables; \S\ref{Sec-Metric} \\
  $\tilde{V}_i$ & Vector Bardeen variable; \S\ref{Sec-Metric} \\
  $v_i$ & Spatial component of fluid velocity; \S\ref{Sec-Fluids-SETensor} \\
  $H^\mu_\nu$ & Projection tensor $\delta^\mu_\nu+u^\mu u^\nu$ onto a hypersurface orthogonal to $u^\mu$; \S\ref{Sec-Fluids-SETensor} \\
  $\chi$ & Coefficient of heat-flow; \S\ref{Sec-Fluids-SETensor} \\
  $Q_\alpha$ & Heat flow tensor; \S\ref{Sec-Fluids-SETensor} \\
  $\xi$ & Coefficient of shear viscosity; \S\ref{Sec-Fluids-SETensor} \\
  $W^\mu_\nu$ & Shear viscous tensor; \S\ref{Sec-Fluids-SETensor} \\
  $\zeta$ & Coefficient of bulk viscosity; \S\ref{Sec-Fluids-SETensor} \\
  $\tilde{\xi}^\mu_\nu$ & Heat-flow component of fluid stress-energy tensor; \S\ref{Sec-Fluids-SETensor} \\
  $\tilde{\chi}^\mu_\nu$ & Shear viscous component of fluid stress-energy tensor; \S\ref{Sec-Fluids-SETensor} \\
  $\tilde{\zeta}^\mu_\nu$ & Bulk viscous component of fluid stress-energy tensor; \S\ref{Sec-Fluids-SETensor} \\
  $\Theta$ & Temperature of matter component; \S\ref{Sec-Fluids-SETensor} \\
  $c_s$ & Speed of sound in a fluid; \S\ref{Sec-Fluids-SETensor} \\
  $S$ & Entropy of a fluid (c.f. the action); \S\ref{Sec-Fluids-SETensor} \\
  $\theta$ & Perturbation to fluid temperature; \S\ref{Sec-Fluids-SETensor} \\
  $dQ$ & Change of heat; \S\ref{Sec-Fluids-SETensor} \\
  $N$ & Number density of a fluid; \S\ref{Sec-Fluids-SETensor} \\
  $m$ & Mass of a particle; \S\ref{Sec-Fluids-SETensor} \\
  $\mathcal{C}_\mu$ & Energy-momentum exchange term; c.f. also $C_{b\leftrightarrow\gamma}$; \S\ref{Sec-Fluids-SETensor} \\
  $\delta$ & Dimensionless perturbation to fluid mass/energy density; \S\ref{Sec-Fluids-SETensor} \\
  $f$ & Distribution function of a collection of particles; \S\ref{Sec-Boltzmann} \\
  $P_\mu$ & Momentum four-vector conjugate to spacetime co-ordinates; \S\ref{Sec-Boltzmann} \\
  $U$ & External potential energy; \S\ref{Sec-Boltzmann} \\
  $C[f]$ & Collisional term in the Boltzmann equation; \S\ref{Sec-Boltzmann} \\
  $m_0$ & Rest-energy of a particle; \S\ref{Sec-Boltzmann} \\
  $p_\mu$ & Proper momentum four-vector defined in a Riemannian frame; \S\ref{Sec-Boltzmann} \\
  $\epsilon$ & Mass-energy of a particle measured by an observer comoving with the universal expansion (c.f. electric field in Minkowski space); \S\ref{Sec-Boltzmann} \\
  $q$ & Amplitude of the comoving momentum; \S\ref{Sec-Boltzmann} \\
  $\nh_i$ & Unit vector in the direction of the momentum; \S\ref{Sec-Boltzmann} \\
  $g_s$ & Statistical weight of a particle; \S\ref{Sec-Boltzmann} \\
  $d\Omega$ & Solid angle differential; \S\ref{Sec-Boltzmann} \\
  $\mu$ & Angle cosine between the particle momentum and Fourier wavemode directions; \S\ref{Sec-Boltzmann} \\
  $F$ & Brightness function of a generic particle; \S\ref{Sec-Boltzmann} \\
  $\Delta$ & Brightness function of a photon fluid; \S\ref{Sec-SachsWolfe} \\
  $\left\{I,Q,U,V\right\}$ & Stokes parameters; \S\ref{SecPhotons} \\
  $\mathbf{e}$ & Electric field observed by a comoving observer; \S\ref{SecPhotons} \\
  $\left\{L,R\right\}$ & Basis axes for the Stokes parameters; \S\ref{SecPhotons} \\
  $\left\{\varepsilon_L,\varepsilon_R\right\}$ & Phase differences of electric field resolved along $L$ and $R$; \S\ref{SecPhotons} \\
  $\left\{I_L,I_R,U,V\right\}$ & Alternative Stokes parameters; \S\ref{SecPhotons} \\
  $\mathbf{P}(\mu,\phi;\overline{\mu},\overline{\phi})$ &
   Scattering matrix for the Stokes parameters for incoming angles
    $\mu,\phi$ and outgoing angles $\overline{\mu},\overline{\phi}$ (c.f. scalar field and electromagnetic scalar potential); \S\ref{SecPhotons} \\
  $\dot{\tau}$ & Differential cross-section of scattering, always Thomson; \S\ref{SecPhotons} \\
  $n_e$ & Density of free electrons; \S\ref{SecPhotons} \\
  $\sigma_T$ & Thomson cross-section; \S\ref{SecPhotons} \\
  $\left\{\Delta_T,\Delta_Q\right\}$ & Photon $I$ and $Q$ Stokes parameters; \S\ref{SecPhotons} \\
  $\mathbf{A}$ & Transformation matrix between bases for Stokes parameters; \S\ref{SecPhotons} \\
  $\Phi_{S,V,T}$ & Collisional terms for scalar, vector and tensor perturbations; \S\ref{SecPhotons} \\
  $\mathcal{M}_{+,\times}$ & Basis for tensor perturbations; \S\ref{SecPhotons} \\
  $\left\{\alpha^*,\beta^*\right\}$ & Basis for tensor perturbations to the brightness function; \S\ref{SecPhotons} \\
  $E$ & Gradient term of spin-invariant polarisation basis (c.f. electric field); \S\ref{SecPhotons} \\
  $B$ & Curl term of spin-invariant polarisation basis (c.f. magnetic induction); \S\ref{SecPhotons} \\
  $\tau(\eta)$ & Optical depth at conformal time $\eta$; \S\ref{SecPhotons} \\
  $g(\eta)$ & Visibility function at conformal time $\eta$; \S\ref{SecPhotons} \\
  $x$ & Dimensionless variable $x=k(\eta-\eta_0)$ (c.f. $y=k\eta$); \S\ref{SecPhotons} \\
  $S^{A}_{S,V,T}$ &
   Source terms in the line-of-sight integrals for the different
   Stokes parameters (superscript) and transformation property (subscript); \S\ref{SecPhotons} \\
  $\Delta_{Al}$ & Moment with index $l$ of the Stokes parameter $A$
    expanded across the Legendre polynomials; \S\ref{SecPhotons} \\
  $\left\{\tilde{\zeta},\breve{\zeta}\right\}$ & Variables characterising statistics of vector photon perturbations; \S\ref{SecPhotons} \\
  $\xi^{1,2}$ & Variables characterising statistics of tensor photon perturbations; \S\ref{SecPhotons} \\
  $C^i_{b\leftrightarrow\gamma}$ & Energy-momentum transfer between baryons and photons; \S\ref{SecBaryons} \\
  $t_c$ & Thomson scattering time $t_c=\dot{\tau}^{-1}$; \S\ref{Sec-TightlyCoupled} \\
  $\delta S_{AB}$ & Entropy perturbation between two fluid species $A$ and $B$; \S\ref{Sec-TightlyCoupled} \\
  $\mathcal{L}$ & Lagrangian density; \S\ref{SecScalarFields} \\
  $\mathcal{L}_{EH}$ & Einstein-Hilbert action for general relativity; \S\ref{SecScalarFields} \\
  $S$ & Action (c.f. entropy); \S\ref{SecScalarFields} \\
  $\phi$ & Scalar field (c.f. azimuthal angle and electromagnetic scalar potential); \S\ref{SecScalarFields} \\
  $\mathcal{K}$ & Curvature parameter of a Robertson-Walker metric; \S\ref{SecScalarFields} \\
  $y$ & Dimensionless variable $y=k\eta$ (c.f. $x=k(\eta-\eta_0)$); \S\ref{Sec-PeriodsInEvolution} \\
  $z$ & Redshift; \S\ref{Sec-PeriodsInEvolution} \\
  $\Omega_A$ & Dimensionless density contrast for a species $A$; \S\ref{Sec-PeriodsInEvolution} \\
  $C_l$ & CMB angular power spectrum; \S\ref{Sec-CMB} \\
  $C(\nhv\cdot\nhv')$ & CMB two-point temperature correlation; \S\ref{Sec-CMB} \\
  $C_{ab,l}$ & CMB angular power spectrum of correlation between $a$ and $b$; \S\ref{Line-of-Sight-CMB} \\
  $F_{\mu\nu}$ & Faraday (electromagnetic field) tensor; \S\ref{FieldAndStressEnergyTensors} \\
  $A_\mu$ & Electromagnetic four-potential; \S\ref{FieldAndStressEnergyTensors} \\
  $\phi$ & Electromagnetic scalar potential (c.f. scalar field and azimuthal angle); \S\ref{FieldAndStressEnergyTensors} \\
  $\mathbf{A}$ & Electromagnetic vector potential; \S\ref{FieldAndStressEnergyTensors} \\
  $\epsilon_i$ & Electric field in Minkowski space; \S\ref{FieldAndStressEnergyTensors} \\
  $\beta_i$ & Magnetic induction in Minkowski space; \S\ref{FieldAndStressEnergyTensors} \\
  $\varepsilon_{\mu\alpha\beta}$ & Levi-Civita tensor density; \S\ref{FieldAndStressEnergyTensors} \\
  $\epsilon_{\mu\alpha\beta}$ & Totally anti-symmetric tensor density; \S\ref{FieldAndStressEnergyTensors} \\
  $\mathbf{b}$ & Magnetic field observed by comoving observer; \S\ref{FieldAndStressEnergyTensors} \\
  $j_\mu$ & Electromagnetic four-current density; \S\ref{Sec-MaxwellEquations} \\
  $\varrho$ & Comoving electromagnetic charge density; \S\ref{Sec-MaxwellEquations} \\
  $\mathbf{j}$ & Comoving electromagnetic current density; \S\ref{Sec-MaxwellEquations} \\
  $\sigma$ & Electromagnetic conductivity; \S\ref{Sec-MaxwellEquations} \\
  $\mathbf{E}$ & Scaled electric field; \S\ref{MHD} \\
  $\mathbf{B}$ & Scaled magnetic field; \S\ref{MHD} \\
  $\varrho_a$ & Scaled charge density; \S\ref{MHD} \\
  $\mathbf{J}$ & Scaled current density; \S\ref{MHD} \\
  $\sigma_a$ & Scaled conductivity; \S\ref{MHD} \\
  $\ra$ & Scaled baryon density; \S\ref{MHD} \\
  $\pa$ & Scaled baryon pressure; \S\ref{MHD} \\
  $\mathbf{L}$ & Lorentz force; \S\ref{MHD} \\
  $\tau^\mu_\nu$ & Scaled electromagnetic stress-energy; \S\ref{MHD-Fourier} \\
  $\mathbf{v}_A$ & Alfv\'en velocity; \S\ref{Sec-DampingScaleMHD} \\
  $l_\gamma$ & Photon diffusion length $(n_e\sigma_T)^{-1}=a/t_c$; \S\ref{Sec-DampingScaleMHD} \\
  $\tilde{\tau}_{ab}$ & Self-convolution of magnetic field; \S\ref{Sec-Tangled} \\
  $\mathcal{H}(k)$ & Power spectrum of helical magnetic field; \S\ref{Sec-Stats-Underlying} \\
  $A$ & Amplitude of magnetic power spectrum; \S\ref{Sec-Stats-Underlying} \\
  $\lambda$ & Scale on which the magnetic amplitude is defined; \S\ref{Sec-Stats-Underlying} \\
  $k_\lambda$ & Wavenumber at which the magnetic amplitude is defined; \S\ref{Sec-Stats-Underlying} \\
  $B_\lambda$ & Field strength at normalisation scale $\lambda$; \S\ref{Sec-Stats-Underlying} \\
  $\mathcal{Q}(k)$ & Magnetic power spectrum normalised to unity, $\mathcal{P}(k)=A\mathcal{Q}(k)$; \S\ref{Sec-Stats-Underlying} \\
  $\mathbf{C}$ & 2-d field underlying 3-d magnetic field in Fourier space; \S\ref{Sec-MagneticRealisations} \\
  $k_c$ & Cut-off scale of magnetic field from genesis scenario; \S\ref{Sec-MagneticRealisations}\\
  $l_{\mathrm{dim}}$ & Size of simulation grid; \S\ref{Sec-MagneticRealisations} \\
  $\mu'_n$, $\mu_n$ & Moments and central moments respectively of a probability distribution; \S\ref{Sec-OnePoint} \\
  $\gamma_1$, $\gamma_2$ & Skewness and kurtosis of a probability distribution; \S\ref{Sec-OnePoint} \\
  $\mathcal{B}_{abcd}$ & Two-point correlation of $\tilde{\tau}_{ab}$; \S\ref{Sec-TwoPoint} \\
  $\mathcal{A}^{abcd}$ & Projection operator recovering a specified power spectrum from the general two-point; \S\ref{Sec-TwoPoint} \\
  $\mathcal{F}_{AB}$ & Angular component of two-point moment; \S\ref{Sec-TwoPoint} \\
  $\gamma$, $\mu$, $\beta$ & Angle cosines for two-point integrations (c.f. $\mu=\khv.\nhv$); \S\ref{Sec-TwoPoint} \\
  $\mathcal{P}^{V,T}(k)$ & Vector and tensor isotropic spectra; \S\ref{Sec-TwoPoint} \\
  $\mathcal{M}_{abcd}$ & Basis for tensor two-point correlations; \S\ref{Sec-TwoPoint} \\
  $\phi$, $r$ & Geometry specification for bispectra (c.f. scalar fields and azimuthal angles); \S\ref{Sec-ThreePoint} \\
  $\mathbf{p}$, $\mathbf{q}$ & Wavemodes forming a closed triangle with Fourier mode $\mathbf{k}$; \S\ref{Sec-ThreePoint} \\
  $\mathcal{B}_{ijklmn}$ & Three-point correlation of $\tilde{\tau}_{ab}$; \S\ref{Sec-ThreePoint} \\
  $\mathcal{A}^{ijklmn}$ & Projection operator recovering a specified bispectrum from $\mathcal{B}_{ijklmn}$; \S\ref{Sec-ThreePoint} \\
  $\theta_{ab}$, $\alpha_a$, $\beta_a$, $\gamma_a$, $\overline{\beta}$, $\overline{\gamma}$, $\overline{\mu}$ & Non-independent set of angles specifying bispectrum geometry; \S\ref{Sec-ThreePoint} \\
  $\xi_{kq}$, $\xi_{pq}$ & Angle cosines between wavemodes; \S\ref{Sec-ThreePoint} \\
  $\mathcal{F}_{ABC}$ & Angular term in bispectrum integral; \S\ref{Sec-ThreePoint} \\
  $B$ & Constant dependent on $k_c$ in large-scale bispectra; \S\ref{Sec-ThreePoint} \\
  $\lambda_S$ & Silk damping scale; \S\ref{Sec-DampingScales} \\
  $k_S$ & Silk damping wavenumber; \S\ref{Sec-DampingScales} \\
  $\overline{B}_{\mathrm{eff}}$ & Effective background field for a cut-off wavenumber $k_c$; \S\ref{Sec-DampingScales}  \\
  $\mathcal{B}_{ABC}(k,p,q)$ & Primordial bispectrum between quantities $A$, $B$ and $C$ and geometry wavenumbers $k$, $p$ and $q$; \S\ref{CMB-Bispectrum} \\
  $J_{ll'l''}(k,p,q)$ & Integral across three spherical Bessel functions; \S\ref{CMB-Bispectrum} \\
  $B^{ABC}_{ll'l''}(k,p,q)$ & CMB bispectrum for multipole numbers $l$,$l'$ and $l''$; \S\ref{CMB-Bispectrum} \\
  $\hat{B}^{ABC}_{l}$ & Reduced CMB bispectrum; \S\ref{CMB-Bispectrum} \\

\hline
\end{longtable}

\bibliographystyle{h-elsevier2}
\bibliography{thesis}

\end{document}